\documentclass{these}
\usepackage[utf8]{inputenc}
\usepackage[section]{placeins}
\usepackage{amsmath}
\usepackage{caption}
\usepackage[final]{graphicx}
\usepackage{amsfonts}
\usepackage{textcomp}
\usepackage{pdfpages}
\newcommand{\ii}{\mathrm{i}}
\title{Des problèmes inverses en Biophysique}
\author{Carlo BARBIERI}
\date{01/09/2011}
\labo{Laboratoire de Physique Statistique de l'\'Ecole Normale Sup\'erieure}
\school{\'Ecole Doctorale de Physique la R\'egion Parisienne --- ED 107}
\speciality{Physique}
\university{l'Universit\'e Pierre et Marie CURIE}{\Large{UNIVERSITE PIERRE ET MARIE CURIE}}
\advisor[M.lle]{Simona Cocco} %Directeur de thèse
\coadvisor[M]{} %Co-directeur de thèse, si nécessaire
\jury{  %définition des membres du jury
\jurymember[M]{Jean-François Joanny}{Pr\'esident du jury}\\
\jurymember[M]{Felix Ritort}{Rapporteur}\\
\jurymember[M]{Massimo Vergassola}{Rapporteur}\\
\jurymember[M]{Christophe Deroulers}{Examinateur}\\
\jurymember[M.lle]{Simona Cocco}{Directeur de Th\`ese}\\

}

\begin{document}
\frontmatter
\maketitle
\resume{Ces dernières années ont vu le développement de techniques expérimentales permettant l'analyse quantitative de systèmes biologiques, dans des domaines qui vont de la neurobiologie à la biologie moléculaire. Notre travail a pour but la description quantitative de tels systèmes à travers des outils théoriques et numériques issus de la physique statistique et du calcul des probabilités.

Cette thèse s'articule en trois volets, ayant chacun pour but l'étude d'un système biophysique.

Premièrement, on se concentre sur l'infotaxie, un algorithme de recherche olfactive basé sur une approche de théorie de l'information proposé par Vergassola et collaborateurs en 2007: on en donne une formulation continue et on en caractérise les performances.

Dans une deuxième partie on étudie les expériences de micromanipulation à molécule unique, notamment celles de dégraffage mécanique de l'ADN, dont les traces expérimentales sont sensibles à la séquence de l'ADN: on développe un modèle détaillé de la dynamique de ce type d'expérience et ensuite on propose plusieurs algorithmes d'inférence ayant pour objectif de caractériser la séquence génétique.

Finalement, on donne une description d'un algorithme qui permet l'inférence des interactions entre neurones à partir d'enregistrements à électrodes multiples et on propose un logiciel intégré qui permettra à la communauté des biologistes d'interpréter ces expériences
a partir de cet algorithme.}{
During the past few years the development of experimental techniques has allowed the quantitative analysis of biological systems ranging from neurobiology and molecular biology. This work focuses on the quantitative description of these systems by means of theoretical and numerical tools ranging from statistical physics to probability theory.

This dissertation is divided in three parts, each of which has a different biological system as its focus.

The first such system is Infotaxis, an olfactory search algorithm proposed by Vergassola et al. in 2007: we give a continuous formulation and we characterize its performances.

Secondly we will focus on single-molecule experiments, especially unzipping of DNA molecules, whose experimental traces depend strongly on the DNA sequence: we develop a detailed model of the dynamics for this kind of experiments and then we propose several inference algorithm aiming at the characterization of the genetic sequence.

The last section is devoted to the description of an algorithm that allows the inference of interactions between neurons given the recording of neural activity from multi-electrode experiments; we propose an integrated software that will allow the analysis of these data.
}
\chapter*{Acknowledments}
First of all I would like to thank my advisor Simona Cocco for the time she has spent mentoring me, the patience she has shown and the countless things I learnt from her.\\
I am also obliged to the members of the committee for having agreed to participate to this occasion and for devoting the time needed to read my manuscript.\\
This dissertation would not have been possible without the interaction with many of the scientits at ENS in Paris and IAS in Princeton. In particular I wish to mention Rémi Monasson for countless hours of help and discussion. I'm also indebted to Francesco Zamponi, Marco Tarzia and Guilhem Semerjian for the scientific and human advice they have provided me with throughout my thesis. Stan Leibler for making the extremely enriching experience at Princeton possible and all the Members of the Simons' Center for System Biology at IAS with a special thought to Arvind Murugan.\\
I really have to thank all the staff at ENS: Annie, Marie and Nora for their professionality and warmth and Eric Perez for always taking the time of asking how things went.\\
I'm obliged to Jean-Pierre Nadal and Jean-François Allemand for making my teaching experience possible, to my teaching colleagues Fréderic Van Wijland, Christophe Mora, Gwendal Fève for their great advice and mentoring. 
I'm truly indebted to all my fellows grad students at ENS: first of all Florent Alzetto with whom I shared two offices and who is a true friend. The guys in DC21: Marc, Antoine, Félix and the two Laetitias. I also need to mention Vitor Sessak which has been of great help throughout my thesis. The geophysics lab: Rana, Penelope, Maya, Laureen, Amaya, Laure and Marianne for our meals and coffees together. The LPS cycling team: Arnaud, Ariel, Xavier, Florent, Clément and especially Sébastien Balibar.\\
I am really grateful to my friends in Princeton: Giulia, Joro, Francesco, Ali, Mathilde, the two Gabrieles, Julien and Daphne. They have made my nine months in Princeton a really pleasent surprise.\\
I wish to thank the various persons who have endured me as a roommate: Filippo, Laetitia, Vitor and Simone and everyone at the ENS college in Montrouge, especially Olivier who is always a good friend and a very stimulating mind.\\
I wouldn't be here without my family and their moral support, I have to thank them for who I am.\\
I would also like to show my gratitude to the countless Italian friends who have visited me during this happy exile in Paris, I hope they haven't forgotten me.\\
This thesis is dedicated to Marie for her loving presence throughout these years.
\tableofcontents
\mainmatter
\chapter*{Introduction}
\section*{Probabilistic models}
Many systems encountered in quantitative biology are best described by probabilistic models. There are essentially three reasons why a probabilistic model would be preferred: either the process is thermally activated, either experimental conditions cannot be controlled in full detail or there are many possible realizations of annealed disorder in some of the involved variables.\\
Systems where the dynamics are thermally activated are widespread at the macromolecular scale (sizes ranging $1-100$ nm), because of this the dynamics of most systems from molecular biology will exhibit stochastic behavior. In this dissertation we will touch such systems in Part II while addressing DNA unzipping experiments.\\
Many biological experiments are performed in conditions where several variables cannot be controlled in detail: organisms which are genetically identical will exhibit different phenotypes, conditions of the medium will vary. In Part I we will observe turbulence can have such an effect in the description of olfactory searches.\\
Thirdly, many biological systems exhibit a characteristic which is similar to that of annealed disorder in condensed matter physics, that is, there are a number of variables which can be treated as random because they are drawn from an ensemble of possible realizations but do not change during experiments. Examples include DNA and RNA (where the variable is the genetic sequence), proteins (amino-acid content) and neural systems (interaction matrix). Such systems will be addressed in Parts II and III.\\
A probabilistic model will assign a proability to the outcome of an experiment. As it is possible to do this, the inverse problem can be of interest, that is we can assign a probability to a model or a set of parameters given the outcome of an experiment. This type of question is at the core of our thesis and of Bayesian inference.\\
\begin{center}
\includegraphics[width=.7\textwidth]{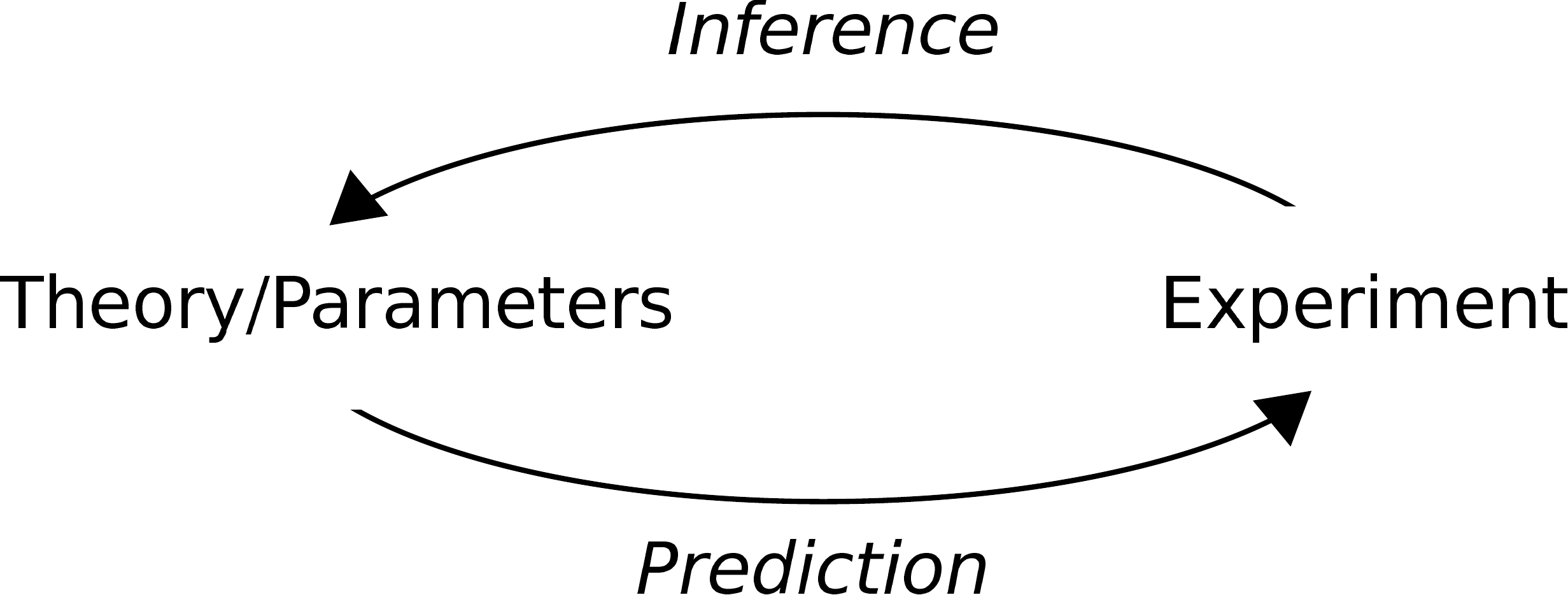}
\end{center}
\section*{Bayes' theorem}
Bayes' theorem was derived by Thomas Bayes and  was only published posthumously in 1763 \cite{BAYES,BAYESR}. It is now regarded as one of the founding pillars of probability theory.\\
By today's standards the name \emph{theorem} is probably a misnomer since its derivation is a straightforward manipulation of the the definition of conditional probability:
\begin{equation}
P(A|B)=\frac{P(A\cap B)}{P(B)}\,
\end{equation}
where $P(A\cap B)$ is the probability of event $A$ and $B$ both happening.\\
If we now switch $A$ and $B$ and redefine combine the two definitions we obtain the classical expression of Bayes' theorem:
\begin{equation}
P(A|B)=\frac{P(B|A)P(A)}{P(B)}\,,
\end{equation}
where $P(A)$ is usually called the \emph{prior}, $P(B|A)$ \emph{likelihood function} and $P(A|B)$ \emph{posterior}.\\
The importance of this theorem in performing statistical inference can only be understated in fact, if one interprets $A$ as the parameters of a model and $B$ as the outcome of an experiment we can see how this theorem relates the predictive power of a model to the inference of the best model or set of parameters. By rewriting the model this way:
\begin{equation}
 P(\text{model}_1|\text{data})=\frac{P(\text{data}|\text{model}_1)P(\text{model}_1)}{\sum_iP(\text{data}|\text{model}_i)P(\text{model}_i)}
\end{equation}
Let us give an example to further clarify this statement. Let us suppose we have two coins: one fair and one which is biased with probability $p$ of heads turning up.\\
While it is straightforward to compute the outcome of an experiment knowing which coin we are handling: say two consecutive heads yield $P(\text{HH}|\text{fair}=1/4$, we wish to know $P(\text{fair}|\text{HH})$.\\
Thanks to Bayes' theorem this can be done in a straightforward manner:
\begin{equation}
P(\text{fair}|\text{HH})=\frac{1/4}{1/4+p^2}
  \end{equation}
\begin{figure}
\begin{center}
\includegraphics[width=.5\textwidth]{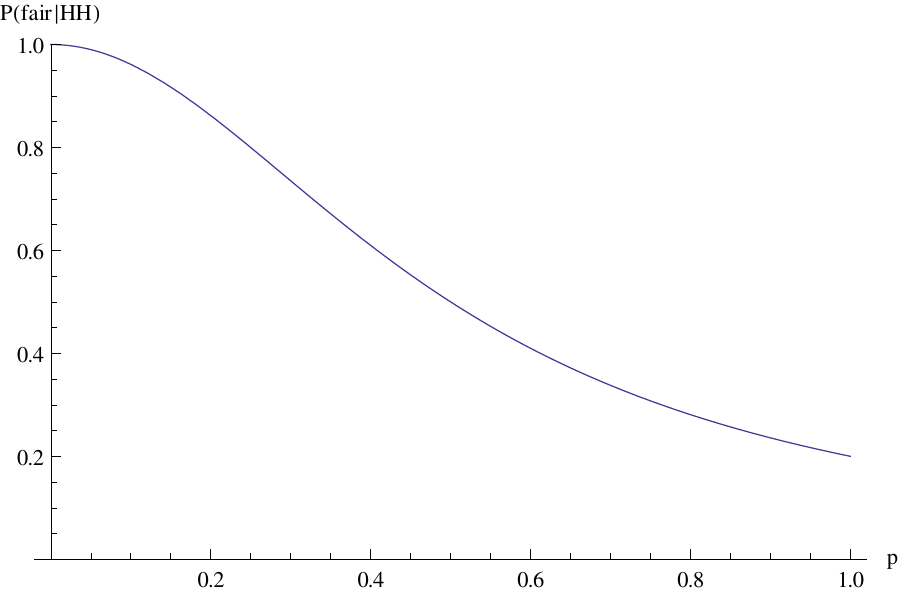}
\end{center}
\caption*{Figure 1: $P(\text{fair}|\text{HH})$ as a function of $p$. Note how the probability is maximum when $p$ vanishes and it's minimum and equal to $1/5$ when the unfair coin always returns heads, that is when $p=1$.}
\end{figure}
The attentive reader will have noticed we have placed ourselves in a very specific situation: we know we only have two coins, and we know the bias of one of them.\\
The problem of testing the hypotesis of whether a coin is biased or not in the most general conditions is a much more complicated one and is illuminating as to the limitations of Bayesian inference.\\
Our toy example had the very compelling feature of defining naturally the \emph{prior distribution}, that is $P(\text{model}_i)$ was $1/2$ for $i=1,2$: both coins were equiprobable. How do we define priors for more general cases?\\
Sometimes some general choices are available, for example one could the maximum entropy probability distribution with given characteristics such has a given support or a given expected value. However this is not always possible especially when the support of the distribution is unbounded.\\
However if we consider successive experiments and we refine the posterior every time we expect the choice of prior to be unimportant asymptotically.
\section*{Bayesian inference}
Bayesian inference is the iterative application of Bayes' theorem to update one's knowledge about a random variable which might be a parameter of our model. It is not the only form of statistical inference, but it has several characteristics which make it more desirable than other techniques such as frequentist inference, where the frequency is interpreted as a probability.\\
First of all Bayesian inference will return a probability distribution, which in general contains a lot more information than an inferred value and a confidence interval.\\
On the other hand, as we have said before, Bayesian inference can depend strongly on the choice of a prior distribution of which there might not always be a natural choice.\\
Let us give an example where a Bayesian approach is much superior: a hunter is hunting with his dog, we can observe the position of the dog but we cannot observe the position of the hunter, we further know the that the dog to be located with a certain probability $p$ in a radius $r$ around the hunter.\\
The frequentist approach would lead to the following reasonment: since I have observed the dog in a given position: the hunter is in a radius $r$ around this position with probability $p$.\\
However relies on several tacit assumptions: the isotropy of the distribution of the dog around the hunter, different directions need not be equiprobable, in fact the dog will prefer to be upwind from the hunter; secondly the uniformity of the distribution of positions of the hunter regardless of where the dog is.\\
To put it in a mathematical form the frequentist approach equates $P(D|H)$ to $P(H|D)$ ignoring $P(H)$, the prior or the distribution of the position of the hunter and ignoring that $P(D|H)$ might depend on more than just the distance between the dog and the hunter.\\
Another classical application of Bayesian inference is the computation of the number of false positive in a medical test: Let us suppose there is a very rare disease which occures only in a tiny fraction $\epsilon$ of the population. A test for this disease returns a false result with probability $p$.
\begin{align*}
P(\text{negative}|\text{sick})=P(\text{positive}|\text{healthy})&=p\\
P(\text{positive}|\text{sick})=P(\text{negative}|\text{healthy})&=1-p\\
P(\text{sick})&=\epsilon
\end{align*}
Bayes theorem tells us that:
\begin{align*}
P(\text{false negative})=P(\text{sick}|\text{negative})&=\frac{p \epsilon}{p \epsilon+(1-p)(1-\epsilon)}\\
P(\text{false positive})=P(\text{healthy}|\text{positive})&=\frac{p (1-\epsilon)}{p (1-\epsilon)+(1-p)\epsilon}\,.
\end{align*}
As you can see these probabilities look much different even if the accuracy of the test is the same for false positives and false negatives. What is happening?
The rarity of the disease determines a very high rate of false positives, in fact it can be shown that more than half of the positives are false unless the probability $p$ of having an inaccurate result is smaller than the prevalence of the disease $\epsilon$.
\section*{Bayesian inference in quantitative biology}
Bayesian inference has an increasingly important role in quantitative biology: the emergence of large data sets coming from molecular biology, neurosciences and molecular biology has increased the need for sophisticated mathematical techniques for their analysis.\\
Examples of biological systems are being successfully investigated through the use of Bayesian inference range from phylogenetics \cite{HUE}, where one wants to reconstruct the most likely evolutionary tree from genetic data to gene regulatory networks where a stochastic approach has been recently shown to be very successful \cite{ELO,ZOU}.\\
Moreover moving away from the molecular scale systems such as neural networks and bacterial motility have greatly benefited by such approaches.\\
In what follows we will concentrate on two main problems and give a brief outline of a third.\\
The first problem we tackled is that of spatial searches with dilute and stochastic information about the location of an object. More precisely we will turn to a strategy originally devised by Vergassola et al. \cite{VERGASSOLA} that makes use of an informational theoretical approach for the location of an odor emitting source.\\
During our thesis we have developed a continuous version of the algorithm and an extensive analysis of its performances and trajectories.\\
The second problem we will turn to regards unzipping experiments of DNA molecules: the force-extension signal that can be measured in these experiments is strongly dependent on the DNA sequence.\\
At first we will describe the direct problem of reproducing experimental traces on a computer and we will describe a software package we have developed with F. Zamponi, R. Monasson and S. Cocco during our thesis, that can simulate the dynamics of such an experiment in a highly modular way.\\
Then we will propose several strategies for the inverse problem of reconstructing the sequence from the  unzipping traces.\\
Lastly we have devoted a section (appendix \ref{APPE}) to a brief technical description of an algorithm for the inference of the interaction matrix of integrate and fire neurons. This algorithm has been developed by Monasson and Cocco and our effort during our thesis has been a translation of the code to the C language, the development of an interface with Matlab and code optimization.\\
\part{Infotaxis}
\chapter{Introduction}
\section{Taxes and the biology of searching}
A \emph{taxis} is the innate directional response of the motility of an organism to a stimulus. On the other hand responses that imply a change in orientation or in the direction of growth are called \emph{tropisms} and those which are not directional are called \emph{kineses}.\\
The term taxis is most commonly found speaking of unicellular organisms, because of its automatic and innate nature, even thought it is sometimes applied to insects and crustaceans. Stereotyped responses in higher organisms are commonly thought to be less reflex-like, they are usually categorized as instincts and are the subject of study of ethology.\\
Taxes can be distinguished according to the nature of the sensory organs implied:
\begin{description}
\item[Klinotaxis] Different successive stimuli are measured by a single sensory organ.\\
\item[Tropotaxis] Well spaced sensory organs measure stimuli on different parts of the organism.
\item[Telotaxis] The perception is mediated by a single directional organ. When the motor response is at an angle to the direction of the source some sources distinguish {\bf menotaxis}.\\
\end{description}
Taxes can also be divided according to the type of stimulus they respond to: chemotaxis (chemical gradients), phototaxis (light sources), geotaxis (gravitational fields), magnetotaxis (magnetic fields) and so on and so forth.\\

\section{Chemotaxis}
The type of taxis which has attracted the most interest in biology is probably chemotaxis, because of its ubiquity in unicellular organisms as inside multicellular organisms.\\
The first observation of bacterial motility date back to the beginnings of microscopy, but we have to wait for the end of the nineteenth century for the first observations of responses to chemical gradients.\\
It is important to distinguish, as we will do in the following, between  bacterial and eukaryotic chemotaxis.\\
Bacteria are very small cells, whose size is of the order of the micrometer, below that of typical fluctuations of chemical fields: this forbids them to be directly sensitive to chemical gradients. Because of this chemosensation must happen through successive intensity assays. According to the preceding section definitions it is a klinotaxis.\\
Eukaryotic cells can be much bigger than bacteria: some species can reach sizes of the order of a millimeter and typical sizes range in the tens and hundreds of micrometers. Because of this in eukaryotes chemosensation happens through the instantaneous differentiation of stimuli coming from different parts of the organisms. In this case chemotaxis can be defined as a tropotaxis.\\
In the light of this distinction and of the differences between motor organs in different organisms, bacterial and eukaryotic chemotaxis must be considered as different phenomena.
\subsection{Chemotaxis in bacteria}
\begin{figure}[htbp]
 \begin{center}
  \includegraphics[width=.5\textwidth]{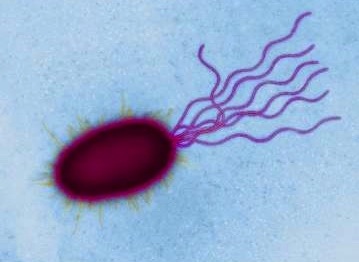}
  \caption{A specimen of \emph{Escherichia coli}. Notice the flagella that enable it to move, now unbundled.}
  \label{fig:ecoli}
 \end{center}
\end{figure}

\begin{figure}[htbp]
\includegraphics[width=.7\textwidth]{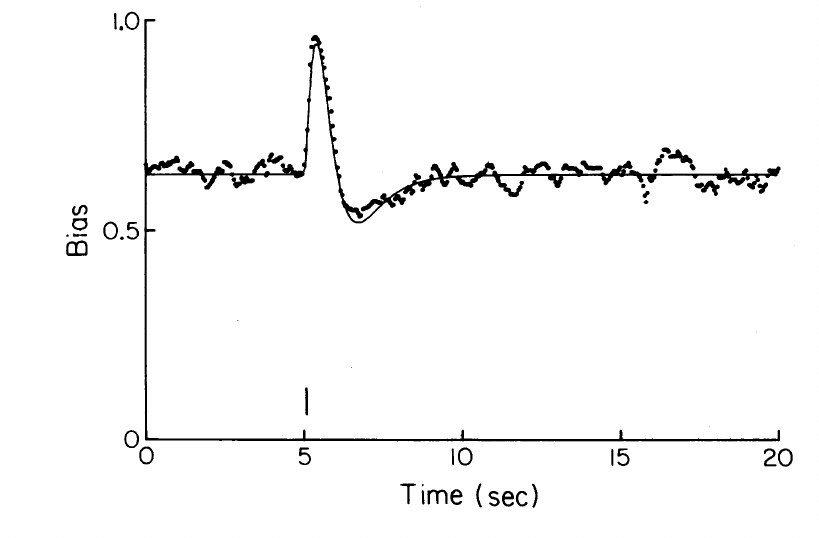}
\caption{The response of bacteria to a chemoattractant in wild type \emph{E. coli}. The dotted curve is the bias in the rate of tumbles after some attractant was pulsed at the vertical bar. From \cite{SEGA}.}
\label{fig:sega}
\end{figure}

Many reviews of bacterial chemotaxis exist in literature, for example the classic Adler's \cite{ADLER} or Berg's \cite{BERG1}, which has an extensive bibliography. Here we will follow another Berg's review \cite{BERG2} which is more focused on theory than on bacterial physiology.\\
Microbiology's workhorse is certainly \emph{Escherichia coli} (pictured in figure \ref{fig:ecoli}), partly for historical reasons, because of it's ubiquity in human guts and certainly for its simplicity.\\
\emph{E. coli} is endowed with about six flagella positioned on its surface. When those turn anti-clockwise they form a bundle and push the bacterium in a definite direction. Flagella can turn clockwise too: when this happens the bundle opens up and the bacteria tumbles on itself in a random fashion.\\
Those two modes of movement are the fundamental components o chemotactic response in flagellates and are called \emph{swims} in the first case and \emph{tumbles} in the second.\\
Swims length is temporally limited by Brownian noise which, at room temperature for a body of size of a micrometer, decorrelates the heading of the bacteria in about ten seconds. Because of this reason bacteria tumble before losing their original heading completely.\\
Tumbles on the other hand are a random event which last about a tenth of a second. The new heading of after a tumble is completely independent of the one before.\\
Up to here the description of the motion of a flagellate does not differ significantly from a random walk; in the absence of chemical gradients the duration of swims is distributed as an exponentially random variable (that is to say that tumbles are a Poisson process).\\
Directional response in the motion of \emph{E. coli} happens through the variation of the average duration of swims: if the bacteria is moving in a favorable direction swims become longer.\\
This observation is compatible with what we have said about the klinotactic nature of bacterial chemotaxis. Because of diffusive reasons, bacteria are not capable of discriminating between favorable and unfavorable directions during a tumble, but it is forced to sample the gradient during the swim. In other words the chemical gradient signal to noise ratio is big enough only on distances of the order of swims, not on the scale of the size of bacteria.\\
\emph{E. coli} temporal response to gradients has been studied thanks to the response to short impulses. Bacteria effectuate time differentiation through an integral of concentration at different times multiplied to a function which has a positive weight for the first second immediately in the past and a negative weight for the three preceding seconds:
\begin{equation}
P(\textrm{tumble})=l-k\int_{-\infty}^0 d t\, c(t) w(t)\,,
\end{equation}
where $k$ and $l$ are positive real constants that ensure normalization and $w(t)$ is a compact support weight function which has the characteristics we have just described and which were measured by Segall et al. in \cite{SEGA} (see Figure \ref{fig:sega}).
This can be rewritten integrating by parts as:
\begin{equation}
P(\textrm{tumble})=l+k\int_{-\infty}^0 d t\, c'(t) W(t)\,,
\end{equation}
where $W$ is a compact support probability distribution which is zero outside the integration domain and $W'(t)=w(t)$.\\
The real world $w$ has been measured by \cite{SEGA} and is shown in figure \ref{fig:sega}, the two lobes have equal area, which is consistent with our definition of $W$.
The fact that the derivative is averaged over a finite period of time is a desirable property, in fact it allows bacteria to average out fluctuations in concentration fields. On the other hand run lengths never get longer than a few seconds, because bacteria aren't able to go in a straight line for long periods of time because of rotational diffusion. 
\subsection{Chemotaxis in eukaryotes}
\begin{figure}[htbp]
 \begin{center}
  \includegraphics[width=.5\textwidth]{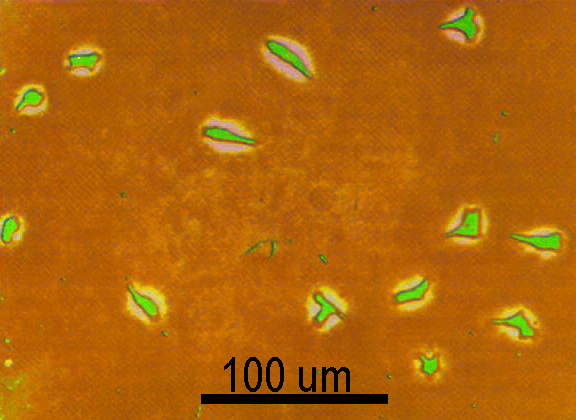}
  \caption{A few specimens of \emph{Dictyostelium discoideum}.}
  \label{fig:disco}
 \end{center}
\end{figure}

As we have previously mentioned, eukaryotes sense chemical gradients in a  way which is much different from bacteria. This difference has an effect on typical trajectories of a chemotactic eukaryote which, being able to sens gradients instantaneously and being much less affected by Brownian effects, is able to climb the chemoattractant gradient directly.\\
Motility in eukaryotic cells happens through ameboid movement (as in slime molds), cilia (as in \emph{Tetrahymena}, or through the eukaryotic flagellum (as in \emph{Chlamydomonas}), all these means of transportation are much more precise than the bacterial flagellum.\\
Eukaryotic chemotaxis is not confined to unicellular organisms: it plays a central role in embryogenesis, in the immune system and also the spread of metastases.\\
As is the case with many biological phenomena eukaryotic chemotaxis has its model organism: \emph{Dictyostelium discoideum} (pictured in figure \ref{fig:disco}), a soil living amoeba which cycles through an unicellular and a multicellular state according to the environmental conditions.\\
When \emph{D. discoideum} undergoes starvation, it starts secreting cyclic AMP which is a chemoattractant, this way cells move towards one another until they stick to each other. When the cells are lumped together they form what is referred to as a pseudoplasmodium, or more colloquially a slug which measures a few millimeters. Some other slime molds can form pseudoplasmodia of sizes of square meters which are commonly found on forest floors.\\
\emph{D. discoideum} we observed for the first time in 1933 \cite{RAPER1}, in the following years its life cycle was described in detail \cite{RAPER2}  and in the fifties cyclic AMP was identified as playing a central role in aggregation \cite{SHAFFER}. Nevertheless it wasn't until the beginning of the seventies that a model for aggregation was proposed \cite{KS1}, and despite some resistance in the microbiology community later accepted.\\
What was novel about this model was that aggregation was described as a truly collective phenomenon, like those found in the statistical physics of phase transitions.\\
%This model, which we will refer to as Keller-Segel model, describes the single cells of \emph{D. discoideum} as particles that follow a Langevin dynamics by climbing the gradient of the concentration field of cyclic AMP:
%\begin{equation}
 %x_i=\alpha \nabla \phi +\xi_i\,,\label{eq:kscell}
%\end{equation}
%where $x_i$ are the coordinates of the $i$th cell, $\phi$ is the chemical concentration of cyclic AMP and the $\xi_i$ are uncorrelated Gaussian noises and $\alpha$ is a dimensional constant.\\
%At the same time each cell secretes cyclic AMP which diffuses through the medium and is degraded into AMP by an enzyme. AMP is chemotactically inactive. This gives:
%\begin{equation}
% \partial_t\phi=D_1\nabla^2\phi-\frac1\tau\phi+\beta\sum_i\delta(x-x_i)\,,%\label{eq:kspot}
%\end{equation}
%where $D_1$ is a diffusion constant, $\tau$ is the mean degradation time, and $\beta$ is the degradation rate of cAMP into AMP.
%{\bf finire}

\section{Discrete infotaxis}
\subsection{Historical models}
\begin{figure}[htbp]
 \begin{center}
  \includegraphics[width=.3\textwidth]{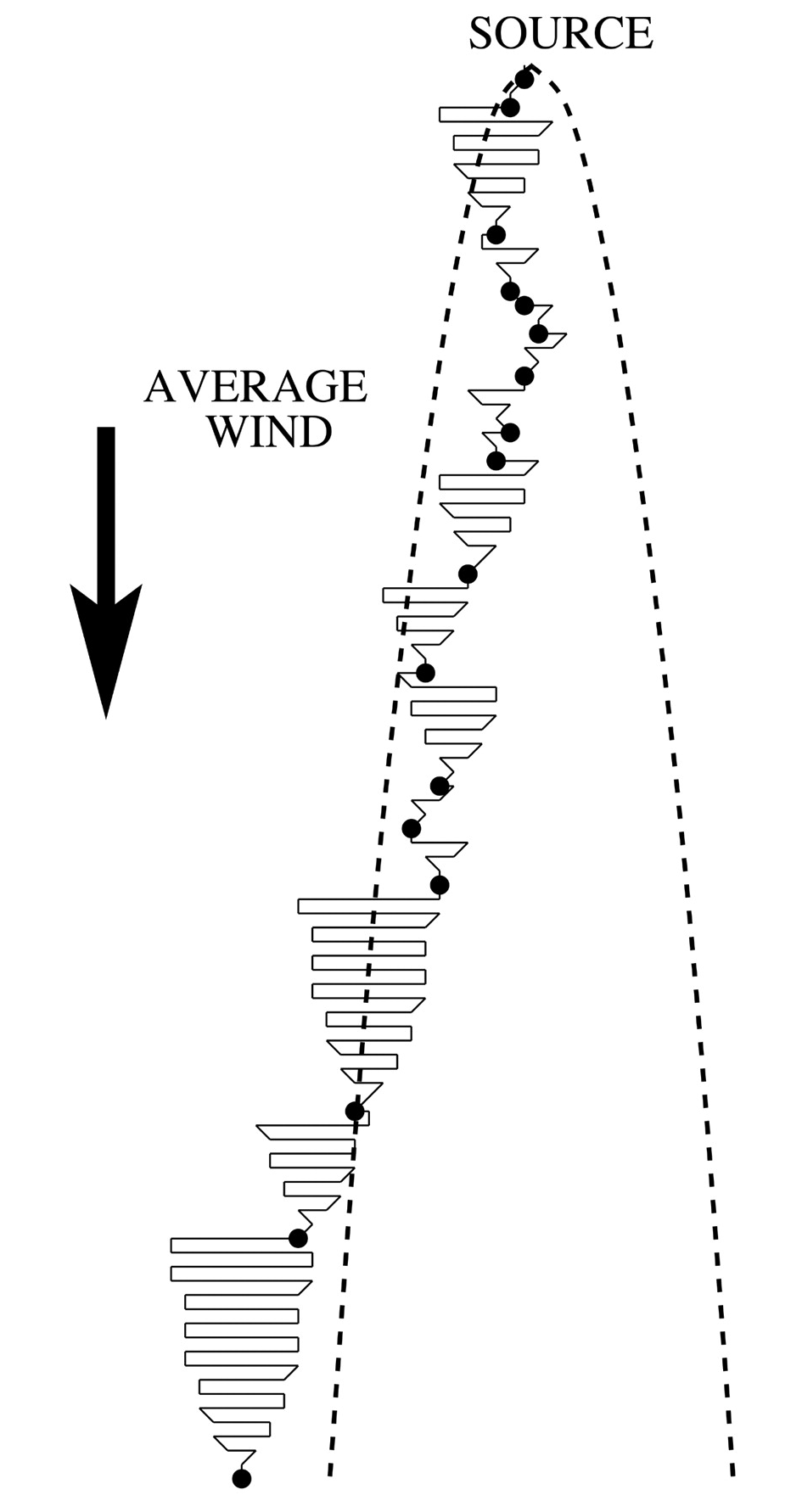}
  \caption{A sample trajectory of the algorithm proposed by Balkovsky and Shraiman. The continuous line is the trajectory, the dashed line the parabola that is the boundary to the area where the probability of encountering odor particles is significantly different from zero and the circles are the odor hits. From \cite{BALKO}}
  \label{fig:shraimann}
 \end{center}
\end{figure}

The description we have given for chemotactic cells relies heavily on the size of cells and on the nature of chemical gradients at their scale. If one wishes to model olfactory search, one has to deal with turbulence, intermittent signals and dilution of fields.\\
First of all most chemoattractants degrade over times and scales which are relevant over the size of a typical search, we will see that this leads to exponentially decaying concentrations and that this has to be taken into account.\\
\begin{figure}[htbp]
 \begin{center}
  \includegraphics[height=7cm]{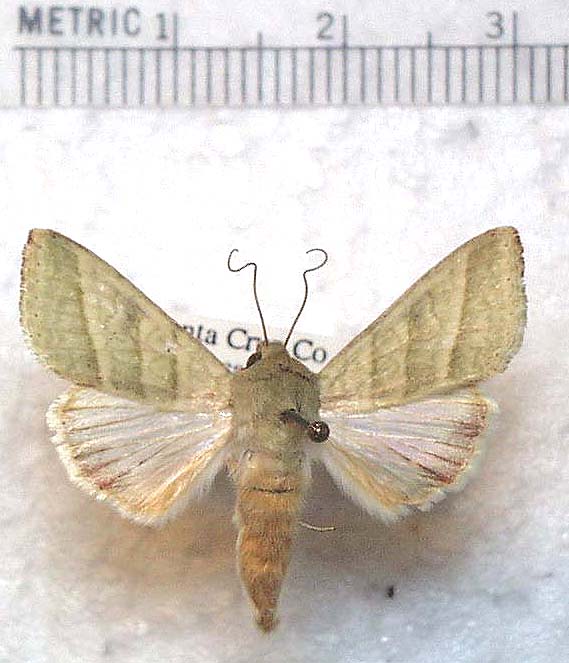}
  \includegraphics[height=7cm]{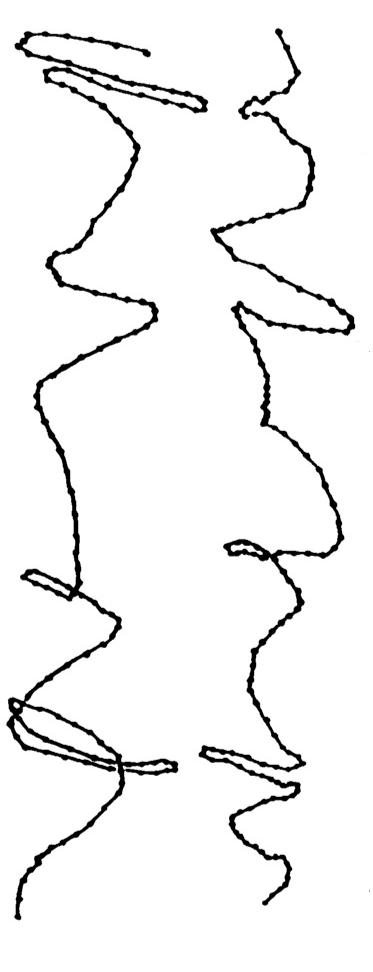}
  \caption{Left: a specimen of tobacco budworm (\emph{Heliothis virescens}), a species of moth. Right: a few recorded trajectories of \emph{H. virescens} from \cite{VICK}.}
  \label{fig:hvire}
 \end{center}
\end{figure}
Moreover the nature of olfactory system is such that it is impossible to instantaneously perceive the spatial derivatives as in a tropotaxis: nostrils are usually very close and even if they were to be as far apart as ears or eyes the spatial information they would get would not be reliable. This is because of the effect of turbulence, local concentrations do not necessarily reflect the distance or direction of the source.\\
In the past there have been a few attempts to define search strategies when information is scarce: one classic reference is Gal's book on search games \cite{GAL}, but the amount of information in classical search games is simply too scarce for our purposes: there is no equivalent of the odor field, that is the source is found when the searcher is close enough and the searcher has no clue whether the source is close by or not unless it has been found.\\
One further development of search strategies was given by Balkovsky and Shraiman \cite{BALKO} who proposed a model for olfactory searches where both the searcher and the odor particles are bounded to move on the sites of a bidimensional discrete lattice. The model supposes an average wind direction, that we can take without loss of generality to be up to down. Odor particles then are made to move down at every time-step and can either move left, right or not move at all on the horizontal axis with equal probabilities. Odor particles don't decay as in more refined models, thus the odor field is never dilute when the searcher is downwind with respect to the source and close to the wind axis.\\
The authors observed that the stationary probability of finding an odor particle in ${x,y}$ when the wind blows in $y$ direction and one particle per time step is emitted is given by:
\begin{equation}
 P({x,y})=\frac1{\sqrt{4\pi D y}}e^{-\frac{x^2}{4\pi D y}}\,,
\end{equation}
where $D=(p_\textrm{r}+p_\textrm{l})/2=1/3$ and $p_\textrm{r}=p_\textrm{l}=1/3$ are the probabilities of moving left and right. That is, being the variance proportional to $y$, most odor particles will be confined to the area $x^2<(4\pi Dy)$.\\
If an encounter has just been made and the searcher has no prior information on the position of the source, it follows from the Bayes' theorem that the source is most probably located in the area defined by a parabola having for vertex the position of the odor encounter. From this observation stems the strategy devised by the authors: once an odor particle has been encountered the searcher explores exhaustively zigzagging the area where the source is most probably located until either the source is found on another particle encountered. For a clearer pictures of what a typical trajectory looks like see figure \ref{fig:shraimann}.\\
The main drawback of this strategy is that it is guaranteed to work only in the case of non-decaying odor particles, that is when the odor concentration does not decrease exponentially with the distance.
\subsection{Definition of the odor detection model}
Recently Vergassola et al. have proposed an algorithm for olfactory searches: here we will describe what is the odor model that underlies their search strategy using the formalism used in the Supplementary informations of their paper \cite{VERGASSOLA}.\\
The stationary concentration of odor particles $c(y)$ in the absence of an average wind is given by:
\begin{equation}
D\nabla c(y) -\frac1\tau c(y)+R\delta(y-y^*)=0\,,
\end{equation}
where $D$ is the diffusion coefficient, that stems from molecular and turbulent diffusion, $\tau$ is the mean decay time,$R$ is the rate of emission of odor particles and $y*$ is the position of the source.\\
This equation has analytic solutions and in two dimensions yields:
\begin{equation}
 c_2(y)=\frac{R}{2\pi D}K_0\left(\frac{|y-y^*|}{\lambda}\right)\,,
\end{equation}
where $K_0$ is the zero-order modified Bessel function of the second kind, $\lambda$ is a characteristic length given by $\lambda=\sqrt{D\tau}$ and can be interpreted as the mean length traveled by an odor particle before decaying. It will be used in the following as the natural unit of lengths.\\
In three dimensions the solution is:
\begin{equation}
 c_3(y)=\frac{R}{4\pi D}\frac{e^{-\frac{|y-y^*|}{\lambda}}}{|y-y^*|}\,.
\end{equation}
The rate of encounter of odor particles per unit for a spherical searcher of radius $a$ is given by relation due to Smoluchowski \cite{SMOLU}:
\begin{equation}
R_3(y)=4\pi D a c_3(y)=R \frac{a}{\lambda} \frac{\lambda e^{-\frac{|y-y^*|}{\lambda}}}{|y-y^*|}\,.
\end{equation}
While in two dimensions the relation is:
\begin{equation}
R_2(y)=\frac{2\pi D}{\ln\left(\frac\lambda{a}\right)}  c_2(y)=R \frac{K_0\left(\frac{|y-y^*|}{\lambda}\right)}{\ln\left(\frac\lambda{a}\right)}\,,
\end{equation}
where $R$ is the number of emitted particles per second.\\
These two equations define the natural unit of time that we will use throughout this work: in three dimensions the unit of time is $\frac{\lambda}{a R}$, while in two dimensions it is $\log\left(\frac{\lambda}{a}\right)/R$. Once the unit of time and length are defined through the actual physical constants of the system we need not worry about those details anymore: the description we will give of the system will be completely independent of them.\\
\begin{figure}[htbp]
\begin{center}
\includegraphics[width=.45\textwidth]{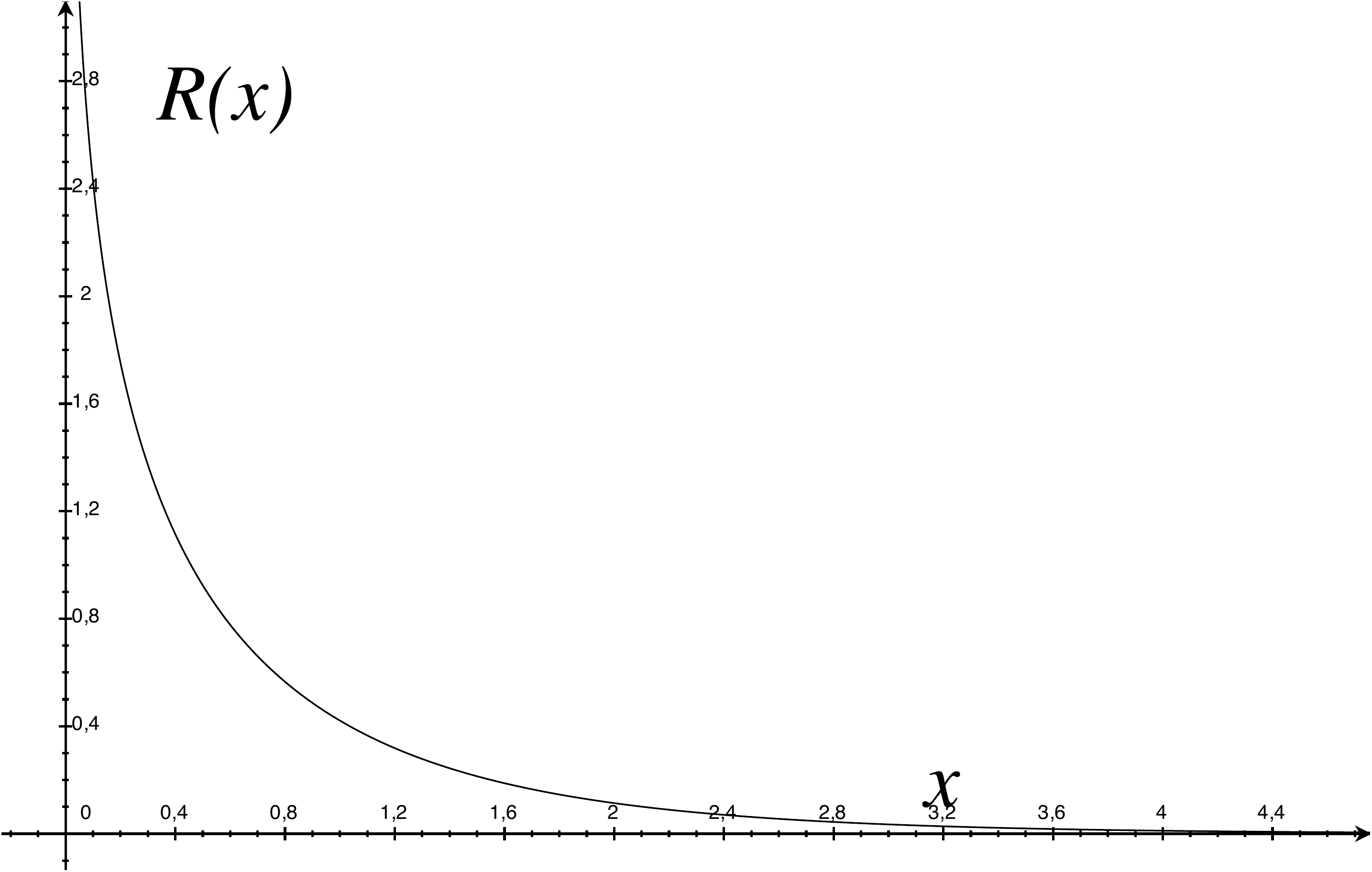}
\includegraphics[width=.45\textwidth]{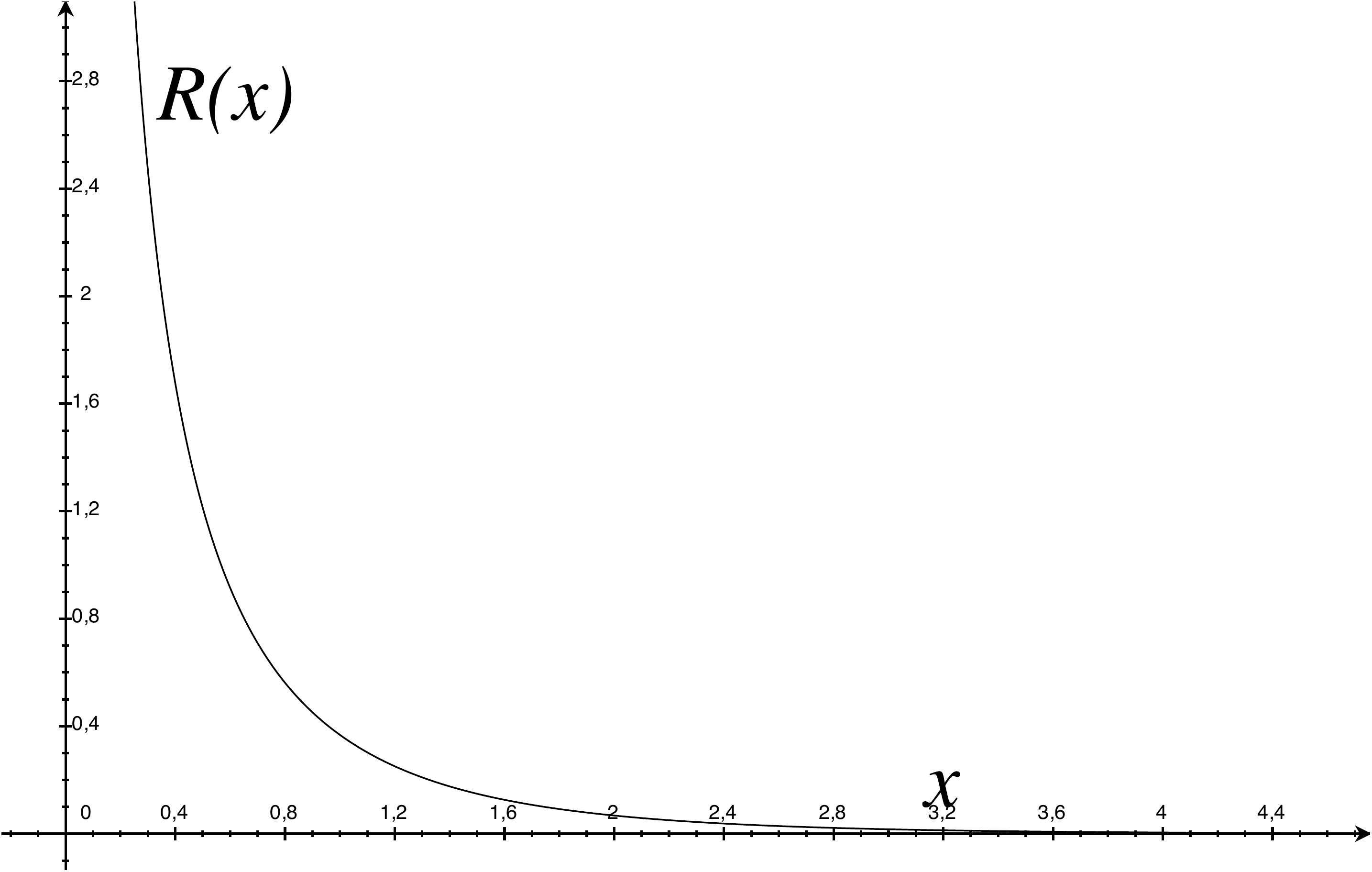}
\caption{The rate of encounter of odor particles, in two (left) and three (right) dimensions. The divergence in the origin is much more abrupt in the three-dimensional case, than in the two-dimensional one, but the asymptotic behavior for large arguments is the same.}\label{fig:R}
\end{center}
\end{figure}Once this relation is known, the idea is to model the erratic nature of odor detection in a turbulent flow as a Poisson process with a rate proportional to this rate of detection. This way odor is perceived through discrete \emph{hits} which vary in frequency as we move closer to the source. Hits contain no information pertaining the direction of the source and are all equal in intensity. The probability of getting $n$ hits during time $\Delta t$ while standing still at coordinates $y$ is:
\begin{equation}
 P_y(n)=\frac{(\Delta t R(y))^n}{n!}e^{-\Delta t R(y)}\,.
\end{equation}
This equation allows us to write the probability of receiving a number of hits n along a trajectory at times ${t_i}$ given the knowledge of the position of the source, that is:
\begin{equation}
 P(x(t),{t_i}|y)=\exp\left(-\int_0^t d t^\prime R(|y-x(t^\prime)|)\right)\prod_{i=1}^H R(|y-x(t_i)|)\,,
\end{equation}
where we have supposed no two hits happen at the same time. While this is reasonable for a continuous time description, in a discrete time framework one has to divide by $n!$ whenever $n$ hits happen during the same time-step, but we will see later this is of no importance.
\subsection{The Bayesian posterior}
Using Bayes' theorem we can write the probability of the source being at position $y$ given the trajectory and the hits' times:
\begin{equation}
 P_t(y|x(t),{t_i})=P_0(y)\frac{\exp\left(-\int_0^t d t^\prime R(|y-x(t^\prime)|)\right)\prod_{i=1}^H R(|y-x(t_i)|)}{\int d y^\prime\exp\left(-\int_0^t d t^\prime R(|y^\prime-x(t^\prime)|)\right)\prod_{i=1}^H R(|y^\prime-x(t_i)|)}\,,
\end{equation}
where $P_0$ is the prior distribution for the position of the source, we will see later how this plays a central role. On the other hand the attentive reader will have noticed how the previously mentioned $n!$ is cancelled out in this expression.\\
This expression has a few interesting features: the exponential term accounts for the vanishing probability of finding the source along the trajectory, that is: if the source was along the trajectory it would be found; it is also responsible for the low probability of points close to the trajectory. On the other hand the terms in the product are diverging and concentrate the probability around the points where most hits have occurred.
\subsection{The expected value of the variation of entropy}
The main idea behind Vergassola et al. algorithm is to exploit the Bayesian posterior as defined in the previous section to define the best movement at the next step.\\
This is done by defining the entropy of the posterior at a given time and by choosing the direction that maximizes its decrease: that is the direction where we expect to gain the most information on the source.\\
We will now compute this quantity in order to analyze the different contributions that make it up.\\
Even if the description given up to now is completely independent of the nature of the space where the searcher moves, be it a discrete lattice or an Euclidean space, and whether the time is discretized or continuous, we will from now on follow the description of the discrete version of the algorithm given by Vergassola et al..\\
Let $P_t(y)$ be the posterior probability distribution at time $t$. It's entropy is defined by:
\begin{equation}
 S(P_t)=-\sum_yP_t(y)\log(P_t(y))\,,
\end{equation}
where the sum runs on all the lattice sites $y$.\\
If our searcher is on one of the site of the lattice, it is now possible to compute the expected variation of entropy of the posterior distribution described above, resulting from a move on one of the adjacent lattice sites $x$:
\begin{equation}
 \left<S(P_{t+\Delta t})-S(P_t)\right>=-P_t(x)S(P_t)+(1-P_t(x))\left(\Delta S_\textrm{norm}+\sum_i\rho_i\Delta S_i\right)\,,\label{eq:discreteentropyvar}
\end{equation}
where the expected value has been taken with respect to the posterior probability distribution at time $t$.\\
Let us analyze the terms one by one:
\begin{description}
 \item [$-P_t(x)S(P_t)$] The source is found to be in $x$ and the entropy vanishes. The probability for this event to happen is given by the posterior $P_t(x)$ and the new value of the entropy is zero, that is the variation is $-S(P_t)$.
 \item[$(1-P_t(x))\Delta S_\textrm{norm}$] The source is not found, the probability of it being at site $x$ is now zero and the whole probability distribution has to be normalized. It can be easily computed as:
 \begin{equation}
 \begin{split}
\Delta S_\textrm{norm}&=-\sum_{y\neq x}\frac{P_t(y)}{1-P_t(x)}\log\left(\frac{P_t(y)}{1-P_t(x)}\right)+S(P_t)\\
&=\frac1{1-P_t(x)}\left(P_t(x)S(P_t)-S_\textrm{b}(P_t(x)\right)\,,
\end{split}
 \end{equation}
where $S_\textrm{b}(p)=-p\log(p)-(1-p)\log(1-p)$ is the binary entropy function.\\
 \item[$(1-P_t(x))\sum_i\rho_i\Delta S_i$] The source is not found, but at site $x$ the searcher receives $i$ hits. $\rho_i$ is the probability of receiving $i$ hits and $\Delta S_i$ is the corresponding entropy variation, that can be calculated remembering that:
 \begin{equation}
 P_{t+\Delta t}^{(i)}(y)= \frac{R(y-x)^i e^{-\Delta t R(y-x)}}{\left< R(y-x)^i e^{-\Delta t R(y-x)}\right>}P_t(y)\,.
  \end{equation}

 \end{description}
The main idea behind Infotaxis is to use this variation of entropy as an instantaneous potential and to move in the direction where the entropy decreases faster. With this in mind different terms play the contrasting roles of exploration and exploitation in the search.
The first term is more negative when the probability of finding the source at site $x$ is larger and can be thought as an exploitation term, where the searcher tries to move greedily where the source is more likely to be found. This term only dominates at the end of the search when the probability is well concentrated.\\
The last two terms favor the collection of new information, through, on one hand, the elimination of possible candidates for the source position, and, on the other, the collection of hits.\\
One of the most compelling features of this algorithm is that the balance between exploration and exploitation seems to be automatic, we will see in the following that this statement needs to be refined, and that one can see the algorithm as greedy on the entropy potential and that a class of more powerful algorithms can be imagined on the basis of this. 
\chapter{Continuous infotaxis}
\section{Derivation of continuous infotaxis}
We now turn to the problem of the derivation of a continuous form for infotaxis which we have done during our PhD. There are a few reasons for doing so: first of all, real organisms experience the world as continuous and a lattice based description of the world seems very artificial.\\
Secondly, the original algorithm poses a very realistic odor propagation model, while retaining a discrete description of the searcher and of its vision of the world. This makes the model anisotropic, in fact if we suppose the source is at a certain euclidean distance, the searcher will experience the same number of hits (on average) regardless of the direction of the source with respect to the axes of the lattice, but the direction of the source might decrease the number of steps needed to reach it of a factor of up to $\sqrt{d}$, where $d$ is the dimension of the space.\\
Another inconvenient of a discrete model is that the lattice is finite and the time needed to sum over all of its sites limits what can be practically done, especially in three dimensions where only a few trajectories on a small lattice were generated \cite{MASSON}.\\
A continuous description on the contrary allows the description of unbounded domains and the use of adaptive techniques to improve precision if needed.\\
One important thing must be stated before we begin: there is not one possible translation of infotaxis in the continuous limit, what we will do is only one of the many options.\\
In the following we will derive our version of continuous infotaxis in two different ways: the first is somewhat lengthy and cumbersome, but it follows closely from the discrete definition, while the second is much more compact but we think showing both might shine different lights on the problem.\\
The first difference between a discrete and a continuous model is the nature of the probabilistic description: from now on we have to distinguish between probabilities and probability densities which we will denote with $p_t(x)$.
In order to complete the discussion of the continuous limit we have to identify three independent scales which are relevant in the spatial part of the limit which are identical in the discrete version of the algorithm. These are: the lattice spacing, the size of the source $\sigma_\textrm{s}$ and the area (or volume) perceived by the searcher in a time-step $\sigma_\textrm{p}$.\\
To rephrase this: in the discrete version of the algorithm during one time step the searcher is able to rule out the presence of the search on one lattice site. The source size is one lattice site. Performing the continuous limit we could, in principle, leave the size of the source and of the searcher perceptions finite for a vanishing lattice spacing.\\
If we analyze one by one the terms of equation (\ref{eq:discreteentropyvar}) we obtain:
\begin{description}
 \item [$-P_t(x)S(P_t)$] While dealing with discrete probabilities the entropy of a sure event is zero, on the other hand for continuous distributions the entropy of a Dirac distribution is negative and divergent. In this case if the source is found the entropy does not diverge because the source has a finite size $\sigma_\textrm{s}$. Therefore this term is $\sigma_\textrm{p}p_t(x)(\log(\sigma_\textrm{s})-S(p_t))$
 \item[$(1-P_t(x))\Delta S_\textrm{norm}$] When the searcher moves the probability in the area $\sigma_\textrm{p}$ around its position $x(t)$ becomes zero, thus the expected value for the variation of entropy due to the new normalization reads $p_t(x)\sigma_\textrm{p}S(p_t)-S_\textrm{b}(p_t(x)\sigma_\textrm{p})$, where we have considered $p_t$ constant in the area $\sigma_\textrm{p}$ and where $S_\textrm{b}(p)$ is the binary entropy, as function defined in the previous chapter.
 \item[$(1-P_t(x))\sum_i\rho_i\Delta S_i$] For what concerns the terms depending on the expected number of hits, we will focus on none or a single hit in a time $\Delta t$, because the probability of having more is negligible when $\Delta t$ is small. That is: $\rho_1=\Delta t \langle R(y-x(t))\rangle+O(\Delta t^2)$ and $\rho_0=1-\rho_1$.\\
Thanks to the definition of the posterior we can write down the probability density at time $t+\Delta t$ if an hit has occurred in the interval $\Delta t$ as:
\begin{equation}
p^{(1)}_{t+\Delta t}(y)=p_t(y) \frac{R(y-x(t))}{\langle R(z-x(t))\rangle}+O(\Delta t)\,,
\end{equation}
or if it hasn't occurred:
\begin{equation}
\begin{split}
p^{(0)}_{t+\Delta t}(y)&=p_t(y) \frac{1-\Delta t R(y-x(t))}{1-\Delta t\langle R(z-x(t))\rangle}\\
&= p_t(y)[1+\Delta t(\langle R(z-x(t))\rangle- R(y-x(t)))]+O(\Delta t^2)\,,
\end{split}
\end{equation}
where we have omitted the vector norms in the argument of the $R$ and where the average is performed over the variable $z$. Notice that we only need the zeroth order in $\Delta t$ for the term for one hit.\\
Omitting all dependencies, the entropy variation for no hits reads:
\begin{equation}
\begin{split}
  \rho_0\Delta S_0&=(1-\Delta t \langle R\rangle)\left(S\left(p^{(0)}_{t+\Delta t}\right)-S(p_t)\right)\\
  &=(1-\Delta t \langle R\rangle)\left(\Delta t\langle (\langle R\rangle -R)\log(p_t) \rangle \right)\\
  &=-\Delta t\left<(\langle R\rangle -R)\log(p_t)\right>\,,
\end{split}
\end{equation}
while that for one hit is:
\begin{equation}
\begin{split}
  \rho_1\Delta S_1&=\Delta t \langle R\rangle\left(S\left(p^{(1)}_{t+\Delta t}\right)-S(p_t)\right)\\
  &=\Delta t \langle R\rangle\left(\left<\frac{R}{\langle R\rangle}\log \left(p_t\frac{R}{\langle R\rangle}\right)\right>+\langle\log p_t\rangle\right)\\
&=\Delta t\left<R\log\left(p_t\frac{R}{\langle R\rangle}\right)+\langle R\rangle\log(p_t)\right>\,.
\end{split}
\end{equation}
 \end{description}
 Putting all the terms together one obtains:
\begin{equation}
\begin{split}
 \left<S(p_{t+\Delta t})-S(p_t)\right>&=\sigma_\textrm{p}p_t(x)\log(\sigma_\textrm{s})+S_\textrm{b}(p_t(x)\sigma_\textrm{p})\\
 &+\Delta t\left< R(y-x)\log\left(\frac{  R(y-x)}{\langle R(z-x)\rangle}\right)\right>\,,
\end{split}
\end{equation}
at first order in $\Delta t$.\\
When the size of the area observed by the searcher vanishes all the terms on the first line vanish (even if the area of the source is zero). One must also observe that in the continuous limit the area $\sigma_\textrm{p}$ must be written as $s v \Delta t$ where $s$ is the cross section of the searcher's perception and $v$ its speed.

On the other hand we can regard the number of received hits as a message on the position of the source. We can compute the mutual information between the random variable $Y$, the position of the source and the random variable $N$, number of hits at first order in $\Delta t$.\\
If one remember the meaning of the rate function $R$, $P(N=1|Y=y)=\Delta t R(y-x)+o(\Delta t)$ and conversely $P(N=0|Y=y)=1-\Delta t R(y-x)+o(\Delta t)$, it follows that:
\begin{equation}
\begin{split}
 I(N,Y)&=\int d y\; P(y)\sum_n P(n|y)\log\left(\frac{ P(n|y)}{P(n)}\right)\\
 &=\left<\sum_n P(n|y) \log\left(\frac{ P(n|y)}{\langle P(n|y)\rangle}\right)\right>\\
 &=\Delta t\left< R(y-x)\log\left(\frac{  R(y-x)}{\langle R(z-x)\rangle}\right)\right>+o(\Delta t)\,,
\end{split}
\end{equation}
where all the terms except for $N=1$ are of higher order in $\Delta t$.\\
The main idea behind discrete infotaxis, that is: to move in the direction that minimizes the entropy of the posterior distribution, here translates into moving in the direction that maximizes the mutual information between the two variables.\\
One of the possible strategies to move in the direction that maximizes the gain in information, and arguably the simplest is that of forcing the searcher to obey Brownian dynamics, where the opposite of the information gain is viewed as a potential to be minimized, that is:
\begin{equation}
V_t(x)=-\left< R(y-x)\log\left(\frac{  R(y-x)}{\langle R(z-x)\rangle}\right)\right>\,,
\end{equation}
And for the searcher:
\begin{equation}
\gamma \dot x=-\nabla_x V_t(x)\,.\label{eq:searcher}
\end{equation}
where $\gamma$ is a friction coefficient that will be considered constant.\\
It can be argued that this equation cannot be considered equivalent to infotaxis, because the velocity is not constant. We have discussed this in detail in \cite{IO2}, and we will not dwell upon the details here.\\
It suffices to say that there is no way to impose a fixed velocity in a continuous framework: suppose for example that we choose $\gamma$ as a function of the right hand side so that the velocity is equal to a constant $V$, we have observed that if we choose too big a $V$ we observe long steps and a lot of backtracking. This is clearly an effect of the finite integration time-step and it is an effect that disappears in the small time-step limit, but we believe it is symptomatic of a system that chooses it's own velocity by changing the direction continuously.
\section{Search strategy before the first hit}
\subsection{Choice of the prior}
Bayesian techniques are usually very powerful, but the choice of a suitable prior can often be difficult. One can hope for the existence of a obvious choice, or that the results do not depend too much on the specifics of the prior.\\
The situation at hand is less clear cut: while all of our quantities have a clear probabilistic interpretation, what we ultimately want is for the algorithm to be performing well.\\
Vergassola et al. chose a prior proportional to the odor propagation function $R$ which has a few desirable properties: it is normalizable, it has a possible interpretation in the framework of our model and it does not define a new, arbitrary length scale while still concentrating most of the probability over a finite area.\\
Another possible choice in the discrete version of the algorithm is the uniform distribution, where every lattice site is given equal weight, even though this was not included in the original infotaxis paper, we have toyed with this prior only to obtain trajectories that go straight until they reach a distance of approximately $\lambda$ from the boundary of the lattice.\\
Unfortunately, this lattice choice does not have any equivalent in unbounded continuous space, because of this we cannot translate our results in this case.\\
We will now concentrate on two priors:
\begin{description}
\item[One-hit prior] Proportional to the right $R$ in the appropriate dimension. It has an integrable divergence at the origin, but for every $\tau$, no matter how small $R(y)\exp(-\tau R(y))$ is finite for $y=0$.\\
\item[Exponential prior] Proportional to $\exp(-y/d_0)$. Choosing $d_0=\lambda$ we have the same asymptotic behavior for large $y$. This can be used to investigate how important the small scale behavior of the prior is. 
\end{description}
In his original paper \cite{VERGASSOLA,PRIVERGA} Vergassola et al. proposed the first prior as a natural choice.\\
As suggested by the name we have chosen, we could consider the one-hit prior as the result of a search process that has started just after the searcher has received the first hit.\\
This of interpretation, however, poses some problems: how can we justify search trajectories that start very far from the source? If we stick to this interpretation they should be considered as very rare events.\\
This can be salvaged by considering only trajectories that start close enough to the source. As we will see in the following, it doesn't make much sense to employ such a sophisticated algorithm when there's effectively no information to gain.\\
\subsection{Spirals}
In \cite{VERGASSOLA} Vergassola and collaborators described logarithmic spirals in discrete infotaxis, before the first hit. After observing several trajectories where the source of odor had been turned off, we have concluded \cite{TESI} that spirals do appear in discrete infotaxis, but they are not logarithmic, but Archimedean in nature. That is the spacing between subsequent arms is constant.\\
In what follows we wish to characterize spirals in two dimensions and their equivalent in three dimensions for continuous infotaxis, the debate over discrete infotaxis having since been settled \cite{MASSON} with further simulations in hexagonal lattices.\\
\subsubsection{One-hit prior}
\begin{figure}[htbp]
\begin{center}
\includegraphics[width=.8\textwidth]{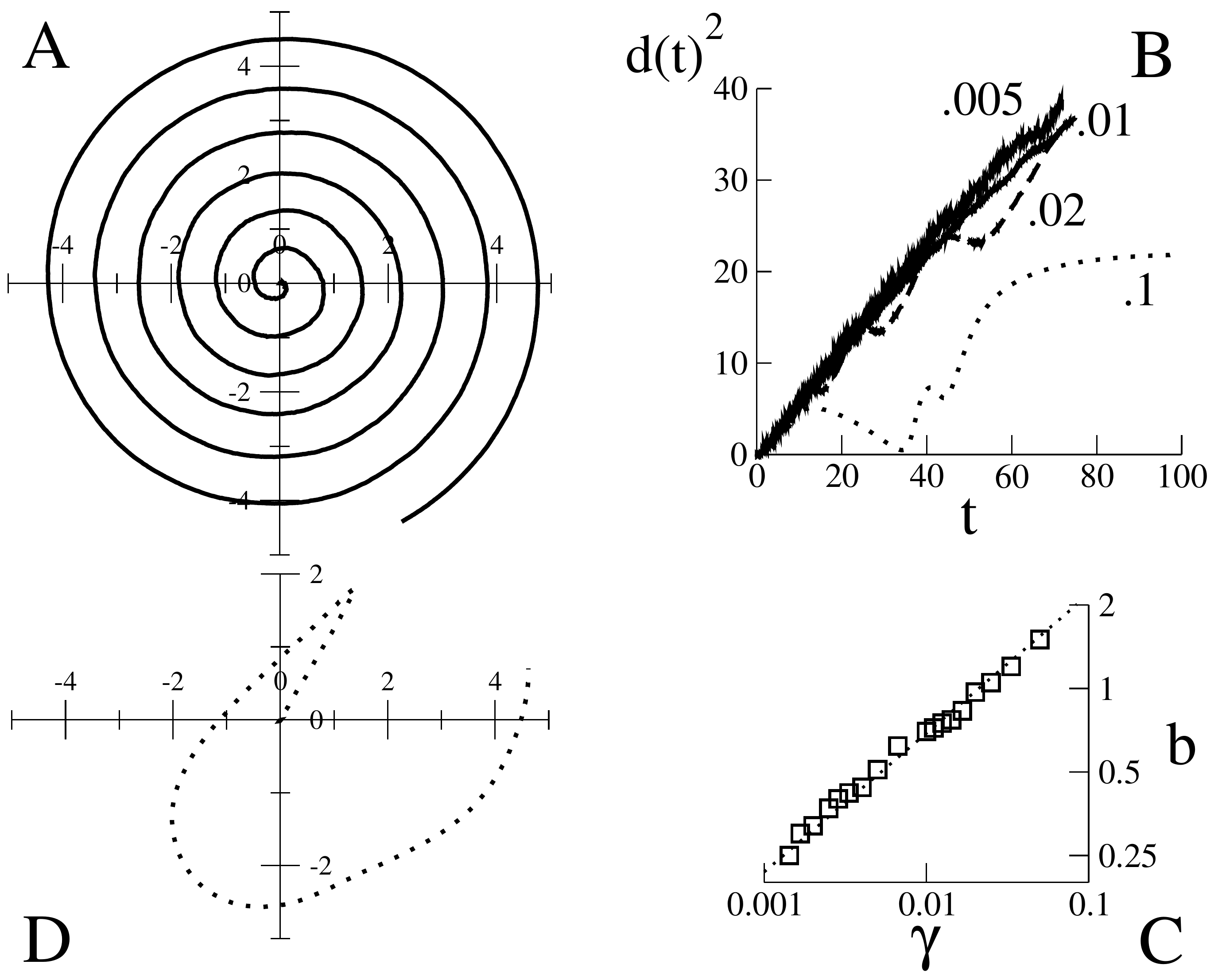}
\caption{{\bf A.} A spiral obtained for $\gamma=0.01$. {\bf D.} A spiral obtained for $\gamma=0.1$. {\bf B.} $d(t)^2$ as a function of time. As this quantity is proportional to the area explored in a given time, we show here that this is proportional to the elapsed time for several values of $\gamma$. For large $\gamma$ this behavior breaks down and the searcher eventually halts. The trajectories for $\gamma=0.01,0.1$ correspond to panels {\bf A} and {\bf D.} {\bf C.} Many values of the spacing $b$ between spiral arms, as a function of $\gamma$. The dotted line corresponds to a slope of $-1/2$.}\label{fig:spiral2d}
\end{center}
\end{figure}

\begin{figure}[htbp]
\includegraphics[width=.5\textwidth]{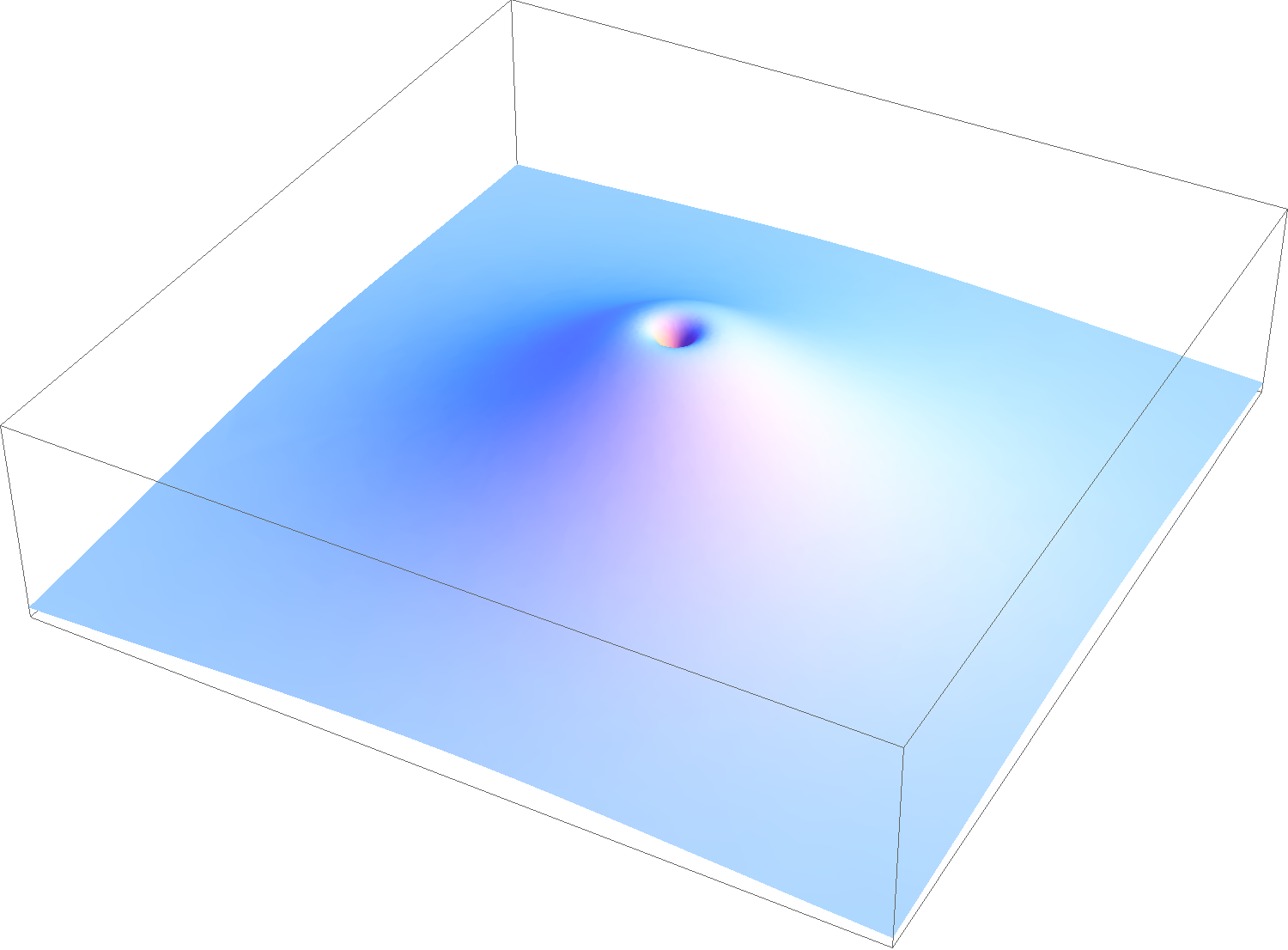}
\includegraphics[width=.5\textwidth]{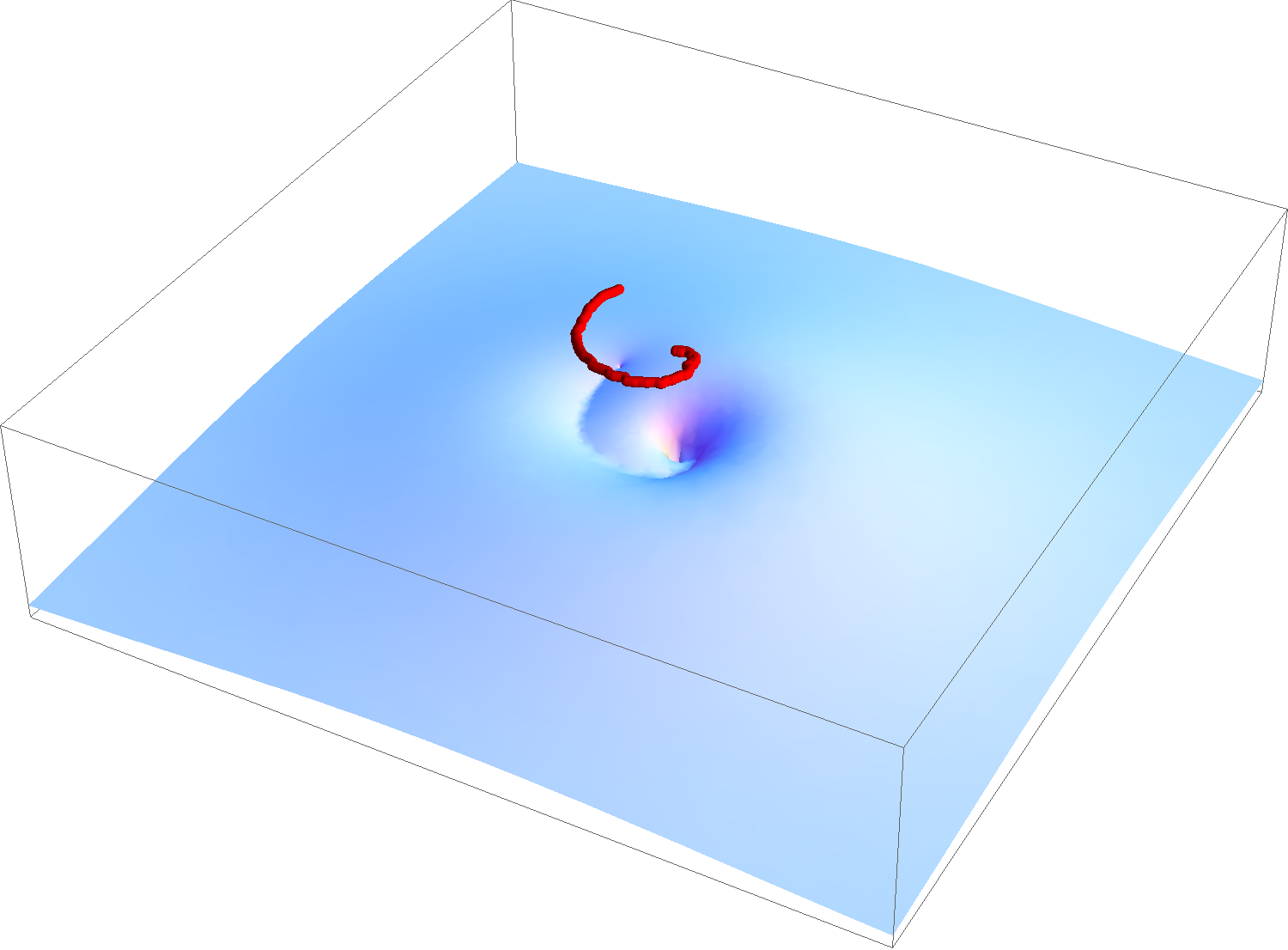}
\includegraphics[width=.5\textwidth]{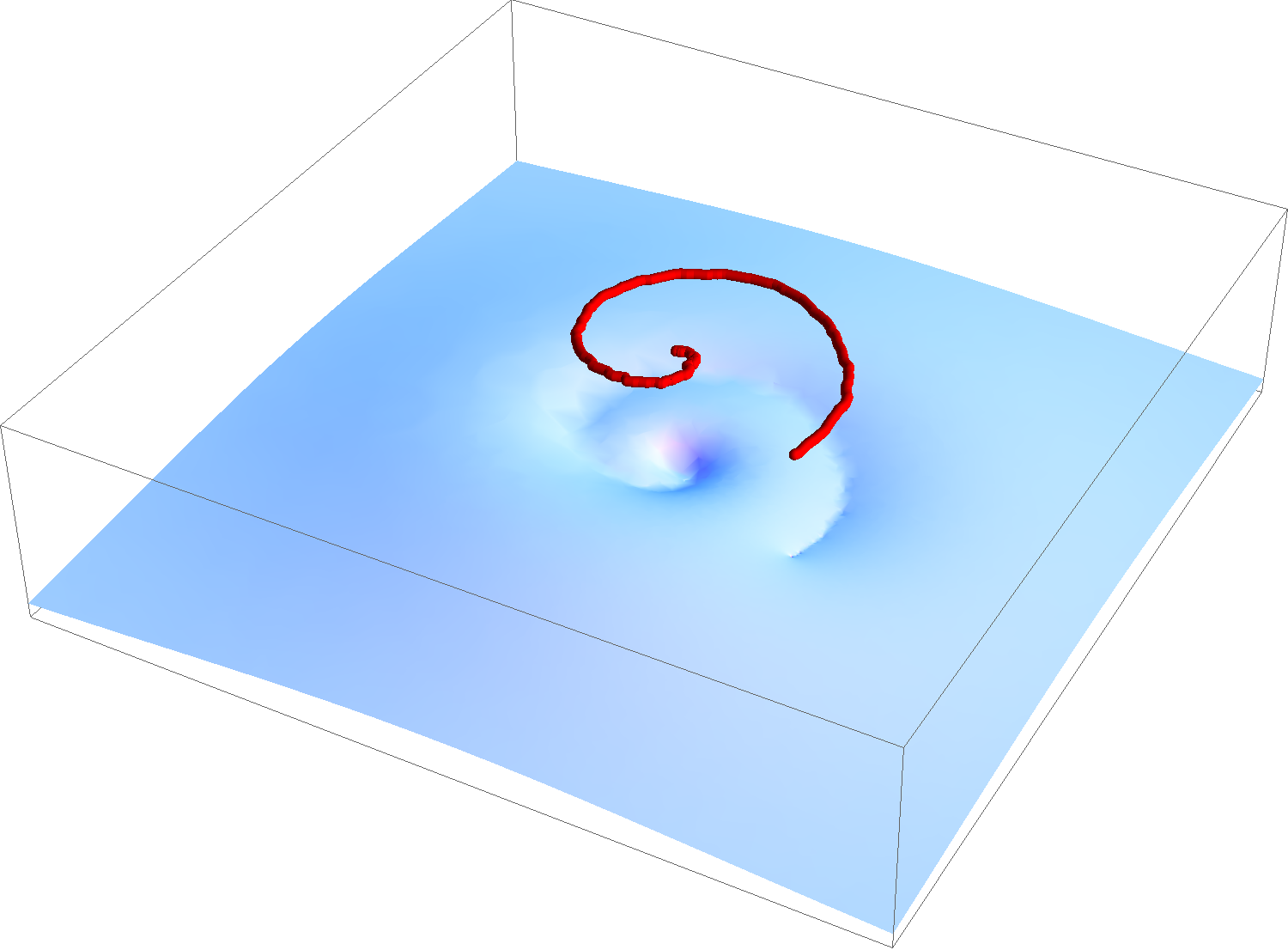}
\includegraphics[width=.5\textwidth]{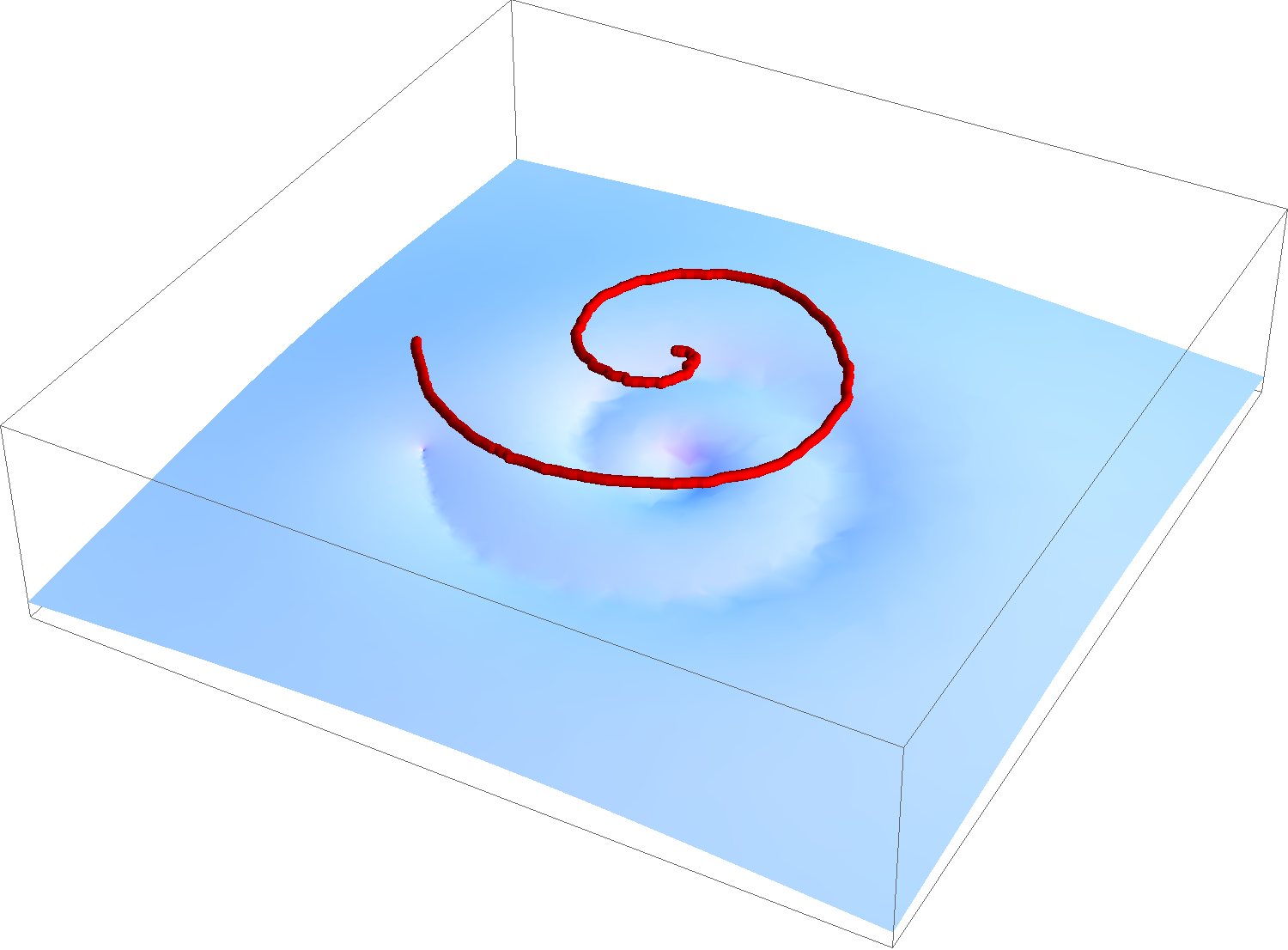}
\caption{The effect of a spiraling trajectory on the probability distributiion at different times.}\label{fig:spiral2dcolor}
\end{figure}

\begin{figure}
\begin{center}
\includegraphics[width=\textwidth]{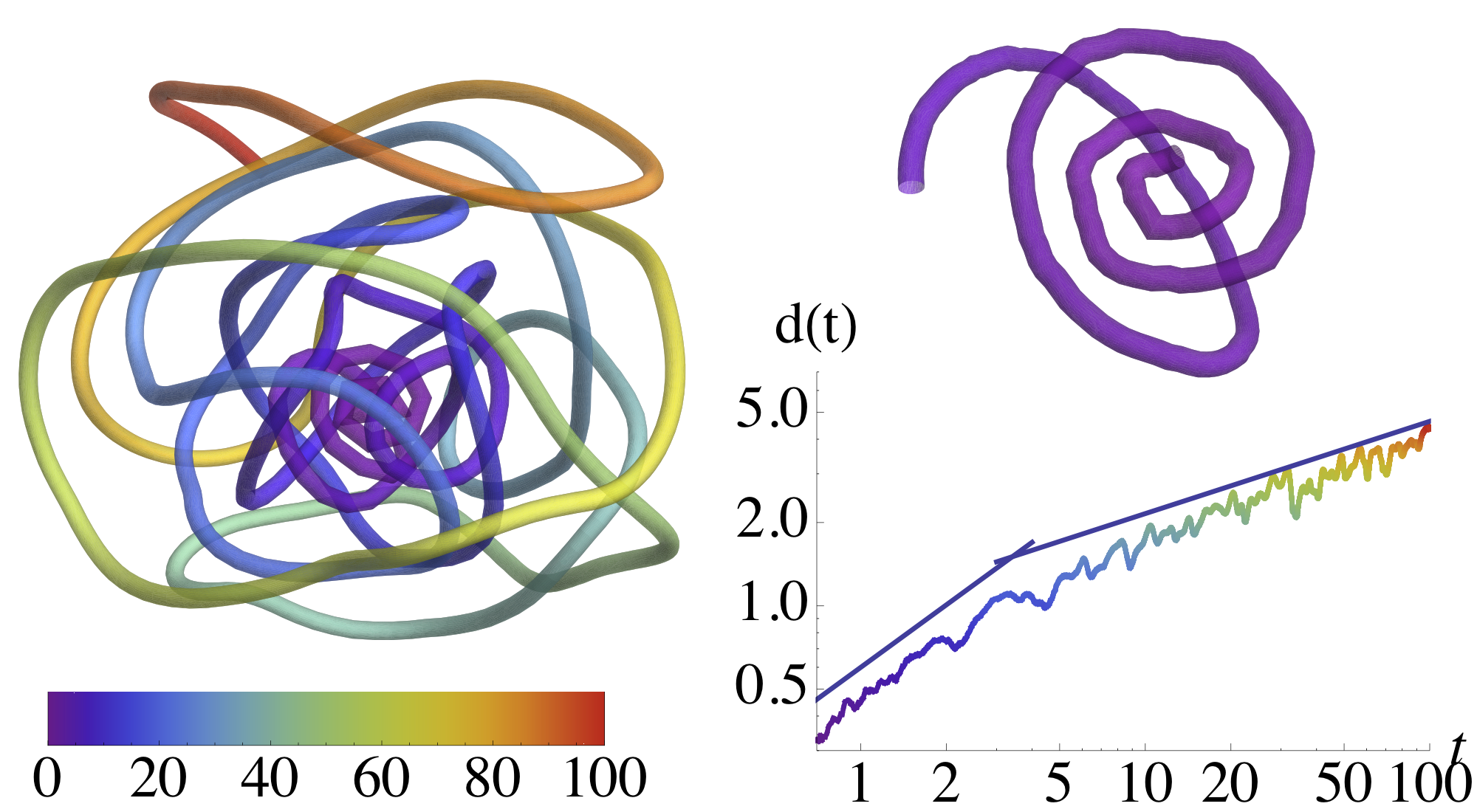}
\caption{A three-dimensional trajectory in the absence of hit for $\gamma=.01$ (left), with its quasi two-dimensional initial portion (top); the time axis is color coded. Bottom: distance to the origin, $d(t)$, compared to the power laws $t^{.75}$, then $t^{1/3}$.}\label{fig:spiral3d}
\end{center}
\end{figure}

In two dimensions the searcher moves in spirals for a wide range of values of $\gamma$, as is shown in figure \ref{fig:spiral2d}. When $\gamma$ is too big spiral behavior breaks down.\\
This behavior can be explained by a very simple argument: for a large range of values of $\gamma$ the searcher effectively visits a region of area proportional to the elapsed time. In a way the probability of finding the source in a given area is discounted in a given time thanks to the negative exponential term in the posterior. Once the source is not found the searcher moves elsewhere. This effect on the prior can be directly observed in figure \ref{fig:spiral2dcolor}.\\
This area does not depend on $\gamma$, while the linear velocity of the searcher does. For this reason this only has an effect on the spacing of the arms. More quantitatively  if $b$ is the spacing between successive arms then what we observe is consistent with $b\sim\sqrt{\gamma}$ and $|\dot x|\sim 1/b$.\\
Spiral behavior is not observed for large $\gamma$ ($>0.08$), we think that this is due to the fact that the $R\log R$ kernel has a range which is proportional to $\lambda$ and for large $\gamma$'s we would expect arm spacing which are larger than this range. In other words the algorithm cannot be sensitive to the probability distribution at large distances.\\
To validate this hypothesis we have run a few simulation with a modified kernel with larger and shorter range, and we have indeed observed that this moves the spiral-breaking-down threshold in the expected direction.\\
In three dimensions there is no exact equivalent of a spiral: the searcher will try to stay as close as possible to where it started as a result of the exponentially decreasing prior, but will move in a self avoiding trajectory, because of the term $\exp\left(-\int^tdt^\prime R(y-x(t^\prime)\right)$ in the posterior probability.\\
We have observed the first part of the trajectory to be quasi-planar and then to break off and start occupying all available space, this is shown in figure \ref{fig:spiral3d} where the dependence of the distance from the origin is plotted as a function of time and compared with the curve $t^{1/3}$ which corresponds to the prediction of space filling trajectories.\\
Three dimensional trajectories look like balls of yarn, compact coiled structures. We think that parallels can be drawn with the solutions of the Thomson problem for polyelectrolites \cite{ANGELESCU,CERDA,SLOSAR}, which has received a lot of attention recently because of its connections to the problem of DNA packing in virus capsides.
\subsubsection{Exponential prior}
\begin{figure}[htbp]
\includegraphics[width=\textwidth]{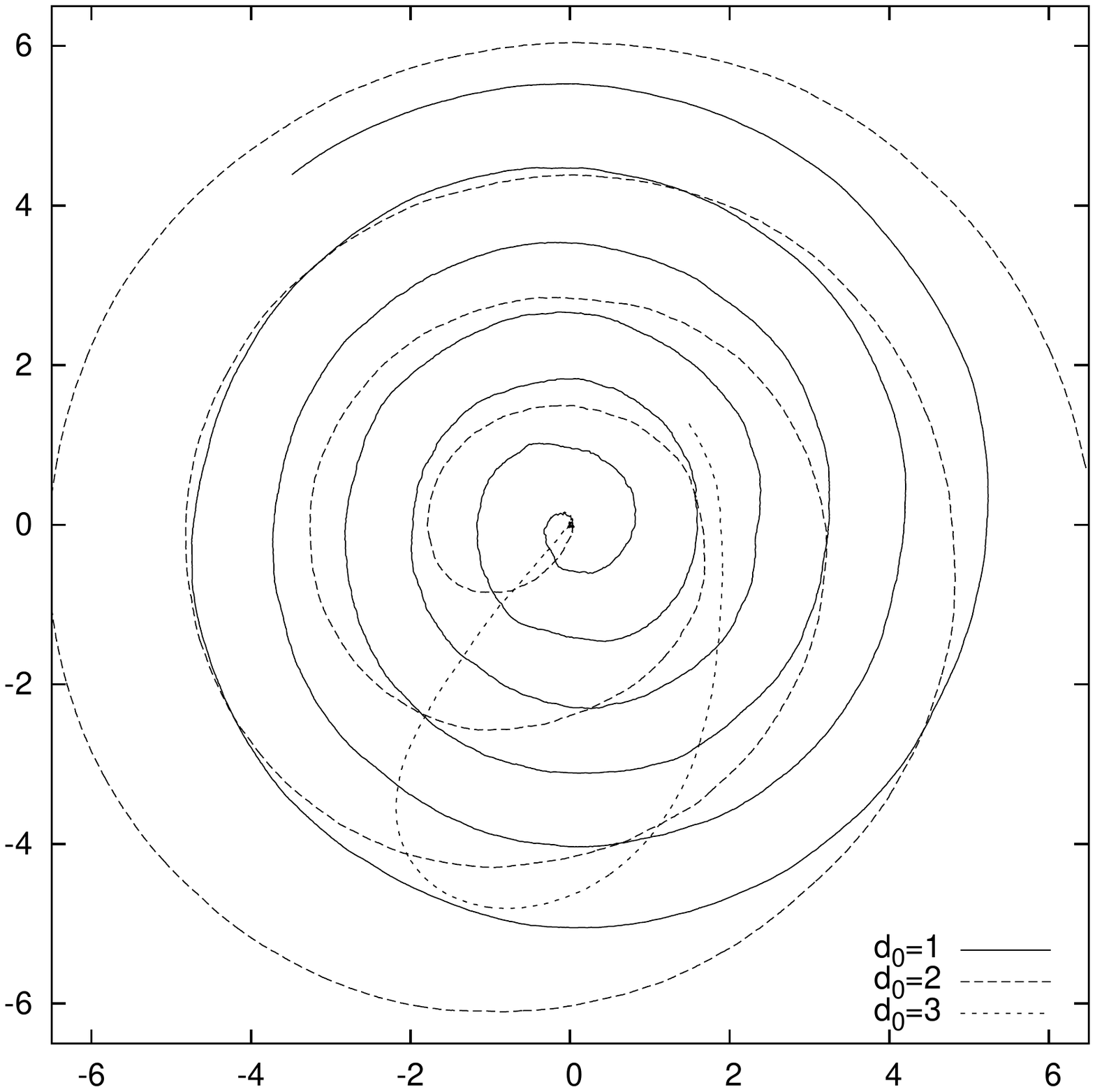}
\caption{Three trajectories without hits for the exponential prior with varying $d_0$. As highlighted in the text, for low enough $d_0$ spirals are observed with spacing $b\propto d_0$. When $d_0$ is too large, as we have observed for varying $\gamma$, spirals break down. Such is the case for $d_0=3$.}\label{fig:dprior}
\end{figure}

Another way of interpreting the choice of the one-hit prior is to consider the details of the prior at short range from the starting point of the searcher as mostly irrelevant and to concentrate on the asymptotic behavior.\\
Ignoring small scale behavior makes a lot of sense in the case of discrete infotaxis, where the scales smaller than the lattice spacing are not accessible, and the probability at the starting point of the searcher is exactly zero regardless of the prior.\\
The exponential prior can be also justified because of its memorylessness property that is: $P(Y>y+d|Y>y)=P(Y>d)$ and furthermore because it is maximum entropy distribution with a fixed mean.\\
This two mathematical properties could be used to justify the Archimedean nature of the spirals, which can be checked in figure \ref{fig:dprior}. The spirals however break down, as discussed before for the case of variable $\gamma$, when the arm spacing $b$ would exceed the range of the kernel $R\log R$.\\
The fact that the behavior of the searcher for both the one-hit prior and the exponential prior produces spirals, suggests that the spirals are a consequence of the asymptotic behavior of the prior at large distances. We will try to verify this with a Taylor series expansion of the right hand side of the equation for the movement of the searcher.\\
\subsection{Small $x$ expansion}
It is possible to characterize the spirals as an instability by performing an expansion for small $x$ of equation \ref{eq:searcher}:
\begin{equation}
\begin{split} 
\gamma  \dot {\vec{x}}(t) =&\alpha _1(t)  \vec{x}(t)+\alpha _2(t) \int_0^t d t^\prime \vec{x}(t^\prime)\\ + &\int_0^t  d t^\prime \vec{x}(t^\prime)  \left[ \beta _1(t) |\vec{x}(t^\prime)|^2 + \beta _2(t) \vec{x}(t^\prime) \cdot \int _0^t d t^{\prime\prime}\vec{x}(t^{\prime\prime}) \right]\\
+& \int_0^t  dt^\prime \vec{x}(t^\prime)  \left[ \beta_3(t)
\int _0^t  dt^{\prime\prime}|\vec{x}(t^{\prime\prime})|^2 + \beta_4(t)\; \left(\int _0^t  dt^{\prime\prime}\vec{x}(t^{\prime\prime}) \right)^2\right]\\+&o(x(t)^3)\,,\label{eq:eqapprox}
\end{split}
\end{equation}
where the $\alpha_i(t)$ and $\beta_j(t)$ are time dependent coefficients, respectively for the first and third degree. All other terms vanish for symmetry reasons. We need to stress that this expansion is only valid for three dimensions.\\
Defining:
\begin{equation}
\label{bracket}
\langle f (y)\rangle_t = \frac{\int d\vec{y} \; \exp(-t R(y)) \; f(y)}{\int d\vec{y} \; \exp(-t R(y))}\,,
\end{equation}
we can express the terms of the development as:
\begin{align}
\alpha_1(t)=&\frac16\left<\frac{(R^\prime(y))^2}R(y)-\left(R^{\prime\prime}(y)+2\frac{R^\prime(y)}{y}\right)\log\left(\frac{R(y)}{\left<R(y)\right>_t}\right)\right>_t
\\
\alpha_2(t)=&\frac13\left<\frac{(R^\prime(y))^2}R(y)\log\left(\frac{R(y)}{\left<R(y)\right>_t}\right)\right>_t
\\
\beta_1(t)=&\frac1{3!5}\left<\left(R^{\prime\prime\prime}(y)+2\frac{R^{\prime\prime}(y)}y+\frac{R^\prime(y)}{y^2}\right)\log\left(\frac{R(y)}{\left<R(y)\right>_t}\right)\right>_t
\\
\beta_2(t)=&\frac1{15}\left<(R^\prime(y))^2\left(R^{\prime\prime}(y)+\frac23\frac{R^\prime(y)}y\right)\log\left(\frac{R(y)}{\left<R(y)\right>_t}\right)\right>_t
\end{align}
\begin{align}
\begin{split}
\beta_3=&\frac1{30}\left<(R^\prime(y))^2\left(R^{\prime\prime}(y)+\frac23\frac{R^\prime(y)}y\right)\log\left(\frac{R(y)}{\left<R(y)\right>_t}\right)\right>_t\\
-&\frac1{18}\left<(R^\prime(y))^2\log\left(\frac{R(y)}{\left<R(y)\right>_t}\right)\right>_t\left<R^{\prime\prime}(y)+2\frac{R^\prime(y)}y\right>_t\\
-&\frac1{18}\left<(R^\prime(y))^2\right>_t\left<R(y)\left(R^{\prime\prime}(y)+2\frac{R^\prime(y)}y\right)\right>_t\\
+&\frac1{18}\frac{\left<R(y)\left(R^{\prime\prime}(y)+2\frac{R^\prime(y)}y\right)\right>_t^2}{\left<R(y)\right>_t}
\end{split}\\
\begin{split}
\beta_4=&\frac1{18}\frac{\left<R(y)
(R^\prime(y))^2\right>_t^2}{\left<R(y)\right>_t}
-\frac1{18}\left<(R^\prime(y))^2\right>_t\left<R(y)(R^\prime(y))^2\right>_t\\
-&\frac1{18}\left<(R^\prime(y))^2\right>_t\left<(R^\prime(y))^2\log\left(\frac{R(y)}{\left<R(y)\right>_t}\right)\right>_t+\frac1{30}\left<(R^\prime(y))^4\log\left(\frac{R(y)}{\left<R(y)\right>_t}\right)\right>_t
\end{split}
\end{align}
If one looks at the equation up to the first order, neglecting the $\beta$ terms, one can already explain the instability that leads to spirals.\\
Since $\alpha_1\simeq \frac {\sqrt 2}{3e}\frac{\log t}t>0$ and $\alpha_2\simeq -\frac{3\sqrt 3}{e^2}\frac{\log t}{t^2}<0$ for large t. $\alpha_1$ is positive so the trajectory starts as a straight line out of the origin, but then the term $\alpha_2$ which is unstable makes it unstable against local bending explaining planar spirals.\\
An analytic solution of this simplified equation is possible if one approximates the coefficients neglecting the logarithmic terms.\\
$\beta_3$ and $\beta_4$ are coefficients to terms that lie in the same plane as the first order ones. Because of this we will only concentrate on $\beta_1$ and $\beta_2$. Those are both positive and lead to the instability of the planar trajectory eventually leading to a full fledged three dimensional structure.
\subsection{Waiting time}
\begin{figure}[htbp]
\includegraphics[width=\textwidth]{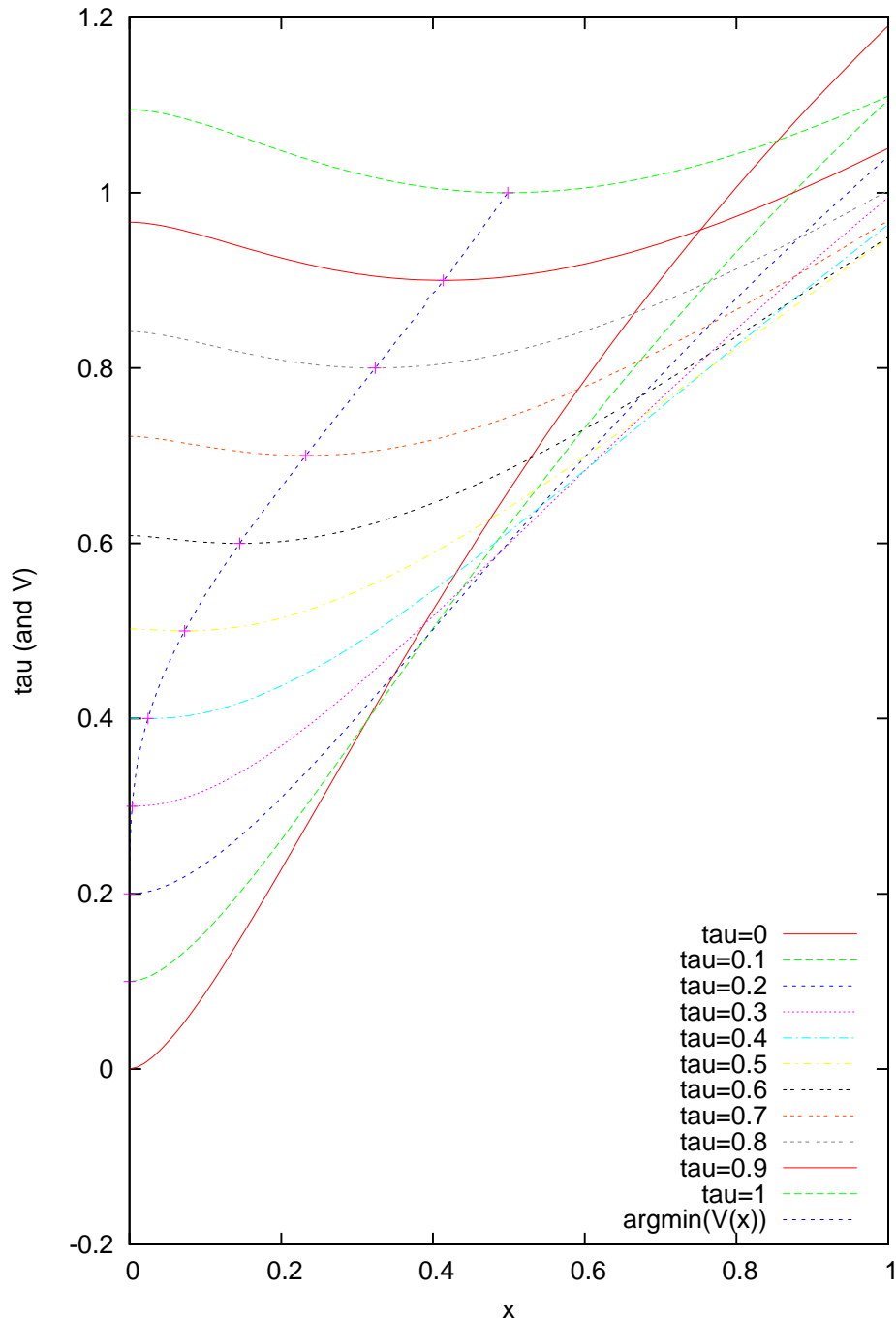}
\caption{Profile of $V(x)$ at varying $\tau$ and position of its minimum. We have subtracted arbitrary constants to the various $V(x)$ in order for the curve of the minima to pass through the minima.}\label{fig:pot}
\end{figure}

One interesting feature of the spirals is that they do not start immediately as in the discrete algorithm. This seems to be at odds with the results obtained in previous section: $\alpha_1$ is always positive, this means that staying in the origin without moving should be unstable.\\
How to reconcile this apparent paradox?\\
Let us define:
\begin{equation}
\tilde V_\tau(x)=-\left< R(y-x)\log\left(\frac{  R(y-x)}{\langle R(z-x)\rangle}\right)\right>_\tau\,,
\end{equation}
where we have sticked to the definition of the brackets of equation (\ref{bracket}) in the previous section.\\
If we plot $\tilde V_\tau(x)$ along a direction for different values of $\tau$ and we compute its minimum, as in figure \ref{fig:pot} we find indeed that there always a maximum in $x=0$, but there is also a non-trivial minimum for every $\tau>0$, albeit this minimum can be very close to the origin for small $\tau$.\\
The curve of the minumum $x_m(\tau)=\arg\min_x\tilde V_\tau(x)$ is well fit by as $x_m(\tau)\simeq6.62\exp(-2.32/\tau)$ in two dimensions.\\
These results can be further substantiated by convolving the $R$ with a Gaussian distribution of width $\sigma$, this way a $\sigma$-dependent crossover in $\tau$ can be shown to exist between waiting and moving. The interpretation of the Gaussian convolution is that, because of the numerical integration, when the searcher is waiting it effectively fluctuates around the origin.\\
All this can be summarized by saying that, if the noise is zero or stays within an acceptable range, the time it takes the searcher to move perceptibly out of the origin is $\simeq0.4$.\\
This is a very important feature of the continuous version of infotaxis which is not present in its discrete counterpart. This is due to the fact that setting a whole lattice site probability to zero creates a very strong repulsive effect, and since the area that is set to zero in the continuous version is infinitesimal there is no inhibition of this effect.\\
The striking feature of this effect is that it reproduces itself whenever there is a new hit: the searcher stops, waits about $0.4$ and then starts moving again. We can think of it as if it were trying to exclude that the source was in its immediate vicinity.\\
There exists a distance from the source when the expected arrival time of two successive hits is smaller than the waiting time, when this happens the searcher will be effectively stuck at this position. We will call this distance $d_\textrm{halt}$: it is dimension dependent. It is $\simeq0.1$ for D=2 and $\simeq0.3$ for D=3.\\
$d_\textrm{halt}$ is more rigorously defined as $0.4R(d_\textrm{halt})=1$. The reason for different values for different dimensions is the different form of $R$.
\section{Numerical integration}
In this chapter we will illustrate the techniques we have employed for numerically integrating the continuous infotaxis equations. We will devote some time to justifying the choice of a technique that increases the complexity of the algorithm in favor of precision.\\
At every time-step we have to compute the integral of the kernel over the probability measure in order to know the velocity of the searcher. The position of the searcher is then updated with a simple Euler integration step, that is:
\begin{equation}
x(t+\Delta t)=x(t)+\Delta t v(t)\,,
\end{equation}
where $v(t)$ is the velocity at time $t$ defined as $-\nabla_x V_t(x)/\gamma$.\\
We have found empirically that a good choice for the integration time-step $\Delta t=\gamma$, this choice ensures precision when $\gamma$ is small and then the searcher is fast and economy when $\gamma$ is big and the searcher is slow.\\
At each time step a Poisson pseudo-random variable is generated for the number of hits, this is recorded in a vector as is the whole trajectory.\\
The whole procedure can be summarized in pseudo-code as:
\begin{verbatim}
searcher=origin
source=d_0/sqrt(dimension)
i=0
while(d_success<distance(source,searcher)<d_fail){
   old_n_hits=n_hits;
   n_hits+=poissonrandom(dt*R(distance(source,searcher)))
   for(j=old_n_hits;j<n_hits;j++)
       hits[j]=searcher
   force=average(force,R,R_prime,x,trajectory,history,hits)
   x+=force*dt/gamma
   trajectory[i]=searcher
   i++
}
\end{verbatim}
An important detail that can't be omitted is the calculation of the averages over the probability distribution. The original discrete infotaxis implementation performed this by storing and updating the complete probability distribution over the lattice. This is clearly impossible in the absence of a lattice. Especially since the search is performed in unbounded Euclidean space.\\
We have, however, tried memorizing the probability distribution at points either on a non-square lattice or randomly picked in order to emulate the behavior of the original algorithm.\\
This approach is plagued by various serious shortcomings: first of all we need to choose the points at the beginning of the search, and it is natural to choose them concentrated around the starting position. After a certain time, however, the searcher will have moved farther away where the points are rarer and numerical precision will start suffering.\\
Another big problem is that the computation of integrals as sums over a set of point that does not change will effectively recreate a lattice, albeit not a regular one. The trajectories will stick to those \emph{lattice} points because visiting them directly is optimal for the information gain.\\
In order to avoid these artifacts, that crippled the simulation even for relatively short run times, we have decided not to store and update the whole probability distribution, but to store the trajectory and the hits and to calculate the probability distribution dynamically at each time-step. It is now possible to perform the integrals by Montecarlo importance sampling around the position of the searcher, and choose a different set of points at each time-step.\\
The procedure is as follows: one performs a change of variable for the  argument $u=|\vec x(t)-\vec y|$ of the functions to integrate. $u=\phi(v)=u_0(1-v)/v$, where $v\in(0,1]$, then angles are sampled uniformly in two or three dimensions.\\
$N_\textrm{MC}$ points are sampled this way (typically $10^4$) for each time step, and summed taking care of the Jacobian of the change of variables $v\mapsto u$.\\
Again in pseudo-code:
\begin{verbatim}
function average(functional,R,R_prime,x,trajectory,history,hits){
    sum=0
    for (i=0; i<MC_steps; i++) {
        y.angle=randomangle()
        y.radius=phi(randomreal())
        jacobian=phi_primep(inverse_phi(point(radius))
        hitscontrib=1
        for(j=0;j<size(hits);j++)
             hitscontrib*=R(distance(y,hits[j]))
         if(dimension==3) jacobian*=rs*rs
         else jacobian*=rs
         sum+=jacobian*priorprob(y)*exp(-history(R,y,trajectory))
             *functional(y,x,R,Rp)
    }
    return sum/MC_steps
}
\end{verbatim}
The only bit left is the computation of the integral over the trajectory at the exponential:
\begin{equation}
\int_0^td t^\prime R(y-x(t^\prime))\,.
\end{equation}
To compute this we have used the classic composite Simpson's rule:
\begin{equation}
\int_o^t f(t^\prime) dt^\prime\approx \frac{\Delta t}{3}\left[f(0)+2\sum_{j=1}^{n/2-1}f(2 j \Delta t)+4\sum_{j=1}^{n/2}f((2j-1)\Delta t))+f(t)\right]\,,
\end{equation}
where $n=t/\Delta t$ needs to be even.\\
Taking extra care to ensure $n$ is even, we get in pseudo-code:
\begin{verbatim}
function history(R,x,trajectory){
    sum=0
    if (size(trajectory)==1)
        return dt*R(distance(trajectory[0],x))
    if (size(trajectory)==2)
        return dt*(R(distance(trajectory[0],x))+R(distance(trajectory[1],x)))
    flag=size(trajectory)%2
    sum=R(distance(trajectory[flag],x))+R(distance(trajectory[size-1],x))
    for (i=1; i<=size/2-1; i++) 
        sum+=2*R(distance(trajectory[2*i+flag],x))
    for (i=1; i<=size/2; i++) 
        sum+=4*R(distance(trajectory[2*i-1+flag],x))
    sum*=dt/3;
    return sum;
}

\end{verbatim}
\section{Results and performances}
\subsection{Typical trajectories}
\begin{figure}[htbp]
\includegraphics[width=.5\textwidth]{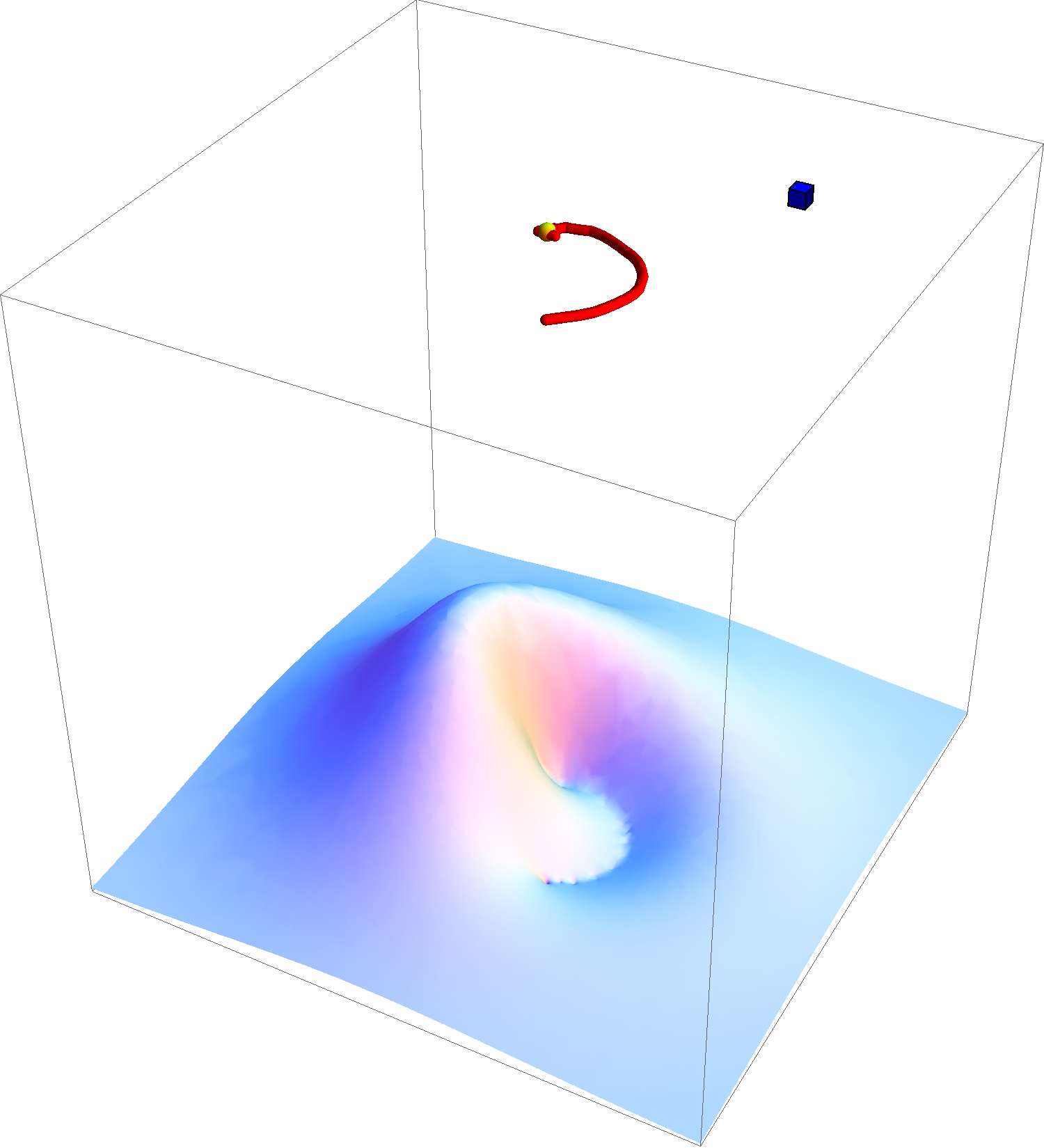}
\includegraphics[width=.5\textwidth]{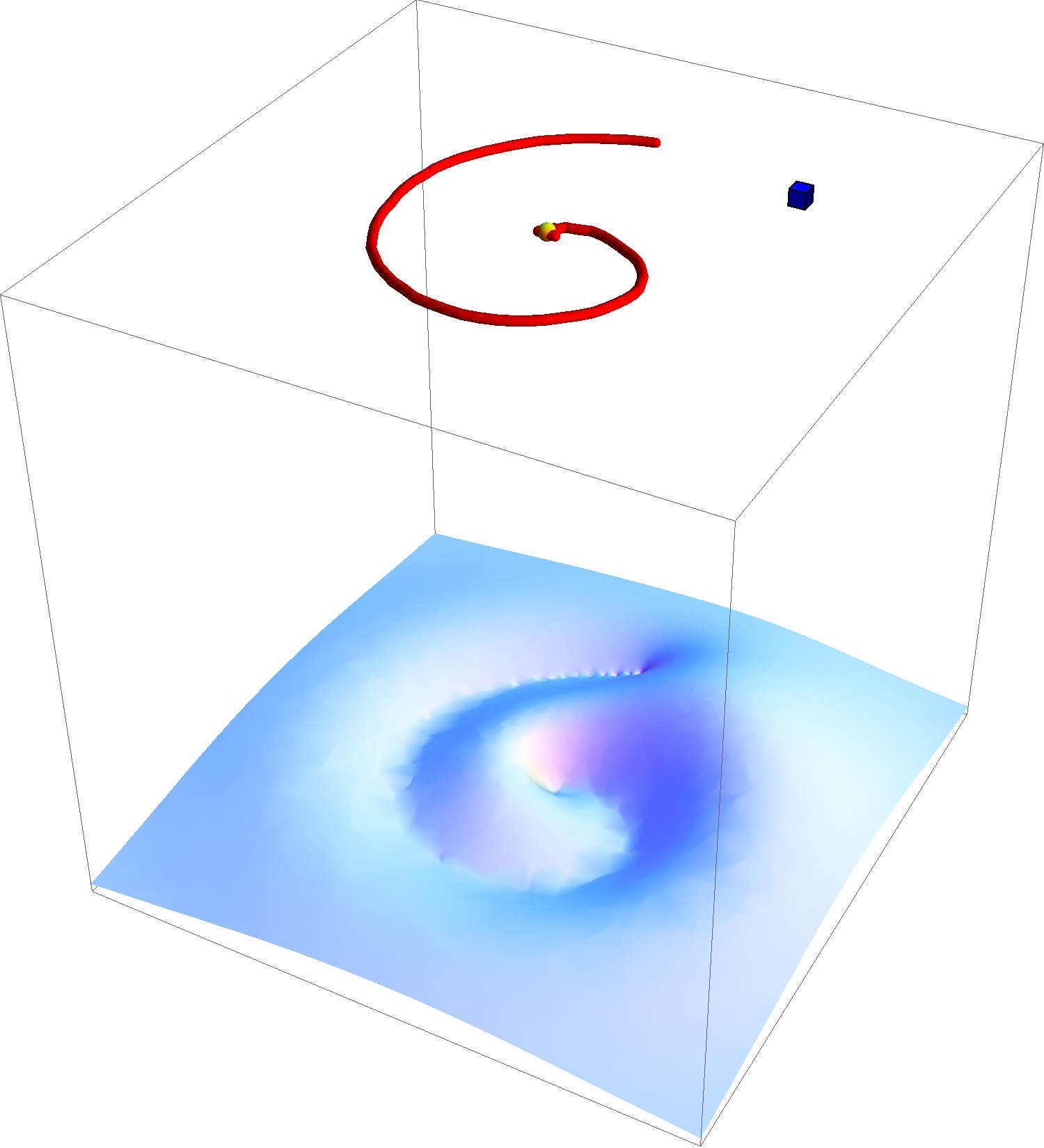}
\includegraphics[width=.5\textwidth]{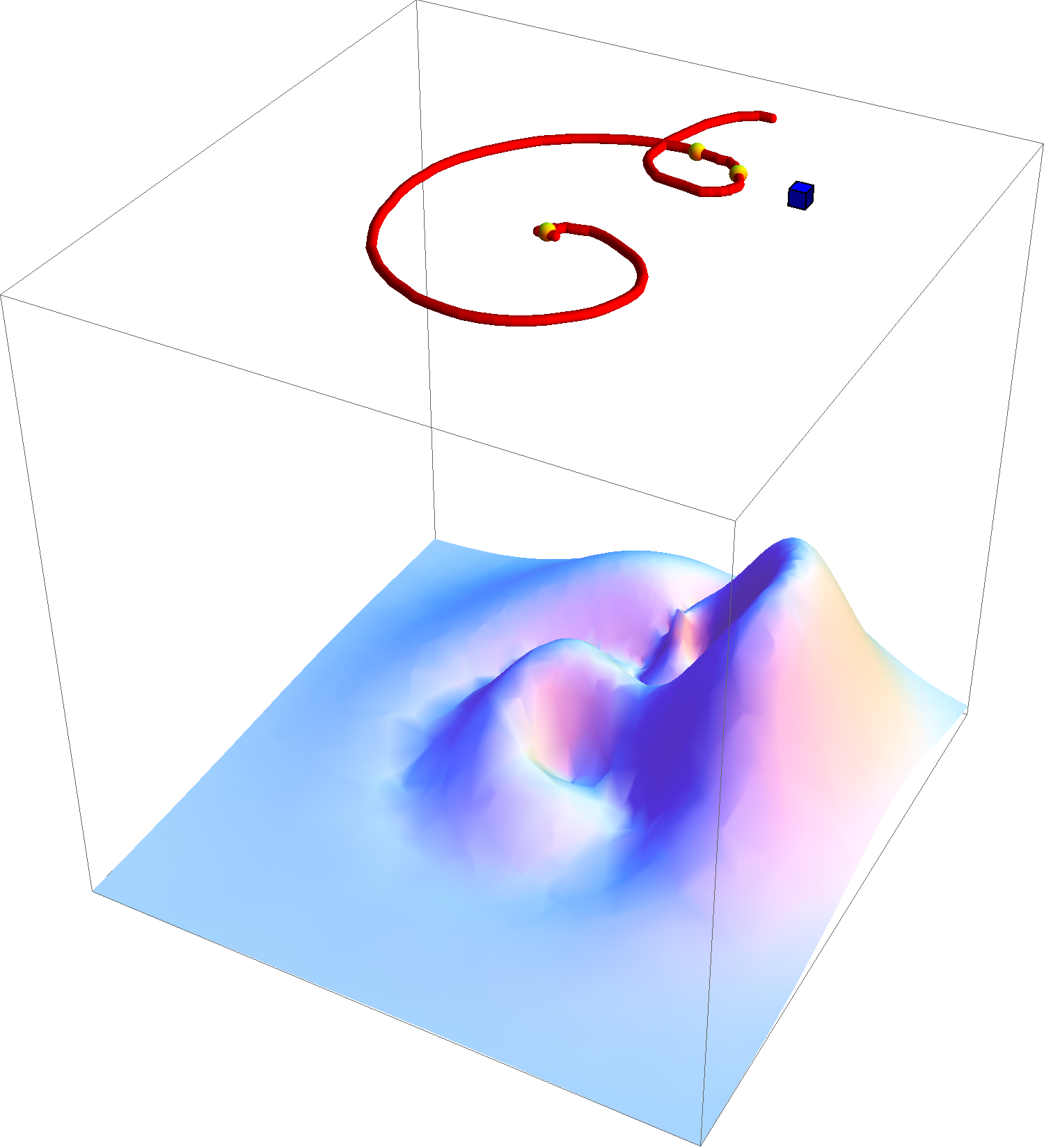}
\includegraphics[width=.5\textwidth]{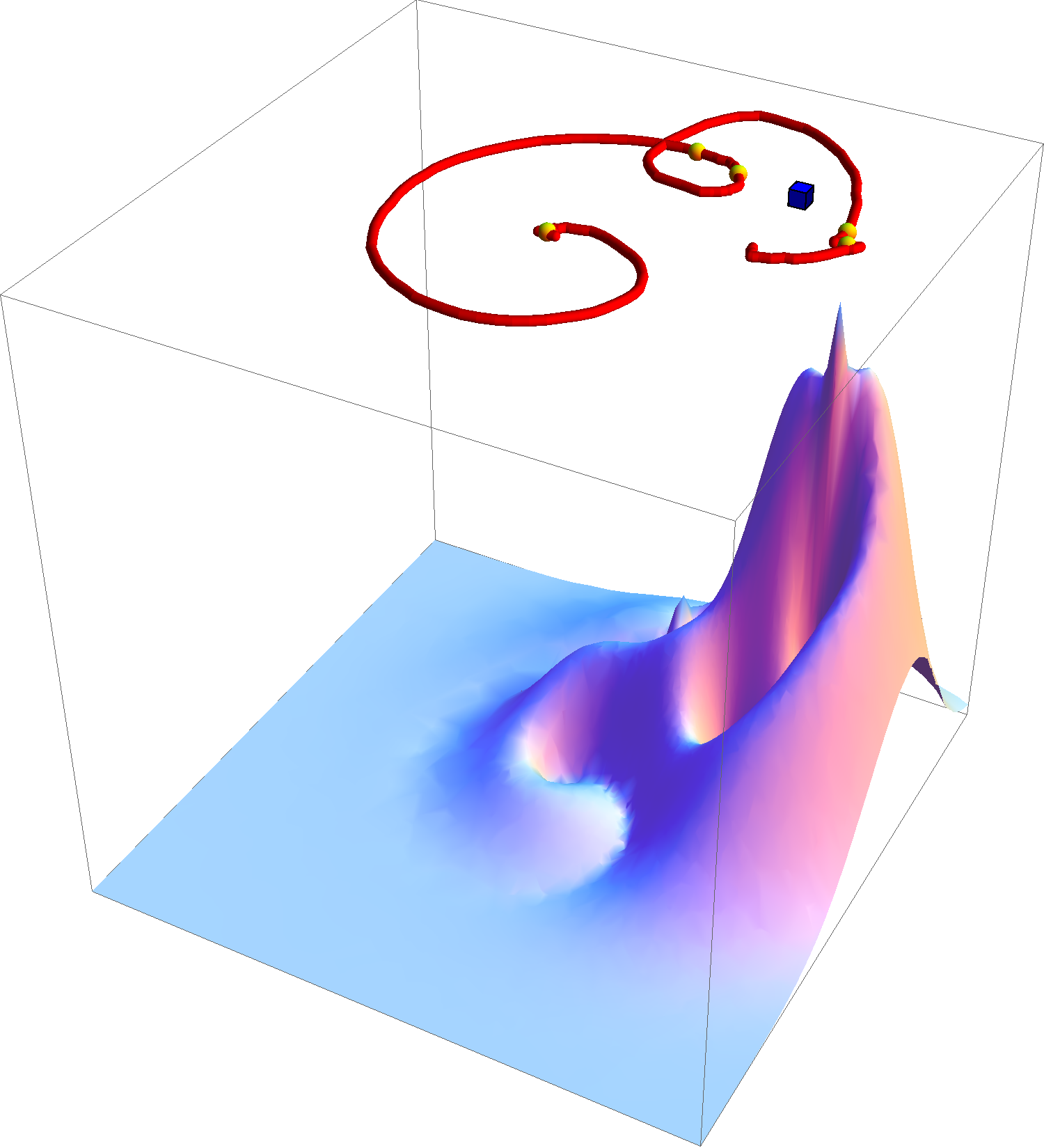}
\caption{The trajectory is plotted in red, superposed to the posterior probability distribution. Along the trajectory hits are displayed as yellow spheres. The source is the blue cube.}\label{fig:hits}
\end{figure}

\begin{figure}[htbp]
\includegraphics[width=\textwidth]{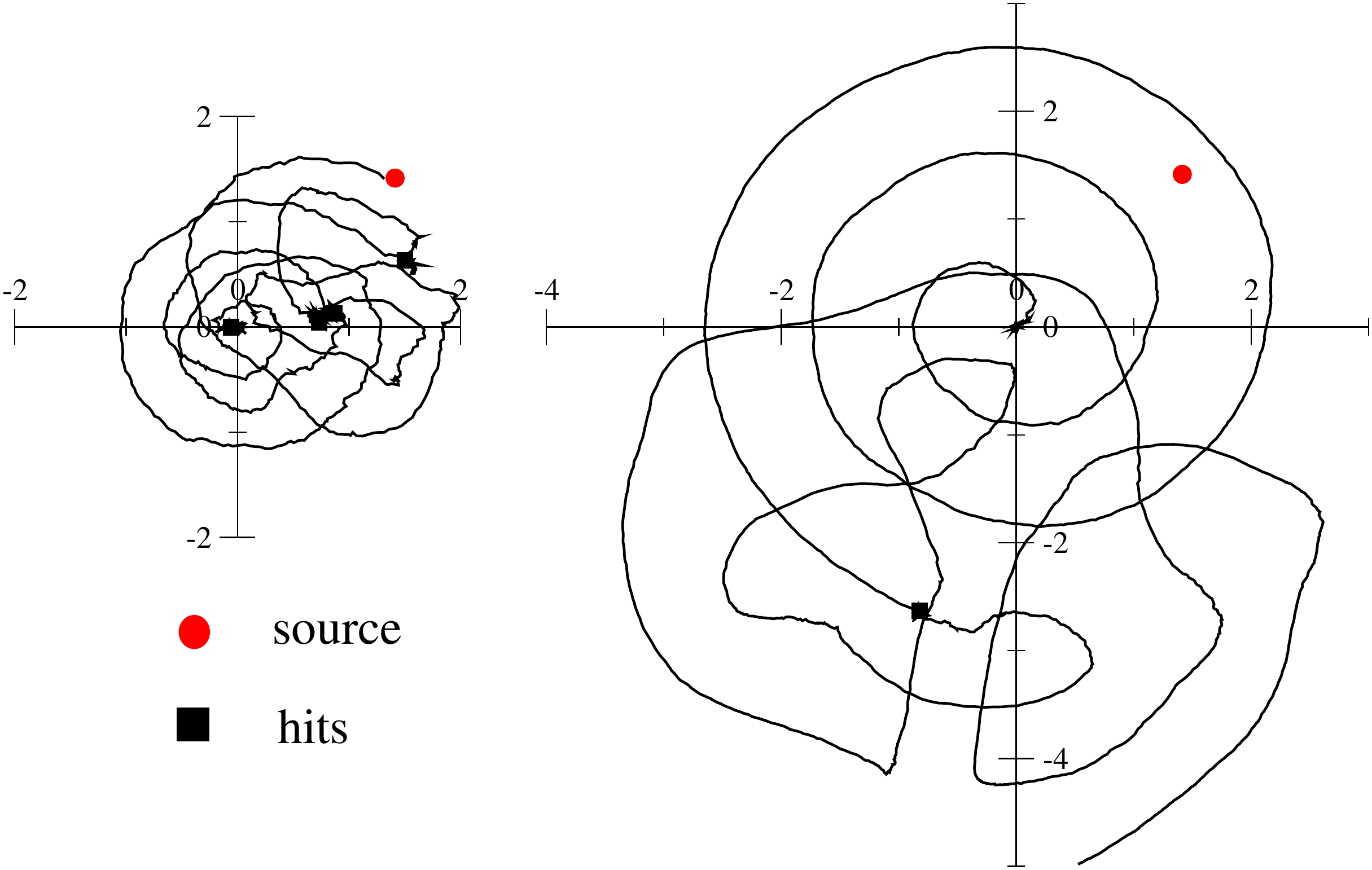}
\caption{Examples of search trajectories with hits two dimensions ($\gamma=.02$). The trajectory on the left finds the source, while the one  on the right is not successful. The initial distance to the source is $d_{0}=2$. The red disk represents points at distance $<d_{\text{halt}}$ to the source. Black squares locate the hits.}\label{fig:succunsucc2d}
\end{figure}

\begin{figure}[htbp]
\includegraphics[width=.5\textwidth]{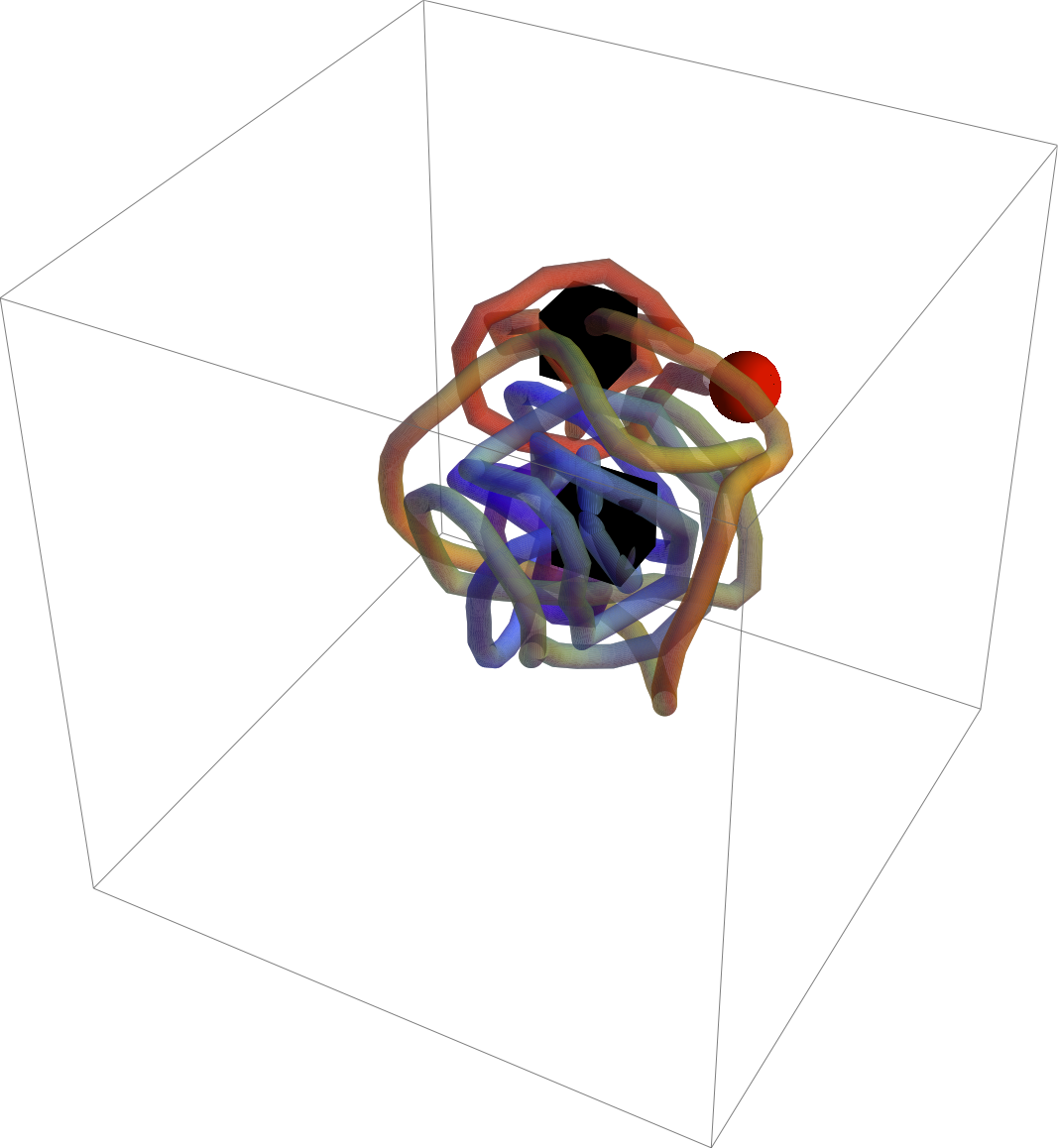}
\includegraphics[width=.5\textwidth]{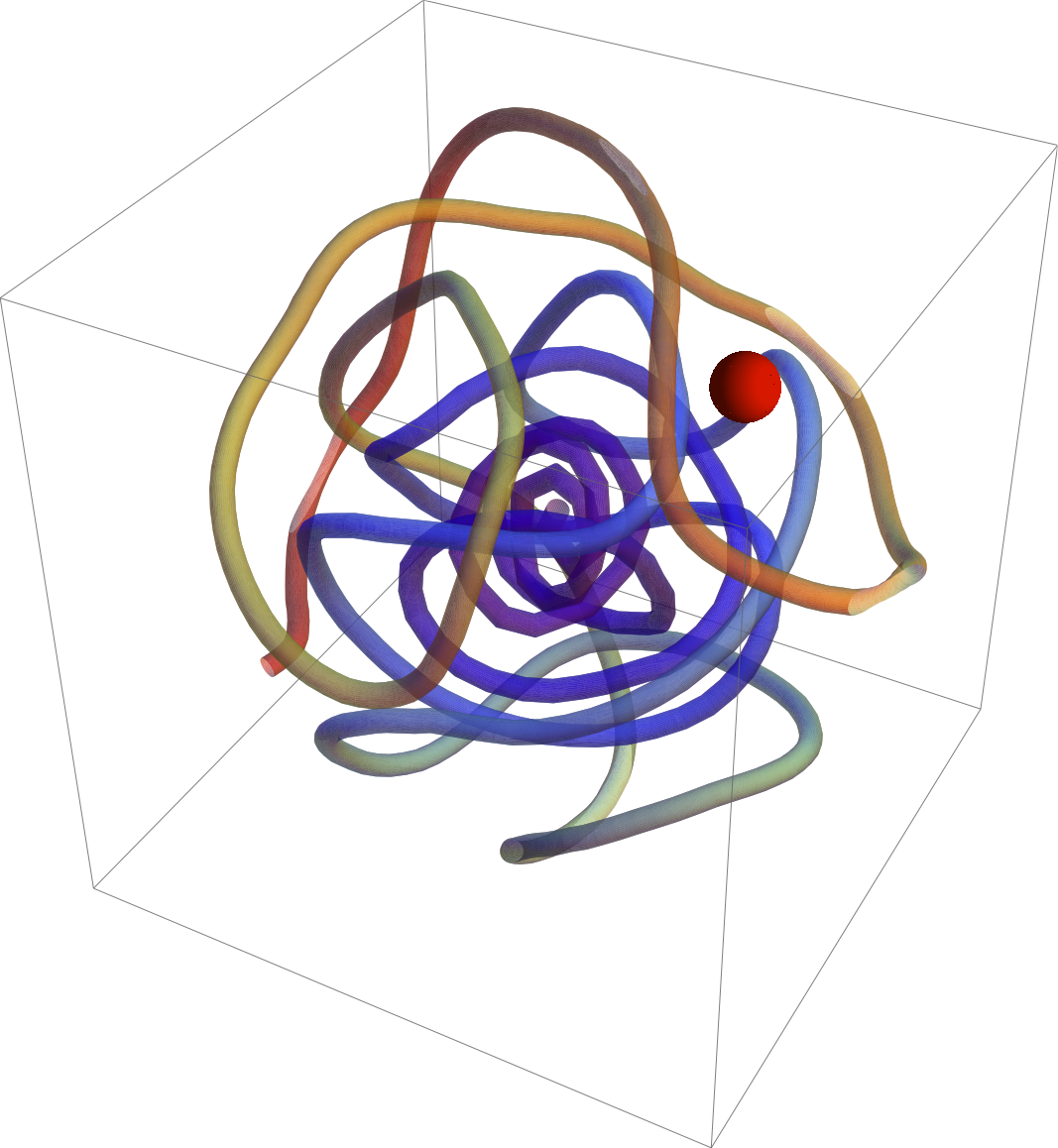}
\caption{Examples of search trajectories with hits three dimensions ($\gamma=.01$). The trajectory on the left finds the source, while the one  on the right is not successful. The initial distance to the source is $d_{0}=2$. The red sphere represents points at distance $<d_{\text{halt}}$ to the source. Black cubes locate the hits.}\label{fig:succunsucc3d}
\end{figure}

In this section we wish to show what the typical trajectories of continuous infotaxis look like in two and three dimensions once we have introduced a source of odor, as the goal for the searcher.\\
In two dimensions one can superimpose the trajectory to the probability and gain some good insights as to how the posterior probability is affected by odor hits.\\
In figure \ref{fig:hits} one can see the searcher starts its trajectory spiraling around its starting position, and how the probability distribution is affected by this: the maximum of the probability is always in front of the searcher, and a valley of minima is dug where it has passed.\\
In the third panel (bottom left) the first hit is received and the probability has a new maximum. If the searcher didn't receive further hits in the last panel (bottom right) it would start spiraling around the position of the new maximum.\\
In the last panel the  probability distribution is very peaked around the real position of the source, which is about to be found.\\
In figure \ref{fig:succunsucc2d} two trajectories are shown for two-dimensional infotaxis: the one on the left is successful in finding the source while the second is not.\\
Notice how the unsuccessful searcher has received a very misleading hit, actually farther away from the source than when it started. We can imagine the probability distribution to be peaked somewhere closer to the position of the hit. This maximum becomes the center of its new spiraling, albeit these new spirals are not as regular as the ones we have observed without hits.\\
Trajectories with hits are much harder to visualize, we try to do so in figure \ref{fig:succunsucc3d}, but the trajectory covers itself. What can be gleaned from these two trajectories is that the searcher seems to be using less information than in two-dimensional searches. In fact the unsuccessful searcher receives no hit at all while the successful one received only two.
\subsection{Average signal}
\begin{figure}[htbp]
\includegraphics[width=\textwidth]{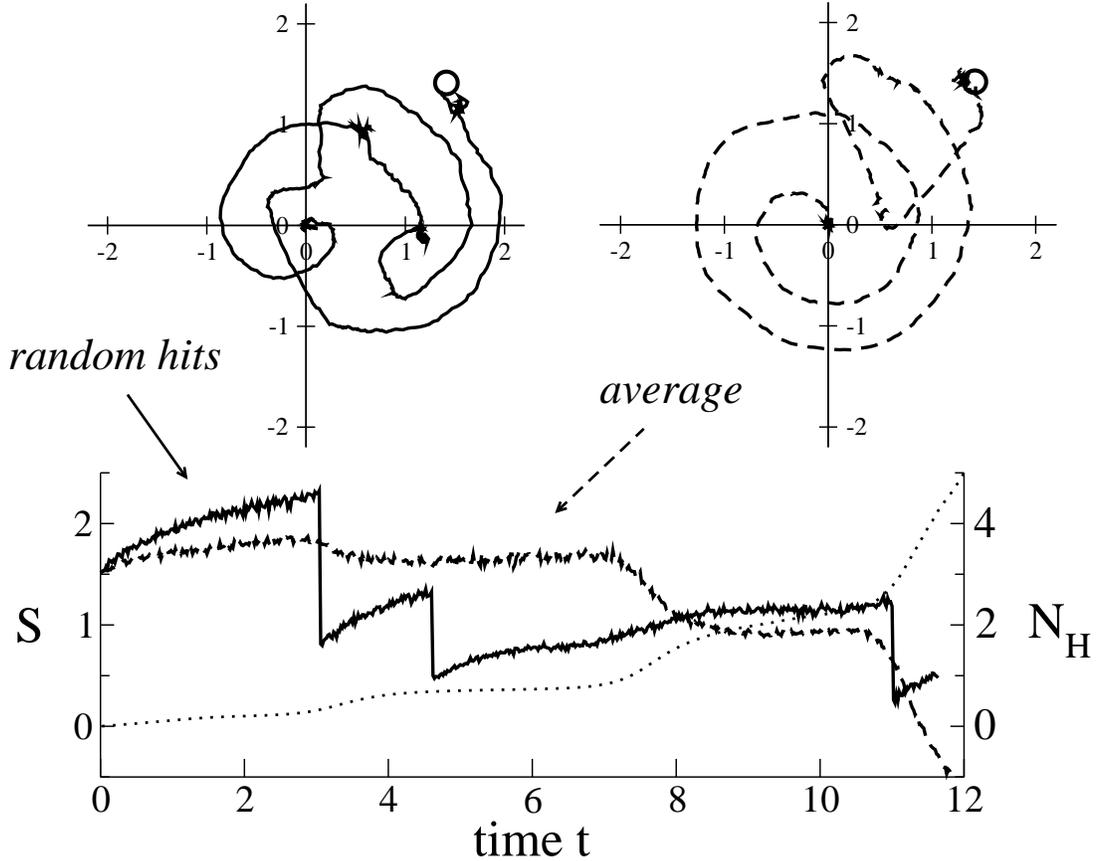}
\caption{Entropy $S(t)$ (bottom, left scale) for one trajectory ${\bf x}(t)$ obtained with random hits (top left, full curve, 3 hits are received) and the average trajectory (top right, dashed curve). The dotted line shows the average number of hits $N_{H}$ (right scale) received along the average trajectory. The source is located in $(\sqrt 2,\sqrt 2)$ (circle).}\label{fig:ave}
\end{figure}

We now wish to define what we think will be a very useful tool for the evaluation of performances: as we will see in the following, a large number of runs are needed in order to sample the probability of success and the time of success. This is due to the fact that the arrival times and positions of hits can vary wildly, and have a very strong influence on the searcher trajectory.\\
If one observes the posterior probability density, one notices that the hits are encoded as the product of $R$ functions centered at the position of each hit. As it is customary with multiplicative processes it is natural to look at the logarithm of the probability distribution.\\
\begin{equation}
\log P_t(y)=-\int_0^tdt^\prime R(y-x(t^\prime)) +\sum_{i=1}^H\log R(y-x(t_i))+\text{const}\,,\label{eq:logprob}
\end{equation}
where $t_i$ are the times at which the hits occur.\\
If the searcher is at time $t$ in position $t$ the probability it will get a hit in the next $\Delta t$ is given by $\Delta t R(y^*-x(t))$ where $y^*$ is the actual position of the source. Having observed this, we can take the expected value of equation (\ref{eq:logprob}) with respect to the probability of receiving a hit at each time-step.\\
This yields:
\begin{equation}
\overline{\log P_t(y)}=-\int_0^tdt^\prime\left[ R(y-x(t^\prime)) +R(y^*-x(t^\prime))\log R(y-x(t^\prime))\right]+\text{const}\\,.\label{eq:logprobave}
\end{equation}
If we now use the exponential of this newly defined quantity as the probability distribution that moves the searcher we obtain trajectories that have features that resemble closely those of trajectories with truly random hits.\\
However, even if we have reduced greatly the variability among trajectories, numerical trajectories obtained for this \emph{average} signal are not completely deterministic. This is due to the stochastic errors involved in Montecarlo integration and how those play an important role in the initial breaking of rotational symmetry.\\
In other words, the searcher starts in a random direction which defines the phase of the turnings of the spiral. This random direction is not a feature of the Poisson noise of the hits, but of the noise coming from Montecarlo integration. If we had access to a perfect integrator, we would need to add noise artificially at least at an initial stage to start the search.\\
In figure \ref{fig:ave} we compare a trajectory with random hits to a trajectory obtained with the \emph{average} signal when those have comparable duration. We also plot the entropy of the posterior distribution. Notice how it plunges in discontinuous jumps for the random signal and how it tapers off gently for the average signal.
\subsection{Performances}
\begin{figure}
\begin{center}
\includegraphics[width=.8\textwidth]{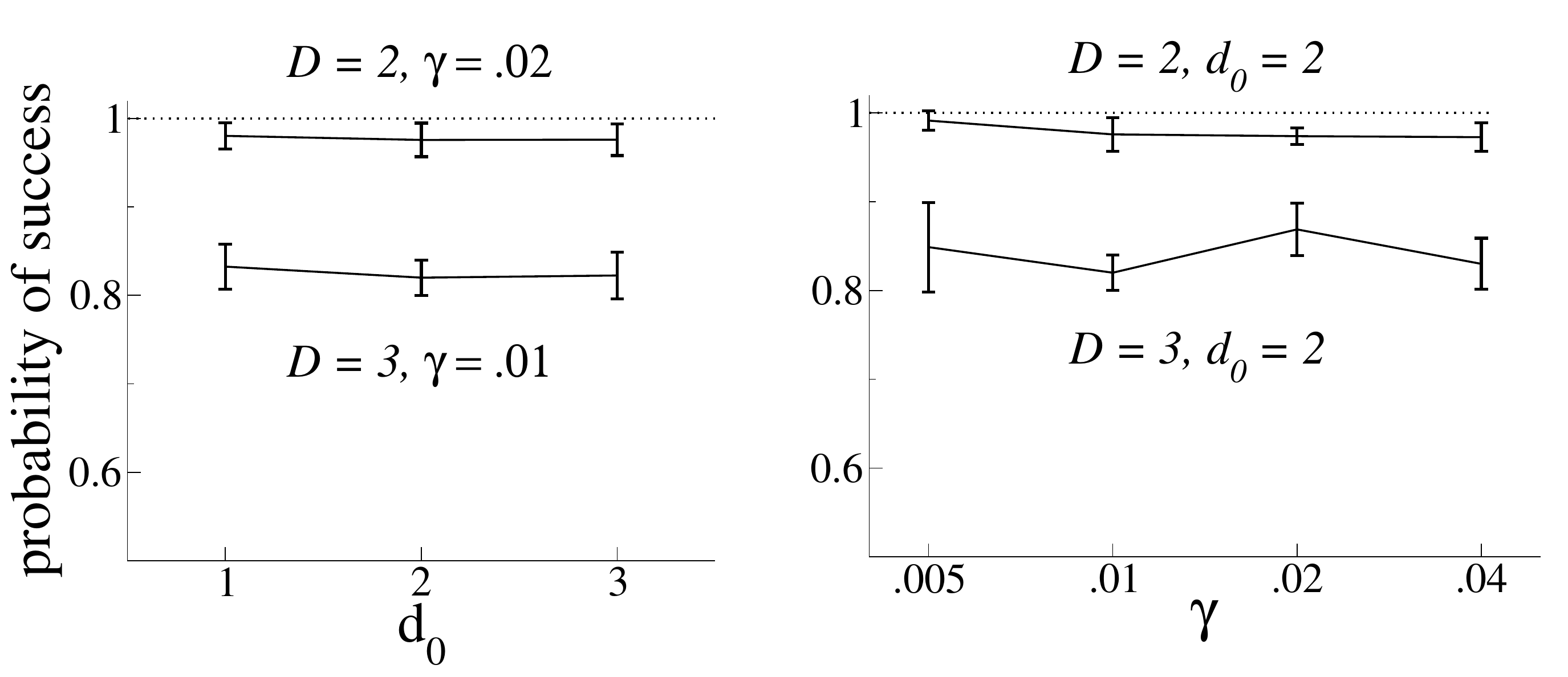}
%fig-proba
\caption{Probability of success of Infotaxis as a function of the initial distance to the source, $d_{0}$ (left), and of the friction $\gamma$ (right). Top points correspond to $D=2$ dimensions, bottom points to $D=3$. The numbers of runs is of about 200 for each point. All probabilities where obtained with $d_{\text{fail}}=8$.}
\label{fig:infoperform}
\end{center}
\end{figure}

\begin{figure}
\begin{center}
\includegraphics[width=.8\textwidth]{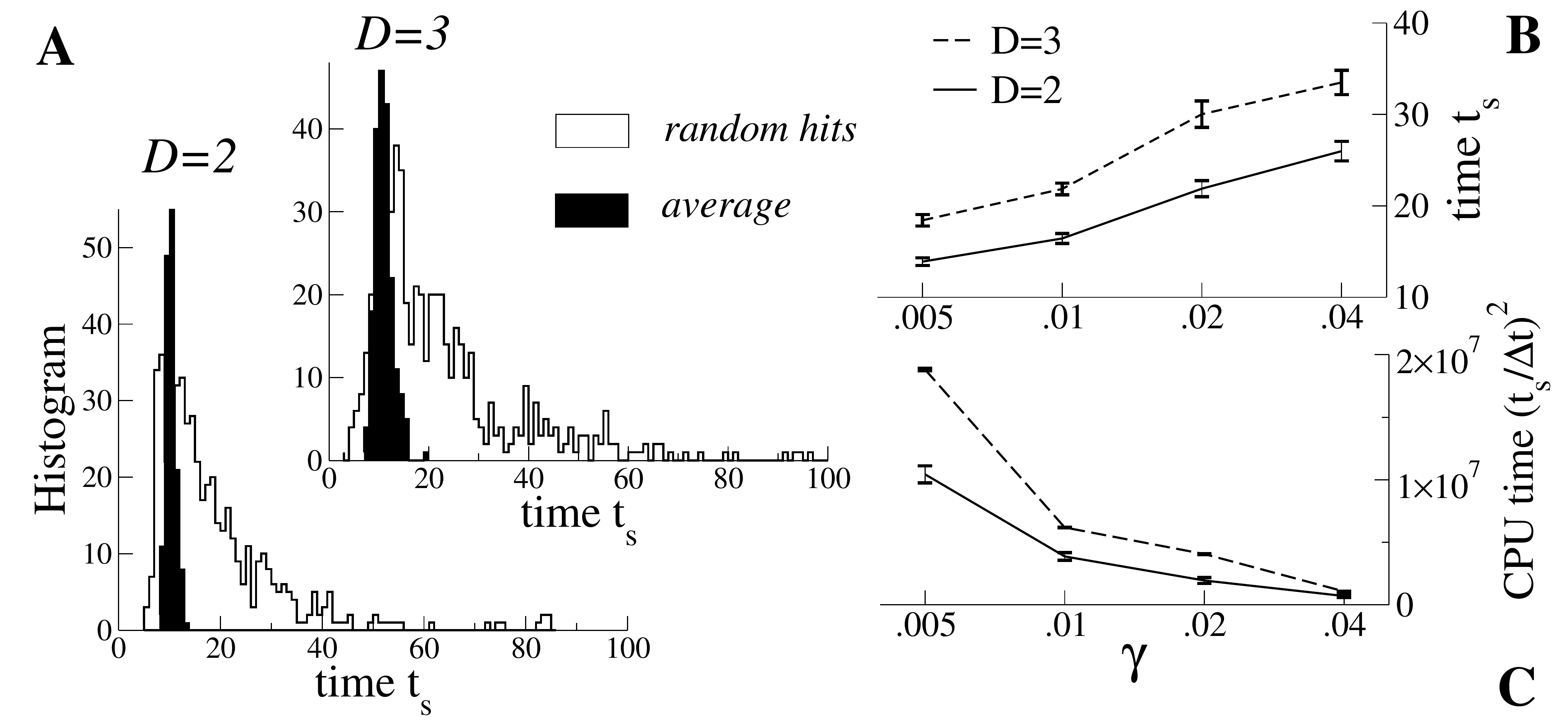}
% fig-histo
\caption{{\bf A.} histograms of the search times $t_\textrm{s}$ in $D=2$ ($\gamma=.02$) and $D=3$ ($\gamma=.01$) dimensions for an initial distance $d_{0}=2$ to the source. Full histograms correspond to the average trajectories, contour histograms to trajectories with random hits. {\bf B.} Average search time $t_\textrm{s}$ as a function of $\gamma$. {\bf C.} total CPU time as a function of $\gamma$, calculated as $(t_s/\Delta t)^2$. }
\label{fig:infotemp}
\end{center}
\end{figure}

In order to evaluate the performance of the algorithm we have to look at the success probability and the time needed to reach the source of odor in case of a success. But first of all we have to give a clear definition of success and failure. This is at odds with the discrete algorithm, where success was obtained when the searcher and the source were at the same position and failure when the searcher wandered out of the lattice.\\
In a continuous, unbounded space these definitions do not apply. However we can define a radius $d_\textrm{fail}\gg1$ from the source that defines the region of space out of which the search has not much hope of ever succeeding. The bigger  $d_\textrm{fail}$, the less our results will depend on it.\\
The definition of a $d_\textrm{success}$ is a bit more delicate since too small a radius would have catastrophic effects because of the pinning phenomenon we have described in the previous section; too big a radius would mean getting a lot of false positives and overestimating the performance of the algorithm. In the end we settled for $d_\textrm{success}=d_\textrm{halt}$.\\
There are two parameters that need to be varied in order to evaluate performance: one is the distance from the source, the other is $\gamma$ that characterizes the dynamics.\\
Another delicate issue is the definition of time: since our algorithm has a complexity per time-step which is linear in the elapsed time, CPU time will not be proportional to simulation time and we would need to optimize one or the other in different scenarios.\\
We have investigated the success probability for different values of the initial distance between the searcher and the source.\\
We have chosen distances between 1 and 3 in units of $\lambda$, because, on one hand, larger initial distances would correspond to vanishing an exponentially vanishing probability of receiving one hit and would only lengthen the spirals without showing any interesting feature of the algorithm.\\
On the other hand distances smaller than 1 are too close with the halting distance especially in three dimensions. Because of these two arguments we believe this is the only region where the behavior of this algorithm might be non-trivial.\\
Another important parameter is the friction coefficient $\gamma$. Overall we have observed that the success probability is affected by neither the starting distance or the friction coefficient. It is compatible with unity in two dimensions and slightly higher than $80\%$ in three dimensions. The results are detailed in figure \ref{fig:infoperform}.\\
This does not surprise us much: searches are easier in two dimensions, where random walks are space filling. The result in three dimensions looks promising and it is much better than any random estimate. The interested reader can refer to the classic reference by Redner \cite{REDNER} for a computation of the probabilities for the associated random phenomena.\\
Let us now define the relevant quantities for the search time: first of all we will restrict ourselves to the successful cases. We define the success time $t_\textrm{s}$ as the time when the algorithm halts because the searcher has entered the disk of radius $d_\textrm{success}$.\\
The CPU time can be defined in a implementation-agnostic form as $(t_\textrm{s}/\Delta t)^2$ since it will be generally proportional to this quantity. It should be noted that in the current implementation, with $10^4$ Monte Carlo sampling points in spatial integrations and on a 2.4 GHz core of an Intel Core 2, $A\simeq 3$ ms.\\
In figure \ref{fig:infotemp} we show, with the $t_\textrm{s}$'s and the CPU times for different $\gamma$'s, an histogram comparing the results obtained with the average equation of the previous section with those obtained with the non-simplified equation.\\
It is interesting to note how if one takes into account only the $t_\textrm{s}$ the algorithm is most efficient at low $\gamma$, however, since lower $\gamma$ call for lower $\Delta t$ in CPU time the algorithm is much faster for high $\gamma$.\\
This can be explained by remembering the dependence of the spiral spacing on $\gamma$: low $\gamma$ means tighter spirals and a searcher that moves much faster linearly: while this behavior turns out to be more effective at exploring the space it is more computationally intensive because the increased scalar velocity calls for a smaller time-step.\\
Overall we think performance can be greatly increased either by reducing the number of Monte Carlo integration points or by reducing the number of points in the time integral.\\
A reduction of the time points stored in memory can be obtained in two ways: the first is to add a finite memory, but if one is not careful one could end up with the searcher very strongly attracted back to the origin after a certain time, because the divergence of the prior is not attenuated by the trajectory anymore.\\
A smarter option would be to add some sort of coarse graining in time: points become much rarer in the distant past, but they have an increasing weight in the discrete sum at the exponent in the posterior probability. We would probably lose some precision this way, but we could recover a linear-time algorithm.\\

\part{DNA unzipping and sequencing}
\chapter[Current sequencing technologies]{Review of current sequencing technologies and their limitations}
In this part we wish to show how micromanipulation experiments on DNA molecules could be exploited to give us better sequencing techniques.\\
In this chapter we will describe current sequencing technologies, then underline what are their current limitation and what is to be gained from single-molecule sequencing. This will be the basis and motivation for our further work.\\
Modern DNA sequencing was developed in the second half of the seventies by Sanger et al. \cite{SANG75,SANG77}, a few other methods were tried in the first part of the decade \cite{MAXGIL}, but since they do not have modern day equivalents we will not discuss them here
\section{Chain-termination method}
\begin{figure}[htbp]
\begin{center}
\includegraphics[width=.65\textwidth,angle=-90]{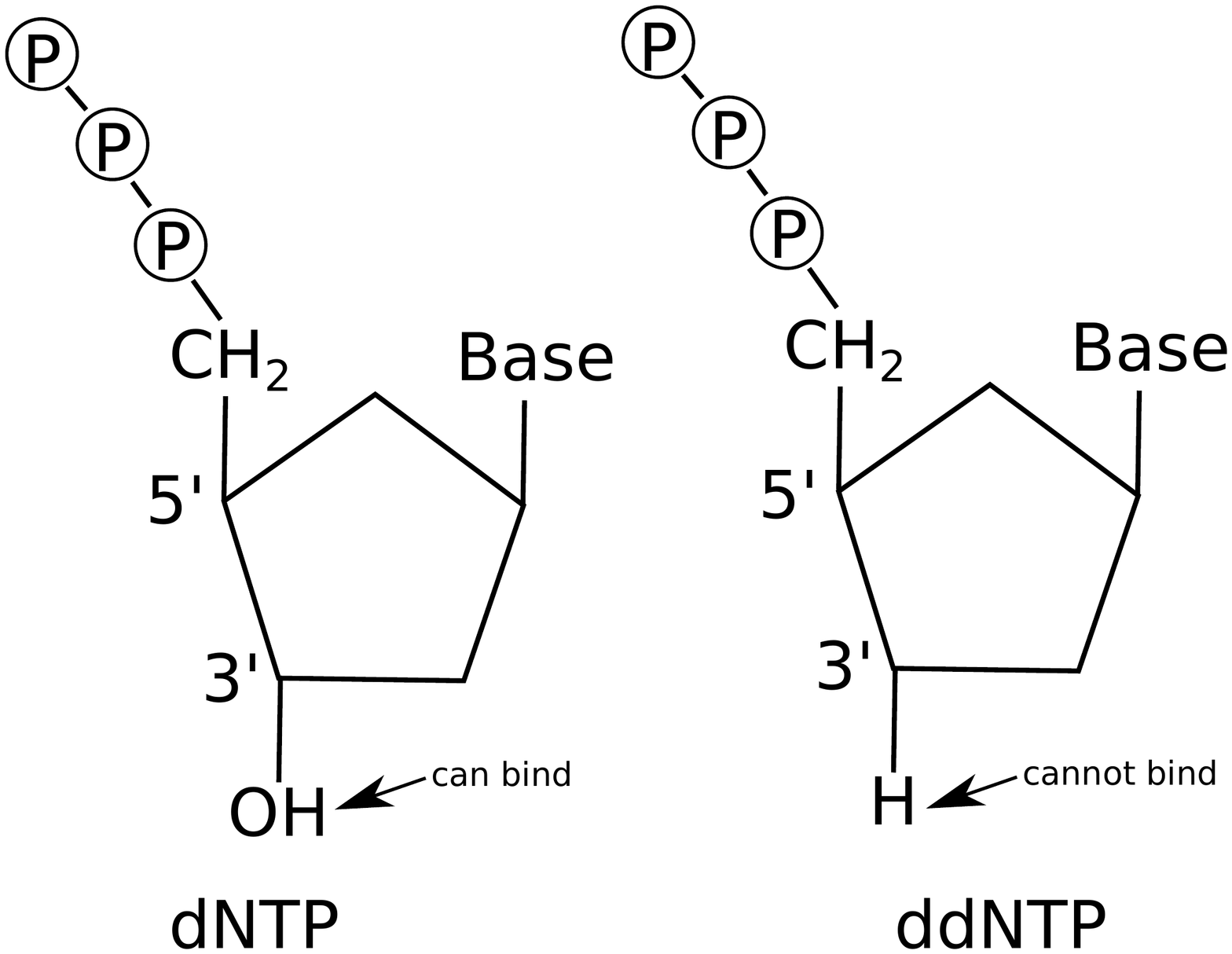}
\caption{Right: a normal Nucleotide TriPhosphate where the sugar is a 2$^\prime$-deoxyribosine. Left: a NTP where the sugar is a 2$^\prime$,3$^\prime$-dideoxyribosine. The absence of the hydroxide on the 3$^\prime$ carbon atom means it no further nucleotide can link to it.}\label{fig:ddNTP}
\end{center}
\end{figure}

The method developed by Sanger is based on the properties of dideoxynucleotides (ddNTPs): these are modified nucleotides: where normal nucleotides would be deoxynucleotides (dNTPs) these lack the 3$^\prime$ hydroxyl group on their dexyribose sugar (see figure \ref{fig:ddNTP}), this means that once they are added to a growing strand of DNA, no further nucleotide can be added because they lack the ability to bind with it \cite{ATKI69}.\\
\begin{figure}[htbp]
\begin{center}
\includegraphics[width=.3\textwidth]{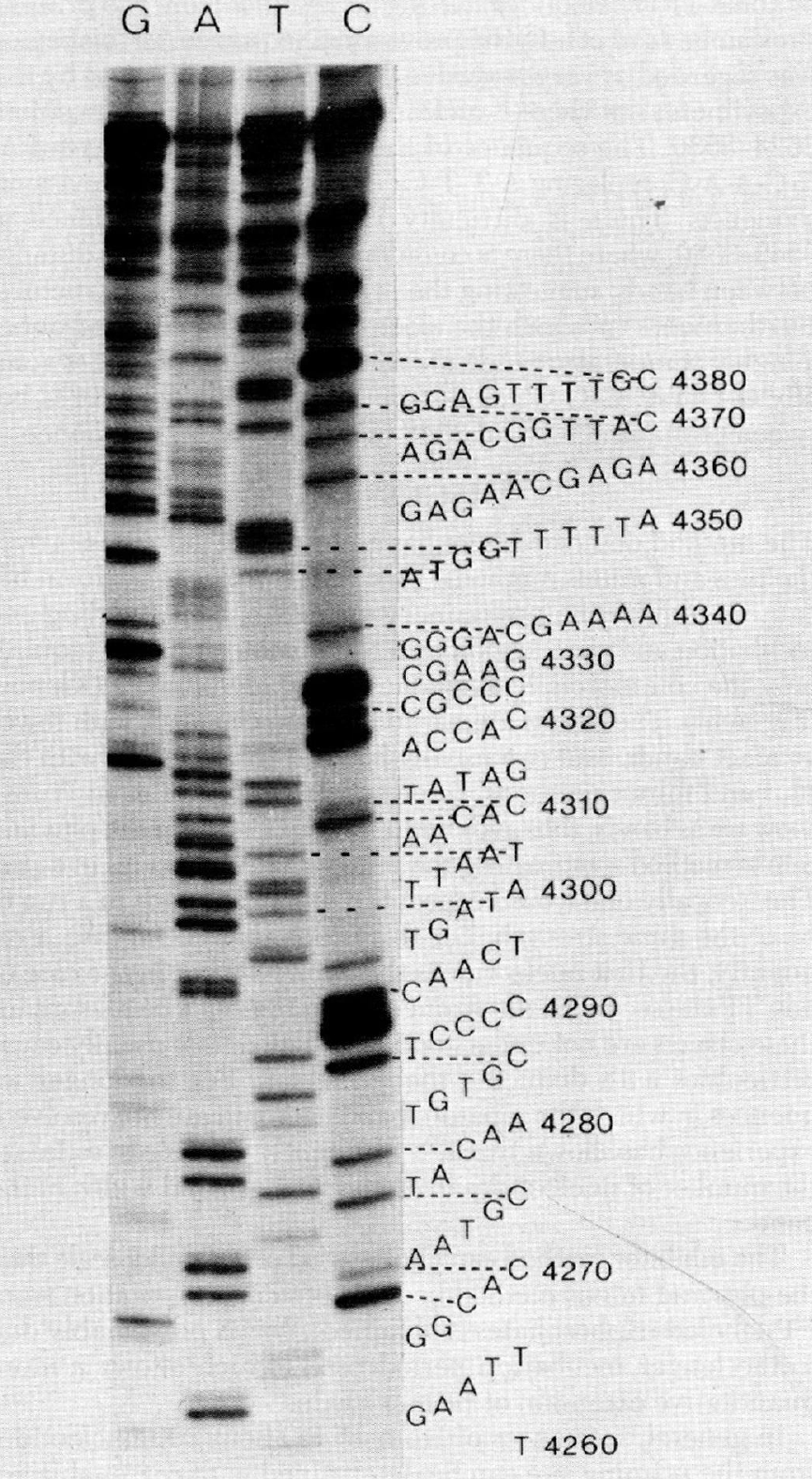}
\caption{One of the figures of Sanger's seminal paper \cite{SANG77} showing an autoradiograph of the four lanes of chain-termination sequencing and how they are used for sequencing.}\label{fig:autoradio}
\end{center}
\end{figure}In order to be sequenced DNA needs to be single-stranded and in multiple copies each of which has a primer attached to the same point. The  copies are then separated in four reactions all of which contain DNA polymerase and all four of the dNTP and only one of the ddNTP in a lower concentration.\\
The DNA polymerase facilitates the binding of the dNTP on the complementary bases, but once in a while a ddNTP will bind to the chain halting the process. At the end of the process we are left with different pieces of DNA all starting at the same point (where the primer was bound) and ending at random points, with the constraint that all the pieces in the reaction that contained only ddATP end at a T basis, all those in the ddCTP reaction end at a G basis and so on and so forth.\\
Now the molecules can be sorted according to their size with gel electrophoresis and photographed on four different lanes (one for each of the basis), a black line will appear in correspondence to each base.\\
Several variation to this technique exist: the ddNTP can be dyed in order for them to fluoresce or tagged with a radioactive substance, but the essential mechanism stays the same.\\
The main problem with this kind of method is that the quality of the sequencing traces degrade after about 1000 bp. This is due to several factors: the first and most important is the nature of the random process involved in the binding of ddNTP. Suppose we are in the ddATP solution and the next base is a T, then the probability $p_{\textrm{dd}}$ of the ddATP binding instead of the dATP binding does not depend on the length of the sequence. On the other hand the probability of still finding a sequence of a certain length after having encountered $n$ T's is $(1-p_{\textrm{dd}})^n$ and thus decreases exponentially.\\
Another source of accumulating errors is the presence of two or more basis of the same kind next to each other, that is to say it is difficult to distinguish four C's in a row from five C's. This type of errors will crop up, making the alignment of the four different lanes difficult.\\
\section{Pyrosequencing}
Another very popular sequencing technique which is behind some current day automated sequencing methods is pyrosequencing. Developed by Ronaghi and Nyrém in the nineties \cite{PYROANA,PYROSCI}, pyrosequencing relies on detecting the activity of DNA polymerase through the use of a chemiluminescent enzyme that will emit light whenever a new bond is formed.\\
A single strand of DNA reacts with DNA polymerase, a chemiluminescent enzyme and solutions of one of the four nucleotides, which are sequentially added and removed. When a nucleotide binds to the next available spot, light is emitted and we know which base has bound because only one type of nucleotide was in solution at that moment.\\
Pyrosequencing is inherently limited to sequences of about 500 bp (more typically less than 100 bp), but it is well suited to being automated and massively parallelized. Because of the limitations in the size of the the fragments it has been  rarely used for \emph{de novo} sequencing, instead it is either used in conjunction with other methods, or for resequencing and for the search for single nucleotide polymorphisms (SNP). Only recently read lengths of about 1000 bp have been attained by a company called 454. This will allow for \emph{de novo} sequencing using pyrosequencing.\\
\section{Sequencing by ligation}
Ligation is the joining of two double stranded DNA segments through the formation of two covalent bonds. This reaction involves an enzyme called DNA ligase. The difference between DNA ligase and DNA polymerase is that DNA polymerase needs one of the two strands to be intact while DNA ligase can repair double stranded DNA.\\
DNA ligase can also be used to join a single strand of DNA to an otherwise intact single strand, but in this case it is very sensitive to mismatches, that is it will hardly ever join two strands which are not complementary.\\
Several techniques are based on this specificity, namely ligase chain reaction (LCR) \cite{LCR1,LCR2} and ligase amplification reaction (LAR) \cite{LAR}, we will not dwell here on the details, it suffices to know that these rely on oligonucleotides (short pieces of ssDNA, here typically 8-9 bases long) and their ligation to a the DNA that is being sequenced.\\
A number of different oligonucleotides is added to the solution where the anchor sequence is. Then the ligase will hybridize two of the bases of the oligonucleotide to the anchor sequence and emit a light signature that allow the two bases to be recognized.\\
Sequences are then reconstructed using two-base encoding, a technique that relies on these superposed two-base reads. Read lengths of up to 25-50 bases have been achieved \cite{LIGATION}.
\section{Limitations}
\begin{figure}[htb]
\begin{center}
\includegraphics[width=.7\textwidth]{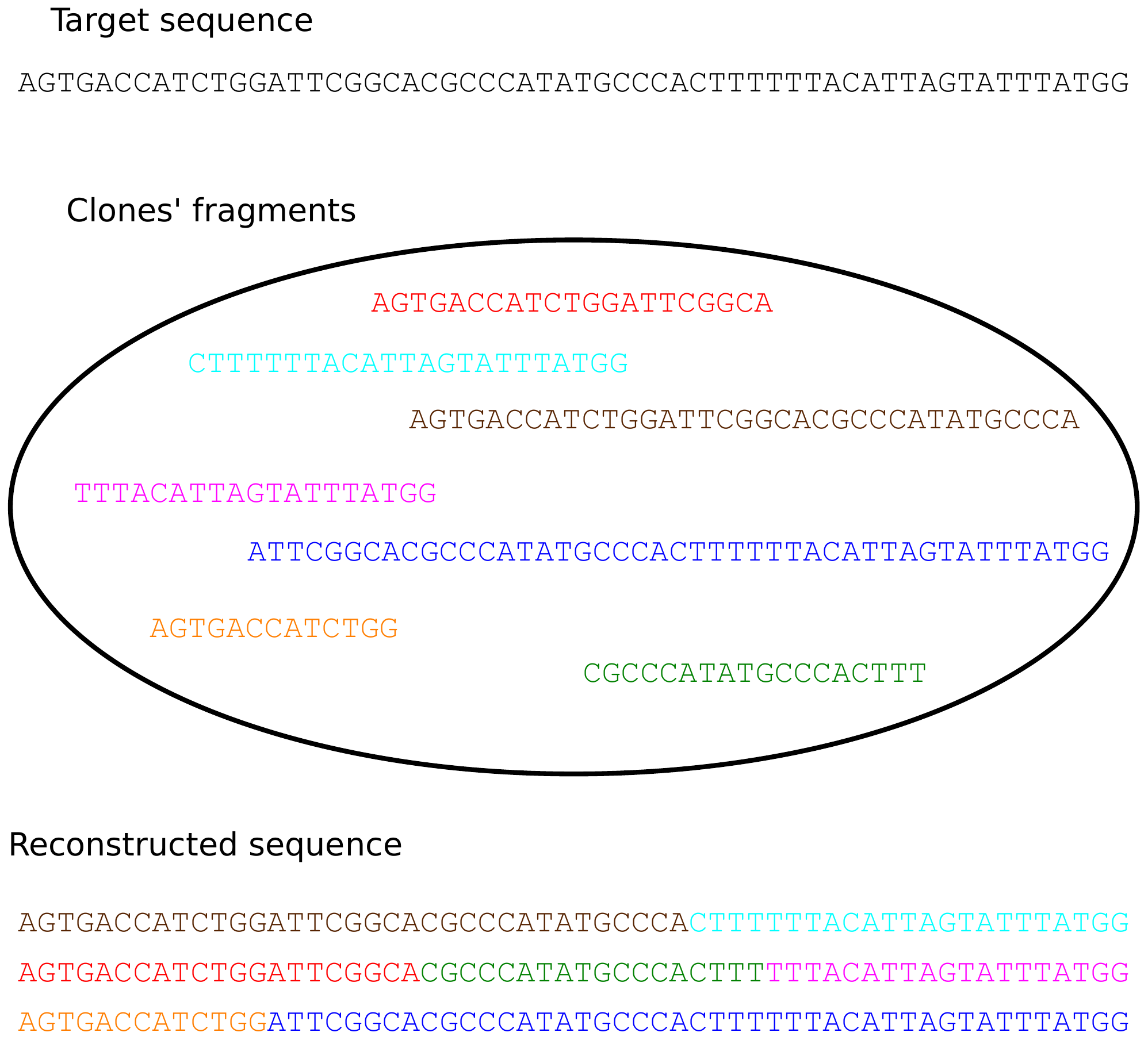}
\caption{Shotgun sequencing: the target sequence is cloned several times and cut at random points. The smaller segments are then sequenced and the sequence is reconstructed thanks to the overlaps.}\label{fig:shotgun}
\end{center}
\end{figure}

As you might have noticed, all of the techniques outlined up to here rely on read lengths of at most 1000 bp, while whole chromosomes and genomes have lengths that exceed this by several orders of magnitude. In order to fill this gap, DNA has to be spliced and amplified to be sequenced. Amplification is usually done through a technique called polymerase chain reaction (PCR) \cite{PCRPROC,PCRBOOK}.\\
DNA can be cut in an ordered way starting from one end and then cutting regularly. This technique is called chromosome walking and it is the best method for sequences which are too long to be sequenced in a go, but still under 10000 bp. The shorter fragments are then sequenced leaving 20 or so superposing bases on each fragment to allow for reconstruction.\\
Longer sequences as whole chromosomes or genomes are usually dealt with a technique developed at the end of the seventies called shotgun sequencing \cite{SHOTGUN}.\\
The name derives from a metaphor: as a shotgun fires a large array of small projectiles in a random pattern, DNA is cut in random points into smaller sequences. The process is repeated multiple times as to have several copies of the same sequence cut in different points. The spliced sequences can then be sequenced one at a time and then recomposed through the use of algorithms that rely on the overlapping between different copies (see figure \ref{fig:shotgun}).\\
Short reads are fine when we are looking for short mutations such as SNPs or anything shorter than the length of the typical read, but genomes are replete with mutations that are much larger in size such as copy number variations (CNV).\\
Copy number variations are mutations that involve the deletion or the duplication of a section of DNA, they have lengths of at least 1 kbp and up to several hundred kbp and are very common throughout the human genome \cite{CNVSCI,CNVNAT}.\\
Copy number variations seem to play a central role in cancer \cite{CNVCANCER}, autism \cite{CNVAUT} and in neurological conditions \cite{CNVNEU1,CNVNEU2,CNVNEU3}. CNV are very hard to find with current sequencing methods, because reconstruction algorithms tend to miss them. The only way to effectively indentify them is to use classic sequencing techniques in conjunction with microarrays for the detection of SNPs and very complex algorithms \cite{CNVMETH}.\\
This is one of the main reasons for developing single molecule techniques for sequencing DNA, but current efforts are not very promising: zero-mode waveguide \cite{ZMWG} seems to be the most advanced but it still offers read lengths of about 1500 bp, that is comparable with chain termination techniques. It is a technique based on holes which are small ($\sim100$ nm) in all of their dimensions compared to the frequency of light used for the observation. Their optical properties allow the observation of the enzymatic activity of a single molecule.\\
On the other hand techniques based on nanopores look promising \cite{NANOPORES}. Nanopores are holes with a diameter of $\sim1$ nm, similar to some proteins found on cellular membranes. DNA can be forced through the nanopore one base at a time. Since each nucleotide obstructs the nanopore in a different way it is possible to distinguish between nucleotides by measuring the electrical properties of the obstructed nanopore. These technique is, however, at a very early stage of development.\\
This is why in the following we will propose a novel approach based on single-molecule experiments of unzipping that could one day be used to sequence DNA.\\
The reader should keep in mind that no single method is free from the trade-off between resolution and scope, that is to say that it is impossible to attain at the same time accuracy at a single base level and very long reads.\\
\chapter{Modeling DNA unzipping}
\begin{figure}[p]
\includegraphics[width=.5\textwidth]{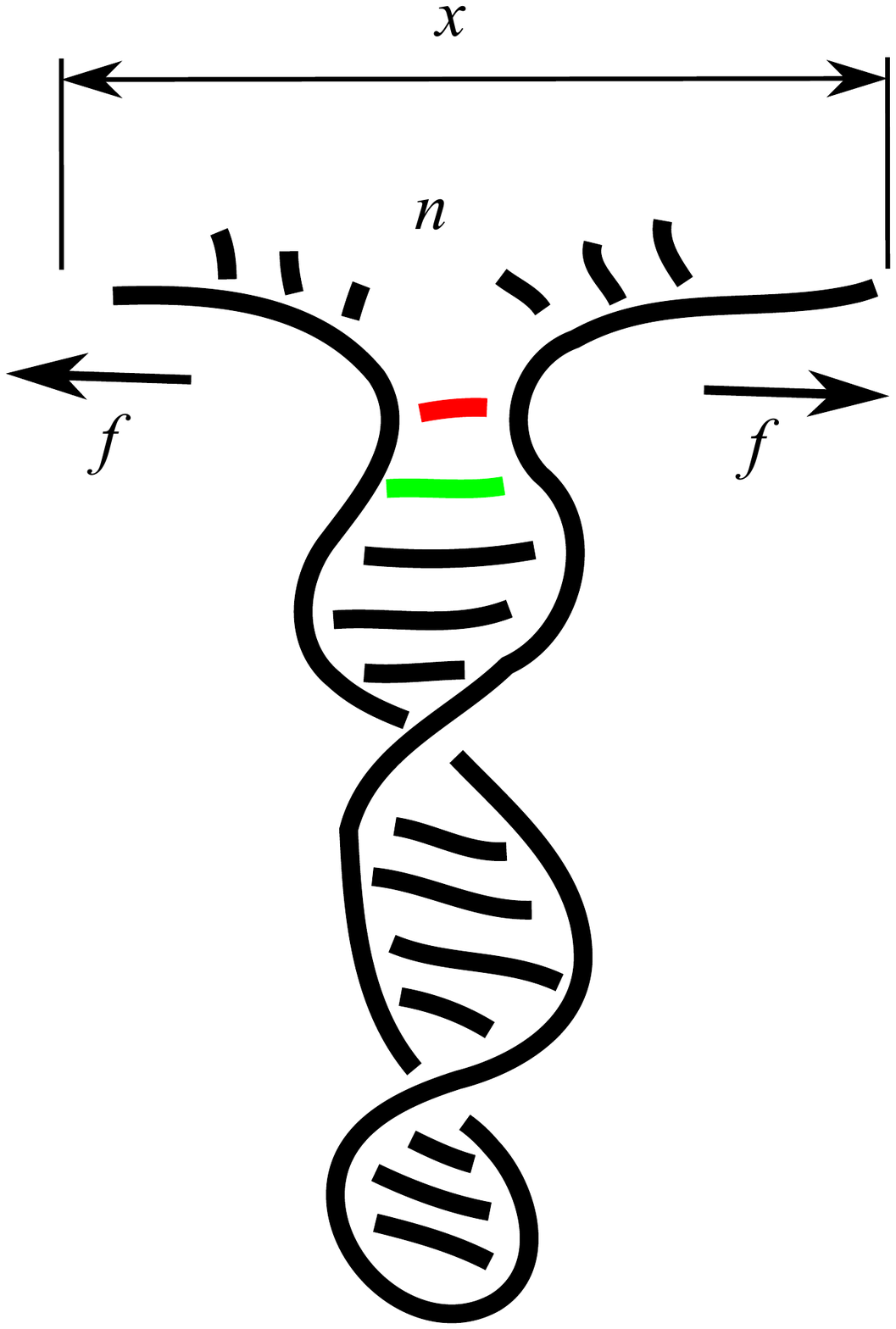}
\includegraphics[width=.5\textwidth]{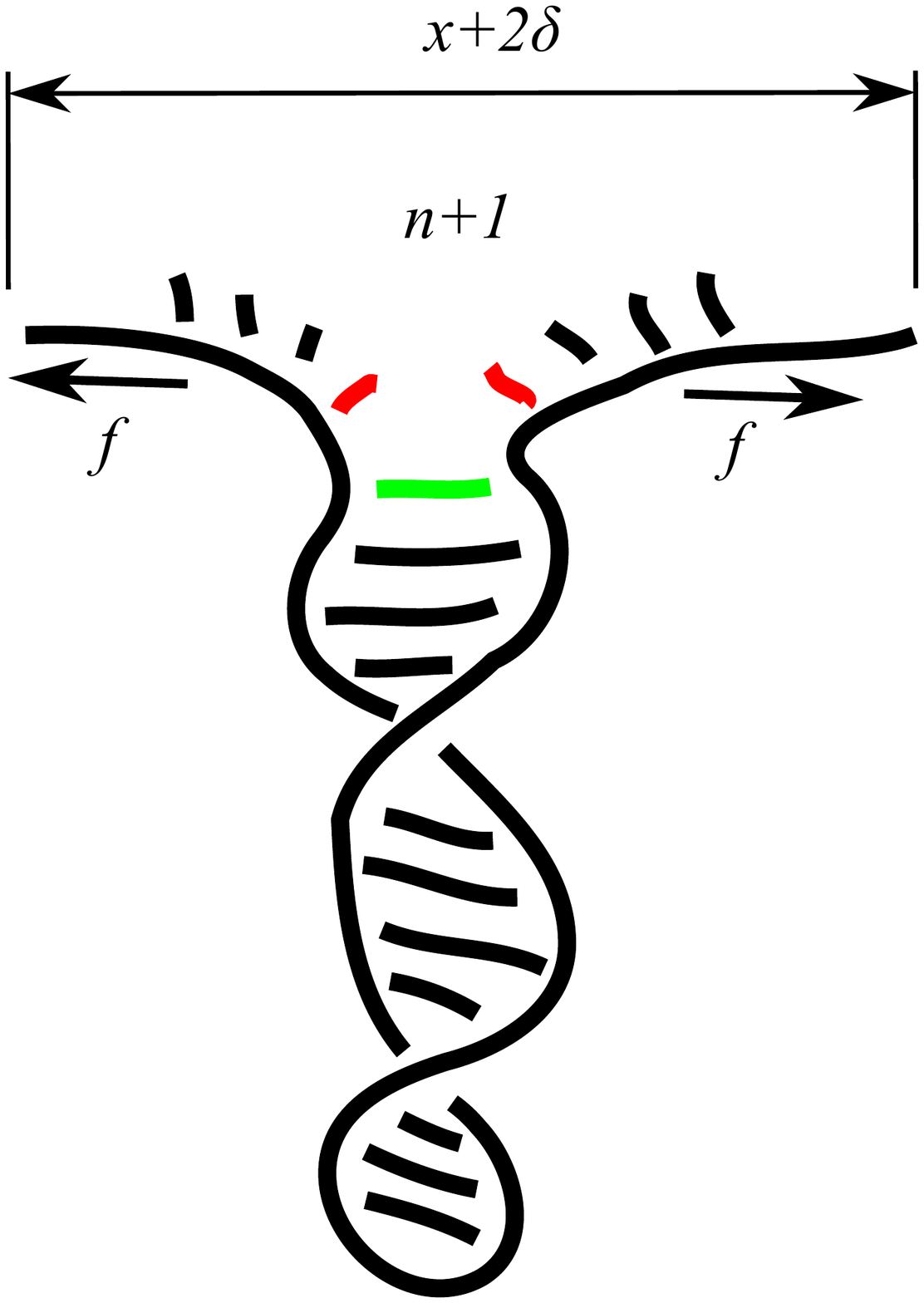}
\caption{Double stranded DNA can be denatured by applying opposite forces to the two strands. In clear analogy with the zipper commonly found in clothing, this type of experiment has benn christened unzipping.}
\label{fig:apertochiuso}\end{figure}

In the past two decades, the development of experimental techniques that allowed the manipulation of single biological macromolecules at the nm and pN scale has afforded us a wealth of experimental data on the physical properties of said molecules.\\
At the same time theoretical models have been devised to predict and model the behavior of said molecules. In particular the elasticity of both single-stranded and double stranded DNA is well know and the phase diagram of dsDNA is well understood. Experiments have permitted to denature dsDNA by applying a mechanical force, those experiments have taken the name of unzipping because the DNA is pulled apart from its two strands as a zipper (see figure \ref{fig:apertochiuso}.\\
These experiments are well understood in their single components: the ds- and ssDNA, the fork where the DNA denatures, what was lacking was a clearer picture how the delicate interplay of these different dynamics.\\
After an introduction to the physics of its single components, we will develop a mesoscopic model for the coupled dynamics and describe a software package for its simulation.\\
The goal here is to see whether the fluctuations an the correlations that compose the dynamics of linkers and beads will affect the unzipping dynamics of the force. This has already been investigated in \cite{MANOSAS}, however this approach is novel and has been published in \cite{IO1}.
\section{Modeling fork dynamics}
\begin{figure}
\begin{center}
\includegraphics[width=.5\textwidth]{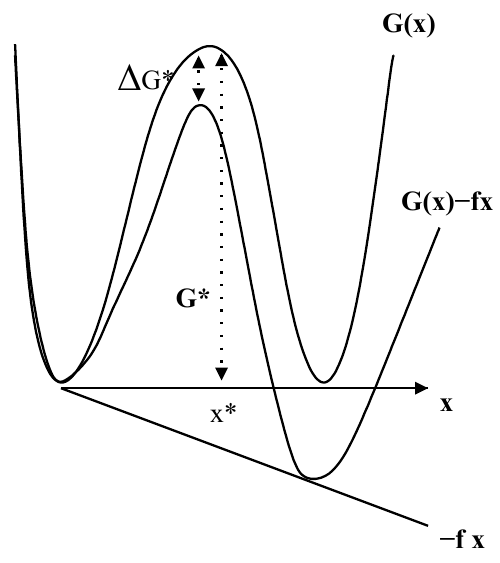}
\caption{The switching rate between the two states is proportional to $\exp(\beta G^*)$, where $G^*$ is the free energy barrier. The application of a force $f$ tilts the distribution and lowers the barrier of $\Delta G^*\simeq -f x^*$. Actual numerical values indicate this can be neglected with respect to $G^*$.}\label{fig:eyring}
\end{center}
\end{figure}

\begin{figure}
\begin{center}
\includegraphics[width=.5\textwidth]{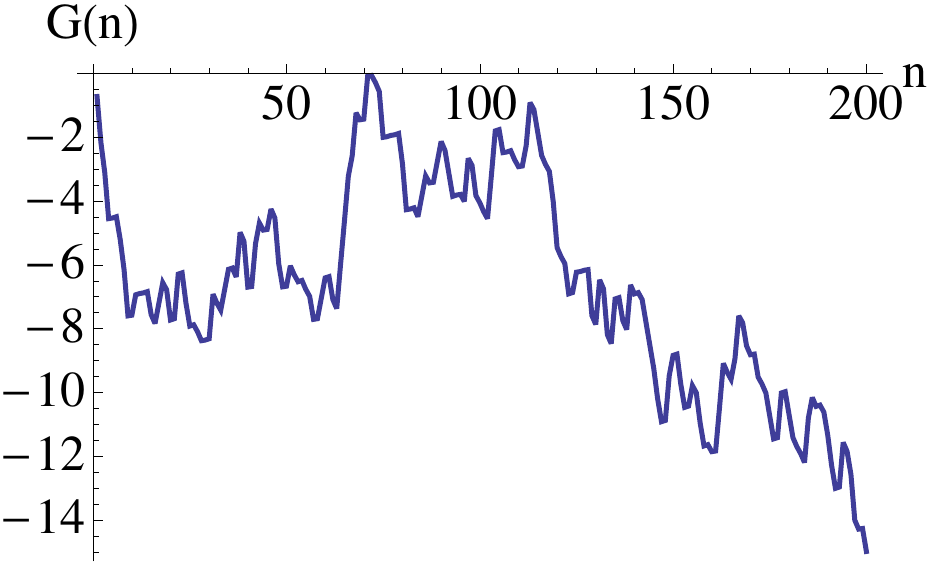}
\caption{Free energy $G$ (units of $k_\textrm{B}T$) to open the first $n$
base-pairs, for 200 randomly selected bases.}
\label{fig:gf50}
\end{center}
\end{figure}

The thermodynamics of DNA pairing is a subject that dates back to before the first sequencing techniques were available: a first model was proposed by Tinoco and collaborators in 1971 \cite{TINO71}, it gave the free energies for the two types of Watson-Crick bonds and it remarked that further study was needed to take into account stacking interactions, which had been known to be the principal cause of DNA stability for some time then \cite{CRO64}.\\
In 1973 the same group published a new letter \cite{TINO73} where new data allowed for the introduction of stacking effects, that is to say that base-pairing free energies now depended, not only on the base itself but on the previous base too. However the results were not very precise and they involved RNA hairpins rather than DNA, it wasn't until the second half of the eighties that reliable data on DNA thermodynamics became available \cite{BRES86}. More recently similar data have been obtained in unzipping experiments. \cite{HUGUET}.\\
The results of all of this studies are that the free energy of a DNA base pair depends on the base pair itself and its nearest neighbor nucleotide content, that is if we now consider a sequence of $N$ bases of dsDNA its free energy will be given by:
\begin{equation}
G(B,N)=\sum_{i=1}^Ng_0(b_i,b_{i+1})\,,
\end{equation}
where $B$ denotes the whole sequence and $b_i=A,T,C,G$ is the $i^\textrm{th}$ base. Typical values of the binding energies are given in table \ref{tableg0}.\\
\begin{table}
\begin{center}
\begin{tabular}{|c|c|c|c|c|} \hline $g_0$ &A&T&C&G \\ \hline A &1.78
&1.55 &2.52 & 2.22\\ \hline T &1.06 &1.78 &2.28 &2.54 \\ \hline C&
2.54 &2.22 &3.14& 3.85 \\ \hline G & 2.28& 2.52 & 3.90&3.14\\ \hline
\end{tabular}
\end{center}
\caption{Binding free energies $g_0(b_i,b_{i+1})$ (units of $k_\textrm{B} T$)
obtained from the MFOLD server for DNA at room
temperature, pH=7.5, and ionic concentration of 0.15 M. The base
values $b_i, b_{i+1}$ are given by the line and column respectively.}
\label{tableg0}
\end{table}What we are interested in is the phenomenon of unzipping under a force, the denaturation of dsDNA when the two strands that compose its double helix are pulled.\\
Let us now suppose for a moment we know the free energy of ssDNA under tension and that this is a linear function of the number of basis and otherwise depends only on the tension $f$ applied to it. At equilibrium we will have that $n$ bases of ssDNA have free energy equal to $ng_\textrm{ss}(f)$. We will focus on the form of $g_\textrm{ss}(f)$ in the following sections, it suffices to say that it needs to be an increasing function of force.\\
If we model only the motion of the  opening fork and we do not include in the  model  the experimental setup (see figure (\ref{fig:setup}): stretching the two strands of DNA away from one another we are able to apply a force and eventually open a base pair. When will this happen? The energy gain from the two new ssDNA bases must be greater than what is lost from the dsDNA energy, that is:
\begin{equation}
\Delta G(i)=g_0(b_i,b_{i+1})-2g_\textrm{ss}(f)\,,
\end{equation}
must be negative for the process to be energetically favored.\\
It is important now to put some numeric values on the quantities involved: the free energies $g_0$ and $g_\textrm{ss}$ are both of the order of a few $k_\textrm{B}T$, forces are expressed in units of pN and distances in units of nm. $k_\textrm{B}T\simeq 4$ pN nm.\\
The typical range of an hydrogen bond is about 0.1 nm, since the critical force needed to break it is of about 15 pN, this works out to an energy of about 0.4 $k_\textrm{B}T$ which can be neglected with respect to the few $k_\textrm{B}T$ of the binding energy which is known thermodynamically. This means that the opening rate will be independent of force.\\
Detailed balance then gives us the closing rate, which depends only on the force fluctuation needed to bring the two strands close enough to form the hydrogen bond. We then have:
\begin{equation}
r_\textrm{o}(n)=r e^{\beta g_0(n)}\,,\qquad r_\textrm{c}(f)=r e^{2\beta g_\textrm{ss}(f)}\,;
\end{equation}
where $\beta$ is the inverse temperature and $r$ gives the timescale of the phenomenon. We will refer to $r$ as the attempt rate; it can be estimated from the rate of self-diffusion for an object the size of a ssDNA base: $r^{-1}=\beta 2 \pi \eta l^3=0.17$ \textmu s, where $l=5$ nm is the size of a base and $\eta$ is the viscosity of water.\\
The interplay of the the stacking and pairing energy $g_0$ and the energy gained from the two newly formed ssDNA bases $2g_{\textrm{ss}}$ is responsible for the formation of a complex energy landscape full of metastable minima, not dissimilar, to a one-dimensional random walk in a random environment, also known as the Sinai model. This suggest the use of methods from the statistical mechanics of disordered systems and from information theory for the description and analysis of such a system.\\
In figure \ref{fig:gf50} we show as an example the free energy derived from the first 50 base-pairs of the $\lambda$-phage DNA at two different forces.\\
Cocco and collaborators first worked out the opening and closing rates in \cite{COCCO01,COCCO03},  as a Eyring-Kramers transition state theory \cite{EYR35,KRA40}. This theory describe the transition with a suitable continuous variable (which here is the separation $x$ between the two bases forming the base pair), $x$ obeys Langevin dynamics over an effective potential that is the free energy $G(x)$. This potential has two local minima at the two equilibrium position that correspond to broken/whole hydrogen bonds (see figure \ref{fig:eyring}).\\

\section{ssDNA as a modified freely jointed chain}
One of the simplest polymer models possible is that of the freely jointed chain. The FJC is composed of $N$ monomers of length $d$, no constraint is put on the angles formed by consecutive segments and the excluded volume is not taken into account.\\
The end-to-end distance is thus given by:
\begin{equation}
\vec{R}=\sum_i^N\vec r_i\,,
\end{equation}
where the $\vec r_i$ are random vectors of length $d$. This length is often referred to as Kuhn length.\\
If we are in the thermodynamic limit we can use the central limit theorem to show that the average end-to-end distance $\langle R\rangle$ vanishes and that it is distributed according to a normal distribution of variance $\langle R^2\rangle=Nd^2$.\\
To get this result \cite{KUHN42,JAMES43} we have assumed that no force was acting on one end of the chain, and it is, in fact, only valid around for end-to-end lengths of the order of $\sqrt{N}d$. In order to get the right result for high extensions we have to add a tensile force $f$ applied in the $x$ direction. We can now compute the average value of the $x$ component of the $i$th link of the polymer as:
\begin{equation}
l_\textrm{FJC}(f)=\langle x_i\rangle=\frac{\int_{-d}^{d}x_i\exp(\beta f x_i)d x_i}{\int_{-d}^{d}\exp(\beta f x_i)d x_i}=d \mathcal{L}(\beta f d)\,,
\end{equation}
Where $\mathcal{L}(x)=\coth(x)-1/x$ is the Langevin function. The total length of the polymer along the $x$ axis is then given by $L_\textrm{FJC}(f)=Nl_\textrm{FJC}(f)$. The interested reader can find further details in a classical reference such as \cite{FLORY53}.\\
At the beginning of the nineties it became possible to measure the elasticity of DNA with magnetic beads \cite{Smi92}. It then became apparent that, up to forces of 20 pN/nm the elasticity of ssDNA is well fitted by a FJC model, but even better results are obtained using a modified FJC where the monomers are extensible at high forces and where the contour length (\emph{i. e.} the total stretched length of the polymer) is not given by the product of the number of momomers and the Kuhn length:
\begin{equation}
l_\textrm{MFJC}(f)=l_\textrm{FJC}\left(1+\frac{f}{\gamma_\textrm{ss}}\right)=d\left(\coth(\beta fb)-\frac1{\beta f b}\right)\left(1+\frac{f}{\gamma_\textrm{ss}}\right)
\end{equation}
where $d=0.56$ nm, $b=1.4$ nm and $\gamma_\textrm{ss}=800$ pN \cite{Smi96}.
\section{dsDNA as an exstensible  worm-like chain}
The worm-like chain (WLC) is one of the simplest continuous models of a polymer: if we define a parametric curve in space $\vec r(s)$ we can define it's tangent vector as $\vec t =\frac{d\vec r (s)}{d s}$ and its curvature vector as $\vec w =\frac{d\vec t (s)}{d s}$, we can further impose that the polymer is inextensible, that is: $|vec t(s)|=1$.\\
Then we can give the internal energy for a polymer stretched by an external force $f$ as:
\begin{equation}
\beta E=\int_0^{L_\textrm{tot}}ds\frac{A}2|\vec w (s)|^2-\beta f \hat t(s) \cdot \hat x\,,
\end{equation}
where $A$ is the persistence length, that turns out to be the correlation length of the direction of the polymer at zero force.\\
The WLC is analytically solvable model, however the solution can only be written as an infinite series \ref{WLCSERIES}. Luckily a very precise numerical fit has been proposed by Marko and Siggia in \cite{MARKO94,MARKO95}:
\begin{equation}
\beta f A=\frac{l_\textrm{WLC}}{l_\textrm{tot}}+\frac1{4(1-l_\textrm{WLC}/l_\textrm{tot})^2}-\frac14\,,
\end{equation}
where $l_\textrm{tot}$ is the contour length of the polymer divided by the number of bases, $A$ is the persistence length and $l_\textrm{WLC}$ is the length of the polymer in the direction of the force $f$.\\
In the following years even more refined fits to the experimental data have been proposed such as the one by Moroz and Nelson \cite{MOROZ97} which used a formula first proposed by Odijk \cite{ODIJK}. Their formula can fit the experimental data for the elasticity of dsDNA for a very large range of forces \cite{BOU99}, thanks to the relaxation of the hypothesis that $|\vec t(s)|=1$, which plays an important role at high forces and the inclusion of torsional effects.\\
However we do not need such a large range of forces for the description of unzipping experiments; because of this that in the following we will use a simplified version of the Odijk formula, namely:
\begin{equation}
l_\textrm{WLC}(f)=l_\textrm{tot}\left[1-\frac12\left(\beta f A\right)^{-1/2}+\frac{f}{\gamma_\textrm{ds}}\right]\,,
\end{equation}
where $l_\textrm{tot}=0.34$ nm $A=48$ nm and $\gamma_\textrm{ds}=1000$ pN.
\section{Two possible ensembles}
\begin{figure}[htbp]
\begin{center}
\includegraphics[width=.7\textwidth]{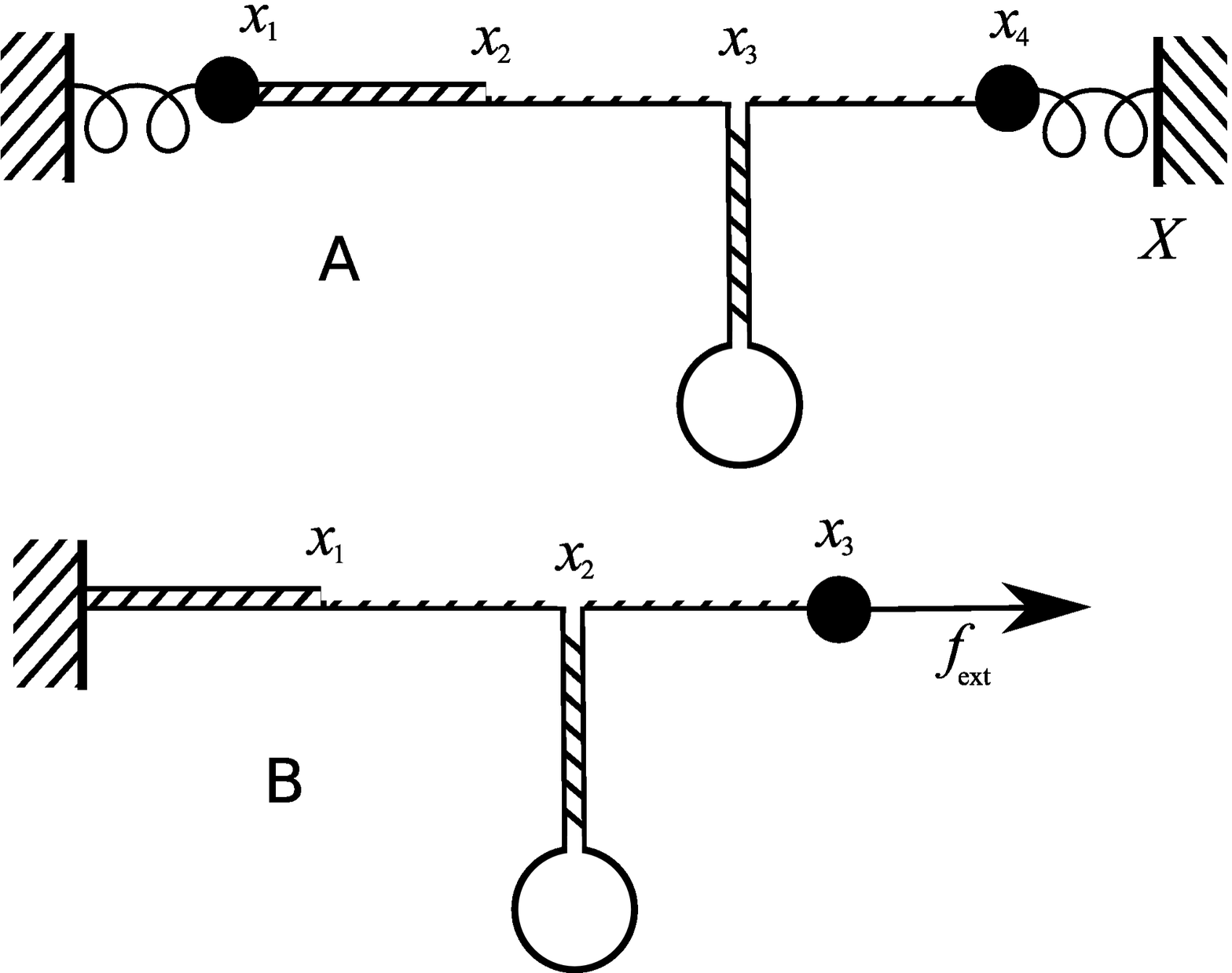}
\caption{Typical experimental setups that will be described in
the following. A) A setup with two optical traps (beads $x_1$ and
$x_4$) drawn as springs and whose centers are the black vertical lines;
B) a setup with a single magnetic bead $x_3$ that applies
a constant force on the molecule attached to a fixed ``wall''. 
In both cases the molecular
construction is made by a DNA molecule that has to be opened
(therefore one should include two single-strand linkers that are
the opened parts of the molecule) and one double-stranded DNA linker.
The coordinates $x_i$ are the distances of the corresponding points
from the left reference position (which is the center of the left optical
trap in case A and the fixed wall in case B).
}
\label{fig:setup}
\end{center}
\end{figure}

The description we have given above does not depend much on the experimental setup, the only time where we have lost some generality is in the description of fork dynamics, where we have assumed the force to be fixed; however both the polymer description and our choice for the dynamics are completely independent of details like this.\\
In the following we will outline two possible experimental setups (pictured in figure \ref{fig:setup}): in the first force is a parameter and the extension of the polymer, which is directly related to the number of open bases, is measured; in the second the distance between two optical traps can be varied as a parameter and the displacement of the beads in the traps can be measured to give a precise measurement of force.\\
The only detail that needs to be sorted out is the change in variable in the thermodynamic potentials that describe different setups, but this can be easily done through a Legendre transform.
Before we go on we should lay out the notation we will use in the following: first of all capital letters denote extensive quantities, while lower case letters correspond to the equivalent intensive quantity. $x$ is the end to end distance of a polymer and $l=x/n$, where $n$ is the number of momomers (bases here). For example $W(x)=nw(l)$\\
Let's lay out all the quantities:
\begin{itemize}
\item $g(f)$ is the free energy per base as a function of force.
\item $l(f)=\frac{\partial g(f)}{\partial f}$ is the length as a function of force.
\item $w(l)=\max_f[fl-g(f)]$ the free energy as a function of length.
\item $f(l)=\frac{\partial w(l)}{\partial l}$ the force as a function of length or the inverse of $l(f)$.
\item $k(l)=\frac{\partial f(l)}{\partial l}$ is the effective spring constant for a given length.
\item $\frac1{k(f)}=\frac{\partial l(f)}{\partial f}$ is the reciprocal effective spring constant as a function of force.\\
\end{itemize}
\subsection{Fixed force, magnetic tweezers}
At the beginning of the 2000s Gosse and Croquette \cite{GOSSE2002} developed a technique called optical tweezing: a superparamagnetic bead with a diameter of the scale of the \textmu m is placed under the two poles of a permanent magnet, which creates a magnetic gradient.\\
The distance between the poles of the magnet (less than 1 mm) is fixed so that on the scale of the typical movements of the bead the gradient of the magnetic field is almost constant and so is the force applied to the bead.\\
Magnetic beads have a preferred direction. This is at the same time an advantage and a disadvantage: the advantage is the possibility of applying a torque to the bead, which has opened the door to experiments involving the coiling and uncoiling of DNA; on the other hand the DNA will bind on a random point of the surface and it is impossible to say exactly where. Given the relative size of the bead and of a single base of DNA, this means that the unzipping experiment can start up to 1000 bp away in two different runs.\\
The position of the bead can be recorded optically. This type of experiments are relatively easy to set up, and the modellization of fork dynamics at fixed force is perhaps more intuitive. On the other hand fixed force experiments tend to be ill suited for sequencing purposes, since it is difficult to control the position along the energy landscape where the fork will stop.\\
For a given portion of the sequence there exists an average critical opening force. When the critical force is exerted the fork will fluctuate around a given number of open bases for a long time because it is in a potential well. On the other hand the top of the potential barriers that separate these wells are very hard to sample, because very little time will be spent there.\\
Another reason why this method is not very well suited for sequencing through unzipping is that the position of the fork for a given force depends strongly on the sequence and it is very hard to generate an unzipping protocol with varying force without a prior knowledge of the sequence.
\begin{figure}[htbp]
\begin{center}
\includegraphics[width=.7\textwidth]{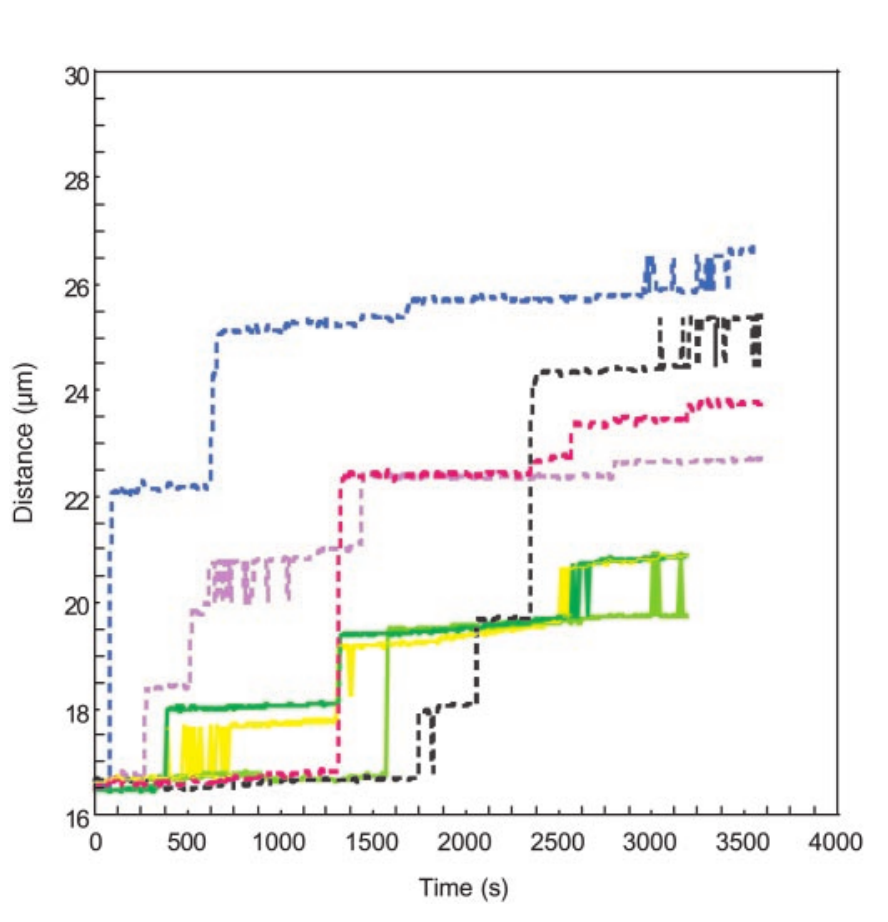}
\caption{Several typical fixed force unzipping traces from \cite{DANI}. Solid lines correspond to a force of 15 pN, while dashed lines correspond to a force of 20 pN. The measured quantity is the distance between the center of the magnetic trap and the surface of a glass micropipette which the DNA is attached to. Horizontal plateaux correspond to minima of the free energy.}\label{fig:danilo}
\end{center}
\end{figure}
\FloatBarrier
\subsection{Fixed distance, optical tweezers}
\begin{figure}[htbp]
\begin{center}
\includegraphics[width=\textwidth]{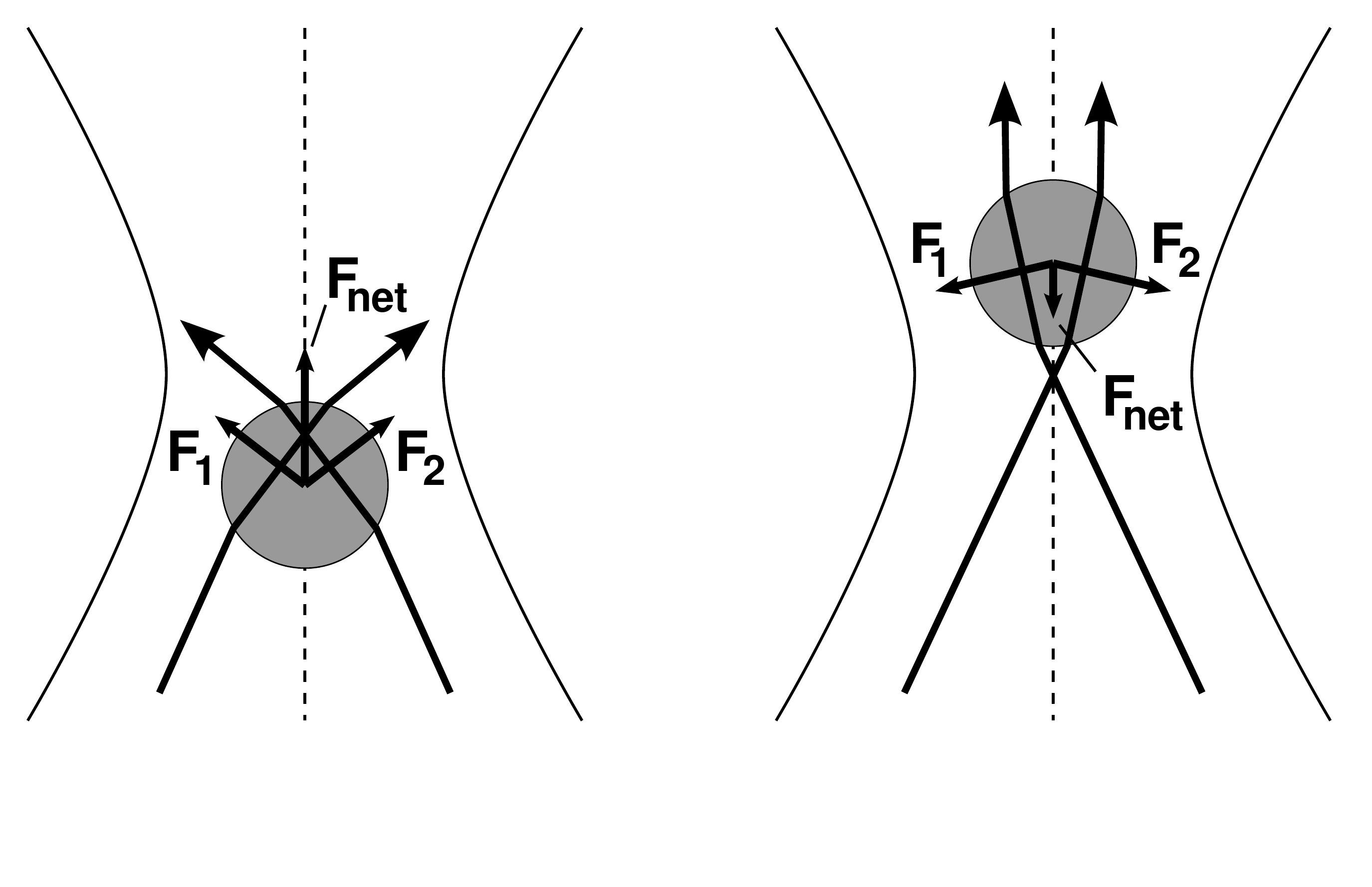}
\caption{The effect of the refraction of light on a dielectric bead in the ray optic approximation. The light propagates from bottom to top. $F_1$ and $F_2$ are the forces acting on the bead because of the concentration of momentum, $F_\textrm{net}$ their resultant.}
\label{fig:opticaltrap}
\end{center}
\end{figure}

Pioneering studies on the effect of radiation pressure from laser light on micrometer-sized dielectric beads were performed at the beginning of the seventies by Ashkin \cite{ASH70}. A few years later the same Ashkin developed a single beam technique for trapping dielectric beads \cite{ASH86}.\\
In optical tweezers a tightly focused laser beam passes through a dielectric sphere which has an optical index higher than that of the surrounding fluid. The incoming light from the laser is refracted by the bead causing a change in the momentum of the outgoing light; because of the conservation of momentum, the bead will experience a change of momentum of opposite sign.\\
For high enough numerical aperture of the laser there exists a stable position of the bead along the axis of propagation of laser light, on the other hand stability along the transversal directions is due to the intensity profile of the laser, which is most of the times Gaussian.
In order to give a precise description of the phenomenon for the conditions most often used in micromanipulation experiments, we should take into account the full Mie theory of light scattering, since the bead size (1 \textmu m) is very close to the wavelength of the laser employed (see for example \cite{MANG08}, where the laser wavelength is 1.064 \textmu m).\\
On the other hand we can give an hand-waving argument for the stability of the trap using ray optics: a particle with a refractive index higher than water will act as a positive lens, roughly speaking if the bead is placed before (after) the focal point the rays will diverge (converge).\\
If the lens converges the ray the light will have more momentum in the direction of propagation of the beam, conversely, if the beams have been diverged the light will lose momentum. See figure \ref{fig:opticaltrap} for a schematic picture. The interested reader should refer to Kerker's book \cite{KERKER} for a full treatment of the Rayleigh and Mie regimes.\\
 The use of optical traps for the manipulation of biopolymers is compelling because it allows to fix the position of the beads and to measure the force exerted on the molecule. This is very attractive for unzipping experiment because it gives us a chance to focus on a specific region of the sequence, while in fixed force experiments the region of DNA where the fork will spend most of the time depends on the sequence itself.\\
The measurement of force is obtained by the observation of the displacement of the bead with respect to the center of the bead. The optical trap is well approximated by an harmonic potential around its equilibrium position. The displacement of the bead can be measured either by direct observation of the diffraction pattern through a microscope, or by measuring the deflection of the laser beam with a PSD \cite{PSD}.\\
\begin{figure}[htbp]
\begin{center}
\includegraphics[width=.7\textwidth]{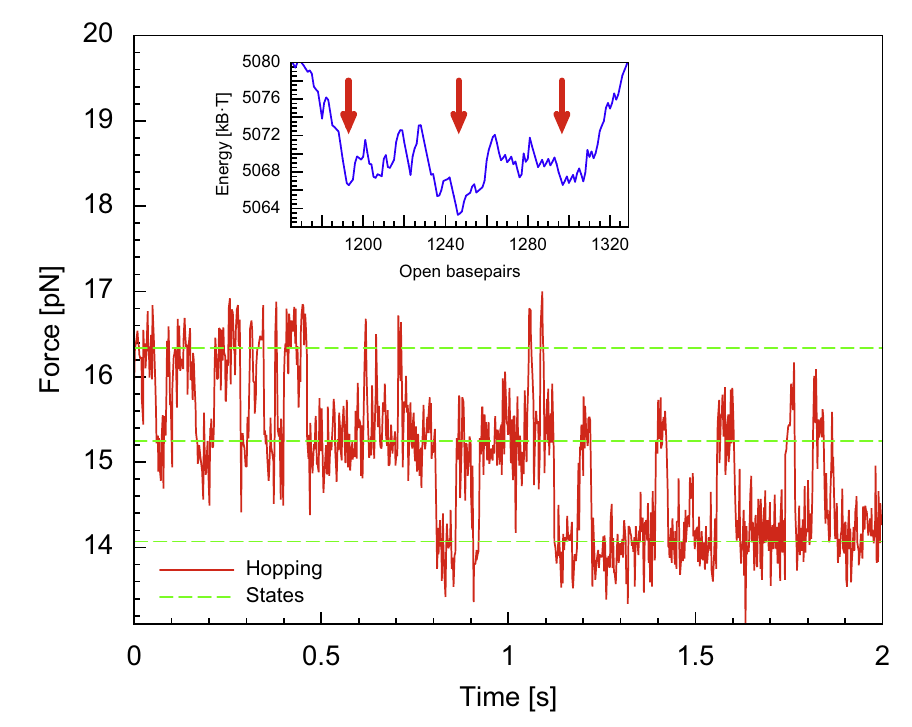}
\caption{A typical fixed distance unzipping trace from \cite{MOSSA}. The force, measured as the displacement of the bead in the optical trap, is measured as a function of time. Notice how the three minima of the free energy correspond to the three green lines where the bead spends most of its time.}\label{fig:mossa}
\end{center}
\end{figure}
\section{Overdamped dynamics}
The motion of very small objects suspended in a liquid does not resemble much to that of objects in everyday life. The most striking features are the absence of inertial effects and Brownian noise.\\
A common way to quantify the ratio between inertial and viscous effects is the Reynolds number Re, which was introduced by Stokes \cite{STOKES}, several years before Reynolds popularized it.\\
It is given by:
\begin{equation}
\mathrm{Re}=\frac{V l \rho}{\eta}\,,
\end{equation}
where $V$ is the mean velocity of the object with respect to the fluid, $\rho$ is the density of the fluid, $l$ is the linear size of the object and $\eta$ is the viscosity of the fluid.\\
It appears in the dimensionless Navier-Stokes\footnote{Please note that this is not the only way to rescale the variables in order to make the adimensional: $\rho V^2$ has the dimensions of a pressure and can be used to the same effect, it turns out this latter is the right scaling for high Reynolds numbers, while the one in the main text is the right one for the limit of low Re.} equation for an object immersed in a Newtonian fluid as:
\begin{equation}
\mathrm{Re}\left(\frac{\partial \mathbf{v}}{\partial t} + \mathbf{v} \cdot \nabla \mathbf{v}\right) = -\nabla p + \nabla^2 \mathbf{v} + \mathbf{f}
\end{equation}
where $\mathbf{v}$ is the speed of the object divided by $V$, $p$ is the pressure of the fluid divided by $\eta V/l$, $\mathbf{f}$ are the external forces per unit volume divided by $\eta V/l^2$, $\nabla$ stands for the the space partial derivatives vector multiplied  by $l$ and finally $\partial/\partial t$ is the time derivative multiplied by $l/V$.\\
The importance of the Reynolds number is that it is the only quantity needed to describe the flow of a fluid, that is to say that once the variables have been properly rescaled sistems of different size, viscosity and density will behave the same way.\\
It is customary to categorize the characteristics of the flow according to the Reynolds number:
\begin{itemize}
\item $\mathrm{Re}\gg 1$: Turbulent flow. Inertial forces are dominant.\\
\emph{E.g.} man swimming, the wing of a plane.
\item $\mathrm{Re}\sim 1$: Laminar flow. Viscous forces dominate. \emph{E.g.} water in a pipe.\\
\emph{E.g.} blood flow, fish swimming, man swimming in glycerol.
\item $\mathrm{Re}\ll 1$: Creeping flow. Inertial forces are completely negligible.\\
\emph{E.g.} Bacteria in water, \textmu m-sized beads in optical traps, macromolecules in solution.
\end{itemize} 
Among the objects that we will consider in the following those who have the largest Reynolds number are the beads in the optical traps; for them $\mathrm{Re}\sim10^{-6}$, because of this, the remarks we will make on their dynamic behavior will be all the more valid for objects with lower Reynolds number.\\
Let us suppose that a bead of diameter $d$, is suspended in water by an optical trap of stiffness $k$. Let us also suppose for the moment that the bead has the same density as the water surrounding it.\\
The bead obeys the Langevin harmonic oscillator equation:
\begin{equation}
m\ddot{x}+\gamma \dot{x}+kx=\xi(t)\,,
\end{equation}
where $m=1/6 \pi\rho d^3$, $\gamma=\beta 6 \pi \eta d$ and $\xi(t)$ is Gaussian noise obeying
\begin{equation}
\langle\xi(t)\rangle=0,\qquad\langle\xi(t)\xi(t^\prime)\rangle=2\gamma k_\textrm{B}T \delta(t-t^\prime)\,.
\end{equation}
We now consider the frequency response by performing a Fourier transform obtaining
\begin{equation}\label{eq:lange}
A(\omega)=\frac1{\sqrt{(k-m \omega^2)^2+(\gamma \omega)^2}}\,.
\end{equation}
The question is: when can this be approximated by its Brownian counterpart, neglecting the mass term?\\
The new equation for the frequency response would read:
\begin{equation}\label{eq:brown}
A(\omega)=\frac1{\sqrt{k^2+(\gamma \omega)^2}}\,.
\end{equation}
Now this response has a cutoff frequency of $\gamma/k$, what we want, in order for our approximation to be valid, is for this frequency to be much smaller than the one at which mass effects become important, that is $m/\gamma$.
Summing up we want
\begin{equation}
\frac{mk}{\gamma^2}=\frac{l\rho k}{(6\pi)^3\eta^2}\gg1\,.
\end{equation}
Plugging in realistic values for the stiffness of the trap $k=0.5$ pN/nm, for the density $\rho=1$ g/cm$^3$, the diameter of the bead $l=1$ \textmu m and the viscosity of water $8.9\,10^{-4}$ Pa s; we find the ratio to be very small: $\frac{mk}{\gamma^2}=1.9\,10^{-4}$.
\begin{figure}
\begin{center}
\includegraphics[width=.7\textwidth]{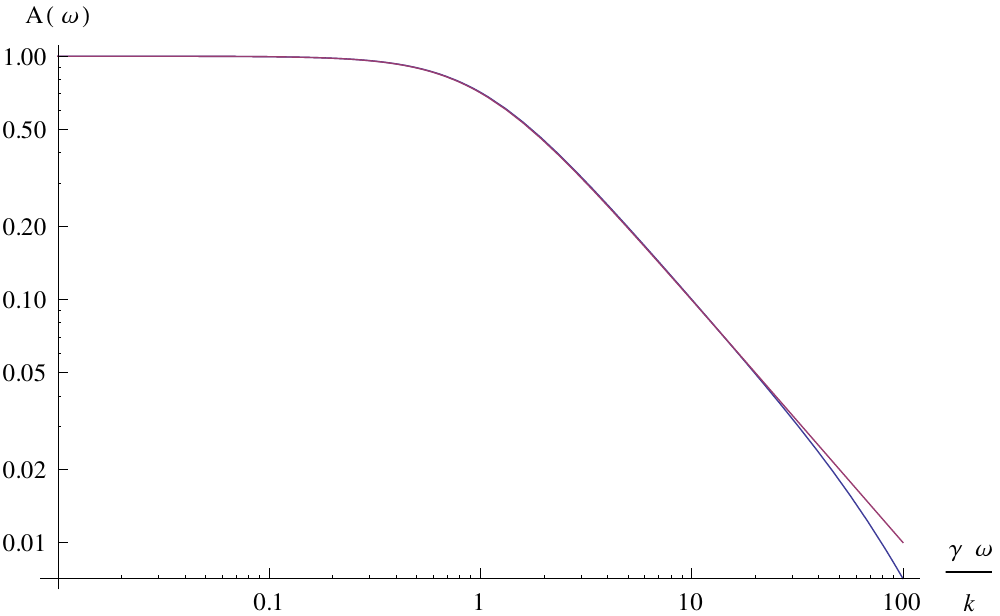}
\caption{The amplitude response as a function of frequency: the blue curve corresponds to eq. (\ref{eq:lange} while the violet one corresponds to eq. (\ref{eq:brown}). For this plot we have chosen $\frac{mk}{\gamma^2}=100$, this way the cutoff frequency is well below the frequency where the mass effects become dominant. }\label{fig:langebrown}
\end{center}
\end{figure}This is in accordance with what we would have expected by using the Reynolds number, in fact \textmu m-sized beads are well within the creeping flow range for speeds up to 10 cm/s.\\
In the following discussion we will be well justified in leaving out the mass terms from our equations and considering all of the dynamics as Brownian or overdamped.\\
\section{Coupling all the dynamics together}
In this section we will derive an effective mesoscopic dynamical equation for coupled heteropolymers. The details of the calculation are somewhat technical, but they offer a different insight from the derivation published in the appendix of \cite{IO1}.\\
Because of the preceding discussion on overdamped dynamics we will ignore all inertial effects. In addition to this simplification we will consider the simplest polymer model: a chain of simple Hookean springs, also known as the Rouse model \cite{Rouse}. We will also consider the model to be effectively one dimensional.\\
What we hope to understand better here is how movement propagates along a heteropolymer, what kind of fluctuations and correlations are important and how. It is also of interest to know whether the polymer can be considered at equilibrium and what are the relaxations times.\\
We will show that in a mesoscopic description where we do not describe single monomers the noise is not decorrelated and we will propose a way to implement these characteristics in computer experiments.
\subsection{Scaling of a homogeneous Rouse polymer}
\begin{figure}[htpb]
\includegraphics[width=\textwidth]{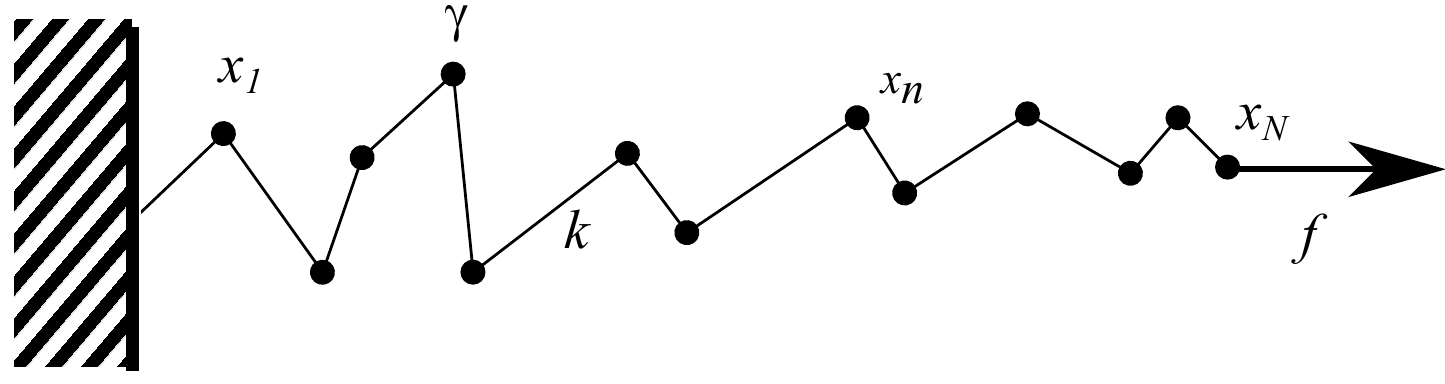}
\caption{A homogeneous Rouse polymer composed of $N$ identical springs and beads each having spring constant $k$ and viscous drag coefficient $\gamma$. The coordinates $x_i$ are taken along the direction of the pulling force $f$.}\label{fig:rouseuniform}
\end{figure}

Let us now derive the equations for the simplest case: that of a homogeneous polymer. At first we will derive the equation for the free end of the polymer and then we will concentrate on a midpoint to see how the dynamics are coupled, we will see of this leads to a viscous drag matrix on the left hand side and how this translates into fluctuation dissipation relations for correlated noise.\\
Each monomer is characterized by its spring constant $k$ and its viscous drag coefficient $\gamma$. Let us suppose that a chain of $N$ identical springs is connected to a non moving wall on one end and that a constant force $f$ is exerted on the other end. The setup is shown in figure \ref{fig:rouseuniform}.\\
The monomers will then obey this system of simultaneous equations:
\begin{equation}
\begin{split}
\gamma \dot{x}_1&=-2kx_1+kx_2+\eta_1\label{first}\\
&\;\,\vdots\\
\gamma \dot{x}_n&=-2kx_n+kx_{n-1}+kx_{n+1}+\eta_n\\
&\;\,\vdots\\
\gamma \dot{x}_N&=-kx_N+kx_{N-1}+f+\eta_N\,,
\end{split}
\end{equation}
where $x_n$ is the coordinate of the $n^\textrm{th}$ link. $\eta_n$ are uncorrelated Gaussian noises of zero average and autocorrelation function:
\begin{equation}
 \langle\eta_i(t)\eta_j(0)\rangle=2\gamma k_\textrm{B} T\delta_{ij}\delta(t)\,.
\end{equation}
Equation (\ref{first}) can be solved formally for $x_1$ in terms of integrals of $x_2$. Thus:
\begin{equation}
 x_1(t)=\frac{1}{\gamma}\int_0^\infty e^{-\frac{2k}{\gamma}(t-t^\prime)}\left[k x_2(t^\prime)+\eta_1(t^\prime) \right] \,.\label{firstint}
\end{equation}
This doesn't bring us any closer to solving the system of equations, in fact iterating this procedure will only produce an integro-differential equation of order $N$. To make the problem tractable we have to solve it in the limit in which the ratio $\gamma/k$ is small, which is reasonable given that for ssDNA at typical conditions it has the value of approximately $10^{-10}$ s, many orders of magnitude below experimental resolution.\\
Equation (\ref{firstint}) thus becomes:
\begin{equation}
 x_1(t)=\frac{1}{2}\left[x_2(t)+\eta_1(t)\right]-\frac{\gamma}{4k}\dot{x}_1(t)+o \left(\frac{\gamma}{k}\right)\,.
\end{equation}
Substitution of this into the equation for $x_2$ yields:
\begin{equation}
 \frac{5}{4}\gamma\dot{x}_2=-(k +\frac{1}{2}k)x_2 +kx_3+\eta_2+\frac{1}{2}\eta_1\,,\label{eq:brownx2}
\end{equation}
The fluctuation-dissipation theorem for Brownian dynamics of the form $\gamma\dot x= -\nabla V(x) +\eta$ states that, in order for Boltzmann equilibrium to be attained the following relation must be verified:
\begin{equation}
 \langle\eta(t)\eta(0)\rangle=2\gamma k_\textrm{B} T\delta(t)\,.
\end{equation}
For equation (\ref{eq:brownx2}) this translates to:
\begin{equation}\begin{split}
 \left<\left(\eta_2(t)+\frac{1}{2}\eta_1(t)\right)\left(\eta_2(0)+\frac{1}{2}\eta_1(0)\right)\right>&= \langle\eta_2(t)\eta_2(0)\rangle+\frac14\langle\eta_1(t)\eta_1(0)\rangle\\&=2 \frac54\gamma k_\textrm{B} T\delta(t)\,.
\end{split}
\end{equation}
This means that our approximation is consistent with the fluctuation-dissipation theorem. It is obvious that the iteration of this procedure will define renormalised $k$ and $\gamma$ and new $\eta_n$ which will be correlated.
We can write:
\begin{equation}
 \gamma_{n-1} \dot{x}_{n-1}=-(k+k_{n-1})x_{n-1}+kx_{n}+\eta^\prime_{n-1}\,.
\end{equation}
Then solve this equation with the usual approximation finding:
\begin{equation}
 x_{n-1}=\frac{k}{k+k_{n-1}}x_n-\frac{ \gamma_{n-1}k}{(k+k_{n-1})^2}\dot{x}_n+\frac{1}{k+k_{n-1}}\eta^\prime_{n-1}+o \left(\frac{\gamma}{k}\right)\,,\label{approxn}
\end{equation}
which must be inserted in the equation for $x_n$:
\begin{equation}
\begin{split}
\left(\gamma+\frac{\gamma_{n-1}k^2}{(k+k_{n-1})^2}\right)\dot{x}_n&=-\left(k+\frac{kk_{n-1}}{k+k_{n-1}}\right)x_n+kx_{n+1}\\
&+\eta_{n}+\frac{k}{k+k_{n-1}}\eta^\prime_{n-1}\,,\label{equan}
\end{split}
\end{equation}
thus defining recurrence relations for the coefficients:
\begin{align}
 k_n&=\frac{kk_{n-1}}{k+k_{n-1}}\,;\\
\gamma_n&=\gamma+\frac{\gamma_{n-1}k^2}{(k+k_{n-1})^2}\,;\\
\langle\eta^\prime_n(t)\eta^\prime_n(0)\rangle&=\langle\eta_n(t)\eta_n(0)\rangle+\frac{k^2}{(k+k_{n-1})^2}\langle\eta^\prime_{n-1}(t)\eta^\prime_{n-1}(0)\rangle\,.
\end{align}
Applying the fluctuation dissipation theorem to the last equation shows that we have chosen the only approximation consistent with the preceding equation. That is to say that the $\gamma$'s on the left hand side obey the same recurrence relations as the Brownian noises.\\
As we have already calculated the values of the constants for $n=2$ we can easily solve the recurrences:
\begin{align}
 k_n&=\frac{1}{n}\,;\\
\gamma_n&=\frac{(2n+1)(n+1)}{6n}\gamma\,.\\
\end{align}
This way we can rewrite equations (\ref{equan}) and (\ref{approxn}) as:
\begin{align}
\frac{(2n+1)(n+1)}{6n}\gamma\dot{x}_n&=-\left(k+\frac{k}{n}\right)x_n+kx_{n+1}+\eta^\prime_{n}\,;\\
x_{n-1}&=\frac{n-1}{n}x_n-\frac{(2n-1)(n-1)}{6n}\frac{\gamma}{k}\dot{x}_n+\frac{n-1}{nk}\eta^\prime_{n-1}\label{ennemenouno}\,.\
\end{align}
The recurrence can be completely closed with the help of the equation for $x_N$ as:
\begin{equation}
\frac{(2N+1)(N+1)}{6N}\gamma\dot{x}_N=-\frac{k}{N}x_N+f+\eta^\prime_{N}\,.
\end{equation}
Not surprisingly we recover the scalings of the Rouse model when it is solved in the continuous $n$ limit (see for example \cite{DoiEdwards}) in that it gives:
\begin{equation}
\frac{N}3\gamma\dot{x}_N=-\frac{k}{N}x_N+f+\eta^\prime_{N}\,,
\end{equation}
What we would like to explore now is what happens to a subpolymer, \emph{i. e.} write down the evolution of one end of the polymer and of a midpoint, integrating out all other degrees of freedom. To do so we need to start from the $(N-1)^\textrm{th}$ link of the polymer.
\begin{equation}
 \gamma \dot{x} \\_{N-1}=-2kx_{N-1}+kx_{N-2}+kx_N+\eta_{N-1}\,.
\end{equation}
which gives, after the usual procedure:
\begin{equation}
 x_{N-1}=\frac{1}{2}\left[x_{N-2}+x_N\right]+\frac{\gamma}{4k}\left[\dot{x}_{N-2}+\dot{x}_N\right]+\frac{\eta_{N-1}}{2k}+o \left(\frac{\gamma}{k}\right)\,,\label{nmenouno}
\end{equation}
which can now be used in the $(N-2)^\textrm{th}$ and $N^\textrm{th}$ equations yielding:
\begin{align}
\frac{5}{4}\gamma\dot{x}_{N-2}+\frac{1}{4}\gamma\dot{x}_{N}&=-\left(k +\frac{k}{2}\right)x_{N-2}+\frac{}{2}x_{N-2}+kx_{N-3}+\eta_{N-2}+\frac{\eta_{N-1}}{2}\\
\frac{5}{4}\gamma\dot{x}_{N}+\frac{1}{4}\gamma\dot{x}_{N-2}&=-\frac{k}{2}x_{N}+\frac{k}{2}x_{N-2}+f+\eta_{N}+\frac{\eta_{N-1}}{2}\,.
\end{align}
If we define:
\begin{align}
\gamma_n^a\dot{x}_{N-n}+\tilde{\gamma}_n\dot{x}_{N}&=-(k+\tilde{k}_n)x_{N-n}+\tilde{k}_n x_{N}+kx_{N-n-1}+\tilde{\eta}_{N-n}\label{alto}\\
\gamma_n^b\dot{x}_{N}+\tilde{\gamma}_n\dot{x}_{N-n}&=-\tilde{k}_nx_{N}+\tilde{k}_n x_{N-n}+f+\tilde{\eta}_{N}^{(n)}\label{basso}\,,
\end{align}
we can solve the first to get recurrence equations:
\begin{equation}
\begin{split}
x_{N-n}&= \frac{\tilde{k}_n}{k+\tilde{k}_n}x_N+\frac{k}{k+\tilde{k}_n}x_{N-n-1}-\left(\frac{\gamma_n^a\tilde{k}_n}{(k+\tilde{k}_n)^2}+\frac{\tilde{\gamma}_n}{k+\tilde{k}_n}\right)\dot{x}_{N}\\&-\frac{\gamma_n^ak}{(k+\tilde{k}_n)^2}\dot{x}_{N-n-1}+\frac{\tilde{\eta}_{N-n}}{k+\tilde{k}_n}+o \left(\frac{\gamma}{k}\right)\,,
\end{split}
\end{equation}
and deriving:
\begin{equation}
\dot{x}_{N-n}=\frac{\tilde{k}_n}{k+\tilde{k}_n}\dot{x}_N+\frac{k}{k+\tilde{k}_n}\dot{x}_{N-n-1}+O\left(\frac{\gamma}{k}\right)\,.
\end{equation}
These last two expressions need to be used in the equation for the $(N-n-1)^\textrm{th}$ link and in equation (\ref{basso}) to define the recurrence relations:
\begin{align}
\begin{split}
& \left(\gamma+\frac{\gamma_n^ak^2}{(k+\tilde{k}_n)^2}\right)\dot{x}_{N-n-1}+\left(\frac{\gamma_n^a\tilde{k}_nk}{(k+\tilde{k}_n)^2}+\frac{\tilde{\gamma}_nk}{k+\tilde{k}_n}\right)\dot{x}_{N}=\\
-&\left(k+\frac{k\tilde{k}_n}{k+\tilde{k}_n}\right)x_{N-n-1}+\frac{k\tilde{k}_n}{k+\tilde{k}_n}x_N+kx_{N-n-2}+\left(\eta_{N-n-1}+\frac{k}{k+\tilde{k}_n}\tilde{\eta}_{N-n}\right)\,;\end{split}\\
\begin{split}
&\left(  \gamma_n^b+\frac{2\tilde{\gamma}_n\tilde{k}_n}{k+\tilde{k}_n}+\frac{\gamma_n^a\tilde{k}^2_n}{(k+\tilde{k}_n)^2}\right)\dot{x}_{N}+\left(\frac{\gamma_n^a\tilde{k}_nk}{(k+\tilde{k}_n)^2}+\frac{\tilde{\gamma}_nk}{k+\tilde{k}_n}\right)\dot{x}_{N-n-1}=\\
-&\frac{k\tilde{k}_n}{k+\tilde{k}_n}x_N+\frac{k\tilde{k}_n}{k+\tilde{k}_n}x_{N-n-1}+f+\left(\tilde{\eta}_{N}^{(n)}+\frac{\tilde{k}_n}{k+\tilde{k}_n}\tilde{\eta}_{N-n}\right)\,;\end{split}
\end{align}
and then:
\begin{align}
\tilde{k}_{n+1}&=\frac{k\tilde{k}_n}{k+\tilde{k}_n}\,;\label{rec1}\\
\gamma^a_{n+1}&=\gamma+\frac{\gamma_n^ak^2}{(k+\tilde{k}_n)^2}\,\label{rec2};\\
\tilde{\gamma}_{n+1}&=\frac{\gamma_n^a\tilde{k}_nk}{(k+\tilde{k}_n)^2}+\frac{\tilde{\gamma}_nk}{k+\tilde{k}_n}\,;\label{rec3}\\
\gamma^b_{n+1}&=\gamma_n^b+\frac{2\tilde{\gamma}_n\tilde{k}_n}{k+\tilde{k}_n}+\frac{\gamma_n^a\tilde{k}^2_n}{(k+\tilde{k}_n)^2}\,;\label{rec4}\\
\begin{split}
\langle\tilde{\eta}_{N-n-1}(t)\tilde{\eta}_{N-n-1}(0)\rangle&=\langle\eta_{N-n-1}(t)\eta_{N-n-1}(0)\rangle\\&+\frac{k^2}{(k+\tilde{k}_n)^2}\langle\tilde{\eta}_{N-n}(t)\tilde{\eta}_{N-n}(0)\rangle\,;
\end{split}\\
\begin{split}
\langle\tilde{\eta}_{N-n-1}(t)\tilde{\eta}_{N}^{(n+1)}(0)\rangle&=\frac{\tilde{k}_nk}{(k+\tilde{k}_n)^2}\langle\tilde{\eta}_{N-n}(t)\tilde{\eta}_{N-n}(0)\rangle\\&+\frac{k}{k+\tilde{k}_n}\langle\tilde{\eta}_{N-n}(t)\tilde{\eta}_{N}^{(n)}(0)\rangle\,;
\end{split}\\
\begin{split}
\langle\tilde{\eta}_{N}^{(n+1)}(t)\tilde{\eta}_{N}^{(n+1)}(0)\rangle&=\langle\tilde{\eta}_{N}^{(n)}(t)\tilde{\eta}_{N}^{(n)}(0)\rangle+\frac{2\tilde{k}_n}{k+\tilde{k}_n}\langle\tilde{\eta}_{N}^{(n)}(t)\tilde{\eta}_{N-n}(0)\rangle\\&+\frac{\tilde{k}^2_n}{(k+\tilde{k}_n)^2}\langle\tilde{\eta}_{N-n}(t)\tilde{\eta}_{N-n}(0)\rangle\,.
\end{split}
\end{align}
Which are quickly solved as:
\begin{align}
\tilde{k}_{n}&=\frac{k}{n}\,;\\
\gamma^a_{n}&=\frac{(2n+1)(n+1)}{6n}\gamma\,;\label{altouni}\\
\tilde{\gamma}_{n}&=\frac{(n+1)(n-1)}{6n}\gamma\,;\\
\gamma^b_{n}&=\frac{(2n+1)(n+1)}{6n}\gamma\,.
\end{align}
This enables us to rewrite equations (\ref{alto}) and (\ref{basso}) as:
\begin{align}
\begin{split}
\frac{(2n+1)(n+1)}{6n}\gamma\dot{x}_{N-n}&+\frac{(n+1)(n-1)}{6n}\gamma\dot{x}_{N}=-\left(k+\frac{k}{n}\right)x_{N-n}\\
&+\frac{k}{n} x_{N}+kx_{N-n-1}+\tilde{\eta}_{N-n}\,;\label{alto1}
\end{split}
\\
\begin{split}
\frac{(2n+1)(n+1)}{6n}\gamma\dot{x}_{N}&+\frac{(n+1)(n-1)}{6n}\gamma\dot{x}_{N-n}=-\frac{k}{n}x_{N}\\&+\frac{k}{n} x_{N-n}+f+\tilde{\eta}_{N}^{(n)}\label{basso1}\,.
\end{split}
\end{align}
Substitution of equation (\ref{ennemenouno}) in equation (\ref{alto1}) yields:
\begin{equation}
\begin{split}
\frac{2Nn(N-n)+N}{6n(N-n)}\gamma\dot{x}_{N-n}&
+\frac{(n+1)(n-1)}{6n}\gamma\dot{x}_{N}=-\left(\frac{k}{N-n}+\frac{k}{n}\right)x_{N-n}\\&+\frac{k}{n}x_{N} +\frac{N-n-1}{N-n}\eta^\prime_{N-n-1}+\tilde{\eta}_{N-n}\label{alto2}
\end{split}
\end{equation}
This defines a system of two coupled equations for $x_N$ and $x_{N-n}$ which cannot in general be decoupled because the coefficient matrices of $\left( \begin{array}{c}
\dot{x}_{N-n}\\
\dot{x}_{N}
\end{array}\right)$ and  $\left( \begin{array}{c}
x_{N-n}\\
x_{N}
\end{array}\right)$ are not proportional to one another.
\begin{equation}
\begin{split}
&\left( \begin{array}{cc}
\frac{2Nn(N-n)+N}{6n(N-n)}&\frac{(n+1)(n-1)}{6n}\\
\frac{(n+1)(n-1)}{6n}&\frac{(2n+1)(n+1)}{6n}
\end{array}\right)\gamma
\left( \begin{array}{c}
\dot{x}_{N-n}\\
\dot{x}_{N}
\end{array}\right)=\\-&
\left( \begin{array}{cc}
\frac{1}{N-n}+\frac{1}{n}&-\frac{1}{n}\\
-\frac{1}{n}&\frac{1}{n}
\end{array}\right)k
\left( \begin{array}{c}
x_{N-n}\\
x_{N}
\end{array}\right)+\left( \begin{array}{c}
0\\
f
\end{array}\right)+\left( \begin{array}{c}
\bar{\eta}_{N-n}\\
\tilde{\eta}_{N}^{(n)}
\end{array}\right)\,,\label{matriceuni}
\end{split}
\end{equation}
where $\bar{\eta}_{N-n}=\frac{N-n-1}{N-n}\eta^\prime_{N-n-1}+\tilde{\eta}_{N-n}$.\\
These equations can be rewritten in the large $N$ limit as
\begin{equation}
\begin{split}
&\left( \begin{array}{cc}
\frac{1}{3}&\frac{\alpha }{6}\\
\frac{\alpha }{6}&\frac{\alpha }{3}
\end{array}\right)N\gamma
\left( \begin{array}{c}
\dot{x}_{N(1-\alpha)}\\
\dot{x}_{N}
\end{array}\right)=\\-&
\left( \begin{array}{cc}
\frac{1}{1-\alpha}+\frac{1}{\alpha}&-\frac{1}{\alpha }\\
-\frac{1}{\alpha }&\frac{1}{\alpha }
\end{array}\right)\frac{k}{N}
\left( \begin{array}{c}
x_{N(1-\alpha)}\\
x_{N}
\end{array}\right)+\left( \begin{array}{c}
0\\
f
\end{array}\right)+\left( \begin{array}{c}
\bar{\eta}_{N(1-\alpha)}\\
\tilde{\eta}_{N}^{(\alpha N)}
\end{array}\right)\,,
\end{split}
\end{equation}
where we have defined $\alpha=\frac{n}{N}$.
\subsection{Scaling of a non-homogeneous Rouse Polymer}
\begin{figure}[htpb]
\includegraphics[width=\textwidth]{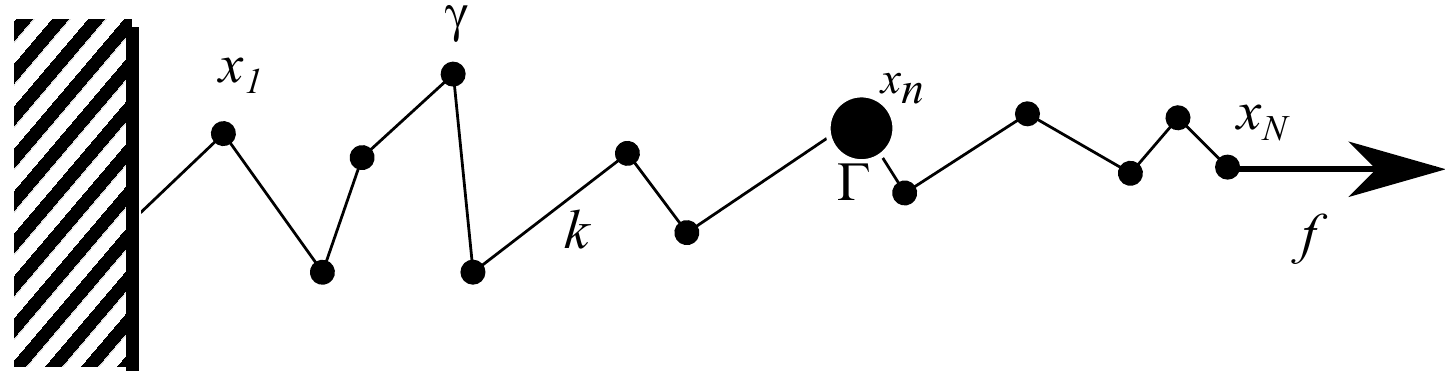}
\caption{A non-homogeneous Rouse polymer composed of $N$ identical springs and $N-1$ beads each having spring constant $k$ and viscous drag coefficient $\gamma$. The $n^\textrm{th}$ bead is taken to have viscous drag coefficient $\Gamma$. The coordinates $x_i$ are taken along the direction of the pulling force $f$. The node with higher viscosity represents the DNA hairpin to be opened in a typical experiment.}\label{fig:rousehairpin}
\end{figure}

\paragraph{The effect of a  single intermediate dishomogeneity}
Let us now suppose that one of the links that compose our polymer has a much greater viscosity than its neighbours, which we leave homogeneous. We wish to investigate this kind of setup because it will give us some insight on how the attached DNA hairpin affects the fluctuations of the linkers and whether or not it decorrelates them.\\
What we are planning to do is to write two coupled equations as in equation (\ref{matriceuni}), namely for the $N^\textrm{th}$ and the $(N-n)^\textrm{th}$ links when the $(N-n+1)^\textrm{th}$ has a much greater viscosity than the others. In what follows we will indicate with $\Gamma$ as opposed to $\gamma$ the viscosity of the different link. The setup is shown in figure \ref{fig:rousehairpin}.\\
Looking at equations (\ref{rec1}-\ref{rec4}) it is immediately apparent that the only one which involves the viscosity of an intermediate link is the one for $\gamma^a_{n}$, namely equation (\ref{rec2}). Retracing the steps that brought us to equations (\ref{matriceuni}), we have to correct equation (\ref{altouni}) as:
\begin{equation}
 \hat{\gamma}^a_{n}=\frac{(2n-1)(n-1)}{6n}\gamma+\Gamma\,.\\
\end{equation} 
This change propagates in equation (\ref{alto1}) but not in equation (\ref{basso1}), and in turn equation (\ref{alto2}) becomes:
\begin{equation}
 \begin{split}
&\left(\frac{2n(N-3)(N-n)+N}{6n(N-n)}\gamma+\Gamma\right)\dot{x}_{N-n}
+\frac{(n+1)(n-1)}{6n}\gamma\dot{x}_{N}=\\-&\left(\frac{k}{N-n}+\frac{k}{n}\right)x_{N-n}+\frac{k}{n}x_{N}+\frac{N-n-1}{N-n}\eta^\prime_{N-n-1}+\hat{\eta}_{N-n}\,;\end{split}
\end{equation}
we have to underline that the second noise term has also changed in order to fullfill fluctuation-dissipation relations.\\
The correlation matrix of the noise terms in equation (\ref{matriceuni}) becomes then:
\begin{equation}
\left( \begin{array}{cc}
\frac{2n(N-3)(N-n)+N}{6n(N-n)}\gamma+\Gamma&\frac{(n+1)(n-1)}{6n}\gamma\\
\frac{(n+1)(n-1)}{6n}\gamma&\frac{(2n+1)(n+1)}{6n}\gamma
\end{array}\right)\,,
\end{equation}
which can be rewritten in a clearer form in the limit of $N\to\infty$ with $\frac{n}{N}=\alpha$ as:
\begin{equation}
\left( \begin{array}{cc}
\frac{N}{3}\gamma+\Gamma&\frac{\alpha N}{6}\gamma\\
\frac{\alpha N}{6}\gamma&\frac{\alpha N}{3}\gamma
\end{array}\right)\,.
\end{equation}
\paragraph{Block polymers}
Suppose we have a polymer composed of two sections: the first composed of $n_1$ links of viscosity $\gamma_1$ and elasticity $k_1$, the second of $n_2$ links of viscosity $\gamma_2$ and elasticity $k_2$. In close resemblance with what we did before we ask ourselves how this modifies the equation for the effective evolution of the floating extremity and of the point where the two sections are linked.\\
It is important to know this because most DNA unzipping experiments so far have relied on linkers of both single- and double-stranded DNA bonded in heteropolymers of various lengths.\\
\begin{figure}[htpb]
\includegraphics[width=\textwidth]{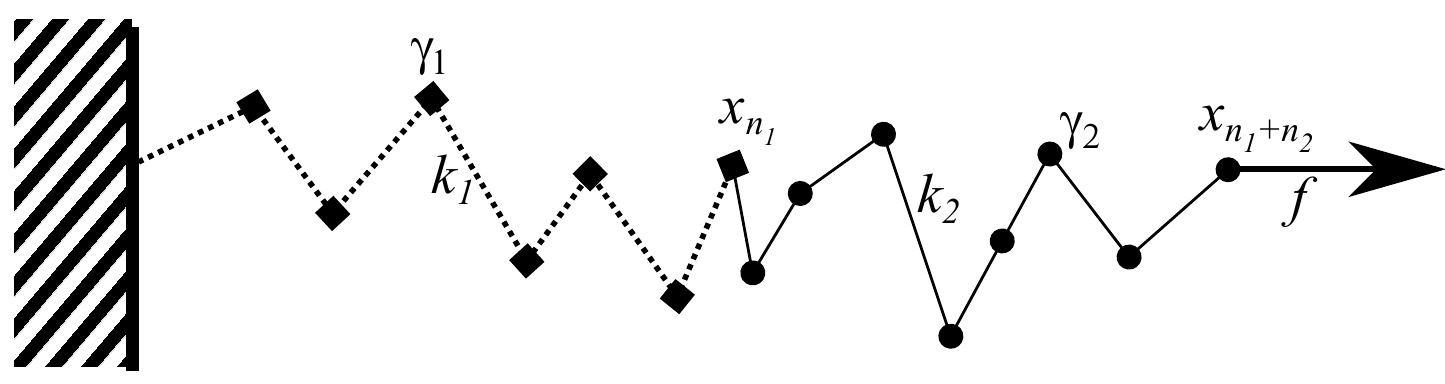}
\caption{A non-homogeneous Rouse polymer composed of $n_1$ identical springs and beads each having spring constant $k_1$ and viscous drag coefficient $\gamma_1$and $n_2$ more identical springs and beads each having spring constant $k_2$ and viscous drag coefficient $\gamma_2$. The coordinates $x_i$ are taken along the direction of the pulling force $f$. One can imagine momoers of type 1 to be ssDNA and those of type 2 to be dsDNA.}\label{fig:rouseblock}
\end{figure}
Equation (\ref{ennemenouno}) involves only links of the first tipe and can thus be easily rewritten with the additional index. We may think that equations (\ref{alto1}) and (\ref{basso1}) share the same fate but a different index; unfortunately this is true only of the elasticities. The viscosity of the link that connects the first and the second section (that is the $n_1^\textrm{th}$) is of the first type. Equations (\ref{alto1}) and (\ref{basso1}) then become:
\begin{align}
\begin{split}
\left(\frac{(2n_2-1)(n_2-1)}{6n_2}\gamma_2+\gamma_1\right) \dot{x}_{n_1}+\frac{(n_2+1)(n_2-1)}{6n_2}\gamma_2\dot{x}_{n_1+n_2}&=\\-\left(k_2+\frac{k_2}{n_2}\right)x_{n_1}+\frac{k_2}{n_2} x_{n_1+n_2}+k_1x_{n_1-1}&+\tilde{\eta}_{n_1}\label{alto3}
\end{split}\\
\begin{split}
\frac{(2n_2+1)(n_2+1)}{6n_2}\gamma_2\dot{x}_{n_1+n_2}+\frac{(n_2+1)(n_2-1)}{6n_2}\gamma_2\dot{x}_{n_1}&=\\-\frac{k_2}{n_2}x_{n_1+n_2}+\frac{k_2}{n_2} x_{n_1}+f&+\tilde{\eta}_{n_1+n_2}^{(n_2)}\label{basso3}\,.
\end{split}
\end{align}
Putting all back together gives us two matrices: one for the $\gamma$'s and the other for the $k$'s:
\begin{equation}
\left( \begin{array}{cc}
\frac{(2n_1+1)(n_1+1)}{6n_1}\gamma_1+\frac{(2n_2-1)(n_2-1)}{6n_2}\gamma_2&\frac{(n_2+1)(n_2-1)}{6n_2}\gamma_2\\
\frac{(n_2+1)(n_2-1)}{6n_2}\gamma_2&\frac{(2n_2+1)(n_2+1)}{6n_2}\gamma_2
\end{array}\right)\,,
\end{equation}
\begin{equation}
\left( \begin{array}{cc}
\frac{k_1}{n_1}+\frac{k_2}{n_2}&-\frac{k_2}{n_2}\\
-\frac{k_2}{n_2}&\frac{k_2}{n_2}
\end{array}\right)\,,
\end{equation}
the former can be rewritten in the limit of $n_1,n_2$ large as:
\begin{equation}
\left( \begin{array}{cc}
\frac{n_1}{3}\gamma_1+\frac{n_2}{3}\gamma_2&\frac{n_2}{6}\gamma_2\\
\frac{n_2}{6}\gamma_2&\frac{n_2}{3}\gamma_2
\end{array}\right)\,.
\end{equation}
\paragraph{Validity of the approximation}
In the beginning of this section we have stated that the microscopic time-scale for ssDNA is given by $\gamma/k=10^-{10}$ s. Now by looking at the scaling of the mesoscopic timescale we'll obtain the range of validity of our aproximation, that is the timescale at which a polymer will continue to behave as a single entity and the propagation time along the polymer will be negligible. The scaling of the macroscopic time is proportional to $\gamma/k N^2/3$ where $N$ is the number of monomers.\\
Given that as of today the state of the art in experiments the maximum sampling frequency is of the order of 10 kHz, polymers larger than 1000 base pairs have relaxation times that are observable.\\
To take this into account in mesoscopic simulations we can split long polymer into smaller pieces even though this doesn't appear to have an appreciable effect on measured relaxation times. The interested reader should refer to section 5.1 of \cite{IO1}. \\
Moreover in \cite[Appendix A]{IO1}, a more formal discussion of the normal modes
of a single homogeneous polymer is given. It turns out that the factor $1/3$ that appears in our equations is an approximation of the true factor $\pi
^2/4$ that would appear in an exact treatment. Here we have preferred to give this approximated result is the only one that can yield the scaling for the off-diagonal terms, and extends well to non-homomgeneous polymers.\\
\subsection{Detailed balance}
Now that we have described all the different pieces of the dynamics of DNA unzipping we would like to derive a coupled mesoscopic dynamics that respects detailed balance equations with the right thermodynamic equilibrium distribution:
\begin{equation}
P(n,x)=e^{-\beta W(x,n)-\beta G_0(n;B)}/Z\,,
\end{equation}
where $G_0(n;B)=\sum_i^n g_0(b_i,b_{i+1})$ is the binding energy of the fork, and $W(x,n)$ is the free energy of the linkers, the beads and the traps, but we need not concentrate on the details for now.\\
This is not a trivial task because we have to take into account the coupling between a continuous time Markov chain (the fork dynamics $n$), and the Brownian dynamics of the polymers and the beads.\\
Let us first identify the possible events at each time step, the fork can either open, close or stay where it is at each time step, and the $x$ variable will perform a Langevin step of size $\Delta x$.\\
We have identified three transitions that correspond to three detailed balance equations:
\begin{align}
 P(n,x) H_\textrm{o}(x,n,\Delta x) &= P(n+1,x+\Delta x) H_\textrm{c}(n+1,x+\Delta x ,-\Delta x)\,;\label{eq:open}\\
 P(n,x) H_\textrm{c}(x,n,\Delta x) &= P(n-1,x+\Delta x) H_\textrm{o}(n-1,x+\Delta x,-\Delta x) \,;\label{eq:close}\\
 P(n,x) H_\textrm{s}(x,n,\Delta x) &= P(n,x+\Delta x) H_\textrm{s}(n,x+\Delta x,-\Delta x) \label{eq:stay}\,;
\end{align}
where o, c and s denote respectively open, close and stay, and $H$ are the transition rates.\\
If we now suppose, as we have discussed previously, that the opening rate depends exclusively on the binding energy, and we further impose it to be a product of the opening rate and a Langevin step we get:
\begin{equation}
\begin{split}
H_\textrm{o}(x,n,\Delta x) &= r \Delta t \, e^{\beta G(n;B)-\beta G(n+1;B)}\\&\times\sqrt{\frac{4 \pi T \Delta t}{\gamma_n}} 
\exp\left[ -\frac{\gamma_n}{4 T \Delta t} \left( \Delta x - \frac{f(x,n) \Delta t}{\gamma_n}  \right)^2 \right]\,,
\end{split}
\end{equation}
that is consistent with the definition of $r_\textrm{o}$ defined previously if we integrate over $\Delta x$.\\
This, in conjunction with equation (\ref{eq:close}) gives the closing rate:
\begin{equation}
\begin{split}
H_\textrm{c}&(x,n,\Delta x) = r \Delta t \, e^{\beta W(x,n)-\beta W(x+\Delta x,n-1)} \, \\ &\times \sqrt{\frac{4 \pi T \Delta t}{\gamma_{n-1}}} 
\exp\left[ -\frac{\beta\gamma_{n-1}}{4  \Delta t} \left( \Delta x + \frac{f(x+\Delta x,n-1) \Delta t}{\gamma_{n-1}}  \right)^2 \right]\,.
\end{split}
\end{equation}
The problem is that this rate depends on quantities computed both in $x$ and $x+\Delta x$ and it is not Gaussian for general $f$. On the other hand if we impose $f(x,n)=-\frac{\partial W}{\partial x}$ and we perform a Taylor of the terms that are calculated in $x+\Delta x$ expansion at the exponent we get for the $W$ part:
\begin{equation}
\begin{split}
 &W(x,n)- W(x+\Delta x,n-1)=\\
 &W(x,n)- W(x,n-1)+W(x,n-1)- W(x+\Delta x,n-1)=\\
&W(x,n)- W(x,n-1)+f(x,n-1)\Delta x -\frac{\partial^2 W(x,n-1)}{\partial x^2}(\Delta x)^2+O((\Delta x)^3)\,;
\end{split}
\end{equation}
and for the Brownian step:
\begin{equation}
\begin{split}
-&\frac{\gamma_{n-1}}{4  \Delta t} \left( \Delta x + \frac{f(x+\Delta x,n-1) \Delta t}{\gamma_{n-1}}  \right)^2=\\
- &\frac{\gamma_{n-1}}{4  \Delta t} \left( \Delta x - \frac{f(x+\Delta x,n-1) \Delta t}{\gamma_{n-1}}  \right)^2-f(x+\Delta x,n-1)\Delta x=\\
-& \frac{\gamma_{n-1}}{4  \Delta t} \left( \Delta x -\frac{f(x,n-1) \Delta t}{\gamma_{n-1}}  \right)^2-f(x,n-1)\Delta x+\frac{\partial^2 W(x,n-1)}{\partial x^2}(\Delta x)^2\\+&O(\Delta t \Delta x)+O((\Delta x)^3)\,.
\end{split}
\end{equation}
Now we only have to notice that terms up to and including order $\Delta x$ cancel out and that for Brownian motion $\Delta t\sim (\Delta x)^2$, to see we can rewrite the rate as:
\begin{equation}
\begin{split}
H_\textrm{c}&(x,n,\Delta x) = r \Delta t \, e^{\beta W(x,n)-\beta W(x,n-1)} \, \\ &\times \sqrt{\frac{4 \pi T \Delta t}{\gamma_{n-1}}} 
\exp\left[ -\frac{\beta\gamma_{n-1}}{4  \Delta t} \left( \Delta x - \frac{f(x+\Delta x,n-1) \Delta t}{\gamma_{n-1}}  \right)^2 \right]\,,
\end{split}
\end{equation}
which is now consistent with the definition of $r_\textrm{c}$ by integrating over $\Delta x$.\\
The attentive reader should note that the force in the Brownian step is computed in $n-1$, that is once the base has been closed, this has important consequences on the implementation of the algorithm.\\
Finally, the rate at constant $n$ is obtained by imposing that:
\begin{equation}
\int d\Delta x H_\textrm{s}(x,n,\Delta x)+H_\textrm{o}(x,n,\Delta x)+H_\textrm{c}(x,n,\Delta x) = 1\,,
\end{equation}
that is:
\begin{equation}\begin{split}
H_\textrm{s}&(x,n,\Delta x) = [1-r_\textrm{o}(x,n)-r_\textrm{c}(x,n)] \, \\ & \times  \sqrt{\frac{4 \pi T \Delta t}{\gamma_{n}}}
\exp\left[ -\frac{\gamma_{n}}{4 T \Delta t} \left( \Delta x - \frac{f(x,n) \Delta t}{\gamma_{n}}  \right)^2 \right] \,,
\end{split}\end{equation}
Now the algorithm can be summarized:
\begin{verbatim}
p=randomreal()
if(p<r_open(n,x)){
    x+=f(x,n)*dt/gamma+randomgaussian()*sqrt(2*beta*dt*gamma)
    n++
}
else if(p<r_open(n,x)+r_close(n,x)){
    n--
    x+=f(x,n)*dt/gamma+randomgaussian()*sqrt(2*beta*dt*gamma)
}
else{
    x+=f(x,n)*dt/gamma+randomgaussian()*sqrt(2*beta*dt*gamma)
}
\end{verbatim}
Note how the order of the Brownian step and the opening or closing is reversed, as we have underlined before this is essential to the satisfaction of detailed balance equations.
\section{Results from the dynamical model}
\begin{figure}[htbp]
\begin{center}
\includegraphics[width=.8\textwidth]{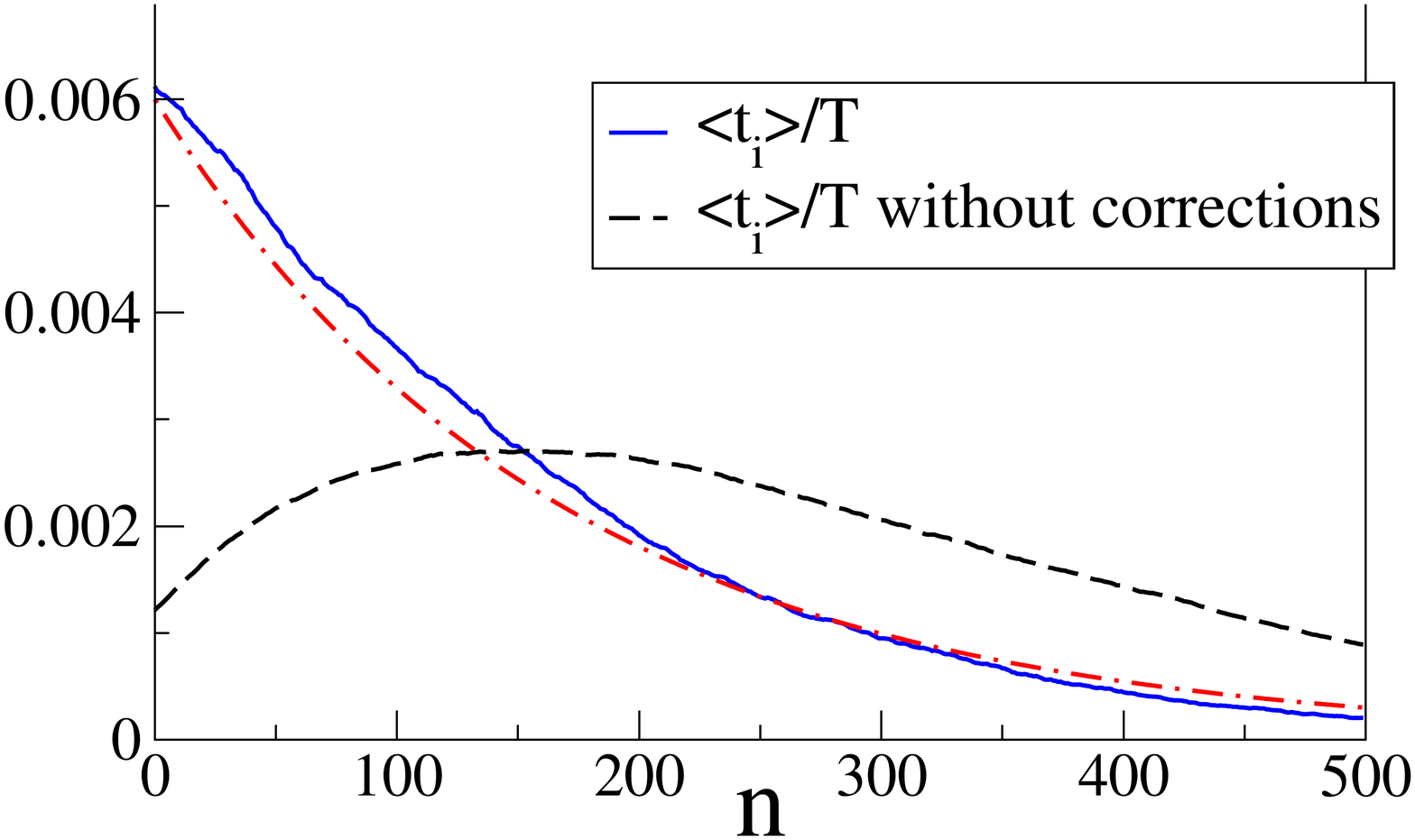}
\caption{Average fraction of 
time spent on each base. 
The full (blue) curve corresponds
to Eq.~(\ref{freefinal}) while the dashed (black) curve
corresponds Eq.~(\ref{freefinal}) without the saddle-point 
corrections (the square-root term).
The dot-dashed (red) line is 
$P_\textrm{eq}(n) \propto \exp [ - n \, \Delta g ]$
with $\Delta g = 0.006$. $n$ is the number of open bases.}\label{fig:correzione}
\end{center}
\end{figure}

\begin{figure}[htbp]
\begin{center}
\includegraphics[width=.6\textwidth]{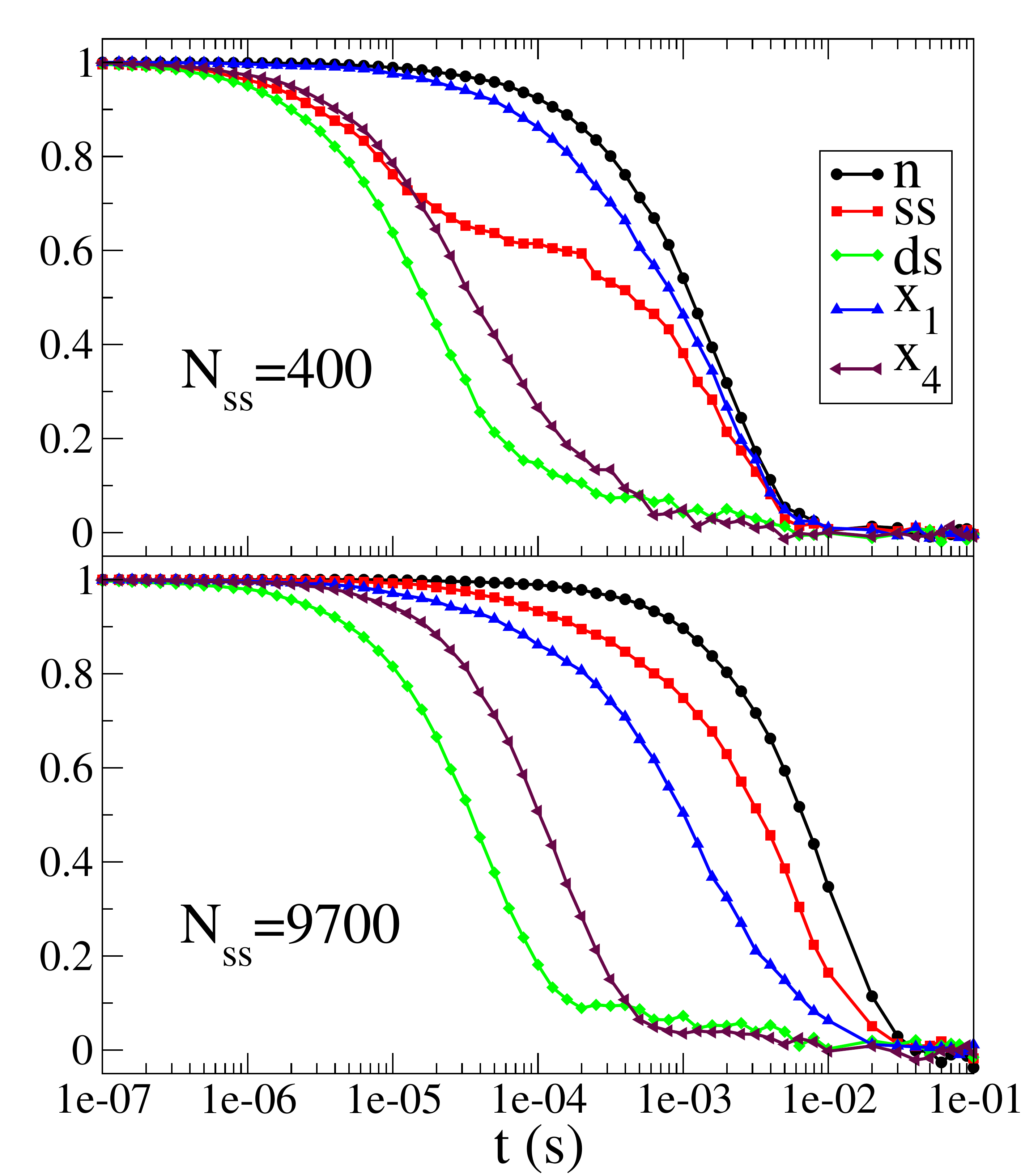}
\caption{
Correlation functions for the setup in
figure~\ref{fig:setup}A at two different values of
the number of open bases, $N_\textrm{eq}=40 +n$. ss and ds indicate the autocorrelation functions of the ssDNA and dsDNA linkers and $x_1$ and $x_4$ are the autocorrelation functions of two optical traps of different stiffnesses ($x_4$ being the stiffest).
}
\label{fig:corren}
\end{center}
\end{figure}

We have spent the best part of the previous chapter defining an effective mesoscopic dynamical model for DNA micromanipulation experiments. Our approach is much more complex than separately simulating fork and polymer dynamics: first because it does not imply equilibrium and secondly because it allows for cross-correlation effects between fork, beads and polymers dynamics.\\
In this section we wish to turn to the novel measurements that we have been able to perform thanks to this software and that were published in \cite{IO1}.\\
First of all we have observed that for complex polymers the expression $W(l)=Nw(l)$ for the free energy breaks down at low $N$. This was immediately clear when we observed that the measured sojourn times did not match the theoretical prediction from the Boltzmann distribution, however, the effect is much smaller even simply adding the nonlinear dependence in $N$ coming from the square root term in:
\begin{equation}
e^{-\beta W(x,n)}=e^{-\beta N w(x/N)}\sqrt{\frac{\beta k(x/N)\ell^2}{2\pi N}}\,,\label{freefinal}
\end{equation}
where $\ell$ is a dimensional constant of no importance, and $k$ was defined previously as the second derivative of $w$ with respect to $l=x/N$.\\
In figure \ref{fig:correzione} we show the effect of the square-root term in the case of an uniform sequence: the time spent on a basis is obtained by simulation with and without the square-root term and by its Boltzmann estimate.\\
Another set of quantities which is in general not available from first principles computations are correlation functions, in \cite{IO1} we have studied in detail the dependence of the correlation functions on the number of open bases and on the length of the linkers in various experimental setups in order to determine the importance of out of equilibrium effects such as propagation times.\\
In the following we will concentrate on a setup similar to that used in Bockelmann's lab at ESPCI: two optical traps of stiffness $0.1$ pN/nm and $0.512$ pN/nm respectively, a dsDNA linker of 3120 bases and a ssDNA linker of 40 bases.\\
We have found that polymers which are shorter than 1000 bases show no appreciable effect due to finite relaxation times. For longer polymers we have devised a way of introducing the propagation effect: we cut up the polymer in pieces which are at most 1000 bases long and we simulate them separately.\\
This is shown in figure \ref{fig:corren}. However we have found this to have an effect on the shape of the correlation function, but not on the correlation time. In fact, if the correlation function is fit with a stretched exponential of the form: $\exp[-(t/\tau)^\beta]$, $\beta$ is slightly smaller.\\
\begin{figure}[htbp]
\begin{center}
\includegraphics[width=.7\textwidth]{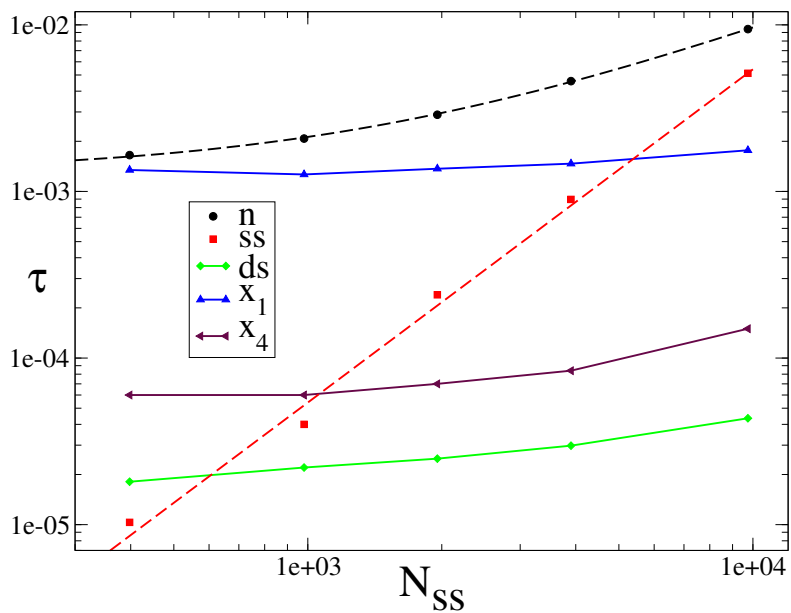}
\caption{
Relaxation times of the correlation functions
in figure \ref{fig:corren} as a function
of the number of open bases. In the case of the single strand (ss), only the fast relaxation time is plotted. 
For the fork and the single strand, 
dashed lines indicate a fit to $\tau_n = A + B N_\textrm{eq}$ 
(with $A=1.3 \cdot 10^{-3}$ and $B=8.4 \cdot 10^{-7}$)  
and $\tau_\textrm{eq} = C N_\textrm{eq}^2$ (with $C=5.4 \cdot 10^{-11}$ s). 
For the others, full lines are guides to the eye.}
\label{fig:times}
\end{center}
\end{figure}We have also been able to study the dependence of the relaxation time for different parts of the setup as a function of the number of open bases and to compare those with theoretical results. This is shown in table \ref{tab:confronto} and in figure \ref{fig:times}.\\
The results are in very good numerical agreement except for the two beads: it turns out that the relationship between the bead and the number of open bases is more subtle than we thought. It appears that there is a very strong correlation between the fork and the bead and this effect is stronger when the optical trap is softer.\\
\begin{table}
\begin{center}
% use packages: array
\begin{tabular}{|l|l|l|}
 \hline& Theoretical (s) & Numerical (s) \\ \hline
Single strand & $4.83\cdot 10^{-11}N_\textrm{eq}^2$ & $5.4\cdot 10^{-11} N_\textrm{eq}^2$\\ \hline
Double strand & $4.96\cdot10^{-5}$ & $\sim 3\cdot 10^{-5}$ \\ \hline
Spring $x_1$ & $1.67 \cdot 10^{-4}$ & $\sim 1.5\cdot 10^{-3}$ \\ \hline
Spring $x_4$ & $3.26 \cdot 10^{-4}$ & $\sim 7 \cdot10^{-5}$ \\ \hline
Fork $N_\textrm{eq}$ & $\propto 14.2 + 0.013 N_\textrm{eq}$& $1.3 \cdot 10^{-3}+8.4 \cdot 10^{-7} N_\textrm{eq}$  \\ \hline
\end{tabular}
\end{center}
\caption{Comparison between the correlation times of the setup in figure~\ref{fig:setup}A
as computed for an isolated element 
and the result of a complete numerical simulation.
In the case of the fork, we reported as theoretical value $1/k_\textrm{eff}$, that must be
multiplied by a viscosity to obtain the relaxation time; it turns out that a viscosity
$\sim 8 \cdot 10^{-5}$ pN s/nm matches the theoretical and numerical results.
}\label{tab:confronto}
\end{table}This effect is desirable, it is in fact the effect that allows us to gain information on the sequence. To better quantify the relationship between the stiffness of the trap and the correlation of the bead with the fork we have defined the quantity:
\begin{equation}
 I(x_4,n)=\sum_n\int dx_4 P(x_4,n)
 \log\left(\frac{P(x_4,n)}{P(x_4)P(n)}\right)\,,
\end{equation}
as the mutual information between the fork and one of the two beads.\\
In figure \ref{fig:entropy} we show the effect on the stiffness of the optical trap on the mutual information between the fork position and the bead position. We find that softer beads yield more information on the sequence. This can be intuitively understood by thinking that a softer trap gives way more easily to the excess length deriving from the opening of a base.\\
It must be stressed, however, that this result holds only per measure, that is if one wanted to know if it were more efficient to have more rigid traps in an experiment one should take into account the autocorrelation times of the bead position. Those are in fact lower for stiffer traps allowing for a larger number of statistically independent measures per unit time.
\begin{figure}[htbp]
\begin{center}
\includegraphics[width=0.7 \textwidth]{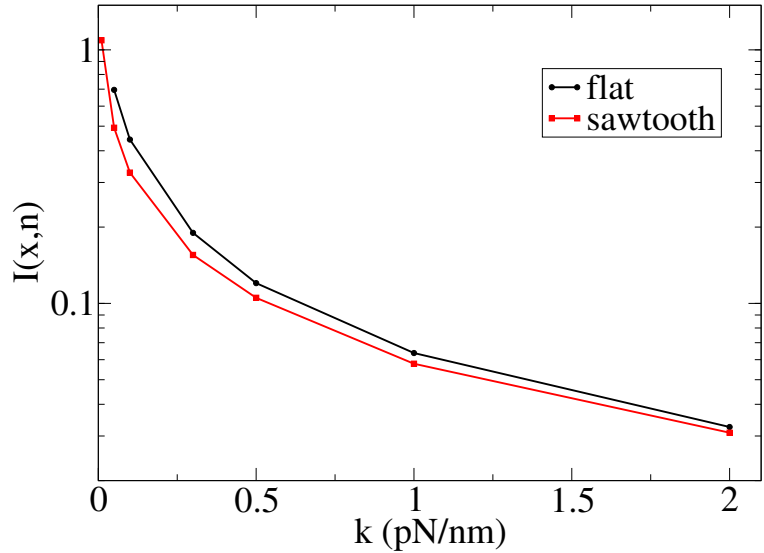}
\caption{Mutual information $I$ between $x_4$ and $n$ as a function of
the trap stiffness,  $k$. Black circles are computed on an uniform
  sequence, while red squares are measured on a sawtooth potential derived from a sequence that alternates stretches of 10 weak bases and stretches of 10 strong bases.}
\label{fig:entropy}
\end{center}
\end{figure}
\chapter{Inferring the DNA sequence}
\label{infedna}
As we have shown in the previous section, DNA unzipping experiments show a remarkable dependence on sequence in the force-extension signal. Several attempts have been made to reconstruct the free energy landscape from different experimental setups \cite{DANI,WOODSIDE,HUGUET2}.\\
In this section we will concentrate on algorithmic and mathematical approaches to solving the inverse problem, that is characterizing the free energy landscape as a function of $n$ and eventually sequencing DNA.\\
Idealized cases, where the number of open bases $n$ is known at all time, are relatively easy to solve, but once we start adding the layers of complexity of real experiments, it becomes really difficult to extract information about the sequence.\\
The first algorithm we will describe is based on the very idealized situation we have just described: infinite sampling frequency and knowledge of the number of open bases.\\
The second supposes we can access the equilibrium value of physical quantities like the position of the beads with arbitrary precision, ignoring fluctuations. This is much more realistic than the first approach, but results are much farther from reconstructing the sequence.\\
Lastly we will take a look at a toy model based on the Ornstein-Uhlenbeck model, that takes into account correlations and finite sampling frequencies.
\section{Infinite bandwidth algorithm}
\begin{figure}[htbp]
\begin{center}
\includegraphics[width=\textwidth]{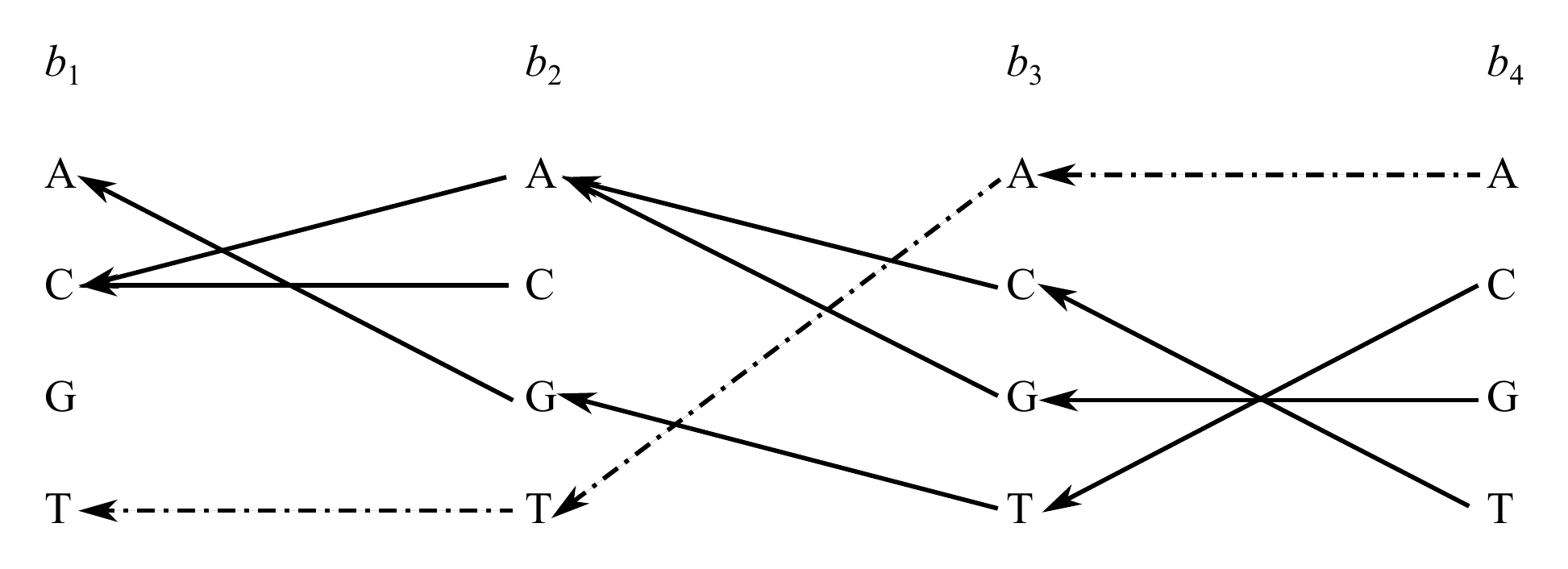}
\caption{We start by choosing $b^{\max}_1(b_2)$ which amounts to choosing the best $b_1$ for each choice of $b_2$ and can be represented by an arrow going from $b_2$ to $b_1$ and then we iterate the procedure until we get to $b_N$ (here $N=4$). It is now possible to compute the optimum $b_N$, in this case A and propagate back to obtain the optimal sequence TTAA.}\label{fig:viterbi}
\end{center}
\end{figure}

In what follows we suppose we have access to the results of a fixed-force experiment where the position of the fork is known at all times. Since this is not a realistic situation, the data on which to perform the inference must be simulated.\\
Let us suppose we have perfect knowledge at all times of the number of open bases. It is clear that this is not realistic at all: first of all because the number of open bases $n$ is not directly measurable and secondly because in order to obtain bandwidths that are large compared to the elementary event time-scale we would need a resolution of the order of the MHz or more and current experimental setups allow for resolutions three orders of magnitude smaller.\\
The details of what follows were first published in \cite{Bal06,BAL07}.\\
In the previous chapter we have defined the opening and closing rate: respectively $r_\textrm{o}(n)$ and $r_\textrm{c}(f)$. For small enough time intervals $\Delta t$ we can write:
\begin{equation}
P(n(t+\Delta t)|n(t))=\left\{\begin{array}{ll}
\Delta t r_\textrm{o}(n(t))&\textrm{for }n(t+\Delta t)=n(t)+1\,;\\
\Delta t r_\textrm{c}(f)&\textrm{for }n(t+\Delta t)=n(t)-1\,;\\
1-\Delta t r_\textrm{o}(n(t))-\Delta t r_\textrm{c}(f)&\textrm{for }n(t+\Delta t)=n(t)\,;\\
o(\Delta t)&\textrm{otherwise.}
\end{array}\right.
\end{equation}
This defines completely the transition probabilities from one state to another, and it can be used to define the probability of a the outcome of an experiment, that is of a complete trace. In order to do so we must define the relevant variables:
\begin{itemize}
\item $t_n$ the total time spent with $n$ open bases;
\item $u_n$ the number of transitions from $n$ to $n+1$;
\item $d_n$ the number of transitions from $n$ to $n-1$.
\end{itemize}
Given those definitions one can write the probability of an experimental trace as $\mathcal{T}$, conditioned on the sequence $\mathcal{B}$ and on the external force $f$, as:
\begin{equation}
\begin{split}
P(\mathcal{T}|\mathcal{B})&=\prod_n(\Delta t r_\textrm{o}(n(t)))^{u_n}(\Delta t r_\textrm{c}(f))^{d_n}(1-\Delta t r_\textrm{o}(n(t))-\Delta t r_\textrm{c}(f))^{t_n/\Delta t}\\
&=C(\mathcal{T})\prod_nM(b_n,b_{n+1};u_n,t_n)\,.
\end{split}
\end{equation}
where we have separated the part that depends on the sequence from that who does not, thus defining:
\begin{align}
C(\mathcal{T})&=(\Delta t)^{u+d}\exp(-t_\textrm{tot} r_\textrm{c}(f));\\
M(b_n,b_{n+1};u_n,t_n)&=\exp\left(g_0(b_n,b_{n+1})u_n-re^{g_0(b_n,b_{n+1})}t_n\right);\\
\end{align}
where we have used the definition of $r_\textrm{o}$ and we have defined $u=\sum_n u_n$, $d=\sum_n d_n$, and $t_\textrm{tot}=\sum_n t_n$.\\
Now we can use Bayes' theorem to compute the probability of a sequence given a trace:
\begin{equation}
P(\mathcal{B}|\mathcal{T})=\frac{P(\mathcal{T}|\mathcal{B})P(\mathcal{B})}{P(\mathcal{T})}\,.
\end{equation}
We can further assume (though it is not generally true) that all sequences are equiprobable that is $P(\mathcal{B})$ is uniform, this will lead us to a first rough estimate of the sequence given a trace.\\
We can maximize the expression we have given for $P(\mathcal{T}|\mathcal{B})$ over the $g_0(b_n,b_{n+1})$ without imposing that it can only take ten values to get a maximum likelihood estimate:
\begin{equation}
g_0(b_n,b_{n+1})=\log\left(\frac{u_n}{rt_n}\right)\,,
\end{equation}
This computation is not bad as a first estimate, but it amounts to searching in a continuous space when we effectively have only 4 possible values for a base. In order to find the most likely sequence $\mathcal{B}^*$ we can use the Viterbi algorithm \cite{VITERBI, MACKAY}.\\
The procedure is as follows: let us consider the first two bases and let us define $P_2(b_2)=\max_{b_1}M(b_1,b_2;u_1,t_1)$, then $b^{\max}_1(b_2)=\arg\max_{b_1}M(b_1,b_2;u_1,t_1)$, and for $n\neq1$ we can write:
\begin{align}
P_{n+1}(b_{n+1})&=\max_{b_n}M(b_n,b_{n+1};u_n,t_n)P_n(b_n)\,;\\
b^{\max}_n(b_{n+1})&=\underset{b_n}{\operatorname{arg\,max}}\,M(b_n,b_{n+1};u_n,t_n)P_n(b_n)\,;
\end{align}
this means that the optimal value for a base depends on the choice for the next base.\\
We can solve these equations up to the last $P_N(b_N)$ which is maximized to obtain $b^*_N=\arg\max_{b_N} P_N(b_N)$ and we can then propagate back to the first value setting $b^*_n=b^{\max}_n(b^*_{n+1})$. The algorithm is explained graphically in figure \ref{fig:viterbi}.\\
What is great about Viterbi algorithm is that its complexity grows linearly in $N$ and one needs to explore only a very small subset of the $4^N$ possible sequences. This is a feature of message-passing algorithms in one dimension.\\
Another interesting feature of this framework is that unzipping experiments can be repeated several times and the different traces can be combined just by computing the product of probabilities:
\begin{equation}
P(\mathcal{T}_1,\mathcal{T}_1,\dots,\mathcal{T}_M|\mathcal{B})=\prod_{i=1}^NP(\mathcal{T}_i|\mathcal{B})\,,
\end{equation}
where $\mathcal{T}_i$ is the trace of the i$^\textrm{th}$ experiment of a series of $M$.\\
Therefore we can combine different experiments to infer the sequence. In \cite{BAL07} it has also been shown that the rate of error decreases exponentially with the number of measurements.\\
As we have said at the beginning of this section, however, this algorithm relies on two unrealistic assumptions: knowledge of the position of the fork, which is never attainable because we actually measure the position of the bead; and an infinite sampling frequency. In the following we will try to come over these two assumptions by building more complex inference algorithms.
\section{Perfect averages algorithm}
\begin{figure} 
 \includegraphics[width=.5\textwidth]{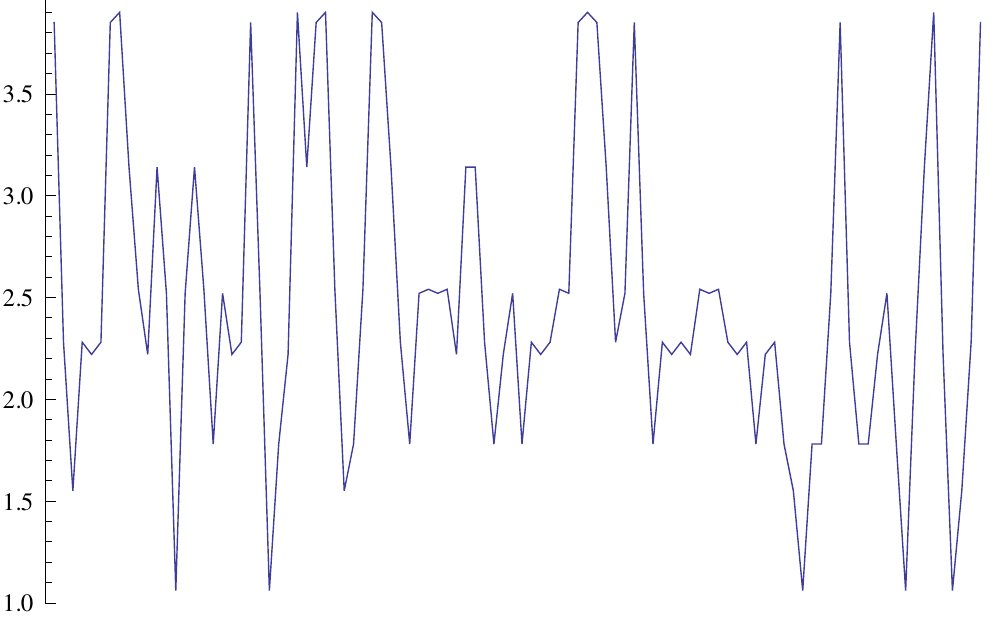}
  \includegraphics[width=.5\textwidth]{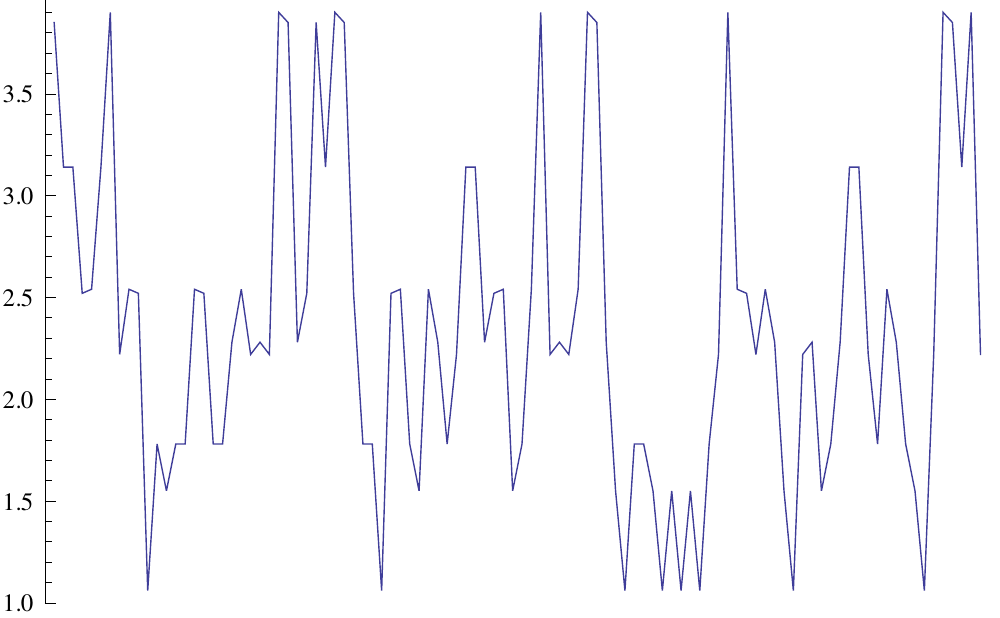}
 \includegraphics[width=.5\textwidth]{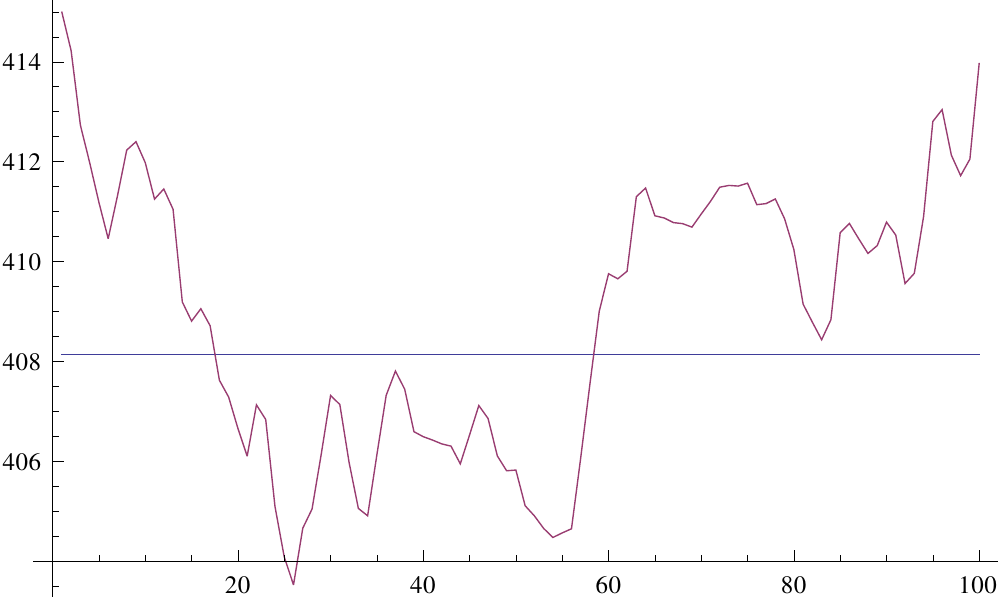}
  \includegraphics[width=.5\textwidth]{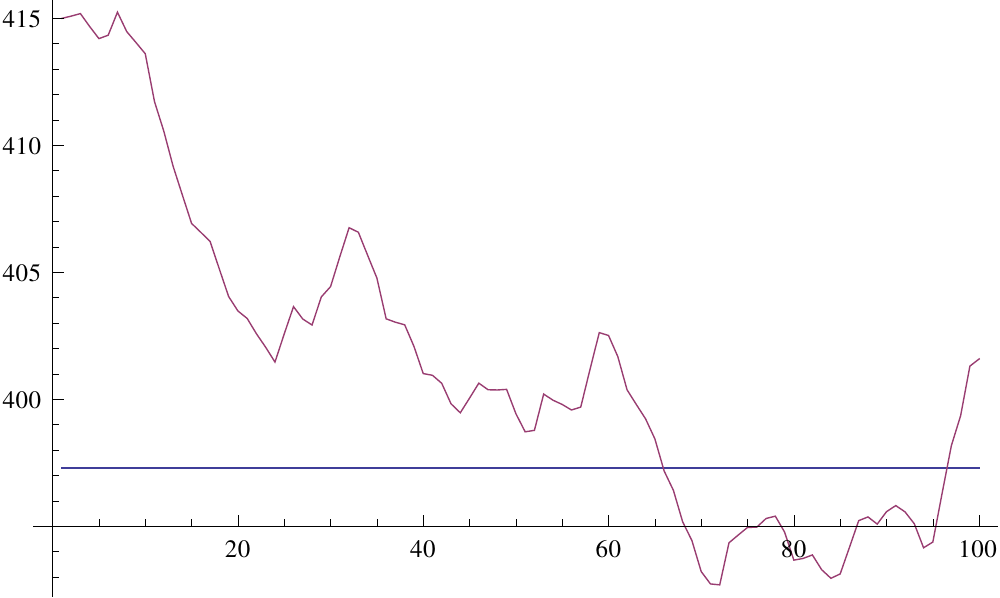}
 \includegraphics[width=.5\textwidth]{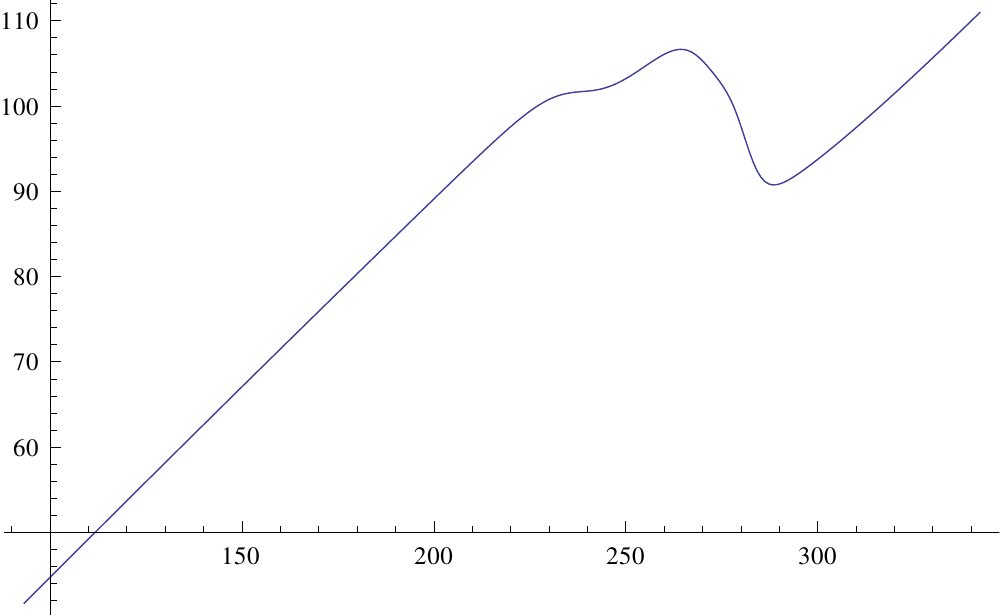}
 \includegraphics[width=.5\textwidth]{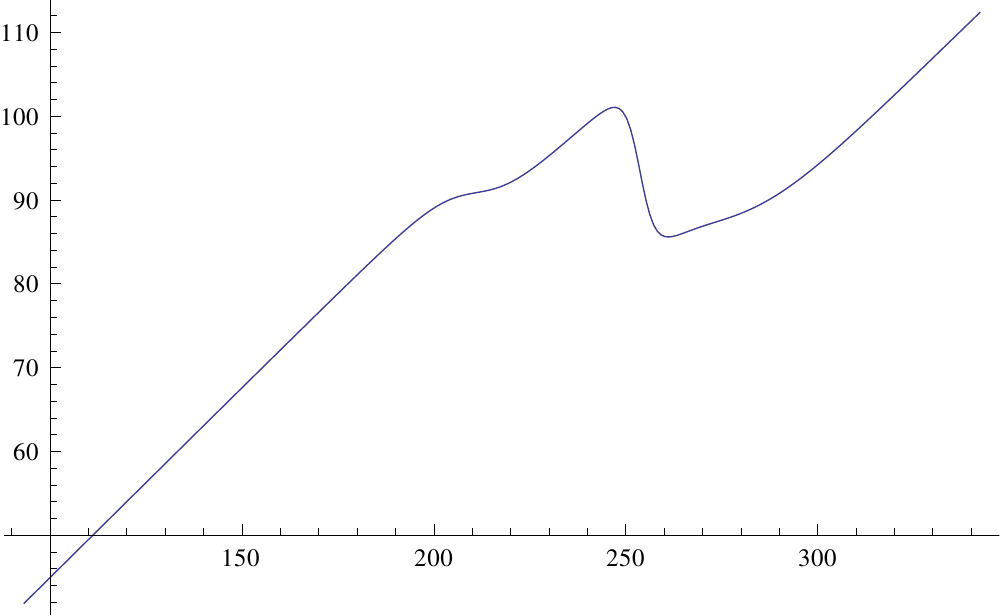}
\caption{Two different random sequences. On top the $g_0(n)$. In the center the free energy defined as $w(n,L)=\sum_j^ng_0(j)+\frac{k_1 k_2 k}{2(k_1 k_2 + k_1k + k_2 k)} (L - n l)^2$ as a function of n for $L=270$, the horizontal line marks the energy level $\tilde E$ such as $\exp(-\beta (\tilde E- E_0))=0.01$, that is sites that are visited (at equilibrium) one hundredth of the time the lowest energy site is. On the bottom the $u(L)$.} \label{energialibera}
\end{figure}

\begin{figure}
 \includegraphics[width=\textwidth]{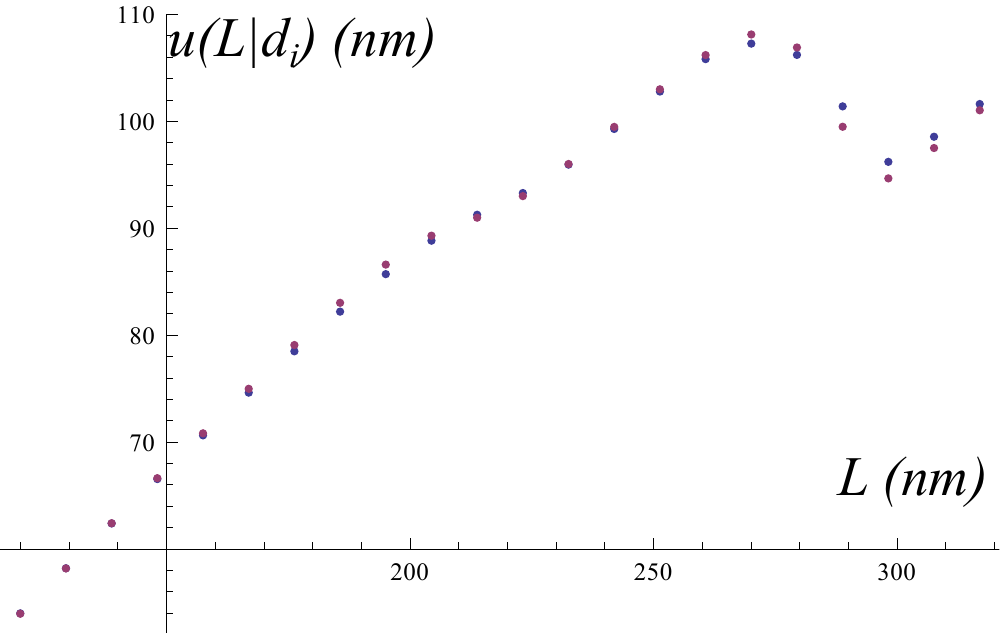}
 \caption{$\bar u(L)$ (blue) and $u_\textrm{trial}(L|d_i)$ (violet)}\label{u2}
\end{figure}

In this section we will perform a few simplifying assumptions in order to keep the equations simple looking. The reader should note, however, that these simplifications are by no mean fundamental and our results will hold even after relaxing those assumptions.\\
The first assumption is that we substitute Gaussian polymers for the complex behavior described in the preceding chapter and the second is that we ignore the $n$ dependence of the spring constants. The first assumption is not of fundamental importance because it amounts to truncating the anharmonic effects in the probabilities; relaxing it would only force us to compute integrals numerically, slowing the computation down.\\
The second assumption is even easier to relax because the $n$ dependence will just change the variance of the different terms in the sum in the next equation.\\
In general we believe that what is most important here is to have a general idea of what can and cannot be done with the spring constants set at realistic values for today's experiments. We will show that even without complex polymers and $n$ dependence we cannot investigate the sequence at a single base level.\\
We define a function $\bar u(L|B)$ as the equilibrium average displacement of one of the beads from the center of its optical trap. $L$ is the distance between the traps and is a parameter of the experiment and $B$ denotes the sequence. The dependence on $B$ will be omitted from now on.\\
The function $\bar u(L)$ has an explicit expression in terms of $g_0(n)$, that is:
\begin{equation}\label{eq:umedio}
\begin{split}
 \bar u (L)&= \frac1{Z(B) }\sum_n^N (L- n l)\frac{k_2 k}{k_1 k_2 +  k_1 k +  k_2 k} \\&\times\frac{\exp\left(-\sum_j^ng_0(j)-\frac{k_1 k_2 k}{2(k_1 k_2 + k_1k + k_2 k)} (L - n l)^2\right)}{\sqrt{k_1 k_2 + k_1 k+ k_2 k}}\,,
\end{split}
\end{equation}
where $k_1=0.025$ nm$^{-2}$and $k_2=0.125$ nm$^{-2}$ are the spring constants of the traps; $k=0.025$ nm$^{-2}$ is the spring constant of the linkers and the open part of the DNA and may depend weakly on $n$; $N$ is the total number of bases. $l=1$ nm is the difference in length when a base is open (two ssDNA bases, one for each side). $g_0$ is the binding energy of the DNA and it's given in table \ref{tableg0}\\
For any given value of $n$, the number of open basis there is a characteristic length of the fluctuations of u, which corresponds to the width of the gaussian in (\ref{eq:umedio}). This length is given by:
\begin{equation}
b=\frac1l\sqrt{\frac1{k_1}+\frac1{k_2}+\frac1{k}}\,,
\end{equation}
the reader should note that spring constants are expressed so that energies are dimensionless, that is as $k=\beta \kappa$ where $\beta$ is the inverse temperature and $\kappa$ a spring constant in the conventional units.\\
In the following (unless otherwise noted), $b=9.38$ as it was calculated from realistic constants from Bockelmann's experiment as described in \cite{IO1}. Other references use different setups that yield different numeric values: Woodside et al. \cite{Woo06} have a setup that would corresponds to $b=6.46$ in the same approximation. Huguet et al. \cite[Supplementary material]{HUGUET} have $b=8.49$ for their setup.\\
In figure \ref{energialibera} we show two sequences and their corresponding free energy landscape at fixed $L$ and the $u(L)$. The reader should note how for a fixed $L$, values of $n$ as far apart as 60 bases can be visited with non negligible probability, and most of the times there exist two or more values as far apart as 20 bases which have a high probability of being visited.\\
If we now define a trial function which depends linearly on a set of coefficients ${c_i}$:
\begin{equation}
 g_\textrm{trial}(n|c_i)=\sum_{i=1}^M c_i \Omega_{b^\prime}(n-b^\prime i)\,,
\end{equation}
where $\Omega_{b^\prime}$ is some one-dimensional function of width $b^\prime$, which we do not need to specify now to keep the discussion as general as possible. We should discuss in the following how $b^\prime$ is related to $b$.\\
We can also define $u_\textrm{trial}$ which is $\bar u$ where $g_0$ has been substituted by $g_\textrm{trial}$, and the cost function:
\begin{equation}
C({c_i})= \frac12 \sum_{i=M_0}^{M+M_0}(\bar u(i b l) - u_{\textrm{trial}} (i b l) )^2\,,
\end{equation}
where $M_0=\min_{b,b^\prime}[g_0(b,b^\prime)]/k_\textrm{eff}$ and $M_0+M=\max_{b,b^\prime}[g_0(b,b^\prime)]/k_\textrm{eff}+N$.  This amounts to taking a measure every $b l$ in the interval where there could be some effect from the sequence, for larger (smaller) $i$ all the bases will be closed (open).\\
The objective in defining this is to find the set of $c_i$ that approximates the best a set of experimental measures.\\
$\min_{b,b^\prime}[g_0(b,b^\prime)]/k_\textrm{eff}=93.28$ nm, and $\max_{b,b^\prime}[g_0(b,b^\prime)]/k_\textrm{eff}=343.2$ nm for the set of parameters specified previously. The reader should notice that the difference between these two numbers is rather large compared to the size of one open base pair (1 nm).\\
In effect most of the times we take many more measures than it is necessary, because for a given sequence the central limit theorem says it is unlikely that such extremes are ever reached, on the contrary the relative fluctuations of the size of the interval of interesting $L$ will scale as $1/\sqrt{N}$.\\
However, this is not a big computational problem because the computation time will not depend as much on the number of measures, as on the number of parameters (the $c_i$) which is fixed. On the other hand taking measures in where the response of the system is purely elastic does not change the landscape over which we are optimizing.\\
It is now possible to minimize the cost function over the ${c_i}$.\\
We will now show some results we have obtained for a random sequence of 50 base pairs and $\Omega_b(x)=\theta(x+b/2)\theta(b/2-x)$ is the boxcar function of width $b$.
There is very good agreement between $\bar u$ and $u_\textrm{trial}$, but if we plot $g_0$ and $g_\textrm{trial}$ the agreement is less good.
\begin{figure} 
 \includegraphics[width=\textwidth]{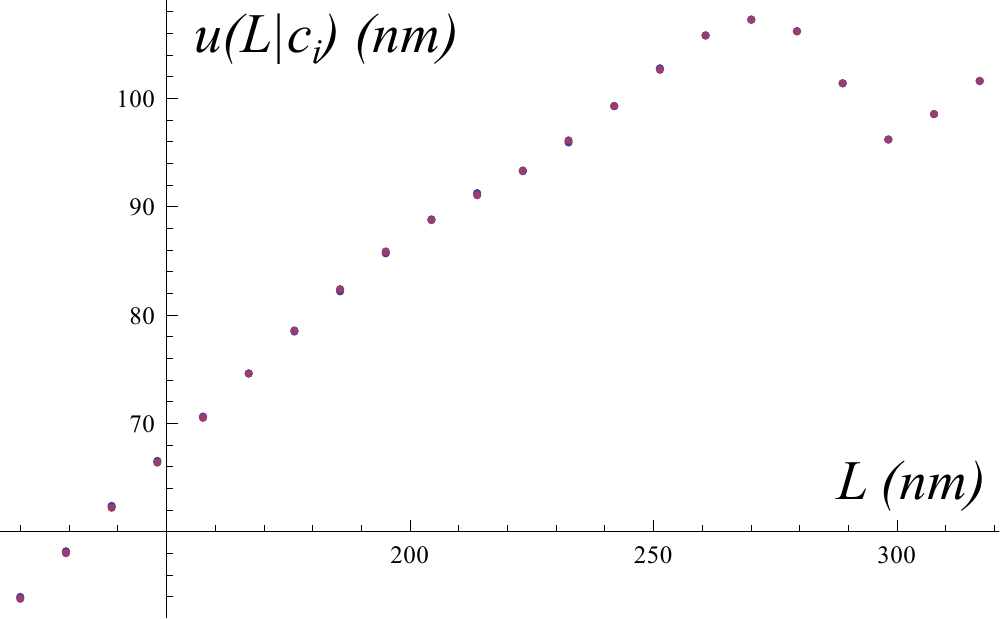}
 \caption{$\bar u(L)$ (blue) and $u_\textrm{trial}(L|c_i)$ (violet)} \label{u1}
\end{figure}
At some points consecutive values of $c_i$, that is $c_i$ and $c_{i+1}$, wander off to values which make it differ greatly from $g_0$. To quantify the difference between $g_0$ and $g_\textrm{trial}(n|c_i)$ we can define another cost function:
\begin{equation}
D({c_i})= \sum_n^N(g_0(n) - g_\textrm{trial}(n|c_i) )^2\,.
\end{equation}
\begin{figure}
 \includegraphics[width=\textwidth]{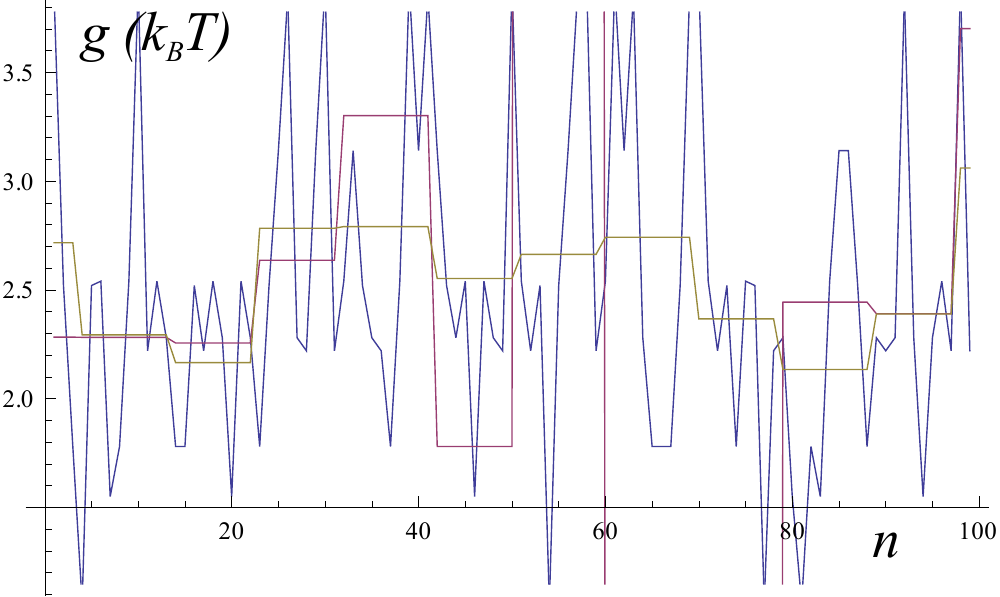}
 \caption{$g_0(n)$ (blue), $g_\textrm{trial}(n|c_i)$ (violet) and  $g_\textrm{trial}(n|d_i)$ (brown)} \label{g0}
\end{figure}
It now seems natural to define the set of parameters that minimize this new cost function as ${d_i}$ and compare the  $g_\textrm{trial}(n|c_i)$ and  $g_\textrm{trial}(n|d_i)$ as we do in figure \ref{g0}. Where $g_\textrm{trial}(n|d_i)$ is given by:
\begin{equation}
 g_\textrm{trial}(n|d_i)=\sum_i^N d_i \Omega_{b^\prime}(n-b^\prime i)\,,
\end{equation}
In practice this amounts to the average of $g_0(n)$ over the step of the trial function, in fact for a given step we have to minimize $\sum_{ j=\lceil ib-b/2\rceil}^{ \lfloor ib+b/2\rfloor}(d_i-g_0(j))^2$, that is:
\begin{equation}
d_i=\frac1{|\omega_i|}\sum_{j\in\omega_i}^Ng_o(j)\,,
\end{equation}
where $|\omega_i|$ is the cardinality of $\omega_i$, the number of bases that make up a step (it can take either $\lceil b\rceil$ or $\lfloor b\rfloor$ as values).\\
This way we have shown that $g_\textrm{trial}(n|d_i)$ is a box average of $g_0(n)$ which is very different from a moving average, and since the $g_\textrm{trial}(n|c_i)$ has the exact same structure it makes sense to compare the two.\\
One might also want to know how $u_\textrm{trial}(L|d_i)$ compares to $\bar u(L)$. We can see that in figure \ref{u2} and the agreement is definitely worse than what it was than when the fit was obtained witht the cost function $C$.

\subsection{Prior}
\begin{figure}
\includegraphics[width=.5\textwidth]{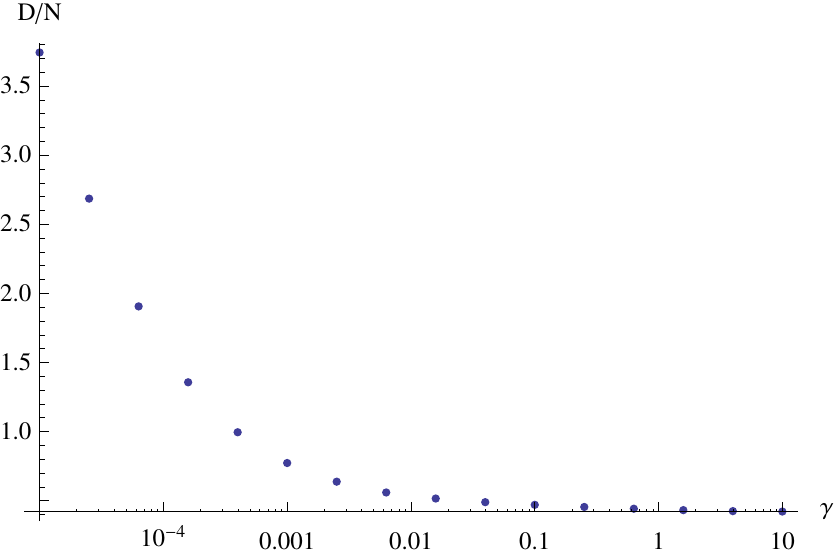}
\includegraphics[width=.5\textwidth]{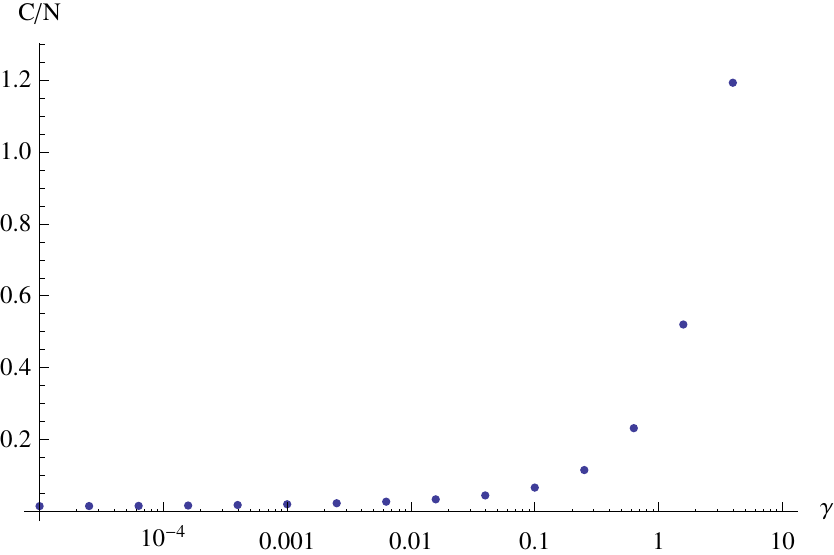}
\includegraphics[width=\textwidth]{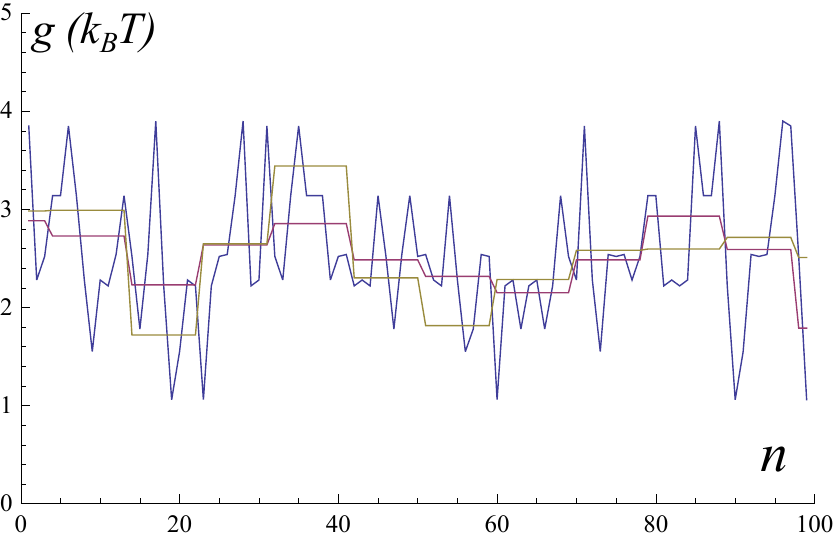}
 \caption{The top two panels show the value of the cost functions $C$ and $D$ as a function of varying $\gamma$. The bottom panel shows the $g_\textrm{trial}(n|c_i)$ for $\gamma=0.0158$ (brown), the real $g_0$ (blue) and the $g_\textrm{trial}(n|d_i)$  (purple). The value of $D/N$ for the $d_i$ is 0.399.}\label{male}
\end{figure}

\begin{figure}
\includegraphics[width=.5\textwidth]{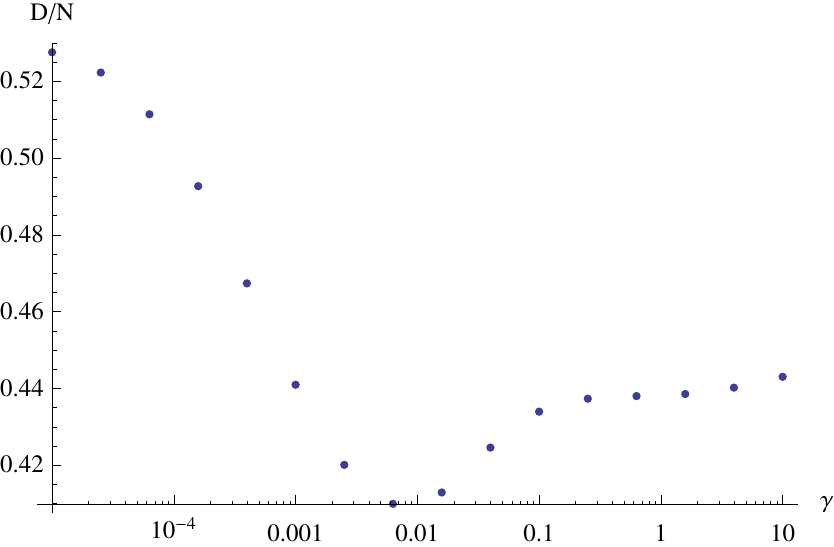}
\includegraphics[width=.5\textwidth]{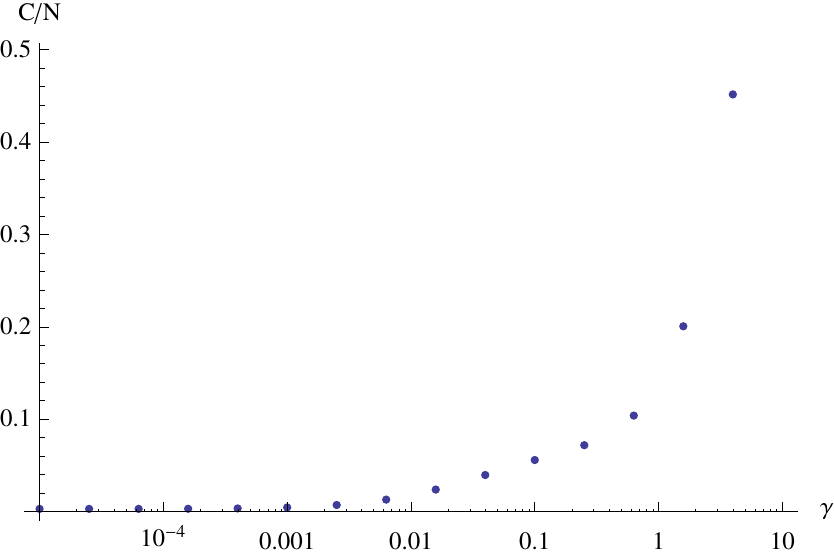}
\includegraphics[width=\textwidth]{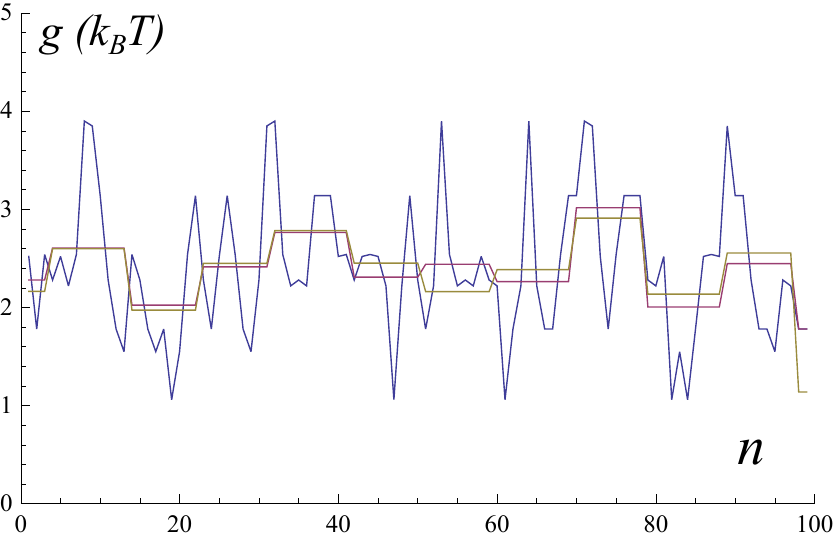}
 \caption{The top two panels show the value of the cost functions $C$ and $D$ as a function of varying $\gamma$. The bottom panel shows the $g_\textrm{trial}(n|c_i)$ for $\gamma=0.0063$ (brown), the real $g_0$ (blue) and the $g_\textrm{trial}(n|d_i)$ (purple). The value of $D$ for the $d_i$ is 0.387.}\label{bene}
\end{figure}

As one can see in figure \ref{g0}: two adjacent steps can sometimes grow in opposite directions to non-physical values.\\
To avoid this kind of problems we have added a prior to center the values of the steps around the average:
\begin{equation}
\tilde C_\gamma({c_i})=\frac12 \sum_{i=M_0}^{M+M_0}(\bar u(i b l) - u_{\textrm{trial}} (i b l) )^2+\gamma\sum_n^N(g_\textrm{trial}(n|c_i) -\bar g_0)^2\,,
\end{equation}
Where $\gamma$ is a constant we use to increase or decrease the effect of the prior. Ideally we hope to obtain a reasonable fit for values of $\gamma$ smaller than the biggest eigenvalue of the Hessian of $C$ when derived with respect to the $c_i$'s.\\
The problem is that sometimes we find no good fit no matter the value of $\gamma$. This is shown in figure (\ref{male}), as the reader can easily see, the best fit for $C$ does not coincide with the best fit for $D$. A decreasing $D$ as a function of $\gamma$ indicates that the best fit is dominated by the prior.\\ prior, to put it in other words: the best fit is the trivial one: $c_i=\bar g_0$ for all $i$.\\
Some other times we have a non trivial minumum over $\gamma$, and things look definitely better as in figure (\ref{bene}).\\
We have tried a prior that would take into account that the potential $g_0$ can only take 10 values, so we have chosen the form:
\begin{equation}
\label{prior10}
-\sum_i^{M}\sum_j^{10}\exp\left(-\frac{(c_i-g_j)^2}{2\sigma^2}\right)\,,
\end{equation}
where the ${g_j}$ are the ten possible values that $g_0$ can take. It is important to note that this strategy makes sens only when the trial function has a stepsize of exactly one.\\
What we have realized is that when $b$ has reasonable values, around those of current state of the art experiments ($\sim 10$), this strategy yields no advantage over the prior we have tried in the preceding section.\\
At the same time one might think that, for smaller values of $b$, say when it's closer to one, this prior might help us reconstruct the original sequence, but the reconstruction is actually just as good.\\
We have yet to find a regime in which this prior makes a difference.\\
In conclusion we have found that most of the times a small value of $\gamma$ (\emph{i. e.} $10^{-4}$) gives pretty good results, otherwise there are very clear signs that the fit has not converged.
\subsection{Optimal value of the step-size}
\begin{figure}
\includegraphics[width=\textwidth]{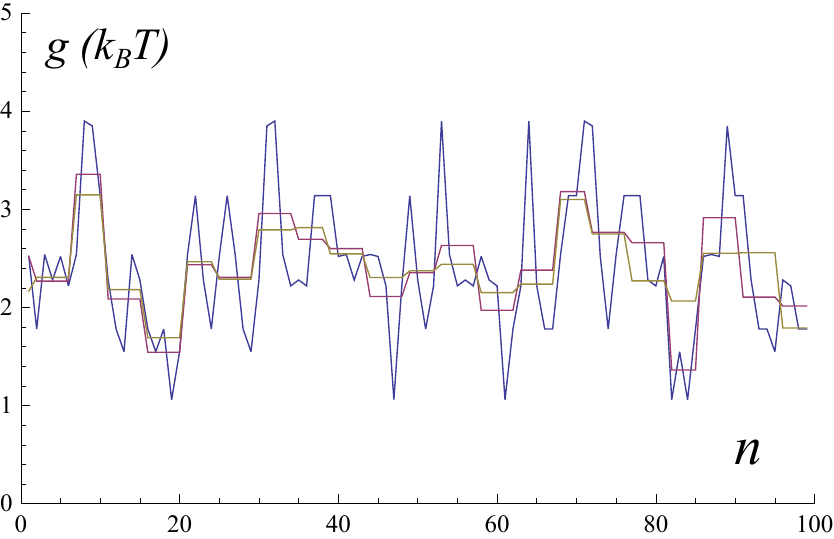}
\caption{The figure shows the $g_\textrm{trial}(n|c_i)$  for $\gamma=0.000399$ (brown), the real $g_0$ (blue) and the $g_\textrm{trial}(n|d_i)$ (purple).}\label{halfb}
\end{figure}

\begin{figure}
\includegraphics[width=\textwidth]{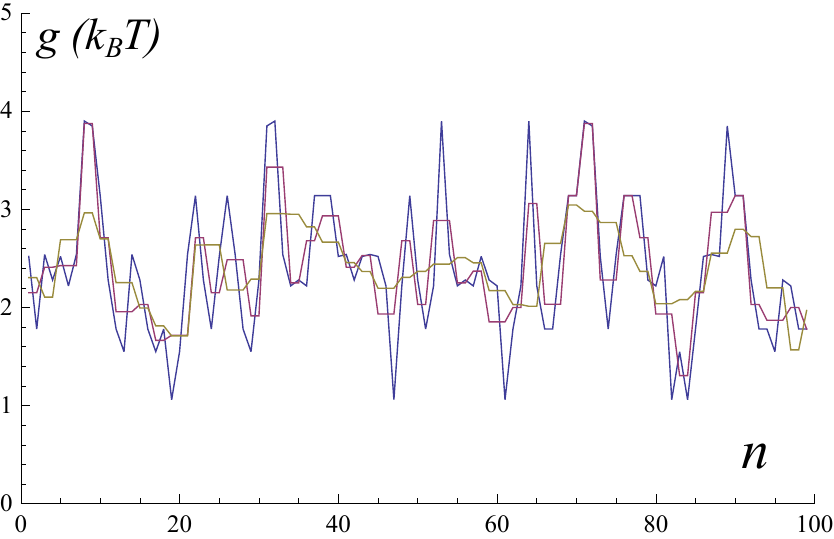}
\caption{The figure shows the $g_\textrm{trial}(n|c_i)$ for $\gamma=0.00016$ (brown), the real $g_0$ (blue) and the $g_\textrm{trial}(n|d_i)$ (purple).}\label{quarterb}
\end{figure}

\begin{figure}
\includegraphics[width=\textwidth]{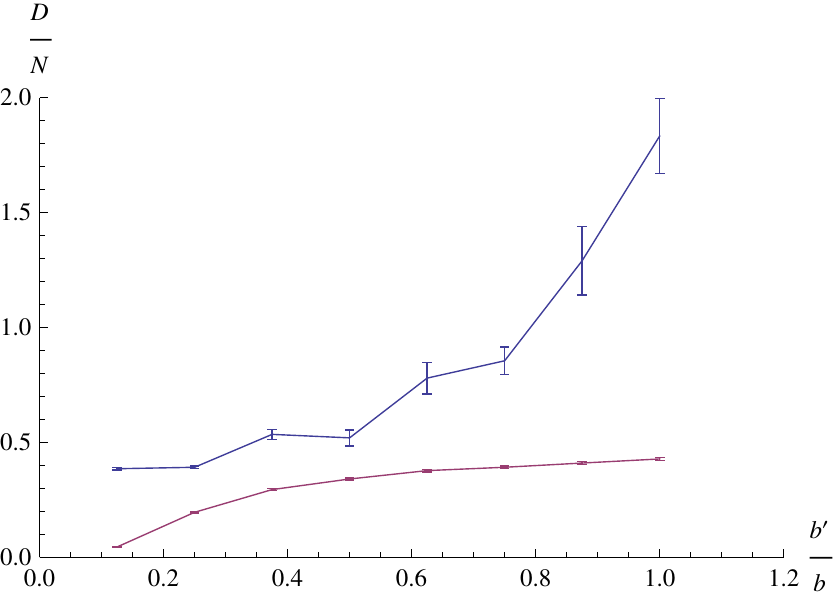}
\caption{$D(c_i)$ (blue) and $D(d_i)$ (purple) as functions of $b^\prime$ averaged over 100 random sequences, the error-bars are the standard deviation of the mean}\label{fig:cd}
\end{figure}

\begin{figure}
\includegraphics[width=\textwidth]{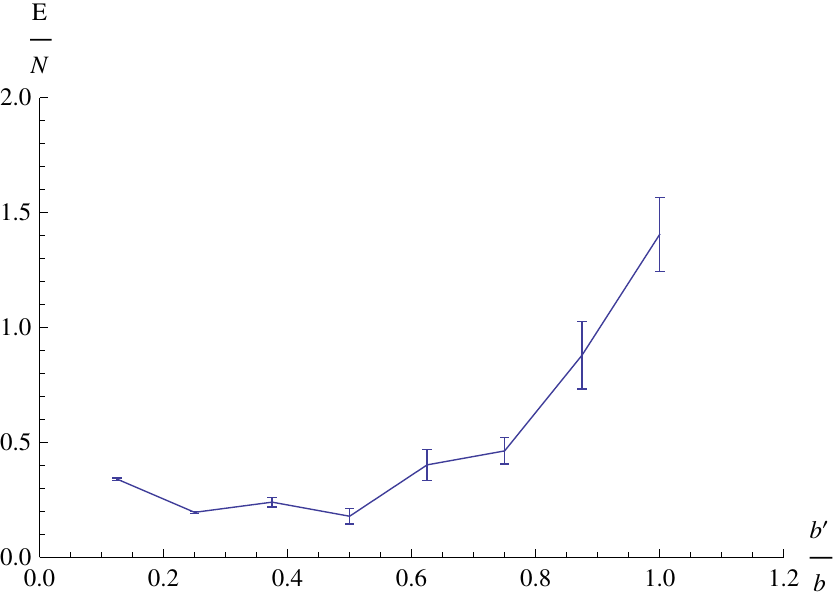}
\caption{$E(c_i|d_i)$ as a function of $b^\prime$ averaged over 100 random sequences, the error-bars are the standard deviation of the mean}\label{fig:E}
\end{figure}

\begin{figure}
\includegraphics[width=.5\textwidth]{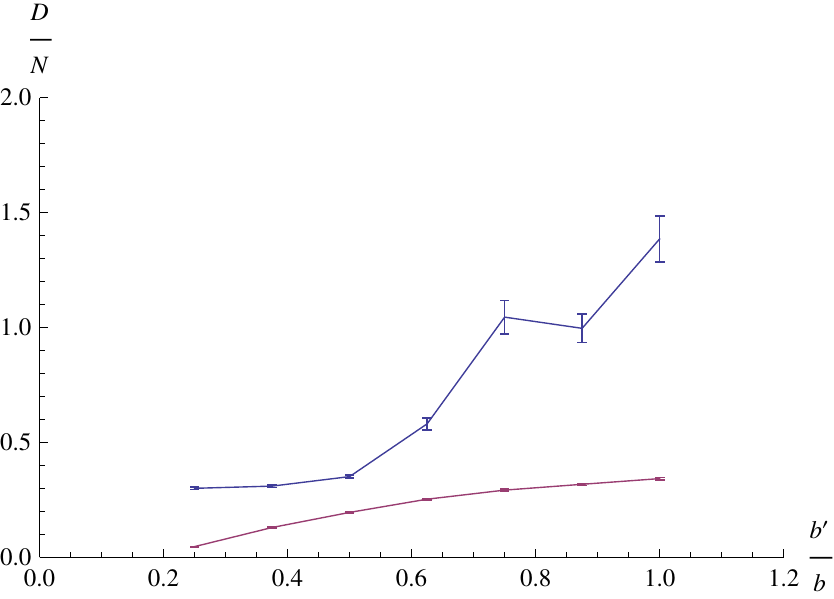}
\includegraphics[width=.5\textwidth]{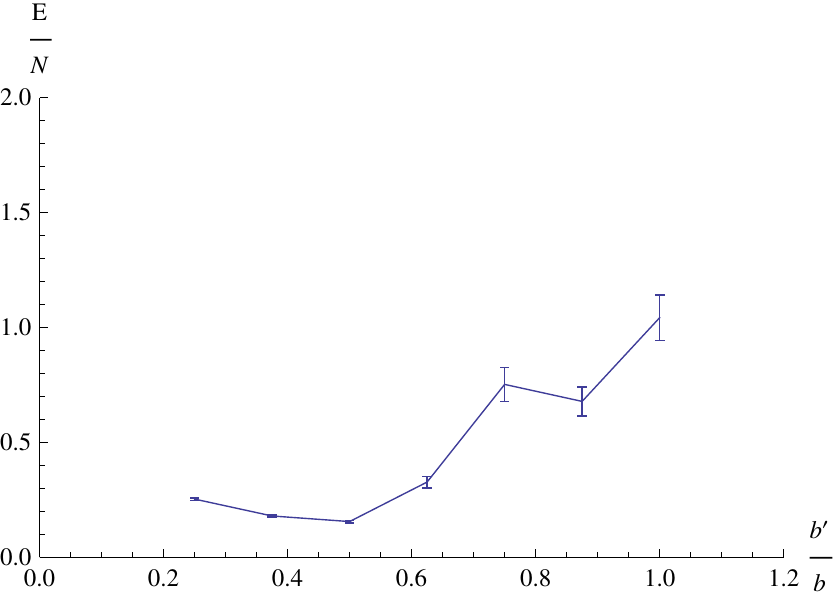}
\caption{Value of cost functions when $b=4.69$. The cost functions are shown as a function of $b^\prime$ averaged over 100 random sequences, the error-bars are the standard deviation of the mean. Left: $D(c_i)$ (blue) and $D(d_i)$ (purple) . Right: $E(c_i|d_i)$}\label{fig:bpiccoli}
\end{figure}

\begin{figure}
\includegraphics[width=.5\textwidth]{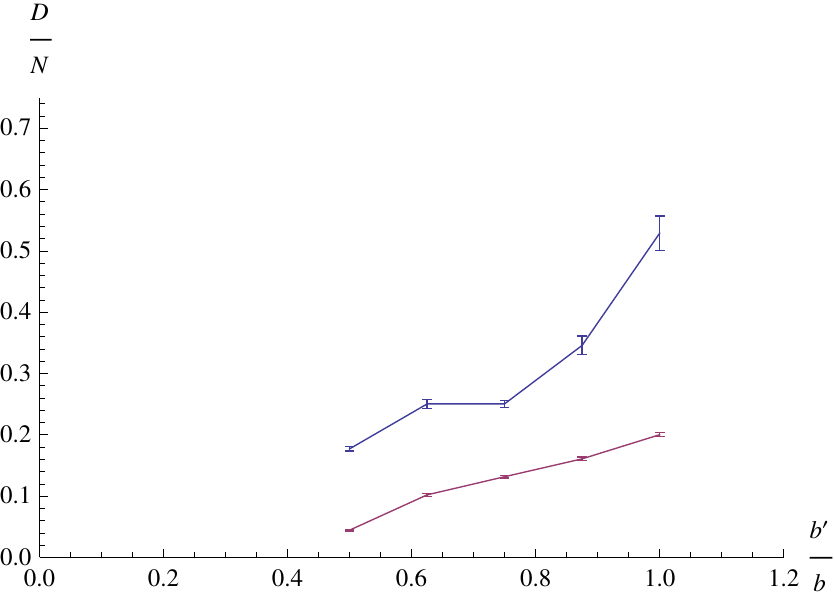}
\includegraphics[width=.5\textwidth]{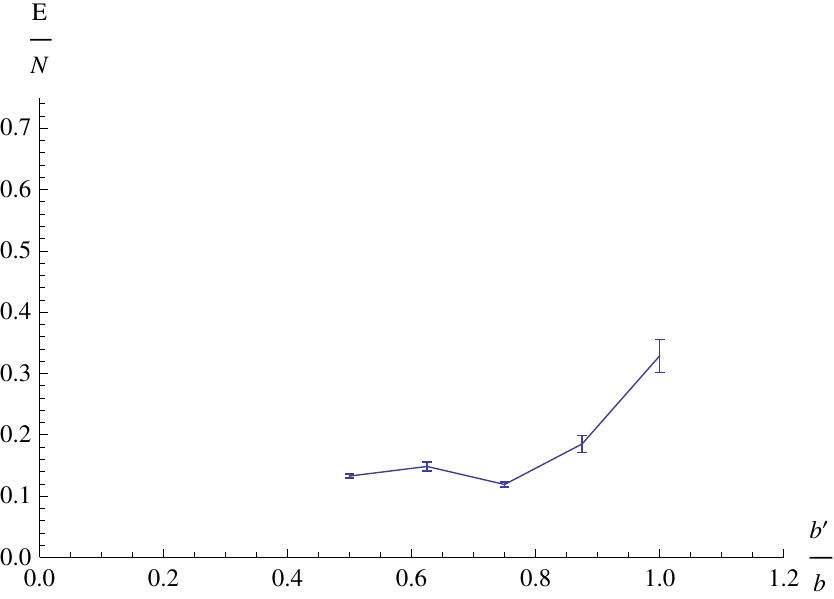}
\caption{Value of cost functions when $b=2.35$. The cost functions are shown as a function of $b^\prime$ averaged over 100 random sequences, the error-bars are the standard deviation of the mean. Left: $D(c_i)$ (blue) and $D(d_i)$ (purple) . Right: $E(c_i|d_i)$}\label{fig:bpiccolissimi}
\end{figure}

\begin{figure}
\includegraphics[width=.5\textwidth]{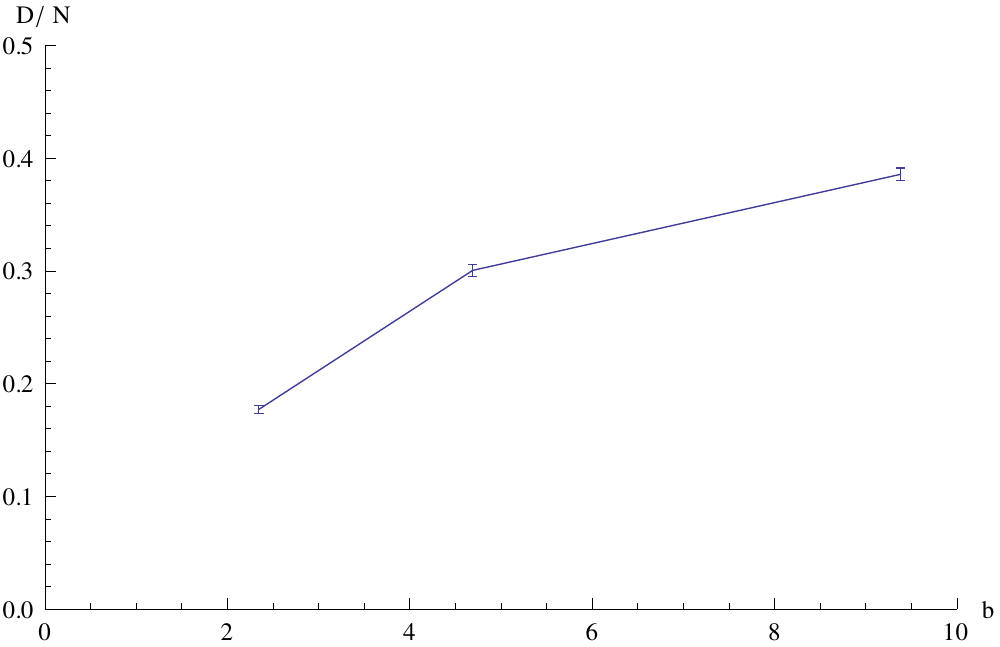}
\includegraphics[width=.5\textwidth]{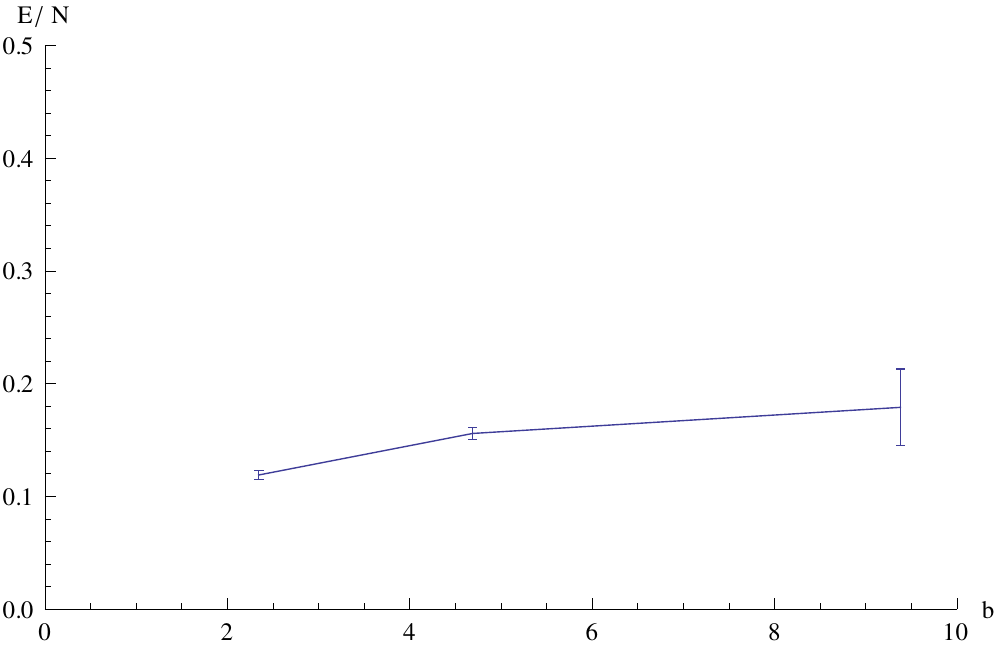}
\caption{Value of the cost functions $D$ (left) and $E$ (right) for different values of $b$. The plotted value is the minimum of the average over $b^\prime$ (see figures \ref{fig:cd}, \ref{fig:E}, \ref{fig:bpiccoli}, \ref{fig:bpiccolissimi}. Error bars are standard deviations over 100 sequences.}\label{minimoDE}
\end{figure}

The question is whether this can be further ameliorated by choosing a smaller stepsize. If we chose a stepsize $b^\prime=b/2$ we obtain the best fit for $\gamma=0.000399$ and a value of $D/N$ of 0.316, while the $d_i$ yield $D/N=0.26$. The results are shown in figure \ref{halfb}.\\
If we further decrease the stepsize to $b/4$ there is not much to be gained: for $\gamma=0.00016$ we obtain $D/N=0.343$ which is larger than what we obtained for $b/2$ while the value for the $d_i$  has further decreased to $D/N=0.16$. The results are displayed in figure \ref{quarterb}.\\
We now wish to study more systematically the optimal value of $b^\prime$, to do so we have computed the optimal $c_i$ and $d_i$ for 100 random sequences of 100 base pairs. The results are shown in figure \ref{fig:cd}: $D(d_i)$ gets better and better with smaller stepsize and for $b^\prime=b/8\simeq  1.17$ it is close to zero. On the other hand $D(c_i)$ seems to taper off to a value of approximately $0.4 N$.
We can now define another function we can use to evaluate the goodness of fit:
\begin{equation}
E(c_i|d_i)= \sum_n^N(g_\textrm{trial}(n|c_i) - g_\textrm{trial}(n|d_i) )^2\,.
\end{equation}
$E$ can be thought of as the distance between the fit of the $u$ ($c_i$) and the boxed average ($d_i$), which is the best attainable fit for a give step-size.\\
We expect $E$ to have a non trivial minimum where $D(c_i)$ starts to saturate, representing the point where the fit obtained through the $u$ is closest to the average over the steps. The results are shown in figure \ref{fig:E}.\\
This kind of metric can be a good gauge of what would happen when $b$ is smaller, we have $b$'s which are a half and a quarter of the original. We have obtained this by making $l$ respectively twice and four times as long.\\
The results for several $b^\prime$ and $b=4.69$ are shown in figure \ref{fig:bpiccoli}and the results for $b=2.35$ in figure \ref{fig:bpiccolissimi}. Please note that we have excluded points where $b^\prime$ would have been less than one.\\
We also include the minimum of the average of $D$ and $E$ over 100 sequences obtained for a given $b$, regardless of the value of $b^\prime$ that corresponds to it in figure \ref{minimoDE}

%\begin{figure}
%\includegraphics[width=\textwidth]{bgrande.pdf}
%\includegraphics[width=\textwidth]{bgrandebpiccolo.pdf}
%\caption{The result of the fit with the prior in eq (\ref{prior10}) for two different values of $b$ and of $\gamma$: on the right $b=9.38$ and $\gamma=$, on the left $b=1$ and $\gamma=$}
%\end{figure}
\subsection{Comparison with the moving average}
\begin{figure}
\includegraphics[width=.5\textwidth]{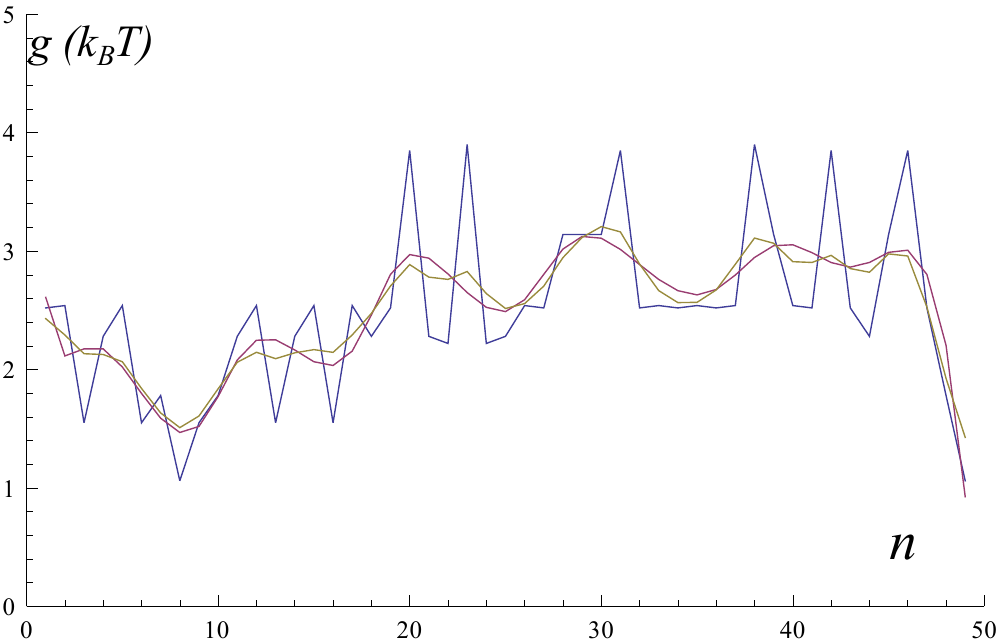}
\includegraphics[width=.5\textwidth]{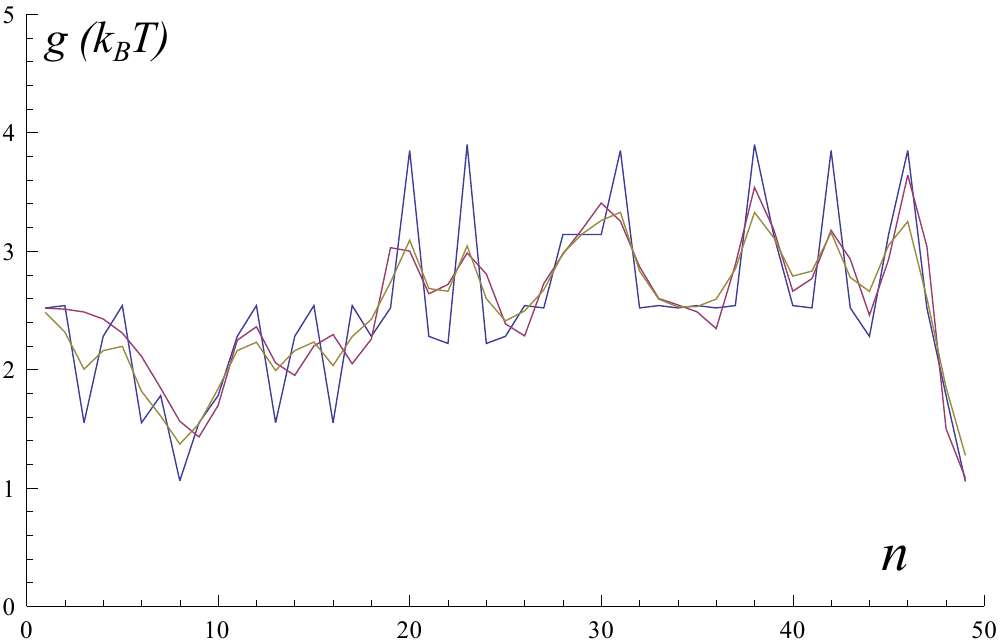}
\caption{Left: $b=9.38$, $g_0(n)$ (blue), $g_\textrm{trial}(n|c_i)$ (violet), $g_\sigma(n)$ (brown) for $\sigma=2.75$. Right: $b=2.35$, same color code, but $\sigma=1.95$.}\label{fig:movingave}
\end{figure}

This part stems from the observation that the $g_\textrm{trial}(n|c_i)$ when $b^\prime=1$ looks a lot like a smoothed version of the $g_0(n)$ we have thus defined $g_\sigma(n)$ a Gaussian filter as the convolution product between the $g_0(n)$ and a Gaussian kernel of width $\sigma$.\\
We then look for the $\sigma$ that minimizes the following cost function:
\begin{equation}
F(c_i,\sigma)=\sum_n^N(g_\sigma(n)-g_\textrm{trial}(n|c_i)^2\,,
\end{equation}
where the $c_i$ are, as usual, the set of parameter that minimize the $C$ cost function.\\
After several runs we have found that the optimal value of $\sigma$ is roughly increasing with increasing $b$, but that different sequences can lead to quite different optimal $\sigma$. One would expect $\sigma$ to be linearly related to the optimal $b^\prime$, but there is too much of a sequence dependence to conclude that. Two examples are shown in figure \ref{fig:movingave}.
\subsection{Difference with the na\"ive prediction}
\begin{figure}
\begin{center}
\includegraphics[width=.45\textwidth]{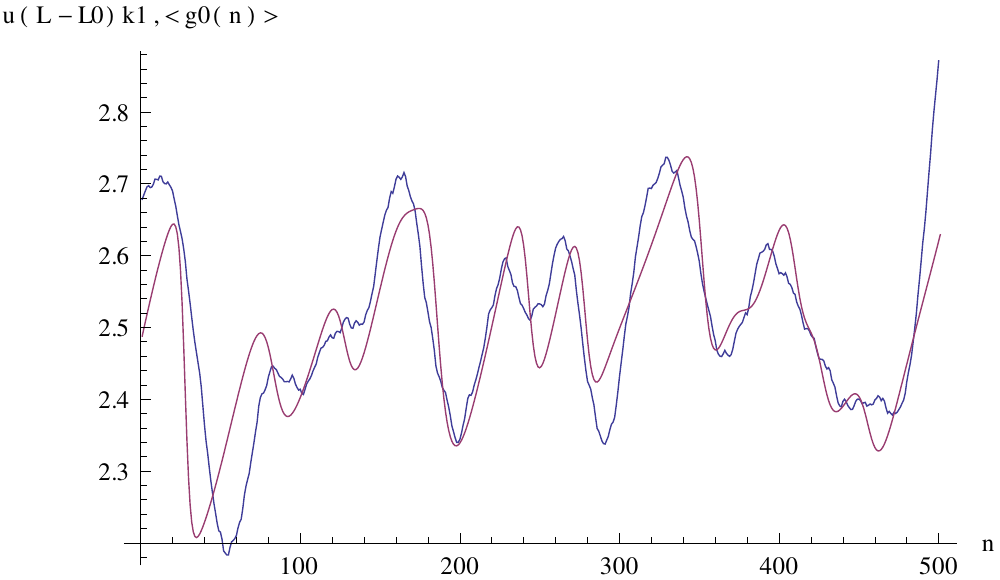}
\includegraphics[width=.45\textwidth]{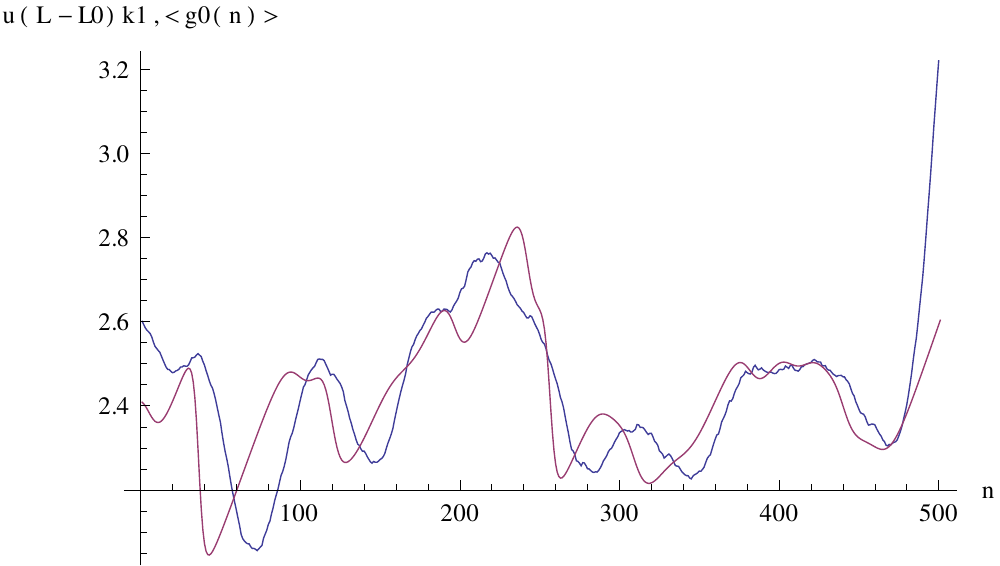}
\includegraphics[width=\textwidth]{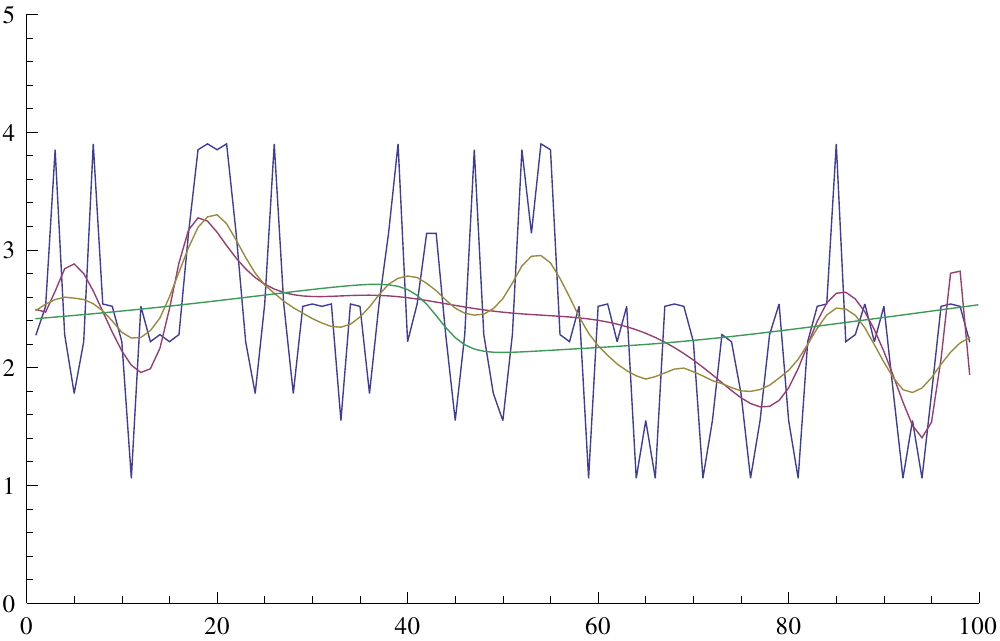}
\caption{Here we show two different sequences 500 bases long and one which is only 100 bases long. The na\"if estimate $u(L+L_0)k_1$ (violet) is compared to a Gaussian running average with $\sigma=25$. On the 100 bases long sequence we compare the real sequence (blue), to the fit obtained for $b^\prime=1$ (violet), the moving average with $\sigma=5.65$ (brown) and the na\"if prediction (bright green)}\label{naif}
\end{center}
\end{figure}

One possible way to perform inference on through the measurement of $u(L)$ at equilibrium is to approximate the expression in equation (\ref{eq:umedio}) through a saddle point. That is to say we find the base $n^*$ that has the biggest contribution for a given length $L$ and neglige all other contributions.\\
\begin{equation}
\bar{u(L)}=\sum_n^Nu(n,L)P(n,L)\simeq u(n^*,L)\,,
\end{equation}
where $P(n,L)$ is the exponential in equation (\ref{eq:umedio}), and $u(n,L)=k_{\textrm{eff}}/k_1(L-n l)$.\\
Now, $n^*$ is given by maximising $P(n,L)$, by solving:
\begin{equation}
g_0(n^*)=k_1l u(n^*,L)\,,
\end{equation}
and this equation looks as though we could use it to infer the $g_0(n^*)$ through the value of $u(n^*,L)$ which should be close to $u(L)$.\
The point where all this doesn't add up is the choice of a suitable $L$: we'd like to find $L(n^*)$ to know which $L$ contributes the most to a given $n^*$. To do so we solve:
\begin{equation}
g_0(n^*)=k_{\textrm{eff}}(L-n^*l)\,.
\end{equation}
Ideally, we'd like the solution of this to depend strongly on $n^*$, but not through $g_0(n^*)$ which is unknown, what we find instead is that with current state of the art experiments $g_0(n^*)/k_{\textrm{eff}}$ is two orders of magnitude larger than $l$.\\
This means that the $L_0(n^*)$ that solves this equation is not a nice, linear function of $n^*$, but instead depends very strongly on the sequence. This translates itself into a wild $n^*$-dependent dephasing between the na\"if prediction and the Gaussian average of the sequence. For short (e.g. 100 bases) sequences this dephasing effect is dominant.\\
What this really points to is that the saddle point approximation is not suitable for a case where $k_{\textrm{eff}}$ is so small, since it should, in principle, diverge.\\
However if we take a constant shift $L_0=\bar g_0/k_{\textrm{eff}}$ we can show how bad this technique is compared to what we have proposed, by comparing it to a Gaussian runnin average with $\sigma=25$ (which is much bigger than the $\sim 2$ we have used before). The results are shown in figure \ref{naif}.\\
This technique can be used to extract rapidly the average of the sequence on $\sim 25$ bases on long sequences, since it is much faster than what we have proposed.
\subsection{Scaling of computational time as a function of sequence length}
\begin{figure}
\includegraphics[width=\textwidth]{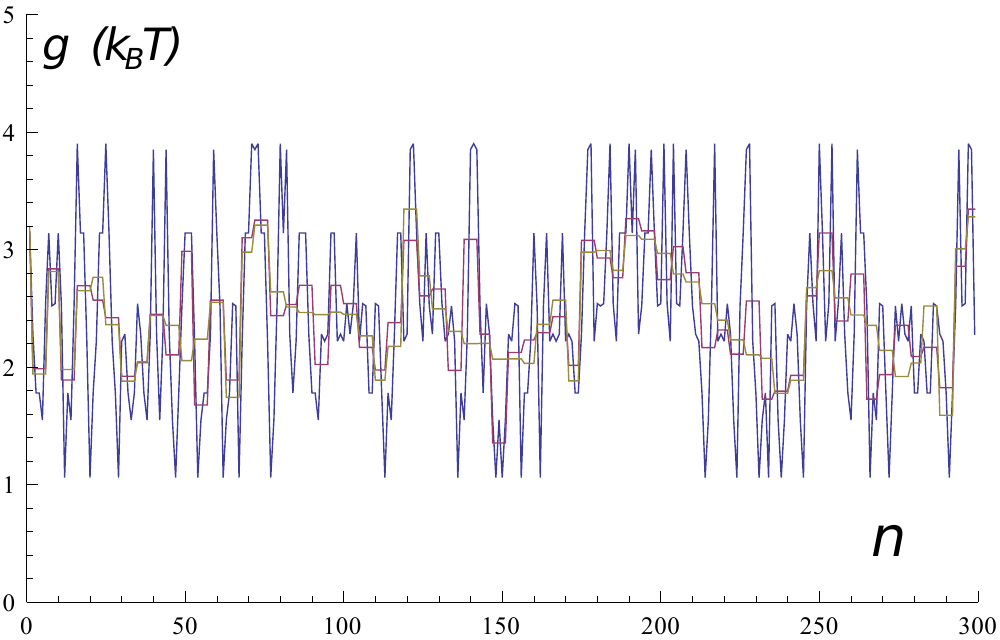}
\caption{For 300 bases, $b=9.38$ and $b^\prime=b/2$: $g_0(n)$ (blue), $g_\textrm{trial}(n|c_i)$ (violet) and  $g_\textrm{trial}(n|d_i)$ (brown)}\label{fig:300}
\end{figure}

As of now the algorithm scales as the cube of the number of basis. In principle it is possible to split a long sequence in smaller batches, and fit the separately, but some practical problems must be adressed.\\
First of all we have to take into account what we have discussed in the previson section, that is: it is impossible to know the relationship between $L$ and $n$ without knowing the $g_0$ with a sufficient degree of precision.\\
So suppose we want to fit a section of the $u(L)$ curve, say from $L_1$ to $L_2$, it is impossible to say what are the $n$'s that correspond to that interval with precision and we may end up adding a few hundreds left and right just to be sure, thus killing any advantage we might have had splitting unless the sequence is some 40 kbp long.\\ 
And here is where the second problem comes into play: up to now we have considered $k$ to be roughly constant, but $k$ really depends on $n$ albeit weakly. A change in $n$ of the order of 1 kbp on the other hand would not be negligible anymore and would lower the value of $k$, and thus of $k_\textrm{eff}$ of an order of magnitude.\\
This is currently a limitation of all current single molecule experiments. whenever opening too long a molecule the linkers become too elastic to yield meaningful insights on the $g_0$.\\
In figure \ref{fig:300} we display a fit of a sequence 300 bp long with $b=9.38$ and $b^\prime=b/2$. This computation takes Mathematica 7 a little more than 20 minutes on a Intel core 2 processor and uses up, about 1.5 GB of RAM. It involves a search in a 64-dimensional parameter space.
\subsection{Estimation of the error bars}
\begin{figure}
\includegraphics[width=\textwidth]{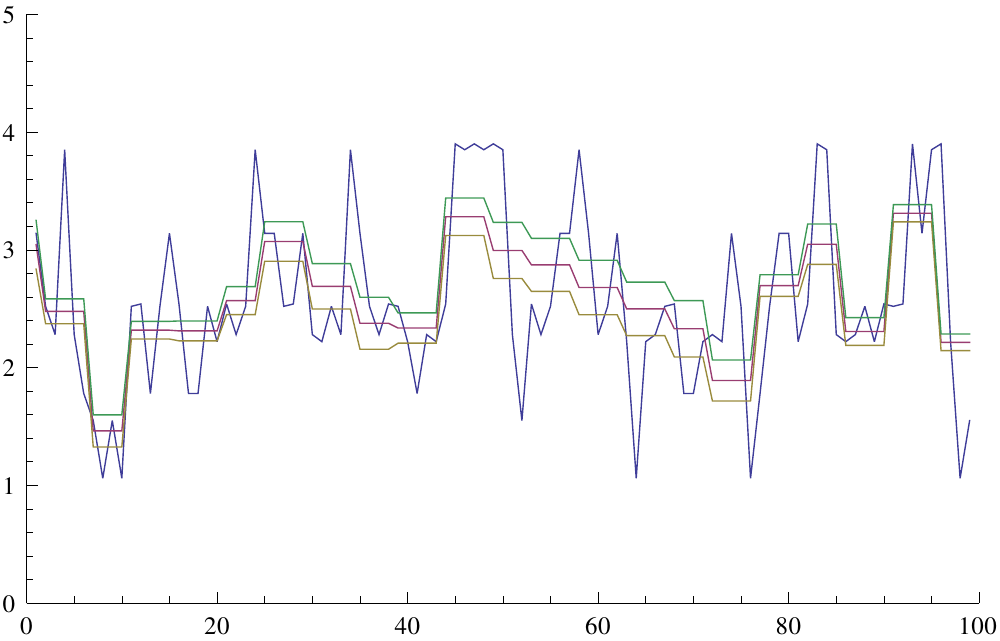}
\includegraphics[width=.5\textwidth]{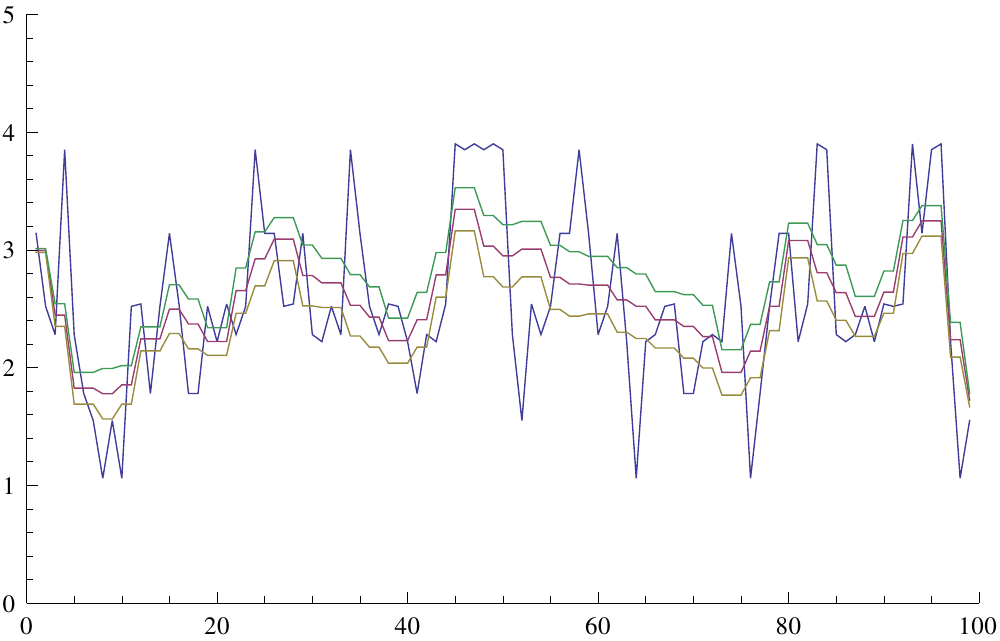}
\includegraphics[width=.5\textwidth]{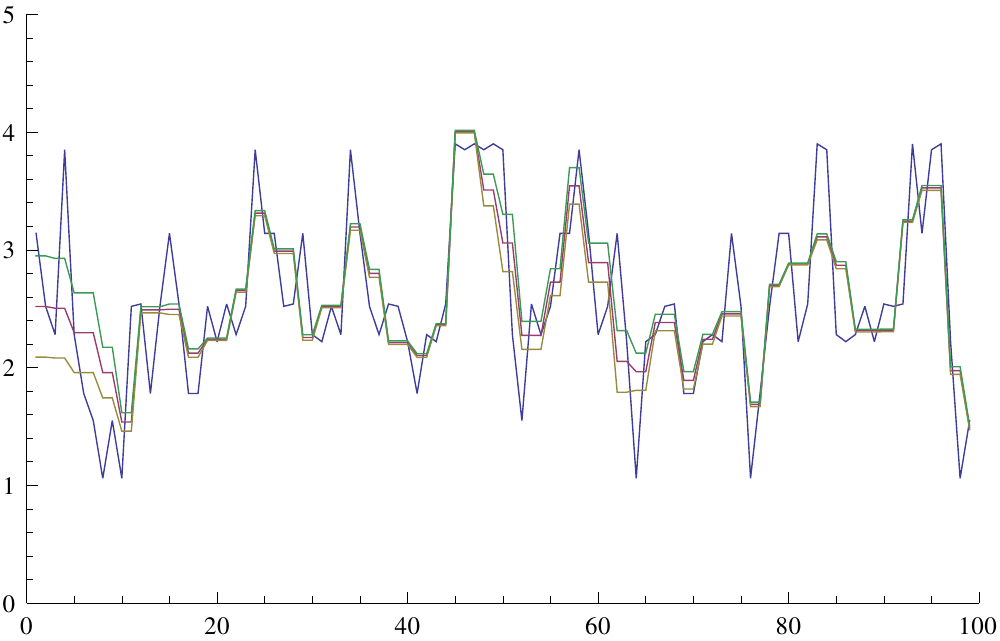}
\caption{For the three panels: $g_0$ in blue. The other three curves are the $g_\textrm{trial}$, and the $g_\textrm{trial}\pm\sigma$. For the top panel we have$b=9.38$, $b^\prime=b/2$, $N=100$. $\gamma=0.1$. On the bottom right we have changed the value of $b^\prime$ to $b/4$ and $\gamma=10^{-4}$. On the bottom right we have changed $l$ to nm so that $b=2.35$ and we have kept $b^\prime=b$, the fit is obtained for $\gamma=10^{-4}$.}\label{fig:errorbars}
\end{figure}

In least squares fitting it is costumary to estimate the variance of the variables through the Hessian of the cost function at the minimum. Let $H_{ij}=\frac{\partial^2 C}{\partial c_i \partial c_j}$ calculated at the minimum. Then the variances are given by
\begin{equation}
\sigma^2_{c_i}=\sigma^2(H)^{-1}_{ii}\,,
\end{equation}
where $\sigma^2$ is the true residual variance, which is unkown, but is usually estimated as $C^*/n$, where $C^*$ is the value of the cost function at the minimum and $n$ is the number of degrees of freedom.\\
If we do so without taking into account the prior $H$ is not positive definite and we end up with negative variances. Because of this we use the full cost function with the prior. Three examples of the results is shown in figure \ref{fig:errorbars}.\\
The reader should note how for an unchanged $b$, there is not much gain in lowering $b^\prime$. On the other hand when $b$ is smaller the fit is much better and this is reflected in the error-bars.
\subsection{Entropy}
Let us suppose we consider the cost functions we have defined in the preceding sections as thermodynamic quantities. As such we can draw a physical analogy and estimate the number of sequences that correspond to a given value of the cost function through the entropy.\\
By defining a pseudo-temperature we can define the partition function as:
\begin{equation}
\begin{split}
Z&=\sum_{\{b_n\}}\exp\left[-\frac{\lambda^2}2E(c_i|d_i)\right]\\
&=\sum_{\{b_n\}}\exp\left[-\frac{\lambda^2}2\sum_n\left( g_\textrm{trial}(n|c_i)- g_\textrm{trial}(n|d_i)\right)^2\right]\\
&=\sum_{\{b_n\}}\exp\left[-\frac{\lambda^2}2\sum_i|\omega_i|\left( c_i- \frac1{|\omega_i|}\sum_{j\in\omega_i}g_0(b_j,b_{j+1})\right)^2\right]\\
&=A\int d\vec x \,e^{-\sum_i\frac1{2|\omega_i|}x_i^2}\sum_{\{b_n\}}\exp\left[-\mathrm{i}\lambda\sum_ix_i\left( c_i- \frac1{|\omega_i|}\sum_{j\in\omega_i}g_0(b_j,b_{j+1})\right)\right]\\
&=A\int d\vec x \,e^{-\sum_i\frac1{2|\omega_i|}x_i^2-\mathrm{i}\lambda x_ic_i}\sum_{\{b_n\}}\exp\left[\mathrm{i}\lambda\sum_n\frac{\Omega_{b^\prime}(n-b^\prime i)}{|\omega_i|}x_ig_0(b_n,b_{n+1})\right]\\
&=A\int d\vec x \,e^{-\sum_i\frac1{2|\omega_i|}x_i^2-\mathrm{i}\lambda x_ic_i}\sum_{\{b_n\}}\prod_n\exp\left[\mathrm{i}\lambda\frac{\Omega_{b^\prime}(n-b^\prime i)}{|\omega_i|}x_ig_0(b_n,b_{n+1})\right]\,,
\end{split}
\end{equation}
where $A=\prod_i(2\pi |\omega_i|)^{-\frac12}$.\\
Using the transfer-matrix method we can recognise $\exp\left[\mathrm{i}\lambda\frac{\Omega_{b^\prime}(n-b^\prime i)}{|\omega_i|}x_ig_0(b_n,b_{n+1})\right]$ as a $4\times4 $ matrix which appears $|\omega_i|$ times identical and then involves a different $x_i$.\\
Because of this we can rewrite the previous equation as:
\begin{equation}
Z=A\int d\vec x \,e^{-\sum_i\frac1{2|\omega_i|}x_i^2-\mathrm{i}\lambda x_ic_i}\sum_{b_0,b_N}\vec b_0 \cdot \prod_i^M [\mathbf{T}(\mathrm{i}\lambda x_i)/|\omega_i|)]^{|\omega_i|} \cdot \vec b_N\,,
\end{equation}
where:
\begin{equation}
\mathbf{T}_{b_n,b_{n+1}}(t)=\exp\left[g_0(b_n,b_{n+1})t\right]=
\left(
\begin{array}{cccc}
 5.93^t & 4.71^t & 12.43^t & 9.21^t \\
 2.89^t & 5.93^t & 9.78^t & 12.68^t \\
 12.68^t & 9.21^t & 23.1^t & 46.99^t \\
 9.78^t & 12.43^t & 49.4^t & 23.1^t
\end{array}
\right)
\end{equation}
By rearranging the terms one can decouple the integrals
\begin{equation}
Z=A\sum_{b_0,b_N}\vec b_0 \cdot\prod_i^M \left[\int d x_i \,e^{-\sum_i\frac1{2|\omega_i|}x_i^2-\mathrm{i}\lambda x_ic_i} [\mathbf{T}(\mathrm{i}\lambda x_i)/|\omega_i|)]^{|\omega_i|} \right]\cdot \vec b_N\,,
\end{equation}
Once we have computed the one dimensional integrals we can multiply the $N$ matrices and sum over the first and last base.\\
We can then change variable ($\beta=\lambda^2/2$) and compute the thermodynamic quantities as:
\begin{align}
E(\beta)&=-\frac{\partial}{\partial \beta}\log Z\\
F(\beta)&=-\frac1{\beta}\log Z\\
S(E)&=\max_\beta(\beta(E-F(\beta)))
\end{align}
The entropy as a function of internal energy can be also obtained with a parametric plot, as it's shown in figure \ref{fig:entropy}. Even through the simplifications obtained thanks to the transfer matrix method, the computation of entropy is very taxing and we didn't have enough memory for computing the entropy for cases where the $c_i$ are not all identical or for longer sequences.\\
\begin{figure}
\includegraphics[width=.7\textwidth]{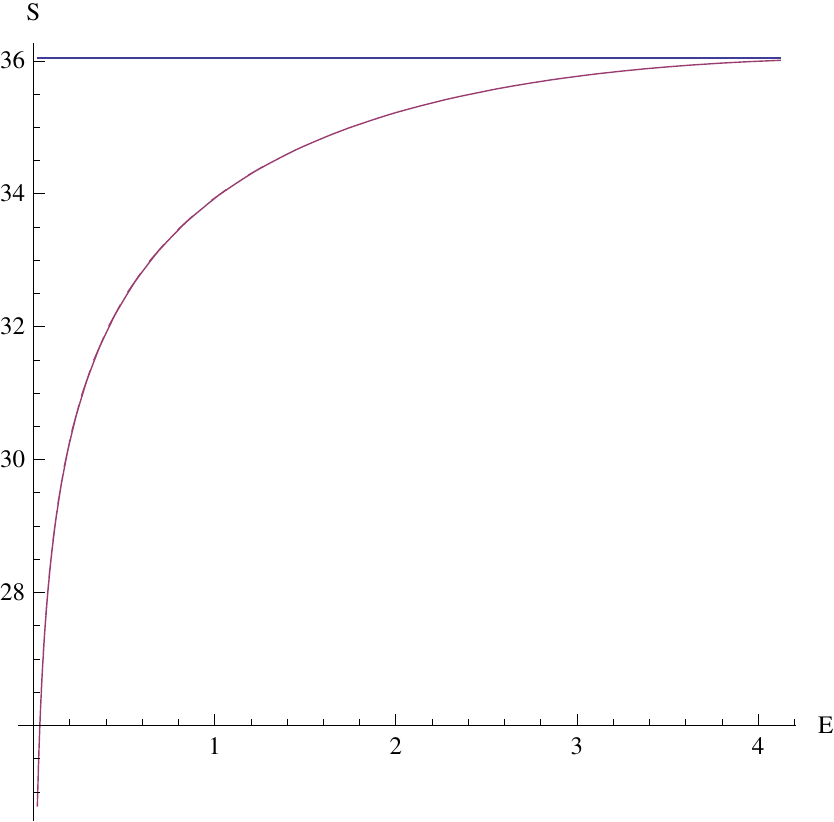}
\caption{Entropy for a sequence of 26 bases (25 values of $g_0$) and 5 measures. The entropy is computed for $c_i=2.52$ for every $i$. At high energies the entropy saturates to $S(\infty)=26\log(4)\simeq36$, the right value which is the logarithm of the number of possible sequences of 26 bases.}\label{fig:entropy}
\end{figure}
\subsection{A different approach}
In this section we will expound a different approach for tackling the same problem as developed by Jörg, Monasson and Cocco and is yet to be published.\\
The main difference is that this formalism allows for the description of the same system in a vector space and defines measures as orthogonality constraints.\\
This formalism has also allowed Jörg et al. to compute interesting statistical properties of this system, but we will not dwell on the details here.\\
Equation (\ref{eq:umedio}) can be rewritten by multiplying both sides by $Z(B)$:
\begin{equation}
  \label{eq:reconstruction_condition}
  \sum_{n} v_n(B) [\alpha(L - n l) - \beta \bar{u}(L)]  \exp{\left[-\frac{\kappa}{2}(L-n l)^2 \right]} = 0\, ,
\end{equation}
where
\begin{equation}
  \label{eq:definition_vn}
  v_n(B) = \exp\left[-\sum_j^ng_o(j)\right]
\end{equation}
\begin{equation}
  \label{eq:definition_alpha}
  \alpha = \frac{k k_2}{\sqrt{(k_1 k_2 + k k_1 + k k_2)^3}} \, ,
\end{equation}
\begin{equation}
  \label{eq:definition_beta}
  \beta =\frac{1}{\sqrt{k_1 k_2 + k k_1 + k k_2}}
\end{equation}
and 
\begin{equation}\label{eq:definition_kappa}
%  \kappa(n)^{-1} = \frac{k(n) k_1 k_2}{k_1 k_2 + k(n) k_1 + k(n) k_2} \, . 
\kappa^{-1} = \frac{1}{k} + \frac{1}{k_1} + \frac{1}{k_2}\, .
\end{equation}
This equation only makes sense if we use for the $\bar u (L)$ the measures we have obtained from an experiment.\\
Equation  (\ref{eq:reconstruction_condition}) can be rewritten as:
\begin{equation}
  \label{eq:reconstruction_condition_perpendicular_vector}
  \sum_{n} v_n(B) p_n(L) = 0 \,,
\end{equation}
where $p_n(L)=[\alpha(L - n l) - \beta \bar{u}(L)]  \exp{\left[-\frac{\kappa}{2}(L-n l)^2 \right]}$ is a vector that depends on a on a given measure $ \bar{u}(L,B)$.\\
This can be thought of as the scalar product $\vec v(B)\cdot\vec p(L)$, suggesting a geometrical interpretation: we have to choose the sequence $B$ so that it is orthogonal to all the vectors given by the measures encoded by the vectors $p(L)$ for different values of $L$.\\
The problem of finding the optimal vector $\vec v(B)$ can be rephrased as a minimization problem over a quadratic form by squaring both sides:
\begin{equation}
  \label{eq:reconstruction_condition_square}
\begin{split}
  \left[\sum_{n } v_n(B) p_n(L)\right]^2 &= \sum_{m ,n}  v_m(B) p_m(L) v_n(B) p_n(L)= \\
  \sum_{m ,n} v_m(B) K_{m,n}(L) v_n(B)&=\vec v^\dagger(B) \mathbf{K}(L) \vec v(B) = 0\, ,
\end{split}
\end{equation}
where $ K_{m,n}(L)=p_m(L)  p_n(L)$. Different measures are easily taken into account by adding the terms obtained for different $L$'s:
\begin{equation}
  \label{eq:chi_square}
 \vec v^\dagger(B) \left[\sum_{i=1}^M W_i\mathbf{K}(L_i)\right]  \vec v(B) = 0 \,,
\end{equation}
where $W_i$ are arbitrary positive weights.
\section{Dynamical algorithm}
\subsection{A toy model: coupled Ornstein-Uhlenbeck processes}
\begin{figure}
\begin{center}
\includegraphics[width=.5\textwidth]{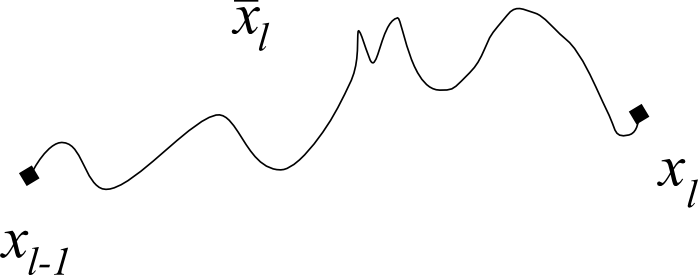}
\caption{The evolution of the Ornstein-Uhlenbeck process during a time step and its average.}
\label{propag}
\end{center}
\end{figure}

\begin{figure}
\begin{center}
\includegraphics[width=.5\textwidth]{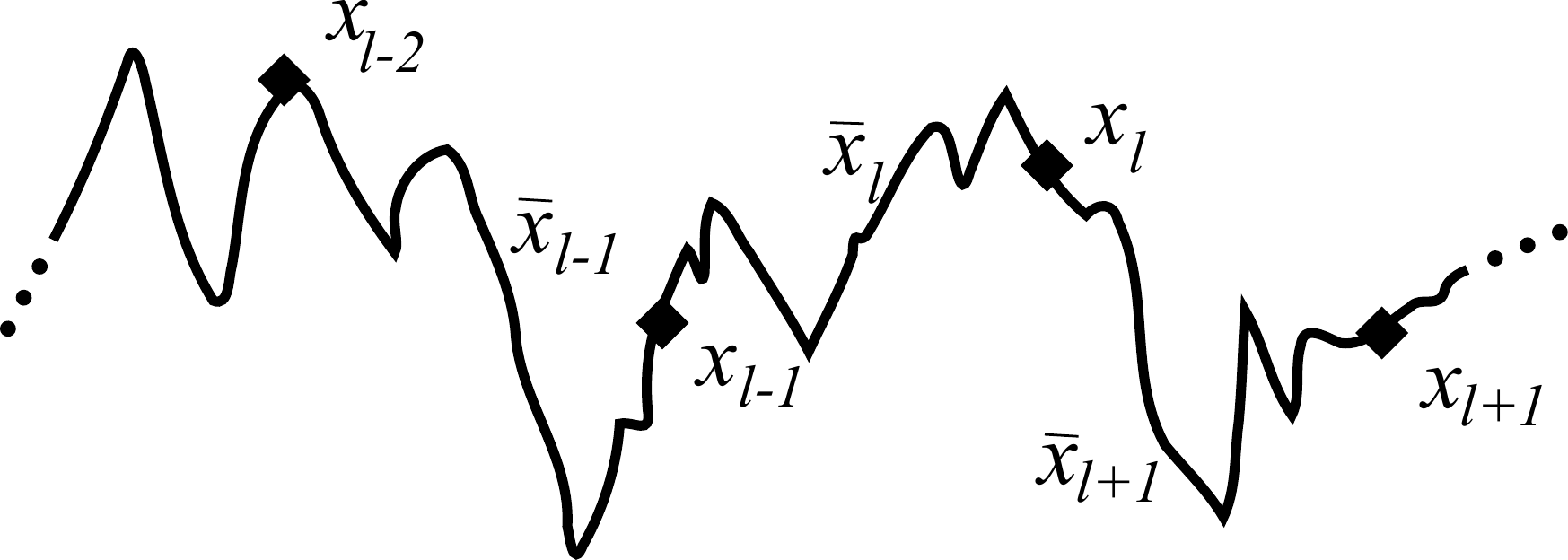}
\caption{The evolution of the Ornstein-Uhlenbeck process during several time-steps.}
\label{tantipropag}
\end{center}
\end{figure}

\begin{figure}[htbp]
\begin{center}
\includegraphics[width=.9\textwidth]{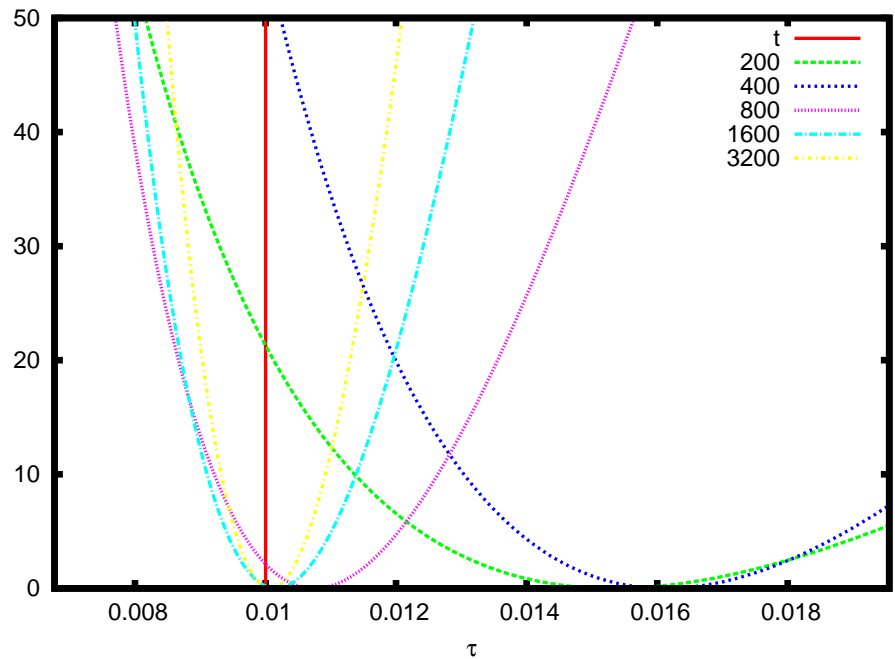}
\caption{The log-likelihood as a function of $\tau$ for different $L$ (indicated in the legend), in red we show the actual value of $\tau$ that generated the data. The log-likelihoods have been offset by a constant value and changed into its opposite for cosmetic reasons. The minimum of the displayed curve is the most likely value of $\tau$.}\label{fig:loglike}
\end{center}
\end{figure}

Real unzipping measurements do not grant us access to the instantaneous force (or displacement) signal. What is actually measured is a signal which is time averaged over a period of a few milliseconds.\\
In this section we wish to explore the effects of time averaging on a simple stochastic system. We will compute the probability of observing a series of time averages given a set of parameters and thanks to the Bayes theorem we will be able to chose the most likely set of parameters given a set of measures.\\
Let us consider an Ornstein-Uhlembeck process \cite{OU}:
\begin{equation}
 \gamma\dot{x}=-k(x-y)+\eta\,,
\end{equation}
where $\eta$ is a Gaussian noise with zero mean and variance $\langle\eta(t)\eta(t^\prime)\rangle=2 k_B T \gamma \delta(t-t^\prime)$.\\
We wish to consider its temporal average $\bar x$ over a certain time and to infer from it the physical quantities $\gamma$ and $k$.\\
The solution of the model is well known and it's the stochastic function:
\begin{equation}
 x(t)=x_0 e^{-\frac{k}{\gamma} t}+y(1-e^{-\frac{k}{\gamma} t})+\frac{1}{\gamma}\int_0^t d t^\prime e^{-\frac{k}{\gamma}( t-t^\prime)} \eta(t^\prime)\,.
\end{equation}
That is a Gaussian process with mean and variance given by:
\begin{align}
 \langle x(t)\rangle&=x_0 e^{-\frac{k}{\gamma} t}+y(1-e^{-\frac{k}{\gamma} t})\,,\\
\langle x(t)^2\rangle-\langle x(t)\rangle^2 &= \frac{ k_B T}{k}\left(1-e^{-2\frac{k}{\gamma} t}\right)\,.
\end{align}
If we now consider the time average over a time $t$ of the same stochastic function we obtain another stochastic function of the form:
\begin{equation}
\begin{split}
 \bar{x}(t)&=\frac{1}{t}\int_0^t d t^\prime x(t^\prime)=\frac{\gamma}{k t}(x_0 -y)(1-e^{-\frac{k}{\gamma} t})+y\\
&+\frac{1}{\gamma t}\int_0^t d t^\prime \int_0^{t^\prime} d t^{\prime\prime} e^{-\frac{k}{\gamma}( t^\prime-t^{\prime\prime})} \eta(t^{\prime\prime})\,.
\end{split}
\end{equation}
That is a Gaussian process with mean and variance:
\begin{align}
 \langle \bar{x}\rangle&=\frac{\gamma}{k t}(x_0 -y)(1-e^{-\frac{k}{\gamma} t})+y\,,\\
\langle \bar{x}(t)^2\rangle-\langle \bar x (t)\rangle^2&= \frac{2 k_B T\gamma}{k^2 t}+\frac{ k_B T\gamma^2}{k^3 t^2}\left(-3+4e^{-\frac{k}{\gamma} t}-e^{-\frac{2k}{\gamma} t}\right)\,,
\end{align}
and additionaly we should consider:
\begin{equation}
 \langle \bar{x}(t) x(t)\rangle - \langle \bar{x}(t)\rangle\langle x(t)\rangle=\frac{ k_B T\gamma}{k^2 t}\left(1-e^{-\frac{k}{\gamma} t}\right)^2\,.
\end{equation}
All this can be summarized defining a covariance matrix as a function of a dimensionless time $\tau=kt/\gamma$:
\begin{equation}
\mathbf{C}=\frac{ k_B T}{k}\left(
\begin{array}{cc}
1-e^{-2\tau} & \frac{\left(1-e^{-\tau}\right)^2}{\tau} \\ 
\frac{\left(1-e^{-\tau}\right)^2}{\tau} & \qquad\frac{ 2}{\tau}+\frac{ 1}{\tau^2}\left(-3+4e^{-\tau}-e^{-2\tau}\right)
 \end{array}
\right)\,,
\end{equation}
But since the process is Gaussian we can write the full probability starting from the means vector and the covariance matrix:
\begin{equation}
P(x(t),\bar x(t)|x_0)=\frac1{2\pi \sqrt{\det \mathbf{C}}}\exp\left(-\frac12\vec x^\dagger\mathbf{C}^{-1}\vec x\right)\,,
\end{equation}
where $\vec x=\left(\begin{array}{c}x(t)-\langle x(t)\rangle\\\bar{x}(t)-\langle\bar{x}(t)\rangle\end{array}\right)$ and $\mathbf{C}^{-1}$ is the inverse of $\mathbf{C}$, that is:
\begin{equation}
\begin{split}
\mathbf{C}^{-1}&=\frac{k}{k_B T\left(\tau\left(1+e^{-\tau}\right)-2\left(1-e^{-\tau}\right)\right)}\\
&\times\left(
\begin{array}{cc}
\frac{2\tau-3+4e^{-\tau}-e^{-2\tau}}{2 \left(1-e^{-\tau}\right)} & -\tau\left(1-e^{-\tau}\right) \\ 
-\tau\left(1-e^{-\tau}\right) & \tau^2\left(1+e^{-\tau}\right)\end{array}\right)
\end{split}
\end{equation}What we have just wrote defines the evolution of the system through an amount of time $t$; let us now just suppose that this is just a step in the evolution of the sistem, that is, at time $(l-1)\Delta t$ the system is in $x_{l-1}$ and it evolves to $x_{l}$ in $l\Delta t$ as shown in figure \ref{propag}. In this time interval its time average is defined as:
\begin{equation}
 \bar{x}_{l}=\frac{1}{\Delta t}\int_{(l-1)\Delta t}^{l\Delta t} d t^\prime x(t^\prime)\,.
\end{equation}
If we set: $x_{l-1}=x_0$, $x_{l}=x(t)$, $ \bar{x}_{l}=\bar{x}(t)$ and $\tau=k \Delta t/\gamma$ we can recycle the previous expression to define a \emph{propagator}:
\begin{equation}
\begin{split}
P(x_l,\bar{x}_l|x_{l-1})&=\frac{1}{2\pi\sqrt{\det\mathbf{C}}}\\
&\times\exp\left( -\frac{1}{2}\left(x_l-\langle x_l\rangle,\bar{x}_l-\langle\bar{x}_l\rangle\right)\mathbf{C}^{-1}\left(\begin{array}{c}x_l-\langle x_l\rangle\\\bar{x}_l-\langle\bar{x}_l\rangle\end{array}\right) \right)\,.
\end{split}
\end{equation}
So, as long as the Ornstein-Uhlenbeck process is a Markov process, we can write the joint probability of the process as:
\begin{equation}
 P(\{\bar{x}_l,x_l\}_{l=1}^L|x_0)=\prod_{l=1}^L P(x_l,\bar{x}_l|x_{l-1})\,.
\end{equation}
This is pictured in figure \ref{tantipropag} and can easily rewritten as a single exponential:
\begin{equation}
P(\{\bar{x}_l,x_l\}_{l=1}^L|x_0)=\frac1{(2\pi\sqrt{\det\mathbf{C}})^L}\exp\left(-\frac{k}{2k_\textrm{B}T}Q\right)\,,
\end{equation}
where $Q$ is:
\begin{equation}
\begin{split}
Q&=\sum_{l=1}^L\bigg[ A x_l^2+B x_lx_{l-1} +C x_{l-1}^2-D(x_l+x_{l-1})\bar{x}_{l}-(x_{l}-x_{l-1})y-\tau\bar{x}_{l}y\\
&+E\bar{x}^2_{l}+\tau \frac{y^2}{2}\bigg]\,,
\end{split}
\end{equation}
where:
\begin{align}
 A&=\frac{2\tau-3+4e^{-\tau}-e^{-2\tau}}{2 \left(\tau\left(1+e^{-\tau}\right)-2\left(1-e^{-\tau}\right)\right)\left(1-e^{-\tau}\right)}\\
 B&=\frac{2-4\tau e^{-\tau}-2e^{-2\tau}}{2 \left(\tau\left(1+e^{-\tau}\right)-2\left(1-e^{-\tau}\right)\right)\left(1-e^{-\tau}\right)}\\
C&=\frac{2\tau e^{-2\tau}+1-4e^{-\tau}+3e^{-2\tau}}{2 \left(\tau\left(1+e^{-\tau}\right)-2\left(1-e^{-\tau}\right)\right)\left(1-e^{-\tau}\right)}\\
D&=\frac{2\tau\left(1-e^{-\tau}\right)}{2 \left(\tau\left(1+e^{-\tau}\right)-2\left(1-e^{-\tau}\right)\right)}\\
E&=\frac{\tau^2\left(1+e^{-\tau}\right)}{2 \left(\tau\left(1+e^{-\tau}\right)-2\left(1-e^{-\tau}\right)\right)}\,.
\end{align}
We would like now to integrate out the $x_i$'s in order to obtain the joint probability distribution for the time averages only:
\begin{equation}
 P(\{\bar{x}_l\}_{l=1}^L|x_0)=\int_{-\infty}^{\infty}\prod_{l=1}^L dx_l P(x_l,\bar{x}_l|x_{l-1})\,.\label{eq:ouavprob}
\end{equation}
In order to perform this integral in full generality we need to change variables in order to diagonalize the quadratic form $Q$ and factorize the integrals.\\
$Q$ can be diagonalised by a discrete Fourier tranform, provided we force periodic boundary conditions (by imposing $x_0=x_L$), that is:
\begin{align}
 X_q&=\frac{1}{\sqrt{L}}\sum_{l=1}^L x_l e^{\frac{-2\pi \ii q l}{L} }\\
 x_l&=\frac{1}{\sqrt{L}}\sum_{q=1}^L X_q e^{\frac{2\pi \ii q l}{L} }\,,
\end{align}
the choice of prefactors ensures the unitarity of the transformation which is known to be orthogonal.
$Q$ is thus transformed into:
\begin{equation}
\begin{split}
Q&=\sum_{q=1}^L\bigg[ \left( A+C+B\cos\left(\frac{2\pi  q}{L}\right)\right)X_q^2-D\left(1+\cos\left(\frac{2\pi  q}{L}\right)\right) X_q\bar{X}_q\\
 &+E \bar{X}_q^2\bigg]-\sqrt{L}\tau y \bar{X}_0+L \tau \frac{y^2}{2}\,.
\end{split}
\end{equation}
Thus integrating over the $X_q$ yields:
\begin{equation}
\begin{split}
\tilde{Q}&=\sum_{q=1}^L\left[E-\frac{D^2\left(1+\cos\left(\frac{2\pi  q}{L}\right)\right)^2}{4\left( A+C+B\cos\left(\frac{2\pi  q}{L}\right)\right)}\right]\bar{X}_q^2-\sqrt{L}\tau y \bar{X}_0+L \tau \frac{y^2}{2}\\
&=\sum_{q=1}^L\frac{\tau^2\left[1+e^{-\tau}-\frac{\left(1-e^{-\tau}\right)^3\cos^4\left(\frac{\pi  q}{L}\right)}{\tau\left(1+e^{-2\tau}\right)-\left(1-e^{-2\tau}\right)+\cos\left(\frac{2\pi  q}{L}\right)\left(1-2\tau e^{-\tau}-e^{-2\tau}\right)}\right]}{2 \left(\tau\left(1+e^{-\tau}\right)-2\left(1-e^{-\tau}\right)\right)}\bar{X}_q^2\\
&-\sqrt{L}\tau y \bar{X}_0+L \tau \frac{y^2}{2}\,.
\end{split}
\end{equation}
We can now use Bayes' theorem to interpret the probability in eq (\ref{eq:ouavprob}) as the likelihood of a set of measures being generated by a given $\tau$.\\
With standard computational techniques one can compute the log-likelihood in time $O(L^2)$. In figure \ref{fig:loglike} we show the results for simulated runs of different lengths. It is easily shown how the prediction of $\tau$ improves with more points, but it's already reasonably good with only 200 points.\\
One could also use the width of this curve to compute $k/(k_\textrm{B}T)$ and by knowing the value of the temperature compute $\gamma$ and $k$.\\
What is compelling about this algorithm is that we are exploiting all the information available: fluctuations, correlations and not only the averages.\\

\chapter*{Conclusions and outlook}
\section*{Infotaxis}
In the part devoted to infotaxis we have developed a continuous version of the algorithm and analyzed its behavior and performance in two and three dimensions.\\
We have shown the probability of success not to depend on the distance from the source when this latter is of the order of magnitude of $\lambda$, the characteristic length of the odor advection phenomenon.\\
Furthermore we have shown the search time to grow with $\gamma$, the parameter that regulates the speed of the searcher in response to the gradient. However, we have observed simulations with small $\gamma$ to be computationally more taxing.\\
The computational time needed to perform a single step is still very large: this is due to the need of computing many Monte Carlo integrations over the whole space, but also from the strategy we have chosen that increases the time complexity of the algorithm to $O(t^2)$.\\
In order to bring back the complexity of the algorithm to $O(t)$ we have toyed with finite memory, as in forgetting earlier events, but this has the effect of removing the exponential term that discounted the probability at the starting point and at the early hits. This is to be avoided because it will attract the searcher very strongly back to where it started.\\
We think the solution to this is to coarse-grain past events by decimating older events and increasing the weight of the points left. This could leave us with a constant number of points and a precision in integration that's only slightly reduced. We think that the coarse-graining could be performed on the fly according to the position of those points compared with the most recent position of the searcher.\\
Another exciting new direction we think could be explored is to think infotaxis as the first and simplest strategy in the class of those based on information theory: infotaxis performs choices by looking at the immediate next step, what would happen if we looked several steps ahead?\\
Such a strategy would translate to adding higher derivatives to the differential equation that regulates the movement of the searcher: the first such step adding an inertial term:
\begin{equation*}
\tau^2(\nabla_x \nabla_x V_t(x))\ddot +x\gamma \dot x=-\nabla_x V_t(x)\,.
\end{equation*}
This inertial term with a mass tensor proportional to the local curvature of the potential at the position of the searcher could have beneficial effects to the performance of the searcher.\\
Finally we think another interesting direction to take would be to build a meta-heuristic for searches that mimics the behavior we have observed in infotaxis without the need of performing the full entropy calculation. For example we could rethink a technique such as the one developed in \cite{BALKO} to work in continuous space and three dimensions.
\section*{DNA unzipping and sequencing}
In chapter \ref{infedna} we have shown several approaches to the inference of DNA sequencing through micromanipulation experiments. The first issue that stands between us and a successful algorithm is the fact that the number of open bases is not directly known, but acts as a hidden variable, while the position of the bead can be measured directly; the second problem is that the temporal resolution in experiments is very low compared to the time-scale of the opening and closing of the fork.\\
The second section of this chapter deals with the first problem: the fact that the fork position $n$ is unknown. It does so in a limit which is not completely realistic by imposing that equilibrium is perfectly attained and that we can sample the equilibrium distribution up to an arbitrary precision. In a real experiment there will be many sources of noise and if we take averages for a long enough time we will end up measuring drifts in temperature and trap position which will change the equilibrium distribution.\\
The approach of the third section, on the other hand takes into account the fact that we could in principle be out of equilibrium and that an infinite sampling frequency is out of the question, but it does so by relying on a very simple model, arguably the simplest non-trivial stochastic process in continuous time and space.\\
In order to devise a more realistic algorithm we should combine this two approaches, but a few difficulties stand in our way: suppose we took the dynamic approach we have used with the Ornstein-Uhlenbeck and tried to apply it to a more complicated system, our experience tells us that even relaxing the periodic boundary conditions in time makes it hard to diagonalize the covariance matrix analytically.\\
On the other hand we could try to adapt the idea developed for the perfect averages approach and use them in conjunction with the dynamic algorithm: we could describe the potential on the hidden variable by a simple potential that depends only on a few parameters, that can in turn be fitted, but it is hard to say how a numeric approach can be combined to the dynamical algorithm.\\
Ultimately we think that many improvements can be brought into play for the experimental procedure if one bears in mind sequencing by unzipping as the ultimate goal. One example are advances in manipulation techniques through holography, allow for the manipulation of multiple beads with a single laser beam \cite{CURTIS} which could allow for the simultaneous rotation of complex objects. Setups similar to a microscopic bobbin or spindle could one day become feasible if one could find a way to prevent ssDNA from forming secondary structures when confined.\\
Another idea suggested to us by Prof. A. Libchaber is the use of proteins that bind to ssDNA stiffening it, bringing us somewhat closer to the measurement of the actual fork position.\\
\part{Publications}
\addcontentsline{toc}{chapter}{Dynamical modeling of molecular constructions and setups for DNA unzipping}
\includepdf[openright=false,pages=-]{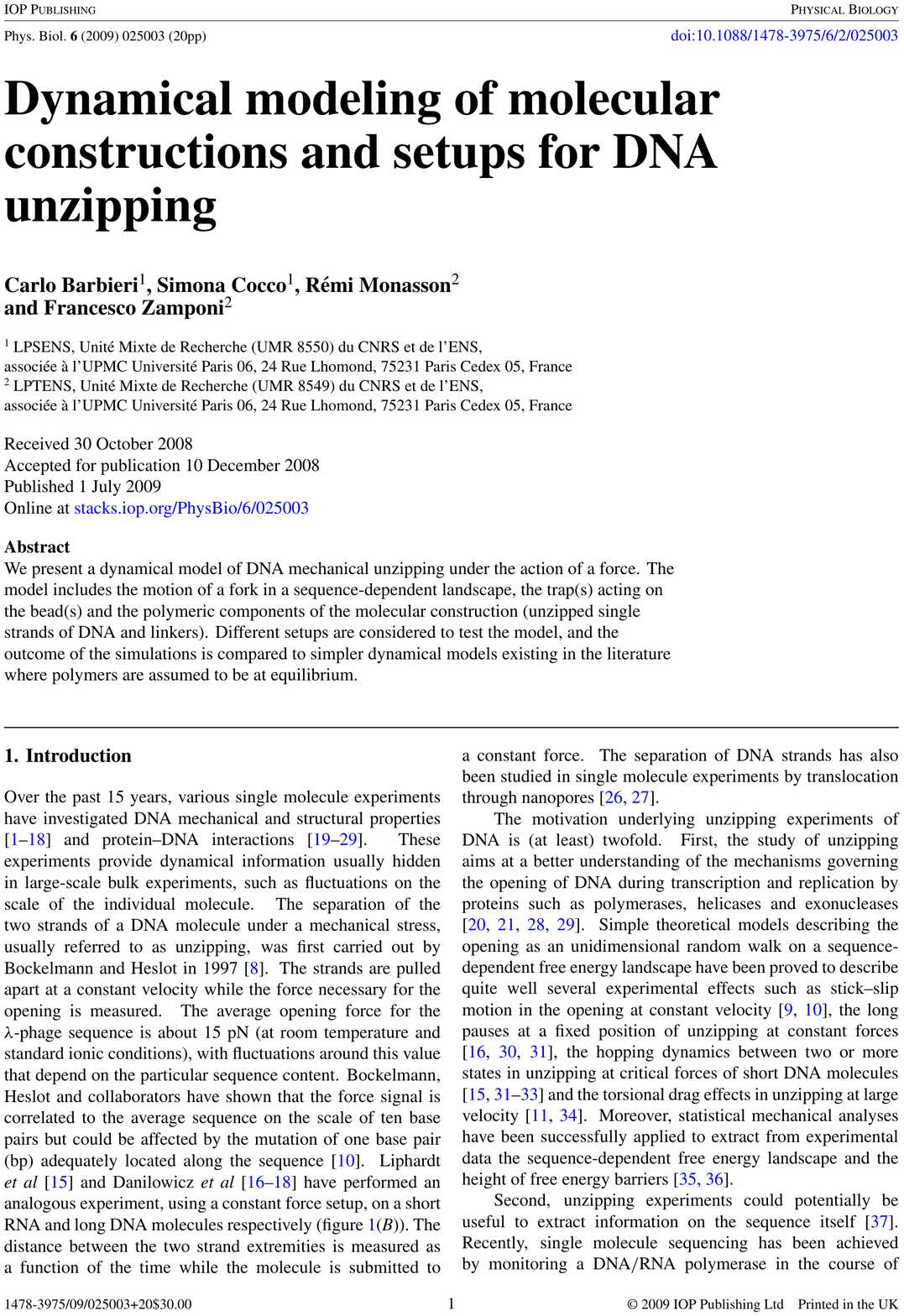}
\addcontentsline{toc}{chapter}{On the trajectories and performance of Infotaxis, an information-based greedy search algorithm}
\includepdf[openright=false,pages=-]{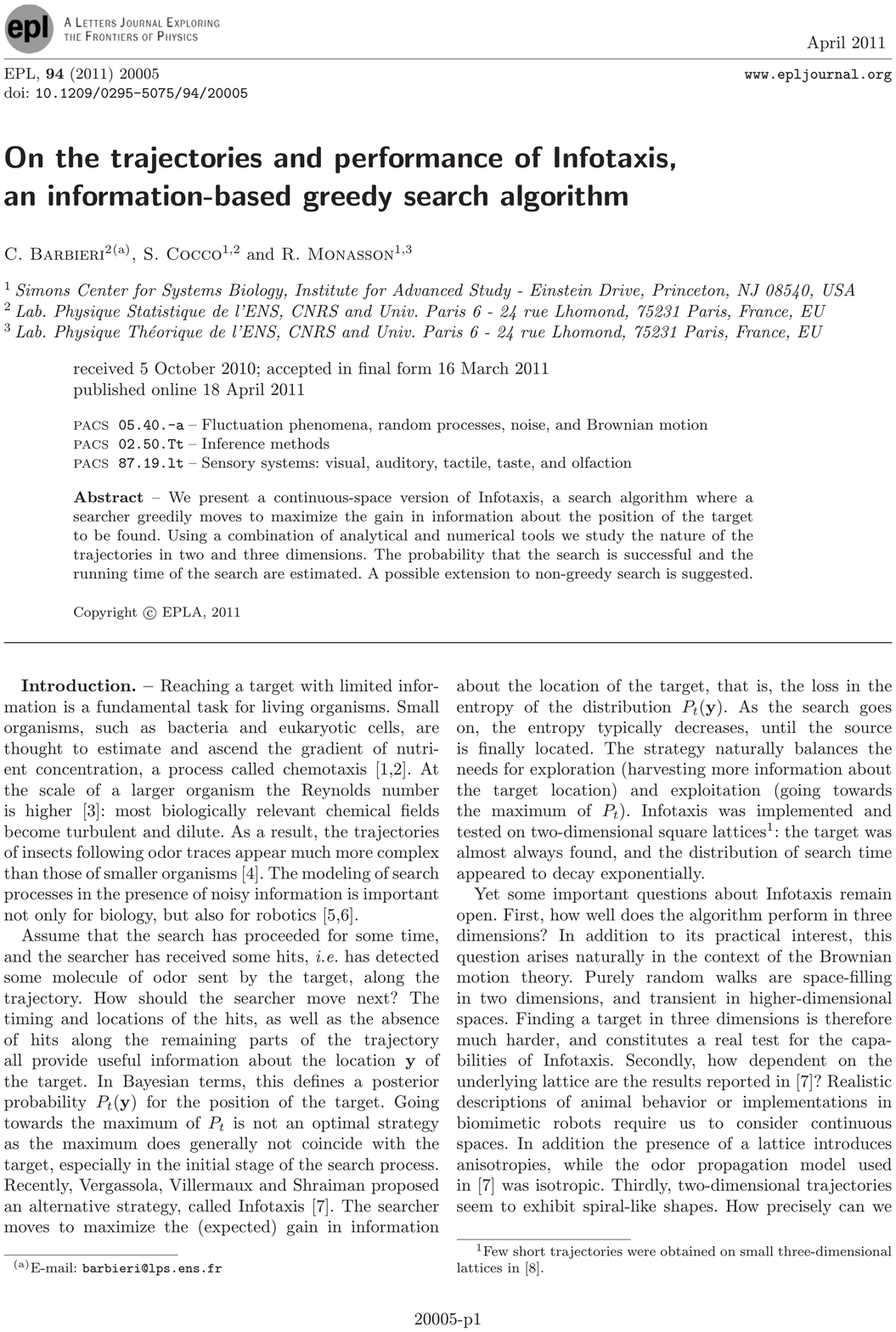}
\appendix
\chapter{Inference of couplings for a set of leaky integrate and fire neurons}
\label{APPE}
\section{Introduction} 
\begin{figure}
\begin{center}
\includegraphics[width=.7\textwidth]{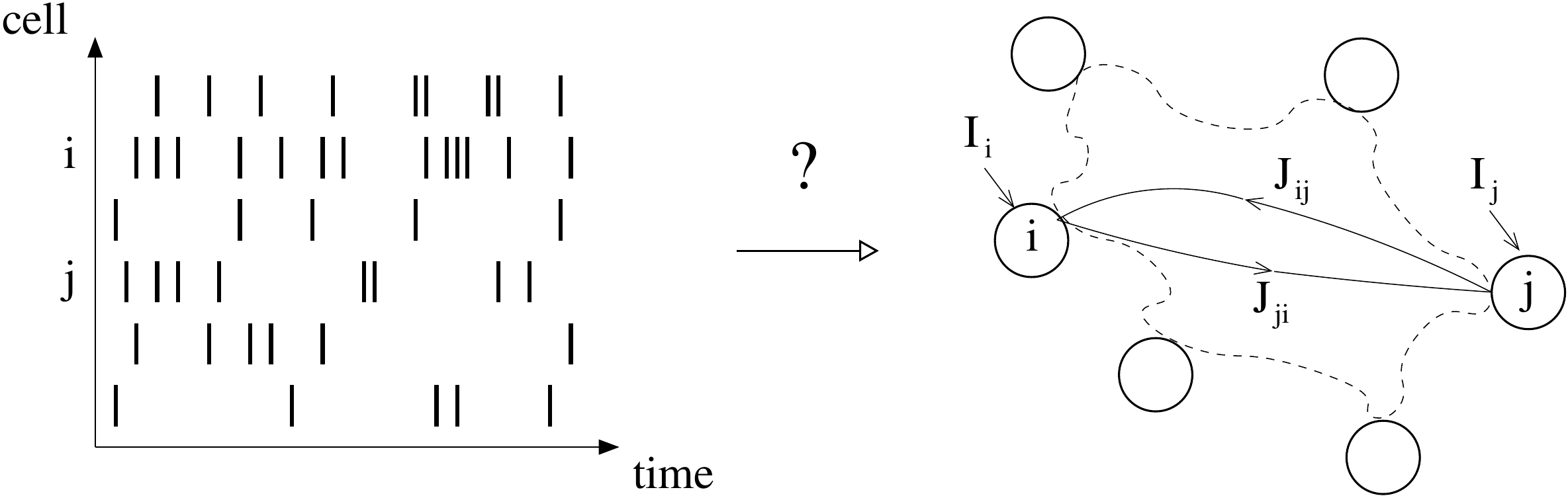}
\caption{Left: times $t_{i,k}$ of spikes emitted by a set of neurons (raster plot). Right: network of LIF neurons with couplings $J_{ij}$ and external currents $I_i$. Given the set of spikes we want to infer the values of the couplings and currents.}
\label{fig-1}
\end{center}
\end{figure}

Recent advances in experimental techniques and in the miniaturization of components have permitted the recording of the activity of several neurons at the same time thorugh the use of multi-electrode recordings \cite{MEA}.\\
The observation of substantial correlations in the firing activities of neurons has raised fundamental issues on their functional role \cite{aver06}. However the problem of inferring the structure of the network and the interaction between different neurons has only recently been attacked (Fig.~\ref{fig-1}). The problem is not easy to tackle, because data sets are already quite big and can contain millions of spiking events from up to a hundred neurons.\\
A classical approach to infer functional neural connectivity is through the analysis of pairwise cross-correlations. The approach is versatile and fast, but cannot disentangle direct correlations from common or correlated inputs. Alternative approaches assume a particular dynamical model for the spike generation, such as the generalized linear model, which represents the generation of spikes as a Poisson process with a time-dependent rate, and the Integrate-and-Fire (IF) model, where spikes are emitted according to the dynamics of the membrane potential \cite{Joli}.\\
While the problem of estimating the model parameters (external current, variance of the noise, capacitance and conductance of the membrane, ...) of a single stochastic IF neuron from the observation of a spike train has received a lot of attention \cite{papisi,lan}, few studies have focused on the inference of interactions in an assembly of IF neurons. 
Cocco and Monasson have recently proposed a Bayesian algorithm to infer the intereactions and the external currents of a set of leaky integrate and fire neurons\cite{noi1,noi2}. They applied their approach to data coming from real experiments on salamander retinas and validated their results with artificial data and cross-checking with another algorithm based on the Ising model.\\
An interesting problem they've come across is the disentanglement of the correlations already present in the stimulus and the correlations that come from the topology of the network itself. This has been discussed by comparing datasets coming from the same retina with different stimuli.\\

\vskip .3cm \noindent
\section{Integrate and fire neurons}
Each neuron is represented by the Leaky Integrate-and-Fire (LIF) model (see \cite{Joli} and references therein). The membrane potential obeys the differential equation,
\begin{equation} \label{ode}
C \frac{dV_i}{dt}(t) = - g\, V_i(t) + \sum _{j (\ne i)} J_{ij} \; \sum _{k} \delta( t - t_{j,k})
+ I_i +\eta_i(t) \ ,
\end{equation}
where $C,g$ are, respectively, the membrane capacitance and conductance. $J_{ij}$ is the strength of the connection from neuron $j$ onto neuron $i$ and $t_{j,k}$ the time at which cell $j$ fires its $k^{th}$ spike; we assume that synaptic inputs are instantaneously
integrated {\em i.e.} the synaptic integration time is much smaller than the membrane leakage time, $C/g$, and the typical inter-spike interval. $I_i$ is a constant external current flowing into cell $i$, and $\eta_i(t)$ is a fluctuating current, modeled as a Gaussian white noise process with variance $ \sigma^2$. Neuron $i$ remains silent as long as $V_i$ remains below the threshold potential $V_{th}$ (set to unity in the following). If the threshold is reached at some time then a spike is emitted, and the 
potential is reset to its rest value (which can be set to zero without loss of generality), and the dynamics resumes.\\
The above model (\ref{ode}) implicitly defines the likelihood $P$ of the spiking times $\{t_{j,k}\}$ given the currents $I_i$ and synaptic couplings $J_{ij}$. If we are given the
spiking times $\{t_{j,k}\}$ we will infer the couplings and currents by maximizing $P$. In principle $P$ can be calculated through the resolution of Fokker-Planck equations (one for each inter-spike interval) for a one-dimensional Orstein-Uhlenbeck process with moving boundaries. However this approach, or related numerical approximations \cite{papisi}, are inadequateis too slow to treat data sets with hundreds of thousands of spikes.\\
In Cocco's and Monasson's approach $P$ is approximated from the contribution coming from
the most probable trajectory for the potential for each cell $i$, referred to as $V_i ^*(t)$. 
This semi-classical approximation is exact when the amplitude $\sigma$ of the noise is small. The determination of $V_i^*(t)$ was done numerically by Paninski for one cell in \cite{Pa1}. What they proposed is a  fast algorithm to determine $V_i^*(t)$ analytically in a time growing linearly with the number of spikes and quadratically with the number of neurons, 
which allows them to process recordings with tens of neurons easily. The algorithm is based on a detailed and analytical resolution of the coupled equations for the optimal potential 
$V_i^*(t)$ and the associated optimal noise $\eta _i^*(t)$ through (\ref{ode}), since this work has not been performed during this thesis and a full explanation would take many pages we will not discuss the details of the algorithm. The interest reader can find an explanation of the approach in a recent publication \cite{noi2}. 

Once the optimal paths for the potential and noise have been determined, one cancalculate the log-likelihood of the corresponding couplings and currents through the integral of the squared optimal noise. This log-likelihood is a concave function of the currents and couplings and can be easily maximized using convex optimization procedures. Measure of the curvature of the log-likelihood allows us to estimate the error bars on the inferred parameters.
\section{Limitations of the original implementation}
Even though the conception of the algorithm and the subsequent testing performed by Cocco and Monasson have been very thorough, distribution of software through its source code can scare all but the most tech savy scholars in the field.\\
Moreover the algorithm was originally written in non-standard Fortran 77 that was only compatible with g77 and not with gfortran. As of version 3.4 of GCC (released in May 2006) development of g77 has stopped and users are encouraged to use gfortran.\\
Because of this most modern Linux distributions do not come with g77 preinstalled and users who need g77 have to install it separately sometimes compiling the compiler itself.\\
Another important hurdle to overcome before the public release of this software was the fact that it originally contained parts of code that were covered by copyright \cite{NRF} and could not be reused freely.\\
An important part of the implementation of the original program relied on the implementation of the classical Newton method for multidimensional minimzation. Knowing that the function to minimize is convex in the parameters guarantees the convergence of the algorithm, howver it is not at all clear that among all minimization techniques Newton's would be the faster in all cases.\\
In fact Newton's method relies on the exact computation of the Hessian matrix which is computatinally taxing and sometimes possible only in an approximate form, because of this we have chosen to reimplement the software in a way that permitted a modular change of minimzation techniques.\\
Further limitations included the lack of inferenence of certain parameters of the model such as the leaking constant $g$, which was fixed at the beginning of the inference and considered equal for all neurons.\\
Because of this we have decided to translate the problem in standard C which is a far more widespread programming language, arguably the most common. Compilers in C are available on virtually every architecture, and there are a variety of free open-source numerical libraries that can be effortlessly used for the implementation of minimization techniques and special functions.\\ 
Morevore standard C code is very easily integrated in more higher level computational software such as Matlab which is very widely used in the computational biology community.\\
We believe that these contributions, even though it is not an algorithmic effort, but of a more mundane nature can be of great use in the diffusion of this algorithm and its use.
\section{Description of the software package} The software package is written in standard C and the source code will soon be available for download. C was the natural choice for a program that can be used either as  a stand-alone executable or called from widely used computational software such as Matlab \cite{MATLAB} and Mathematica \cite{MATHEMATICA}.
\vskip .3cm \noindent
The software is composed of several functions:
\begin{itemize}
\item A function that reads data in the form of spike trains, \emph{i.e.} a two column file where the first column is the time at which the spike was emitted and the second is an integer value that identifies the neuron responsible for that spike. Data are  then stored in data structures of variable size, to be conveniently accessed by other functions. 
\item A function that goes through the data and identifies all spikes incoming to a cell in the time interval between two successive spikes of that cell. This function also performs a check on collision, that is, multiple spikes emitted by the same cell at the same time.
\item A function that computes the log-likelihood and its first and second derivatives with respect to the interactions and the current. Note that the time-consuming calculation of second derivatives can be switched off if the minimization algorithm employed does not require them.
\end{itemize}

These functions have appropriate wrappers that allow for use with the minimization routines available from the GNU Scientific Library (GSL) \cite{GSL} and Matlab. The choice of minimization technique and related parameter is left to the user, but we have observed the gnewton technique of the GSL to be the fastest in most cases.

The user chooses the values of the parameters of the LIF model (\ref{ode}); when the leaking constant $g$ is set to zero, that is when the integrator is non-leaky, a specific and much faster program is used. Otherwise the most likely value for $g$ can be inferred from the data for each neuron. Also available to the user specifications is a choice of priors over the coupling values, based on the $L_{1}$ and $L_{2}$ norms, which can be used to ensure convergence to realistic values and/or to eliminate couplings which are very close to zero. The program can be easily modified to add further specific priors.
The user can further improve upon the instantaneous synaptic integration assumption in the model (\ref{ode}). To do so an option allows the user to introduce a synaptic reweighing function, replacing $J_{ij}$ with $J_{ij}\times K(t_{i,k'}-t_{j,k})$, where $t_{j,k}$ is the time of the spike fired by cell $j$ and entering cell $i$, and $t_{i,k'}$ is the next spiking time of cell $i$; $K(x)=0$ for $x=0$ and $K(x) \simeq 1$ for $x > \tau_{s}$, the synaptic integration time.

The output data can be printed to a file or to a Matlab array. The file is composed of three columns, the first two denote the indices of the coupling matrix $J_{ij}$ and the third the value of the coupling constant. The diagonal elements of the matrix $J_{ii}$ are the currents $I_i$.

\backmatter
\bibliographystyle{these} 
\bibliography{biblio}
\end{document}